\documentclass[longauth,a4paper,11pt]{article}
\pdfoutput=1 

\usepackage{jcappub} 

\usepackage[T1]{fontenc} 
\usepackage{mathtools}

\usepackage{array}
\newcolumntype{P}[1]{>{\centering\arraybackslash}p{#1}}
\newcolumntype{M}[1]{>{\centering\arraybackslash}m{#1}}
\usepackage{graphicx}
\usepackage[dvipsnames]{xcolor}
\usepackage{longtable}
\usepackage{url}
\usepackage{verbatim}
\usepackage[utf8]{inputenc}
\usepackage{makecell}
\usepackage{comment}
\usepackage{cleveref}
\usepackage{multirow}
\usepackage{lineno}
\usepackage{hyperref}

\crefname{section}{§}{§§}
\graphicspath{ {Figures/} }

\title{Prospects for \boldmath{$\gamma$}-ray observations of the Perseus galaxy cluster with the Cherenkov Telescope Array}
\author{K.~Abe$^{1}$,}
\author{S.~Abe$^{2}$,}
\author{F.~Acero$^{3,4}$,}
\author{A.~Acharyya$^{5}$,}
\author{R.~Adam$^{6,7,*}$,}
\author{A.~Aguasca-Cabot$^{8}$,}
\author{I.~Agudo$^{9}$,}
\author{A.~Aguirre-Santaella$^{10}$,}
\author{J.~Alfaro$^{11}$,}
\author{R.~Alfaro$^{12}$,}
% ...

\affiliation{$^{1} \ $Department of Physics, Tokai University, 4-1-1, Kita-Kaname, Hiratsuka, Kanagawa 259-1292, Japan}
\affiliation{$^{2} \ $Institute for Cosmic Ray Research, University of Tokyo, 5-1-5, Kashiwa-no-ha, Kashiwa, Chiba 277-8582, Japan}
\affiliation{$^{3} \ $Universit\'e Paris-Saclay, Universit\'e Paris Cit\'e, CEA, CNRS, AIM, F-91191 Gif-sur-Yvette Cedex, France}
\affiliation{$^{4} \ $FSLAC IRL 2009, CNRS/IAC, La Laguna, Tenerife, Spain}
\affiliation{$^{5} \ $University of Alabama, Tuscaloosa, Department of Physics and Astronomy, Gallalee Hall, Box 870324 Tuscaloosa, AL 35487-0324, USA}
\affiliation{$^{6} \ $Universit\'e C\^ote d'Azur, Observatoire de la C\^ote d'Azur, CNRS, Laboratoire Lagrange, France}
\affiliation{$^{7} \ $Laboratoire Leprince-Ringuet, CNRS/IN2P3, \'Ecole polytechnique, Institut Polytechnique de Paris, 91120 Palaiseau, France}
\affiliation{$^{8} \ $Departament de F{\'\i}sica Qu\`antica i Astrof{\'\i}sica, Institut de Ci\`encies del Cosmos, Universitat de Barcelona, IEEC-UB, Mart{\'\i} i Franqu\`es, 1, 08028, Barcelona, Spain}
\affiliation{$^{9} \ $Instituto de Astrof{\'\i}sica de Andaluc{\'\i}a-CSIC, Glorieta de la Astronom{\'\i}a s/n, 18008, Granada, Spain}
\affiliation{$^{10} \ $Instituto de F{\'\i}sica Te\'orica UAM/CSIC and Departamento de F{\'\i}sica Te\'orica, Universidad Aut\'onoma de Madrid, c/ Nicol\'as Cabrera 13-15, Campus de Cantoblanco UAM, 28049 Madrid, Spain}
\affiliation{$^{11} \ $Pontificia Universidad Cat\'olica de Chile, Av. Libertador Bernardo O'Higgins 340, Santiago, Chile}
\affiliation{$^{12} \ $Universidad Nacional Aut\'onoma de M\'exico, Delegaci\'on Coyoac\'an, 04510 Ciudad de M\'exico, Mexico}
% ...
\affiliation{ $^* \ $\textbf{Corresponding authors} (alphabetical order): \\
Rémi Adam (\href{mailto:remi.adam@oca.eu}{remi.adam@oca.eu}), \\
Sergio Hern\'andez-Cadena (\href{mailto:skerzot@ciencias.unam.mx}{skerzot@ciencias.unam.mx}), \\
Moritz Hütten (\href{mailto:huetten@icrr.u-tokyo.ac.jp}{huetten@icrr.u-tokyo.ac.jp}), \\
Judit P\'erez-Romero (\href{mailto:judit.perez@ung.si}{judit.perez@ung.si}), \\
Miguel A. S\'anchez-Conde (\href{mailto:miguel.sanchezconde@uam.es}{miguel.sanchezconde@uam.es})} 

\abstract{
Galaxy clusters are expected to be both dark matter (DM) reservoirs and storage rooms for the cosmic-ray protons (CRp) that accumulate along the cluster’s formation history. Accordingly, they are excellent targets to search for signals of DM annihilation and decay at $\gamma$-ray energies and are predicted to be sources of large-scale $\gamma$-ray emission due to hadronic interactions in the intracluster medium (ICM). 
In this paper, we estimate the sensitivity of the Cherenkov Telescope Array (CTA) to detect diffuse $\gamma$-ray emission from the Perseus galaxy cluster. 
We first perform a detailed spatial and spectral modelling of the expected signal for both the DM and the CRp components. For each case, we compute the expected CTA sensitivity accounting for the CTA instrument response functions. The CTA observing strategy of the Perseus cluster is also discussed.
In the absence of a diffuse signal (non-detection), CTA should constrain the CRp to thermal energy ratio $X_{500}$ within the characteristic radius $R_{500}$ down to about $X_{500} < 3 \times 10^{-3}$, for a spatial CRp distribution that follows the thermal gas and a CRp spectral index $\alpha_{\rm CRp} = 2.3$. Under the optimistic assumption of a pure hadronic origin of the Perseus radio mini-halo and depending on the assumed magnetic field profile, CTA should measure $\alpha_{\rm CRp}$ down to about $\Delta \alpha_{\rm CRp} \simeq 0.1$ and the CRp spatial distribution with 10\% precision, respectively. Regarding DM, CTA should improve the current ground-based $\gamma$-ray DM limits from clusters observations on the velocity-averaged annihilation cross-section by a factor of up to $\sim 5$, depending on the modelling of DM halo substructure. In the case of decay of DM particles, CTA will explore a new region of the parameter space, reaching models with $\tau_{\chi}>10^{27}$s for DM masses above 1 TeV. 
These constraints will provide unprecedented sensitivity to the physics of both CRp acceleration and transport at cluster scale and to TeV DM particle models, especially in the decay scenario.

}

\begin{document}
\maketitle
\flushbottom

\section{Introduction}\label{sec:intro}
\subsection{Diffuse $\gamma$-ray emission from cosmic-rays and dark matter in galaxy clusters}
Clusters of galaxies are the largest virialized structures in the Universe, with masses up to about $10^{15}$ M$_{\odot}$. They are dominated by dark matter (DM; $\sim 85$\% in mass) and permeated by the intracluster medium (ICM; $\sim 10 - 15$\% in mass), whose physical properties are shaped by the hierarchical growth of structures through the merging of subclusters and the smooth accretion of surrounding matter \citep{Voit2005,2012ARA&A..50..353K}. While the ICM is mostly thermal, these energetic merging events do not only dissipate the kinetic energy into heat via shock waves and turbulence, but may also accelerate cosmic-rays (CR) in the ambient magnetic field \citep{Brunetti2014}. Galaxies in galaxy clusters account only up to a few percent of the total mass, yet they can also directly inject CR via active galactic nuclei (AGN) feedback or star formation activity \citep{Blanton2010,McNamara2012,Fabian2012}.

Direct evidence for the presence of CR electrons (CRe) and magnetic fields have been found in a growing number of galaxy clusters, thanks to the radio observations of diffuse synchrotron emission \citep{vanWeeren2019,Knowles2019,Botteon2022}. These sources are classified as radio halos (that roughly follow the thermal ICM) and radio relics (elongated and peripheral \citep{vanWeeren2010}). Radio halos are further classified as giant radio halos ($\sim$Mpc size), associated with cluster mergers \citep{Casano2010}, suggesting that they are powered by the energy dissipated during these events, and mini-halos in more relaxed clusters, which extend on $100-300$ kpc scales and are generally confined in the core of cool-core clusters \citep{Giacintucci2017, Ruszkowski:2023rzd}. Recent observations have made the phenomenology more complex, showing that in a number of cases mini-halos are surrounded by larger scale (usually very steep spectrum) emission, similar to giant radio halos \citep{Savini2018,Savini2019,Biava2021}. Furthermore, it has been discovered that radio emission may extend on scales larger than those of giant radio halos, in the form of radio bridges \citep{Govoni2019,Botteon2020} and mega-halos \citep{Cuciti2022}. These emerging evidences from observations suggest a more complex picture of non-thermal phenomena in galaxy clusters that may require a revision of current classification schemes. Galaxy clusters are also expected to act as storehouses for the CR protons (CRp) and heavier nuclei \citep{Berezinsky1997} due to their long lifetimes, once they are injected in the ICM via several mechanisms (cosmic shocks and galaxies via supernovae explosions, starbursts or AGN) \citep{Brunetti2014,Bykov2019,Wittor2021}. These CRp should interact hadronically in the ICM to produce $\gamma$-ray emission \citep{Dennison1980,Blasi1999,Dolag2000,Hussain_etal2021, Hussain:2022tls}. Such interactions imply the production of high-energy secondary electrons, which will eventually contribute to the cluster-scale radio emission. According to the current theoretical picture for radio halos and relics, the emitting electrons are (re)accelerated by turbulence and shocks, respectively \citep{Ensslin1998,Brunetti2001,Petrosian2001,Markevitch2005,Brunetti2011,Kang2012}. The large volumes that are probed by radio halos in which the ICM is tenuous disfavour an important contribution from secondary electrons in these sources \citep{Brunetti2014}, although secondaries can contribute to the population of the seed particles to reaccelerate \citep{Brunetti2005,Brunetti2011,Pinzke2017}. On the other hand, pure hadronic models may still explain the smaller mini-halos in dense cores \citep{Ignesti2020,Perrott2021}, although a number of evidences suggests that gas motions may play a major role \citep{Mazzotta2008,ZuHone2013}. Cosmological numerical simulations of CR in clusters have obtained quantitative predictions for the expected $\gamma$-ray emission \citep{Ensslin2007,Pfrommer2008,Pinzke2010}, which has proved useful when searching for clusters in $\gamma$-ray observations \citep{Ackermann2014}. Nevertheless, many uncertainties related to acceleration and transport physics affect the expected $\gamma$-ray signal \citep{Ensslin2011,Brunetti2014,Wittor2021} and cosmological simulations including full turbulent reacceleration physics have not been obtained yet. In this context the study of $\gamma$-ray emission from galaxy clusters plays a central role.

In the past, galaxy clusters have provided strong gravitational evidence in favour of the existence of DM \citep{Zwicky:1933gu, 1993MNRAS.262.1023W, Clowe:2006eq}. Thus, being DM-dominated, these objects also represent natural astrophysical targets for current DM search efforts. Since the underlying nature of DM (and thus its potential signatures) is still unknown \citep{Bertone:2016nfn}, galaxy clusters have been used to probe the properties of the DM particle with a variety of techniques (e.g., \citep{2009A&A...498L..33G, Randall:2008ppe, 2011ARA&A..49..409A, Limousin:2022lvv}). One of the most promising ones is the search for DM-induced $\gamma$-ray signals (\citep{Funk:2013gxa, Conrad:2017pms, Doro:2021dzh}, for reviews), expected from the annihilation or decay of the so-called Weakly Interacting Massive Particles (WIMPs), one of the most studied DM particle candidate.
WIMPs (e.g., \citep{Jungman:1995df, Hooper:2009zm, Arcadi:2017kky, Bertone:2018krk}) would be produced thermally in the Universe via the ``freeze-out'' mechanism and would have masses $\mathcal{O}(0.1-100)$ TeV\footnote{There is an on-going discussion within the community with respect to the viability of these models beyond hundreds of TeV (see e.g., \citep{Smirnov:2019ngs, Tak:2022vkb}).}. They can arise from several theoretical frameworks, ranging from minimal extensions of the Standard Model \citep{Jungman:1995df, Beneke:2018ssm, Kowalska:2018toh} to extra-dimensions \citep{Servant:2002aq, Cembranos:2003mr} and others \citep{Battaglieri:2017aum, Dienes:2011ja, Buchmuller:2007ui, Feng:2010gw}. The expected $\gamma$-ray flux from their annihilation or decay in astrophysical objects mainly depends on the target DM density (squared, for annihilation) and its distance to Earth. Thus, for DM decay, local galaxy clusters can yield the highest expected fluxes compared to other possible targets, as they are the most massive structures in the Universe. As for DM annihilation, clusters can provide fluxes comparable to the ones from dwarf spheroidal galaxies (dSphs, \citep{2011JCAP...12..011S}), as long as the DM interactions expected in their substructures are taken into account \citep{PieriEtAl2011, Sanchez-Conde:2013yxa, Moline:2016pbm}. These substructures, usually referred to as subhalos, are a natural result of the $\Lambda$CDM hierarchical growth of structures \citep{SilkEtAl1993, BergstroemEtAl1999a, Madau:2008fr}, and their abundance is expected to be comparatively significant in clusters, as they are the largest exponents of structure formation at present. Despite the optimal characteristics of galaxy clusters to be used for $\gamma$-ray DM searches, the main drawback with respect to other targets is the predicted $\gamma$-ray emission from more conventional astrophysical processes. Indeed, the expected $\gamma$-rays from hadronic interactions of the CRp in the ICM can act as a complex background to search for a DM-induced signal using standard analysis techniques.

The search for diffuse $\gamma$-ray emission from galaxy clusters has been going on for over two decades, both using space-based observations in the GeV band \citep{Reimer2003,Ackermann2010,Huber2013,Ackermann2014,Griffin2014,Prokhorov2014,Zandanel2014,Ackermann2015,Ackermann2016,Colavincenzo2020} and ground-based observations at energies $>100$ GeV \citep{Aharonian2009,Aleksic2010,Aleksic2012,Arlen2012,Ahnen2016}. Yet, such signal remained elusive\footnote{Note that \citep{Xi2018} claimed the detection of the Coma cluster with \textit{Fermi}-LAT. \citep{Adam2021} confirmed the signal detection and showed that it would imply a CRp to thermal energy ratio of about $1\%$. However, they also noted that possible confusion with point sources could not be excluded, so that the first non-ambiguous detection of diffuse $\gamma$-ray emission from galaxy clusters is still to be achieved. See also \citep{Baghmanyan:2021jwg} who claimed the detection of extended emission.}. Nonetheless, these non-detections allowed to constrain the CRp to thermal energy ratio down to few percent \citep{HESS:2023roi}, challenging the understanding of diffusive shock acceleration in the ICM when combined with radio data \citep{Vazza2014,Vazza2015,Vazza2016}, although the large modelling uncertainty affects the predictions \citep{Bykov2019}. The stringent $\gamma$-ray limits were also used to disfavour hadronic models for nearby radio halos \citep{Brunetti2012,Brunetti2017} and to test models based on the reacceleration of secondary particles \citep{Brunetti2017,Adam2021}. 

The non-detection of $\gamma$-ray emission from clusters is in agreement with the lack of DM-induced $\gamma$-ray signals from other promising astrophysical targets, especially dSphs \citep{Fermi-LAT:2015att, Strigari:2018utn, DiMauro:2021qcf}. DM searches in clusters mainly targeted very massive and local objects \citep{Reiprich:2001zv, Jeltema:2008vu, 2011JCAP...12..011S}, since the expected flux is proportional to the mass of the objects and decreases with the distance squared. Yet, the lack of a DM signal in clusters has allowed to provide also strong constraints for annihilating WIMPs from a combined analysis of various clusters \citep{2010JCAP...05..025A, Huang:2011xr, 2012MNRAS.427.1651H, Nezri:2012tu, Lisanti:2017qlb, Tan:2019gmb, Thorpe-Morgan:2020czg, Dugger:2010ys} or single cluster observations \citep{2012JCAP...07..017A, 2012ApJ...750..123A, Ackermann2015}. These DM annihilation limits are nevertheless not at the level of discarding thermal WIMP models\footnote{This is true under the assumption of ``vanilla'' WIMP models.}. The situation comparatively improves for WIMP DM decay. Indeed, lower limits on the WIMP lifetime derived from the observation of clusters \citep{Huang:2011xr, Cirelli:2012ut, MAGIC:2018tuz} are among the most constraining ones at present. Some authors have also studied the possibility that some past hints of detection of $\gamma$-rays in clusters  were due to WIMP DM \citep{Hektor:2012kc, Colafrancesco2011, 2012MNRAS.427.1651H}, however these works were inconclusive due to the faintness of such signals, indeed never confirmed. 

The Cherenkov Telescope Array\footnote{\url{https://www.cta-observatory.org/}} (CTA, \citep{Actis2011}) will be the next generation ground-based $\gamma$-ray observatory. It will be amongst the most sensitive $\gamma$-ray telescope from 20 GeV to 300 TeV. CTA will be based at two sites: La Palma, in the North, and Paranal in the South, allowing us to observe sources in a large fraction of the sky. CTA will provide a major improvement, up to one order of magnitude in sensitivity and up to a factor 2 in angular resolution, with respect to previous instruments\footnote{\url{https://www.cta-observatory.org/science/ctao-performance/}}. The study of CR physics and DM are among the main drivers of CTA science \citep{CTA2019}. In particular, CTA will allow us to search for diffuse $\gamma$-ray emission from galaxy clusters. One of the proposed CTA key science projects consists in the observation of the Perseus galaxy cluster, which is among the most promising targets for such observations. These observations should complement DM searches in the Galactic center \citep{CTA:2020qlo}, dwarf galaxies, the Large Magellanic Cloud \citep{CherenkovTelescopeArray:2023aqu}
or the search of axion-like particles, and should be used to probe fundamental physics \citep{CTA:2020hii}.

\subsection{The Perseus cluster as a promising $\gamma$-ray target}\label{sec:Perseus}
The Perseus cluster (Abell 426) is the brightest cluster in the X-ray sky \citep{Edge1992}. It is one of the most massive nearby clusters and presents the typical properties of a relaxed cool-core cluster with a dense core. Nonetheless, two main cold fronts have been identified and interpreted as the result from the sloshing due to minor mergers \citep{Walker2017,Sanders2020}. The Perseus cluster hosts a radio mini-halo \citep{Miley1975,Noordam1982,Pedlar1990,Burns1992,Sijbring1993,Sijbring1998,Gendron-Marsolais2017,Gendron-Marsolais2020,Gendron-Marsolais2021}, and X-ray cavities associated with the radio lobes of the central AGN, NGC~1275 (3C84), indicate that the feedback is important in the cluster center \citep{Boehringer1993,Fabian2000}. AGN activity may also be responsible for weak shocks and turbulence \citep{Fabian2006,Zhuravleva2014}, which could reaccelerate particles, in addition to direct CR injection from the AGN. The contribution to the radio mini-halo from hadronic interactions, direct CR injection from AGN and the role of turbulence remains unknown.

Two AGN from the Perseus cluster are known $\gamma$-ray emitters: NGC~1275 \citep{Abdo2009} and IC~310 \citep{Neronov2010}. Both sources are variable in time. NGC~1275 is the central galaxy of the cluster and IC~310, located about 0.6 deg southwest from the X-ray peak, is consistent with a narrow-angle tail radio galaxy infalling into the cluster \citep{Gendron-Marsolais2020}. A few other remarkable radio galaxies are present in the cluster volume: NGC~1265, NGC~1272, CR~15 \citep{Gendron-Marsolais2020}, that were not detected at $\gamma$-ray energies\footnote{But they could be included in the sky model as point sources and their contribution accounted for}. While these sources are expected to contaminate the CTA data when searching for diffuse emission, they are also contributing to inject CR into the ICM, which could eventually contribute to large-scale $\gamma$-ray emission.

The Perseus cluster has been recognised as one of the best targets for searches of CR-induced $\gamma$-ray emission \citep{Pinzke2010,Aleksic2010}. This is because the mini-halo traces a dense region where hadronic collisions and the production of secondaries is maximised. While its central $\gamma$-ray bright galaxy, NGC~1275, prohibits reliable constraints on the diffuse ICM component from \textit{Fermi}-LAT, the better angular resolution of CTA and the larger accessible energy range probed is expected to allow us separating the different sources of a possible $\gamma$-ray emission. In fact, the expected mild angular extent of the cluster (about 1 deg) due to its proximity (about 75 Mpc) implies that CTA is expected to resolve the diffuse emission if bright enough, but also that the angular extension of the signal is smaller than the field of view diameter by a factor of 5 to 10, so that systematic effects associated with the background modelling are limited. Along this line, also due to its large mass and proximity, Perseus stands out as one of the best clusters to search for DM-induced $\gamma$-ray emission. Indeed, the expected annihilation/decay flux is comparable to the one from other promising local galaxy clusters \citep{Combet2012}, such as Coma, Fornax, Ophiuchus, Hydra or Centaurus -- see \citep{2011JCAP...12..011S, Jeltema:2008vu}\footnote{The exception is Virgo. The Virgo cluster \citep{Fouque:2001qc} is sometimes considered as the best cluster for DM searches since it is the closest one to Earth. However, its large angular extension, comparable to the field of view of CTA, and its dynamical condition as a not-yet-virialized object, complicate considerably a potential DM analysis.} as well as other traditional DM targets such as dwarf satellite galaxies or nearby galaxies. 

A previous search for diffuse TeV $\gamma$-ray emission towards the Perseus cluster was performed using the MAGIC telescopes \citep{Aleksic2010,Aleksic2012,Ahnen2016}. Not having detected the signal, they reported upper limits on the CR to thermal pressure ratio assuming different models. For instance, using a spectral photon index slope of 2.2 and a relatively compact profile for the CRp, they obtained an upper limit of $\sim 1-2$\% on this ratio. This provided the best limit on the CR content of a cluster obtained from ground-based $\gamma$-ray observations so far, at a similar level to that obtained with \textit{Fermi}-LAT \citep{Ackermann2014}\footnote{Note that a high-significance image of the Perseus diffuse $\gamma$-ray emission was reported in \citep{Sinitsyna2014} using the SHALON telescopes. However, their flux is in strong disagreement with the limits allowed by the MAGIC observations so that these results are very controversial.}. \textit{Fermi}-LAT data from Perseus have been recently used to obtain constraints on the velocity-averaged DM annihilation cross-section as well \citep{Thorpe-Morgan:2020czg}. The obtained limits are more than two orders of magnitude above from the reference value of the thermal relic cross-section. Additionally, MAGIC observations were used to set constraints on the DM decay lifetime \citep{MAGIC:2018tuz}, these being among the strongest constraints for DM masses in the TeV, range up to date reaching the value of $2\times 10^{26}$ s. 

Finally, the location of the Perseus cluster in the sky allows for low zenith-angle observations from the CTA Northern Array, guaranteeing the best sensitivity of the array over its whole energy range. For all these reasons, the Perseus cluster was selected as the prime target for diffuse $\gamma$-ray emission searches from galaxy clusters with CTA, as one of the Key Science Projects (KSP) \citep{CTA2019}. We refer to this previous work for further details about this choice.

The paper is organized as follows. Section~\ref{sec:Modeling_CR} and Section~\ref{sec:DM_modelling} present the cluster modelling of the cluster in the context of CR and DM induced $\gamma$-ray emission, respectively. The observation setup and the background sky modelling are discussed in Section~\ref{sec:Observation_setup}. Section~\ref{sec:CTA_CR_sensitivity} and Section~\ref{sec:CTA_DM_sensitivity} provide the results on the CTA sensitivity to CR and DM physics, respectively. We conclude in Section~\ref{sec:Conclusions}. A few appendices complement the paper.

Throughout this paper, we assume a flat $\Lambda$CDM cosmology with $H_0 = 70$ km s$^{-1}$ Mpc$^{-1}$, $\Omega_{\rm M} = 0.3$, and $\Omega_{\Lambda} = 0.7$. The coordinates of the Perseus cluster center are taken as the ones of its central galaxy NGC~1275, R.A., Dec = 49.9507, 41.5117 deg\footnote{NASA/IPAC extragalactic database, \url{http://ned.ipac.caltech.edu/}.} and its redshift is $z = 0.017284$ \citep{Hitomi2018}, corresponding to a luminosity distance of $d_L \simeq 75$ Mpc. Given our reference cosmological model, 1 deg in the sky corresponds to a 1.265 Mpc distance at the redshift of Perseus. We adopt the characteristic angular radius $\theta_{500} = 59.7 \pm 0.4$ arcmin\footnote{$R_{500}$ is the radius within which the mean cluster density is 500 times the critical density of the Universe at the cluster's redshift, and $M_{500}$ is the mass within this radius. $\theta_{500}$ is the angle on the sky that corresponds to $R_{500}$. Quantities with subscript 500 (or 200) refers to quantities within $R_{500}$ (or $R_{200}$).} obtained by \citep{Urban2014} using the fit of the \textit{Planck} universal pressure profile\footnote{This value is in excellent agreement with the one derived from our thermal model and the assumption of hydrostatic equilibrium, assuming an hydrostatic mass bias of 0.1, in Section~\ref{sec:Modeling_CR}.}. This corresponds to a physical radius $R_{500} = (1.26 \pm 0.01$) Mpc and to a mass $M_{500} = (5.77 \pm 0.12) \times 10^{14}$~M$_{\odot}$. In the paper, International System of Units is used unless specified.

\section{Modelling the $\gamma$-ray emission associated with cosmic-rays}\label{sec:Modeling_CR}
The prediction of the diffuse $\gamma$-ray emission induced by CR in the ICM requires the detailed modelling of the physical components at play, their interactions, and the underlying radiative processes. In this section, we describe such a model based on the {\tt MINOT} software \citep{Adam2020}\footnote{The software is available at \url{https://github.com/remi-adam/minot}.}. {\tt MINOT} is dedicated to compute the observable properties of the ICM (radio synchrotron, thermal Sunyaev-Zel'dovich effect, X-ray thermal bremsstrahlung, inverse-Compton, $\gamma$-rays from hadronic interactions, and neutrino emission) given the user-defined physical state of the cluster. The code describes galaxy clusters as spherically symmetric objects. Here, we discuss essentially the calibration of the input model using external data assuming different scenarios, which we feed to {\tt MINOT}. This includes the thermal gas density and pressure (Section~\ref{sec:modeling-thermal_gas}), the magnetic field strength (Section~\ref{sec:modeling-magnetic_field}), and the CR spatial and spectral distributions (Section~\ref{sec:modeling-cosmic-rays}). This allows us to perform predictions for the $\gamma$-ray signal and estimate model uncertainties using the method implemented in {\tt MINOT}.

\subsection{Modelling the intracluster medium components}
We model the Perseus cluster assuming spherical symmetry, considering only radial profiles to describe its ICM components. This assumption is expected to be fairly accurate given the fact that the Perseus cluster is overall a relaxed system. 

While clusters are diffuse objects with no clear definitions of their extension, we expect density and pressure discontinuities near the virial radius \citep{Molnar2009,Aung2020}. We consider a maximum radial extent of the cluster and define this truncation radius as $R_{\rm tr} = 3 R_{500} = 3.78 \ {\rm Mpc} \simeq 2 R_{200}$ using the measurement from \citep{Hurier2019} as a reference. The exact choice of this value does not significantly affect our results, but this allows us to perform numerical integrations over a well defined region.

\subsubsection{Thermal gas}\label{sec:modeling-thermal_gas}
The modelling of the thermal gas is necessary to compute the CR induced $\gamma$-ray emission for two reasons: 1) the nuclei (protons, helium and heavier elements) are involved in hadronic interactions that lead to the emission of $\gamma$-rays and secondary electrons, which require modelling of the thermal gas density; 2) the thermal pressure, or thermal energy, is used for the relative normalization of the CR energy.

The thermal electron density is modeled as a double $\beta$-model \citep{Cavaliere1976}, following \citep{Churazov2003}:
\begin{equation}
n_e(r) = n_{0,1} \left(1+\left(\frac{r}{r_{c,1}}\right)^2\right)^{-3\beta_1/2} + n_{0,2} \left(1+\left(\frac{r}{r_{c,2}}\right)^2\right)^{-3\beta_2/2}.
\end{equation}
The core parameters are taken from the \textit{XMM-Newton} measurement \citep{Churazov2003,Churazov2004}. While \citep{Churazov2003,Churazov2004} used the \textit{Einstein} telescope results from \citep{Jones1999} for the peripheral outskirts region parameters, we use instead the more recent \textit{Suzaku} measurement \citep{Urban2014}. We have $\left(n_{0,1}, r_{c,1}, \beta_1, n_{0,2}, r_{c,2}, \beta_2\right) = \left(4.6 \times 10^{-2} {\rm \ cm}^{-3}, 57 {\rm \ kpc}, 1.2, 3.6 \times 10^{-3} {\rm \ cm}^{-3}, 278 {\rm \ kpc}, 0.71\right)$, when accounting for the different cosmological models.

We rely on X-ray spectroscopic measurements to describe the gas temperature, as
\begin{equation}
k_{\rm B} T(r) = 7 \times \left(1+\left(\frac{r}{r_{t,1}}\right)^3\right) \left(2.3 + \left(\frac{r}{r_{t,1}}\right)^3\right)^{-1} \left(1 + \left(\frac{r}{r_{t,2}}\right)^{1.7}\right)^{-1} \ {\rm keV},
\end{equation}
with $k_{\rm B}$ the Boltzmann constant. The first terms allow us to describe the core temperature as provided by \citep{Churazov2003}, where $r_{t,1}=73.8$ kpc, given our cosmological model. We introduce the last term in order to account for the temperature drop in the outskirt as measured by \citep{Urban2014}, where we set $r_{t,2} = 1600$ kpc.

Given the thermal electron density and temperature, we compute the electron thermal pressure as
\begin{equation}
P_e(r) = n_e(r) k_{\rm B} T(r).
\end{equation}
Additionally, we assume an ICM helium mass fraction of 0.2735 and we model the metal abundances as constant, with $\frac{Z}{Z_{\odot}} = 0.3$ \citep{Werner2013}, using solar abundances from \citep{Lodders2009}. The electron number density and thermal pressure, together with the abundances are used to compute the full thermodynamic properties of the thermal component following \citep{Adam2020}.

\begin{figure}
	\centering
	\includegraphics[width=0.49\textwidth]{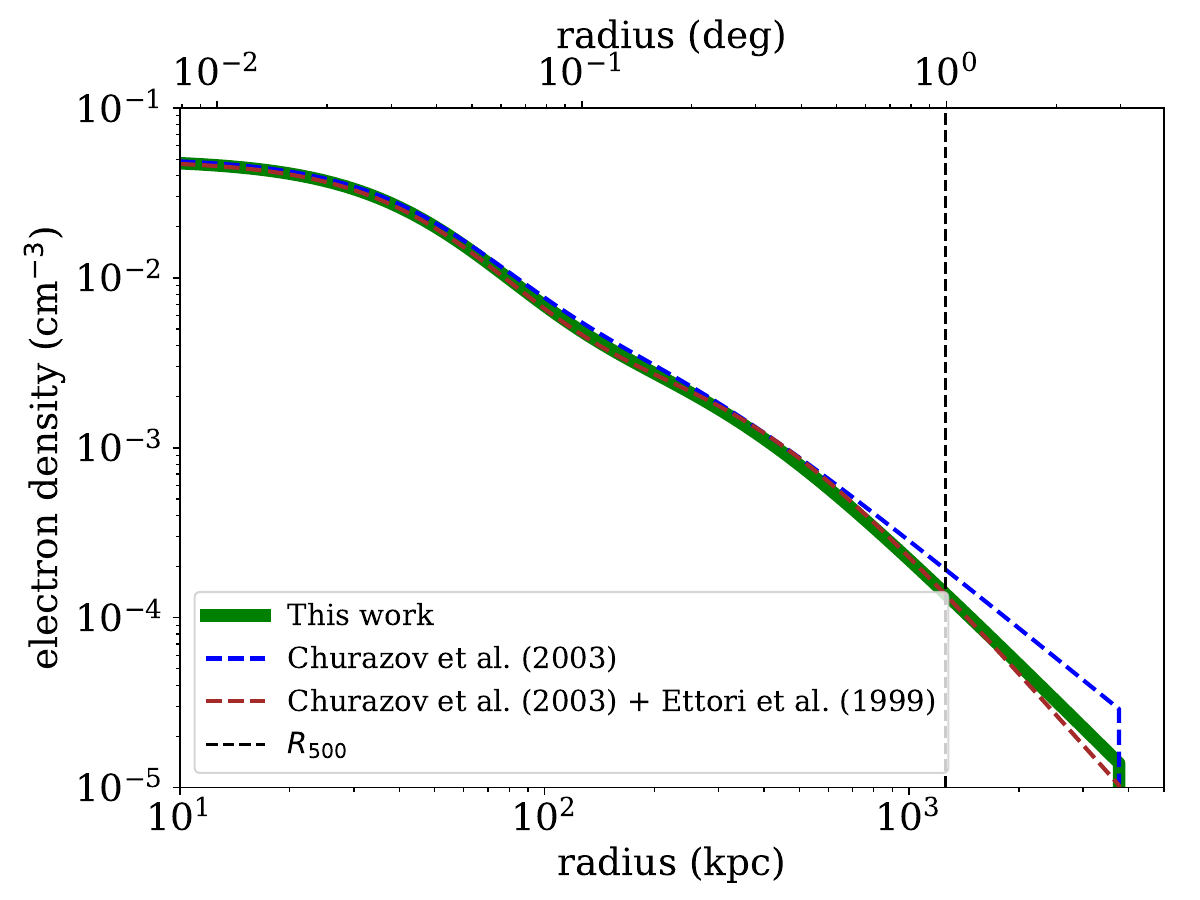}
	\includegraphics[width=0.49\textwidth]{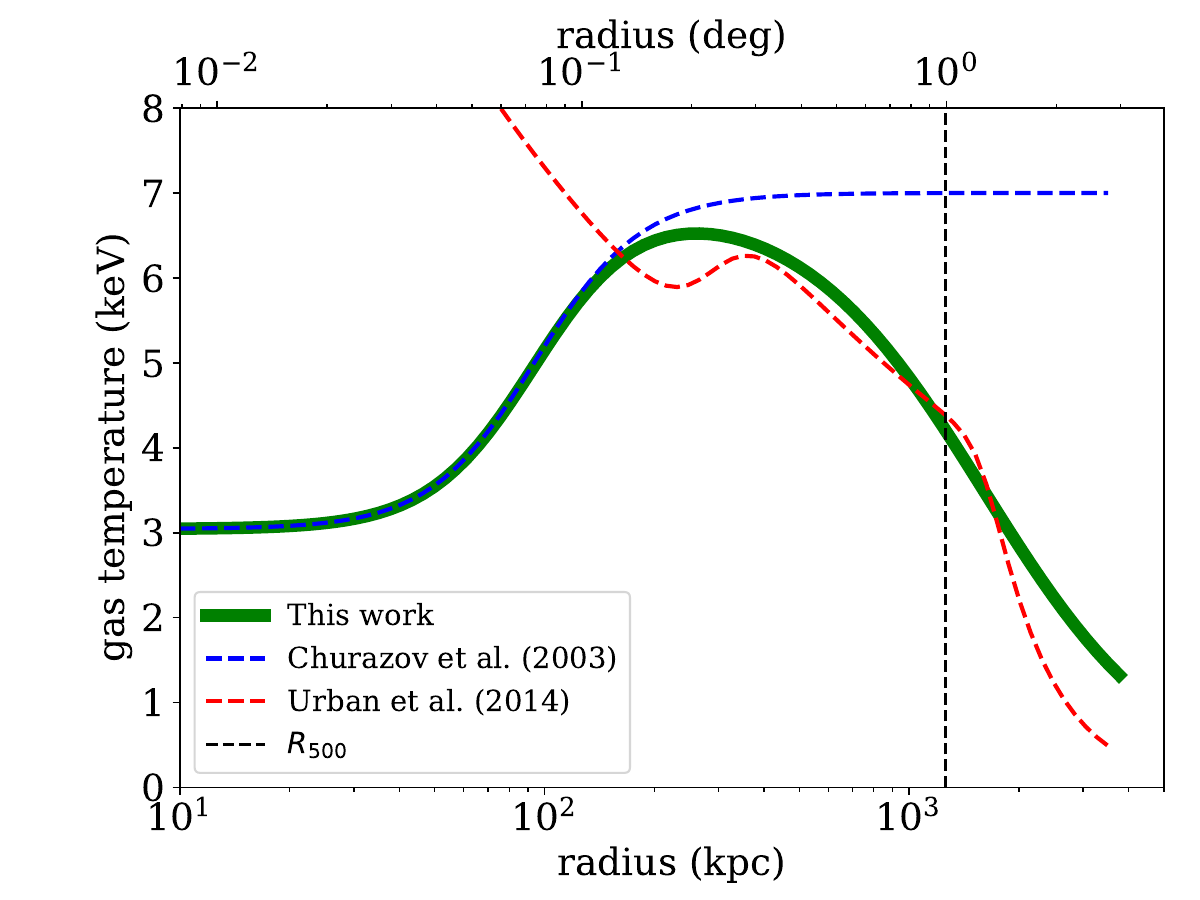}
	\includegraphics[width=0.49\textwidth]{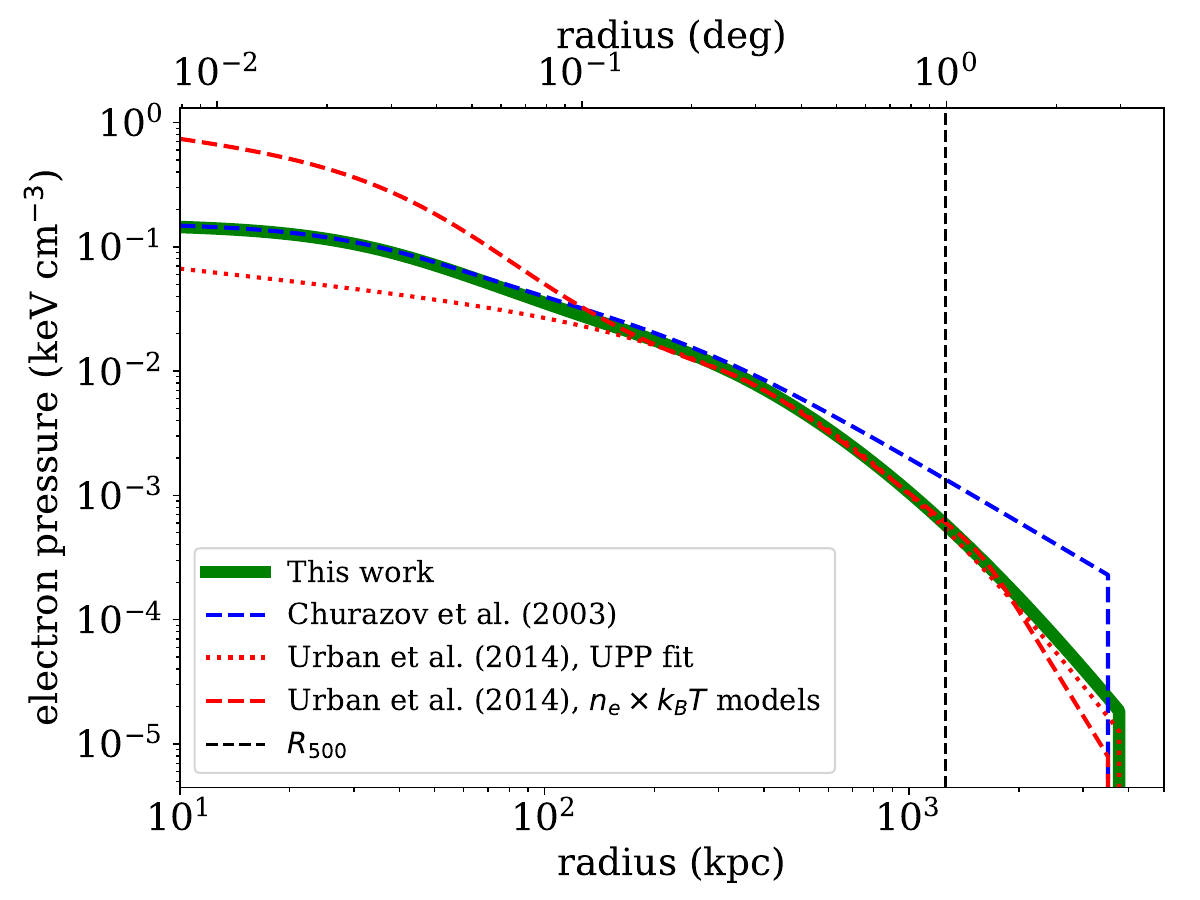}
	\includegraphics[width=0.49\textwidth]{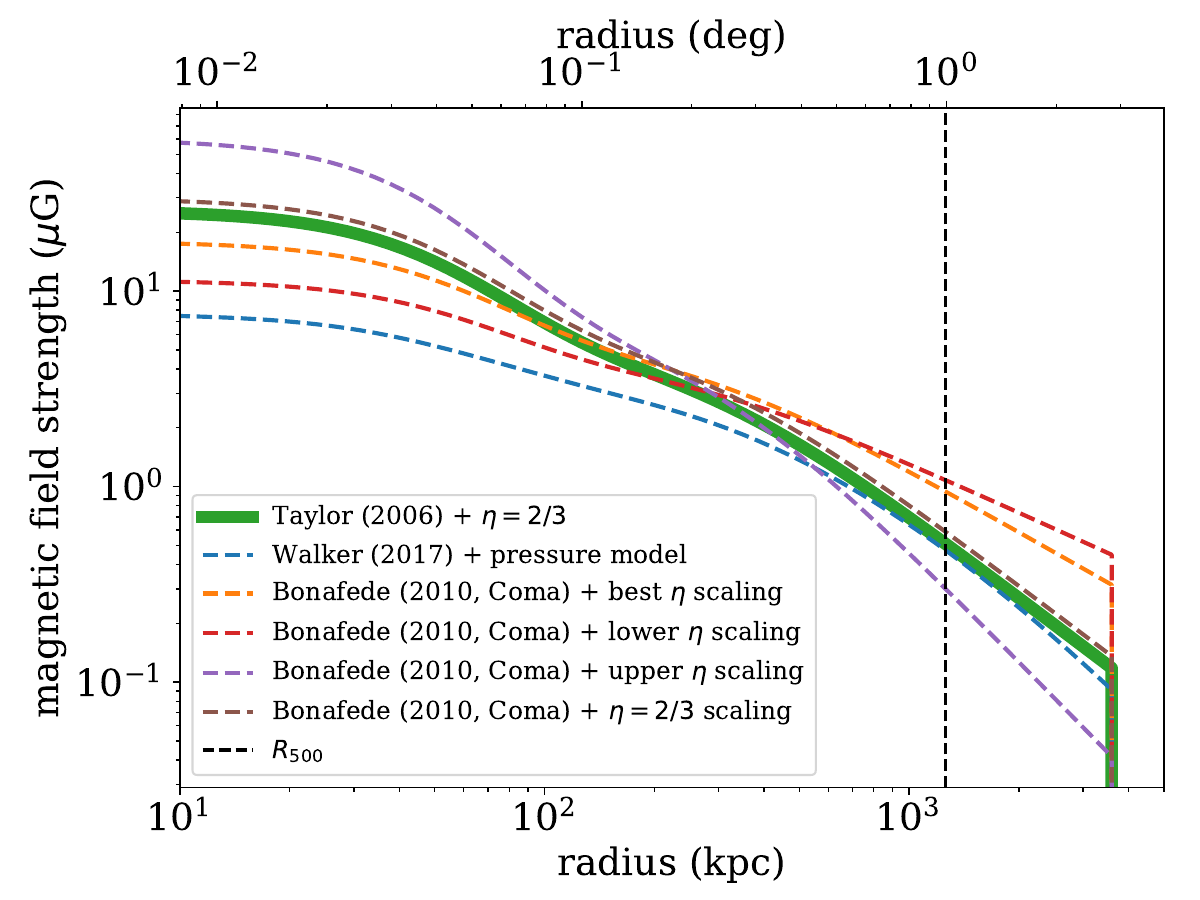}
	\caption{Perseus cluster thermal gas and magnetic field models.
	{\bf Top left panel:} thermal electron number density profile. The model by \citep{Churazov2003} combines \textit{XMM-Newton} observations in the core and \textit{Einstein} observations in the outskirt \citep{Jones1999}. The red curve is similar, but we have replaced the outskirt model by that obtained in \citep{Ettori1998} with ROSAT.
	{\bf Top right panel:} gas temperature profile. The model by \citep{Churazov2003} is constrained by the data up to about 200 kpc. The model by \citep{Urban2014} is constrained by the data down to about 200 kpc.
	{\bf Bottom left panel:} thermal electron pressure profile. In the case of \citep{Urban2014}, both the combination of gas temperature and density, and the \textit{Planck} universal pressure profile fit to the their data are shown.
	{\bf Bottom right panel:} Magnetic field strength models according to scaling from literature measurements.
    The truncation radius is visible at $3 R_{500}$.}
\label{fig:thermal_and_mag_model}
\end{figure}

In Figure~\ref{fig:thermal_and_mag_model}, we present the thermal electron number density, the temperature and the electron pressure radial profiles. We compare our model to other parameterizations available in the literature. In the core, all density profile models agree since they all rely on \textit{XMM-Newton} data \citep{Churazov2003}. In the outskirts, the agreement is good up to $R_{500}$. Beyond, ROSAT \citep{Ettori1998} and Suzaku \citep{Urban2014} agree well but the Einstein telescope model \citep{Jones1999} leads to a larger electron number density. The temperature profile is typical of that of a cool-core cluster. Our model matches the \citep{Churazov2003} model in the core and the \citep{Urban2014} model in the outskirt according to their domain of validity. All the pressure profile models agree well in the outskirt except for the model derived from the extrapolation of the \citep{Churazov2003} temperature in the outskirt, which over-predicts the pressure by a factor of a few. Similarly, in the core, the \citep{Urban2014} models do not agree with the more direct measurement based on \citep{Churazov2003} where they extrapolate the profile with an isothermal component. The observed differences are due to the different scales probed by the respective instruments and the extrapolation of the profiles.

Given the definition of our thermal model and the data used to constrain it, we are able to accurately describe the cluster from its core (10 kpc) to the outskirts (2 Mpc). The uncertainties associated with the thermal model are expected to be negligible compared to the uncertainties associated with the non-thermal component. In appendix \ref{app:validation_thermal_model_planck}, we also present a validation of our thermal model using the \textit{Planck} Compton parameter map, showing that our model accurately describes the pressure profile of the cluster. We note that because the temperature model based on \citep{Churazov2003} is valid only up to about 200 kpc, its use leads to an overestimation of the thermal energy by a factor of a few, depending on the details of the line-of-sight integration of the model, when extrapolated beyond its validity region. This is what is done in \citep{Aleksic2010,Aleksic2012,Ahnen2016} up to the virial radius ($\sim R_{200}$), which should affect their constraints on the CR by a similar amount.

\subsubsection{Magnetic field strength}\label{sec:modeling-magnetic_field}
\label{sec:magnetic_field_model}
The modelling of the magnetic field strength is necessary when considering jointly the $\gamma$-ray emission and the radio synchrotron emission. However, the magnetic field strength and the structure in galaxy clusters remain poorly known to date \citep{Donnert2018}. We thus consider several approaches in order to model the magnetic field distribution in the Perseus cluster, which will allow us to quantify the associated systematic effect.

Our first approach relies on the scaling of the magnetic field to the thermal gas density. In this case, the magnetic field strength is given by
\begin{equation}
	\left<B\right>(r) = \left<B_{\rm ref}\right> \left(\frac{n_e(r)}{n_{\rm e,ref}}\right)^{\eta_B},
	\label{eq:Bfield_scaling}
\end{equation}
where $\left<B_{\rm ref}\right>$ and $n_{\rm e,ref}$ are magnetic field and density normalization parameters, respectively. We first compute the magnetic field strength using the rotation measure estimate from \citep{Taylor2006}, which gives $\left<B\right>(10 \ {\rm kpc}) \sim 25$ $\mu$G, and set $\eta_B=2/3$ assuming magnetic field flux conservation.

The Coma cluster is one of the only clusters for which the magnetic field strength profile was measured \citep{Bonafede2010}. As a second approach, we thus assume that the Coma and Perseus clusters have the same magnetic field strength to gas density ratio. In this case, we use the values $\eta_{B} = [0.4,0.5,2/3,0.9]$, corresponding to $\left<B_{\rm ref}\right> = [3.9,4.7,5.0,5.4]$ $\mu$G with $n_{\rm e, ref}=
3.42 \times 10^{-3} \ {\rm cm}^{-3}$, allowing us to account for the uncertainties in the measurement \citep{Bonafede2010}.

The structure of the Kelvin-Helmholtz instability visible in Perseus was used to infer the thermal to magnetic pressure ratio, $\beta_{\rm pl} = P_{\rm gas}/P_B \sim 200$, for the overall cluster volume prior to sloshing \citep{Walker2017}. In a last approach, we use this measurement to infer the magnetic field strength profile, as
\begin{equation}
	\left<B\right>(r) = \left(\frac{2 \mu_0}{\beta_{\rm pl}} P_{\rm gas}(r)\right)^{1/2},
\end{equation}
where $\mu_0$ is vacuum permeability.

In Figure~\ref{fig:thermal_and_mag_model} (bottom right panel), we present our models of the magnetic field strength profiles of the Perseus cluster. All models provide a similar order of magnitude, but the differences reflect the difficulty in having an accurate description. The scatter between models nearly reaches an order of magnitude in the core and in the outskirt, with the best agreement between all models around $200-600$ kpc. We note that the model with $\eta_{B}=0.9$ and $\left<B_0\right>=5.4$ $\mu$G implies a core magnetic field that reaches about 50 $\mu$G, corresponding to an energy density that is one third of the thermal pressure ($\beta_{\rm pl} \sim 3$) and thus very high compared to physical expectations. Assuming the pure hadronic model, a higher magnetic field will lead to a lower $\gamma$-ray flux for a fixed radio emission. Thus, the highest magnetic field model gives a conservative estimate of the expected $\gamma$-ray emission. In the following, as a baseline, we use the model based on \citep{Taylor2006} with $\eta_{B} = 2/3$ that corresponds to an intermediate estimate.

\subsubsection{Cosmic-ray model}\label{sec:modeling-cosmic-rays}
The $\gamma$-ray emission induced by hadronic interactions is directly related to the spatial and spectral distribution of CRp in the ICM. They are modeled according to a radial profile and a canonical power-law in momentum space,
\begin{equation}
\frac{dN_{\rm CRp}}{dpdV}(E,r) = A_{\rm CRp} \ p^{-\alpha_{\rm CRp}} \  n_e(r)^{\eta_{\rm CRp}}.
\label{eq:crp_modeling}
\end{equation}
The CR radial profile assumes a scaling with respect to the gas density so that only one parameter, $\eta_{\rm CRp}$, is necessary to describe the spatial distribution. Physically, this parameter allows us to account for the CR dynamics and the competition between advection and streaming, which remains poorly known. We also consider the case where the CRp profile is scaled with respect to the thermal pressure profile, and thus related to the thermal energy. The value of the power-law slope $\alpha_{\rm CRp}$ is related to the acceleration of CR such as the associated Mach number distribution \citep{Pfrommer2006}. The normalization $A_{\rm CRp}$ is computed given the value of the CRp to thermal energy ratio, $X_{\rm CRp}(R) = \frac{U_{\rm CRp}(R)}{U_{\rm th}(R)}$, by integrating Equation~\ref{eq:crp_modeling} accordingly. The thermal energy is computed by integrating the pressure profile. The CR energy is computed by integrating the CR distribution over the volume and between $E_{p,{\rm min}} = 1.22$ GeV, i.e., the energy threshold of proton-proton interactions, and $E_{p,{\rm max}} = 10$ PeV, above which some CRp could escape the cluster \citep{Pinzke2010}, although the exact value does not affect our results. In the following, we refer to $X_{\rm CRp}(R_{500})$ as $X_{500}$ and use this reference for normalization. The value of this parameter has been predicted using numerical simulations, being $X_{500} \sim 1$\% \citep{Pinzke2010}.

The CRe are modeled accounting for two contributions: the primary electrons that are independent from the CRp, and the secondary electrons that are produced from hadronic interactions assuming stationarity. The primary electrons, whenever considered, are modeled similarly to the CRp (Equation~\ref{eq:crp_modeling}), but with different spectral model to account for energy losses (e.g., power-law or exponential cutoff power-law). We will show in Section \ref{sec:Calibration_of_the_model_parameters} that in practice, primary CRe are irrelevant for our purpose. Secondary CRe are computed as detailed in \citep{Adam2020}.

\subsection{Non-thermal radiative processes}\label{sec:non-thermal_radiative_processes}
Given the ICM model, we compute the different observables associated with the $\gamma$-ray emission from hadronic interactions, the radio synchrotron, and inverse-Compton emission. The calculations are done using the {\tt MINOT} software \citep{Adam2020}.

\subsubsection{Hadronic interactions and $\gamma$-ray emission}
The hadronic production rate of $\gamma$-rays is computed by integrating the collision rate of proton-proton interactions multiplied by the energy distribution of $\gamma$-rays produced per collision, over the energy of the CR. The computation is based on the parameterization from \citep{Kafexhiu2014}, and its implementation follows the work by \citep{Zabalza2015}. In this paper, we use the {\tt Pythia8} proton-proton interaction model and include corrections for proton-nuclei collision \citep{Adam2020}.

Once the rest frame production rate of $\gamma$-rays is computed, the radial profile and energy spectrum of $\gamma$-rays, as would be observed from Earth, are obtained by line of sight integration and eventually integrating over the energy or the solid angle, respectively. We also account for the Universe opacity as a function of photon energy using the extra-galactic background light (EBL) model from \citep{Dominguez2011}, but this choice does not affect our results given the fact that the corresponding cutoff is at sufficiently high energy, of about 30 TeV \citep{Hussain_etal2021,Hussain:2022tls}.

In proton-proton interactions, in addition to $\gamma$-rays, secondary electrons are obtained following the same scheme. In this case, however, the prescription from \citep{Kelner2006} is used instead of that from \citep{Kafexhiu2014}. The spatial and spectral distribution of secondary CRe is then obtained from the particle injection rate by applying energy losses assuming a steady state scenario.

Modelling uncertainties associated with the particle interaction rates are expected to be accurate to about 30\% at CTA energies, for the same parent CRp population (see Figure~7 from \citep{Adam2020}). This is estimated by comparing the output from the different parametrizations implemented in {\tt MINOT} (from \citep{Kafexhiu2014}: {\tt Pythia8}, {\tt Geant4}, {\tt QGSJET}, {\tt SIBYLL}, and \citep{Kelner2006}). See also \citep{Orusa2023} for another recent determination of the $\gamma$-ray production cross-section.

\subsubsection{Inverse-Compton $\gamma$-ray emission}
Leptonic $\gamma$-ray emission arises via the inverse-Compton scattering of CRe (secondaries, but also primary CRe whenever considered) onto cosmic microwave background (CMB) photons. The inverse-Compton {\tt MINOT} calculations are based on the analytical approximation given by \citep{Khangulyan2014}. We will see later that, in practice, inverse-Compton emission should be negligible (from both primary or secondary electrons) for CTA observations and it will be ignored in the following sections.

\subsubsection{Synchrotron radio emission}
Synchrotron emission should be modeled when considering radio data in order to calibrate our model or check that it does not imply excess radio signal compared to observations. The {\tt MINOT} software computes synchrotron emission following \citep{Aharonian2010}, for which uncertainties are negligible for this work. This assumes that the orientation of the magnetic field is randomized as it is expected for radio halos.

\subsection{Calibration of the model parameters}\label{sec:Calibration_of_the_model_parameters}
In this section, we discuss the different methodologies employed to set our model parameters.

\begin{table}[h]
	\caption{Summary of the parameter values and their explored range, and the $\gamma$-ray flux at CTA energies for the hadronic and inverse-Compton emission (given as: reference value, [min, max]). The flux $F_{500}$ is computed within $\theta_{500}$ by cylindrical integration for energies above 150 GeV and given in units of $10^{-14}$ cm$^{-2}$ s$^{-1}$. In the case of pure hadronic and pure leptonic models, the central value corresponds to the best-fit model and the interval corresponds to the 68\% confidence level. The changes in $\alpha_{\rm CRp}$ for the different magnetic field models is only due to degeneracies in the parameter space and correlations with $\eta_B$. It is not a physical effect and the difference remains well below statistical uncertainties.
 }
	\begin{center}
	\resizebox{\textwidth}{!} {
	\begin{tabular}{|c|ccc|cc|}
	\hline
	Model & $X_{500}$ (\%) & $\alpha_{\rm CRp}$ & $\eta_{\rm CRp}$ & $F^{\rm (had)}_{500, E_{\gamma}>150 {\rm \ GeV}}$ & $F^{\rm (IC)}_{500, E_{\gamma}>150 {\rm \ GeV}}$ \\
	\hline
 	& & & & \multicolumn{2}{|c|}{($10^{-14}$ cm$^{-2}$ s$^{-1}$)} \\
	\hline
	\hline
	Baseline & 1.0, [0.0, 20.0] & 2.30, [2.00, 3.00] & 1.00, [0.00, 1.50] & 70.2, [0, 11373.8] & 2.1, [0, 625.4] \\
	Pure hadronic + Taylor \citep{Taylor2006} & 4.5, [3.2, 14.6] & 2.36, [2.25, 2.65] & 0.78, [0.67, 0.82] & 159.8, [69.1, 234.6] & 4.3, [0.6, 10.7] \\
	Pure hadronic + Bonafede upper \citep{Bonafede2010} & 3.8, [2.6, 13.4] & 2.37, [2.24, 2.66] & 0.79, [0.65, 0.85] & 132.2, [65.2, 217.4] & 3.3, [0.6, 10.9] \\
	Pure hadronic + Walker \citep{Walker2017} & 7.5, [4.9, 15.8] & 2.35, [2.19, 2.54] & 0.76, [0.69, 0.80] & 273.0, [157.8, 421.0] & 9.2, [2.4, 28.1] \\
	Pure leptonic  & \multicolumn{3}{c|}{--} & 0 & $<28.1$ \\
	\hline
	\end{tabular}
	}
	\end{center}
	\label{tab:model_summary}
\end{table}

\subsubsection{Baseline model}
\label{sec:CR_baseline_model}
As a baseline, we consider the results obtained from numerical simulations and cosmic-ray transport description (e.g., \citep{Pinzke2010,Ensslin2011}; see also \citep{Ackermann2014} for an application with \textit{Fermi}-LAT) in order to set the value of the free parameters of the model: $\eta_{\rm CRp}$, $\alpha_{\rm CRp}$ and $X_{500}$ (see Eq.~\ref{eq:crp_modeling}). According to these works, the radial distribution of CR is expected to roughly scale with the thermal gas density ($\eta_{\rm CRp} = 1$) when advection by the turbulent gas dominates, but may flatten if diffusion and streaming become significant. The spectral slope is expected to be $\alpha_{\rm CRp}\sim 2.3$ and a normalization, $X_{500}$, of a few percent is usually expected. However, large uncertainties are associated with these values, such as the transport of CR for which the streaming velocity is poorly known \citep{Ensslin2011}, or the details of acceleration mechanism \citep{Bykov2019}. Our baseline parameter set is defined as $\left(X_{500}, \eta_{\rm CRp}, \alpha_{\rm CRp}\right) = \left(10^{-2}, 1.0, 2.3\right)$, but we explore a large range of values, as given in Table~\ref{tab:model_summary}.

\begin{figure}
	\centering
	\includegraphics[width=0.45\textwidth]{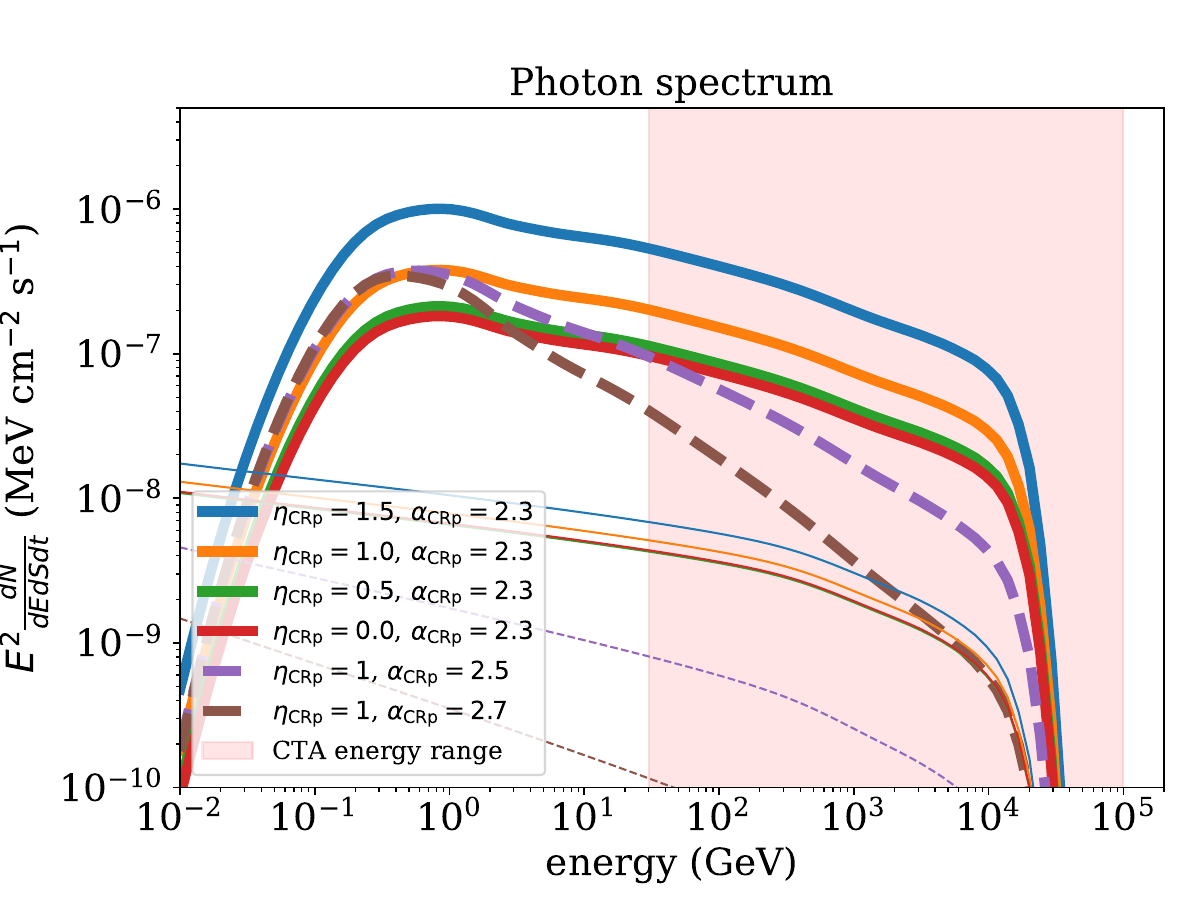}
	\includegraphics[width=0.45\textwidth]{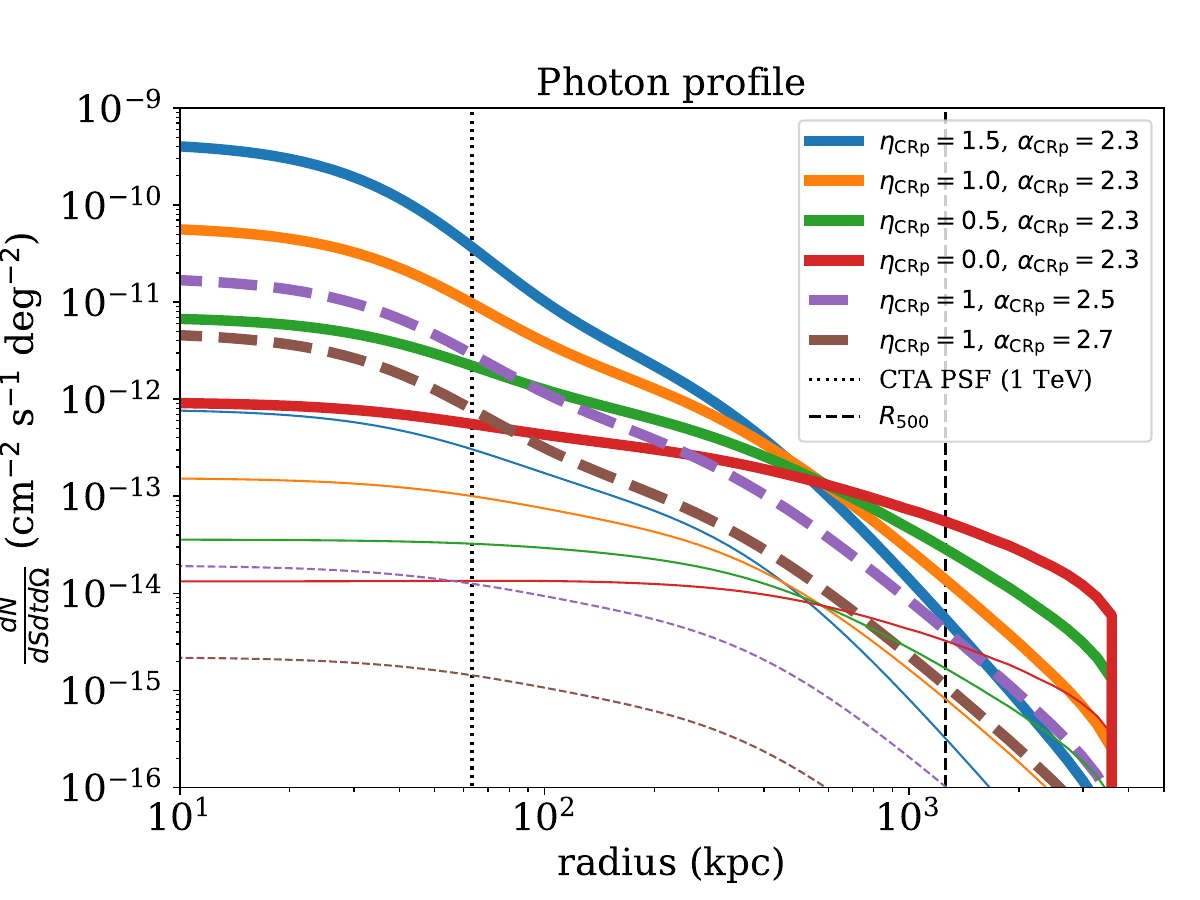}
	\includegraphics[width=0.45\textwidth]{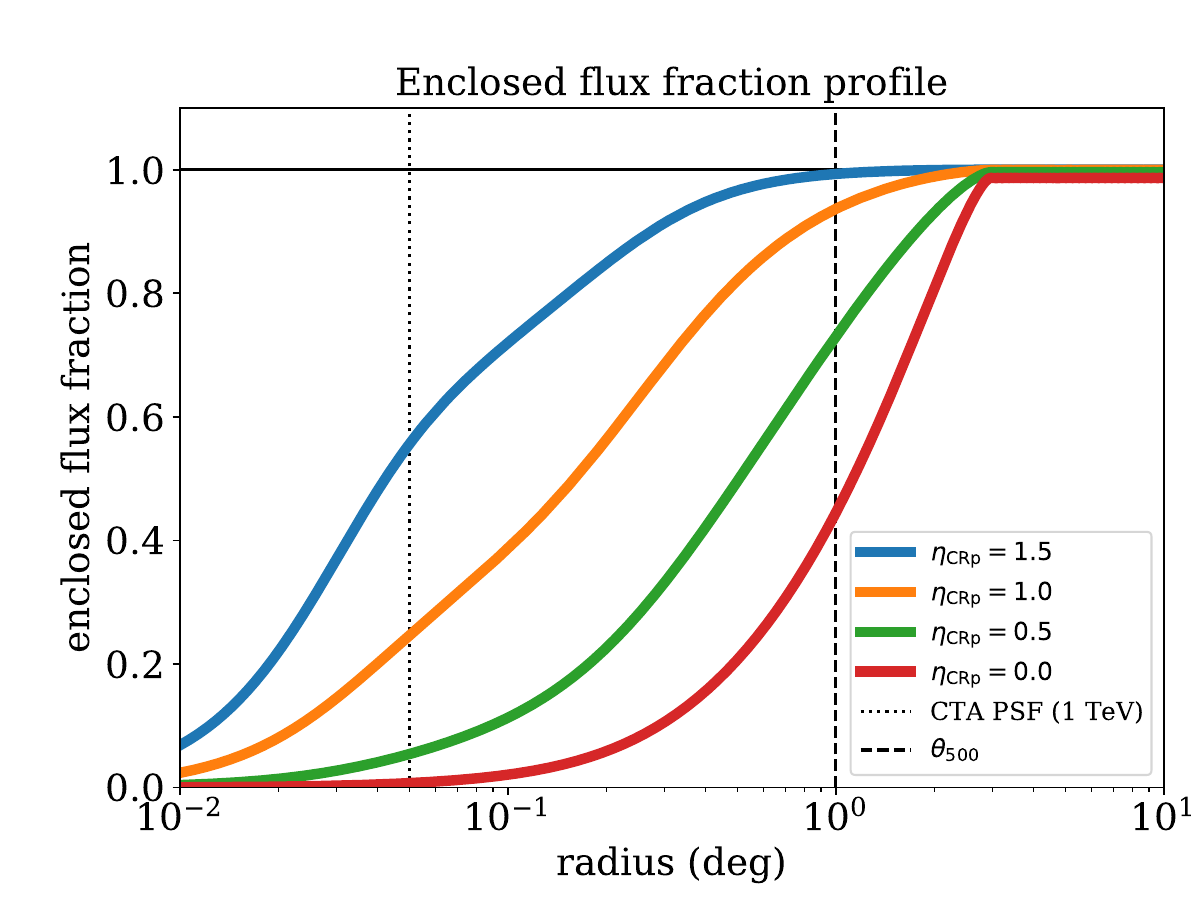}
    \includegraphics[width=0.45\textwidth]{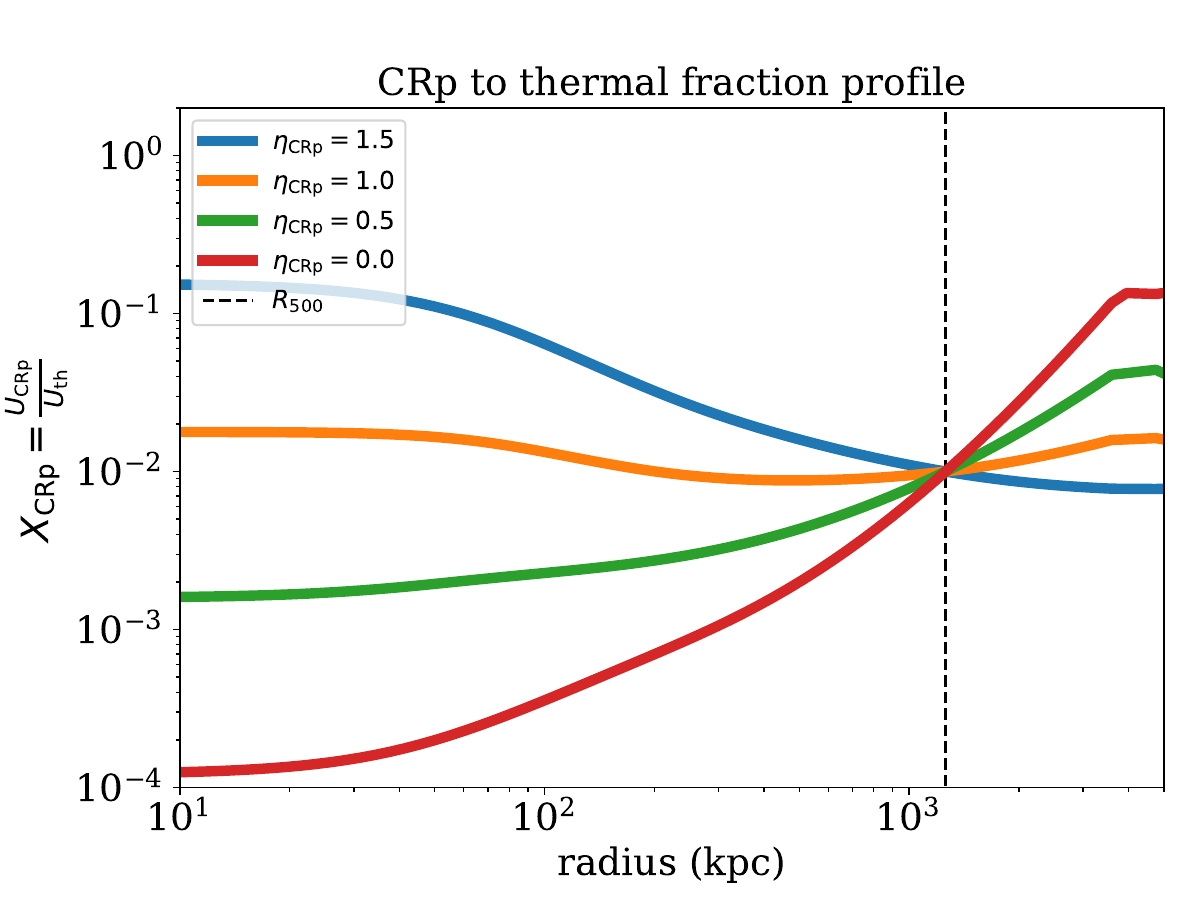}
	\caption{Observable properties of the hadronic $\gamma$-ray emission for different values of the model parameters. The inverse-Compton emission is also reported as a thin line with similar color and style on the top panels.
	{\bf Top left panel}: $\gamma$-ray emission spectrum in the case of hadronic processes (within $R_{500}$).
	{\bf Top right panel}: $\gamma$-ray emission profile, as a function of the projected radius, in the case of hadronic processes (computed within $[0.15,50]$ TeV).
	{\bf Bottom left panel}: fraction of the enclosed $\gamma$-ray flux, as a function of the projected radius (computed within $[0.15,50]$ TeV).
    {\bf Bottom right panel}: enclosed CRp to thermal energy ratio, as a function of physical radius.
    Note that the two bottom panels do not depend on $\alpha_{\rm CRp}$. The magnetic field model used when computing the secondary CR electrons that induce the inverse-Compton signal is the one from \citep{Taylor2006}.}
\label{fig:gamma_ray_observable_sim}
\end{figure}

In Figure~\ref{fig:gamma_ray_observable_sim}, we show the hadronic $\gamma$-ray observables associated with the models that we consider and how they vary as a function of the parameters. We also provide the associated radial profile of the ratio between integrated CRp energy and thermal energy, $X_{\rm CRp}(<R)$. All the models are calibrated to have the same normalization $X_{500}~=~10^{-2}$. In the CTA energy range, the $\gamma$-ray flux decreases when increasing the slope $\alpha_{\rm CRp}$ and the profile gets more compact when increasing the scaling $\eta_{\rm CRp}$. We note that for a fixed normalization at $R_{500}$, the flux computed within $R_{500}$ also increases with $\eta_{\rm CRp}$ because of the increased proton-proton collision rate due to the spatial overlap of the CR and the target gas. We note that the inverse-Compton emission arising from secondary CRe in the baseline model is always subdominant compared to the hadronic emission.

\subsubsection{Pure hadronic scenario}
\label{sec:CR_hadronic_model}
It is also possible to use radio data of the Perseus mini-halo to calibrate our model parameters in the case of a pure hadronic scenario. To do so, we extract the Very Large Array (VLA) surface brightness profile at 1380 MHz from \citep{Pedlar1990} and use the spectrum measured at 327, 609, and 1395 MHz from \citep{Gitti2002}, which were taken from \citep{Sijbring1993}. We use an aperture of 15 arcmin, as it corresponds to the mini-halo extent given in this work. The profile is affected by the presence of NGC~1275 in the core. In addition the VLA observation may suffer from the absence of adequately short baselines that are necessary to properly detect the mini-halo emission at large scales. For this reason, we exclude the data points below 23 kpc, and those above 80 kpc in the fitting. We also assume a 10\% uncertainty on the measured profile \citep{Zandanel2014b}.

We fit the radio data using a Markov Chain Monte Carlo (MCMC) technique with the {\tt emcee} package \citep{Foreman2013}. We use a Gaussian likelihood function and fit simultaneously the spectrum and the profile. In addition to our three physical parameters, we consider a nuisance parameter to account for the inter-calibration uncertainty between the spectrum and the profile, which were extracted from different instruments, and marginalise over this value using a Gaussian prior centered on unity and with a standard deviation of 10\%. Flat priors are used for the other parameters: $\eta_{\rm CRp} \in [0,5]$, $\alpha_{\rm CRp} \in [2,4]$ and $X_{500} \in [0,0.2]$. These limits are defined according to realistic physical expectations of the parameter values. The upper limit of 0.2 on $X_{500}$ corresponds to the expected hydrostatic mass bias of galaxy clusters, itself dominated by turbulent motions plus magnetic fields, so that $X_{500}>0.2$ would be unrealistic (see \citep{Pratt2019} for a review about the mass of galaxy clusters). For a given model allowed by the radio data, we compute the $\gamma$-ray observables associated with hadronic interactions and inverse-Compton emission. We reproduce the constraint for some of the considered magnetic field models discussed in Section~\ref{sec:magnetic_field_model}. The magnetic field model based on \citep{Taylor2006} with $\eta_{B}$ is used as a reference and we quantify the implications of this choice in Appendix \ref{app:impact_of_magnetic_field}. 

In Figure~\ref{fig:gamma_ray_observable_hadronic}, we show the pure hadronic best-fit model, its 68\% enclosed confidence interval, 100 model realizations randomly sampled from the MCMC chains, and the considered data points. The pure hadronic model provides a good description of the data that we have considered. Given the reference magnetic field model, it favors parameter values of $\left(X_{500}, \eta_{\rm CRp}, \alpha_{\rm CRp}\right) \sim \left(5 \times 10^{-2}, 0.8, 2.5\right)$, which is compatible with the MAGIC upper limit \citep{Ahnen2016}, but we stress that uncertainties are large and that the parameters are degenerate. Accordingly, the $\gamma$-ray observables are affected by large uncertainties, especially on the CR spectral slope, which leads to large uncertainties for the expected flux in the CTA energy range. We also stress that systematic effects may affect the radio data that we used. Moreover, this approach relied on a magnetic field model, which limits our prediction. Accounting for different parameterization of the magnetic field, we have bracketed the associated systematic uncertainty. Overall, our estimate is expected to be accurate within a factor of about two in this scenario (see Appendix~\ref{app:impact_of_magnetic_field}).

\begin{figure}
	\centering
	\includegraphics[width=0.49\textwidth]{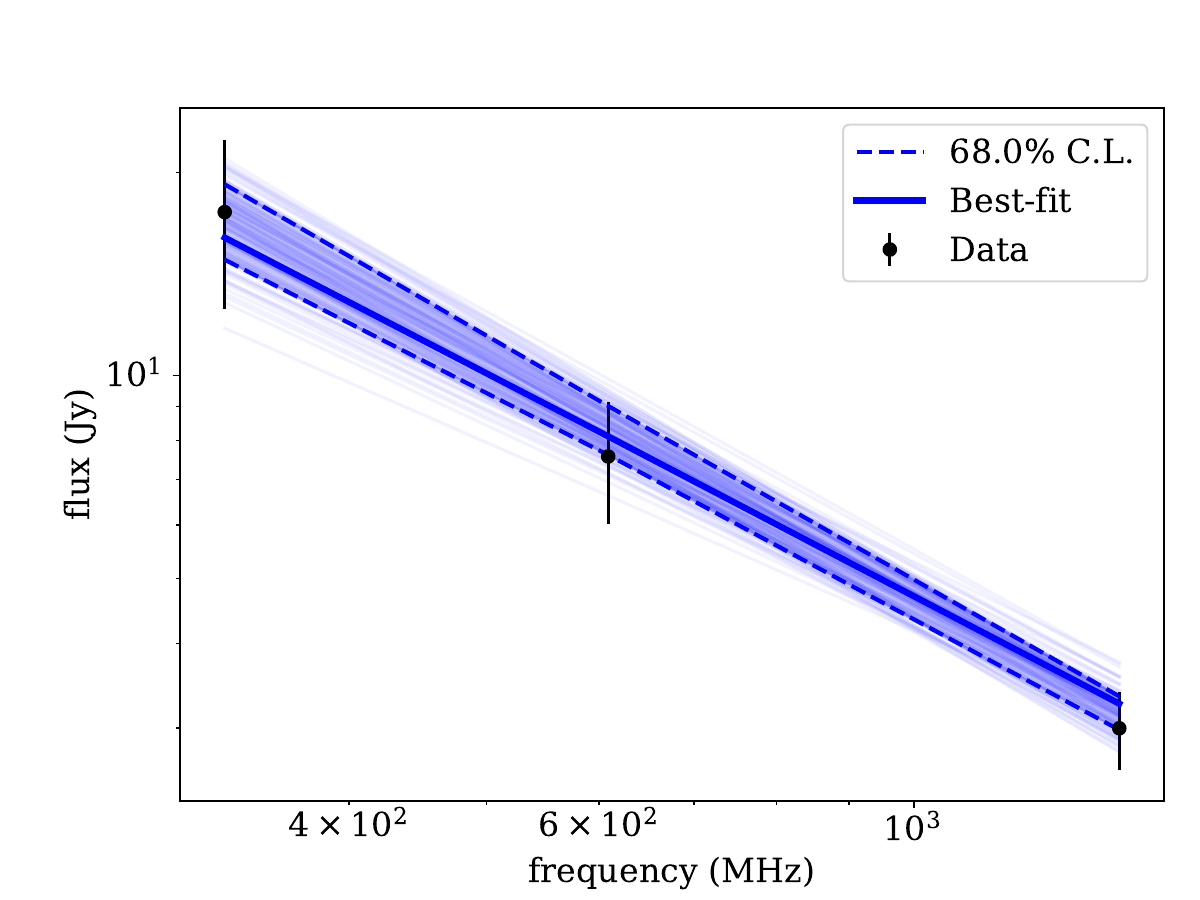}
	\includegraphics[width=0.49\textwidth]{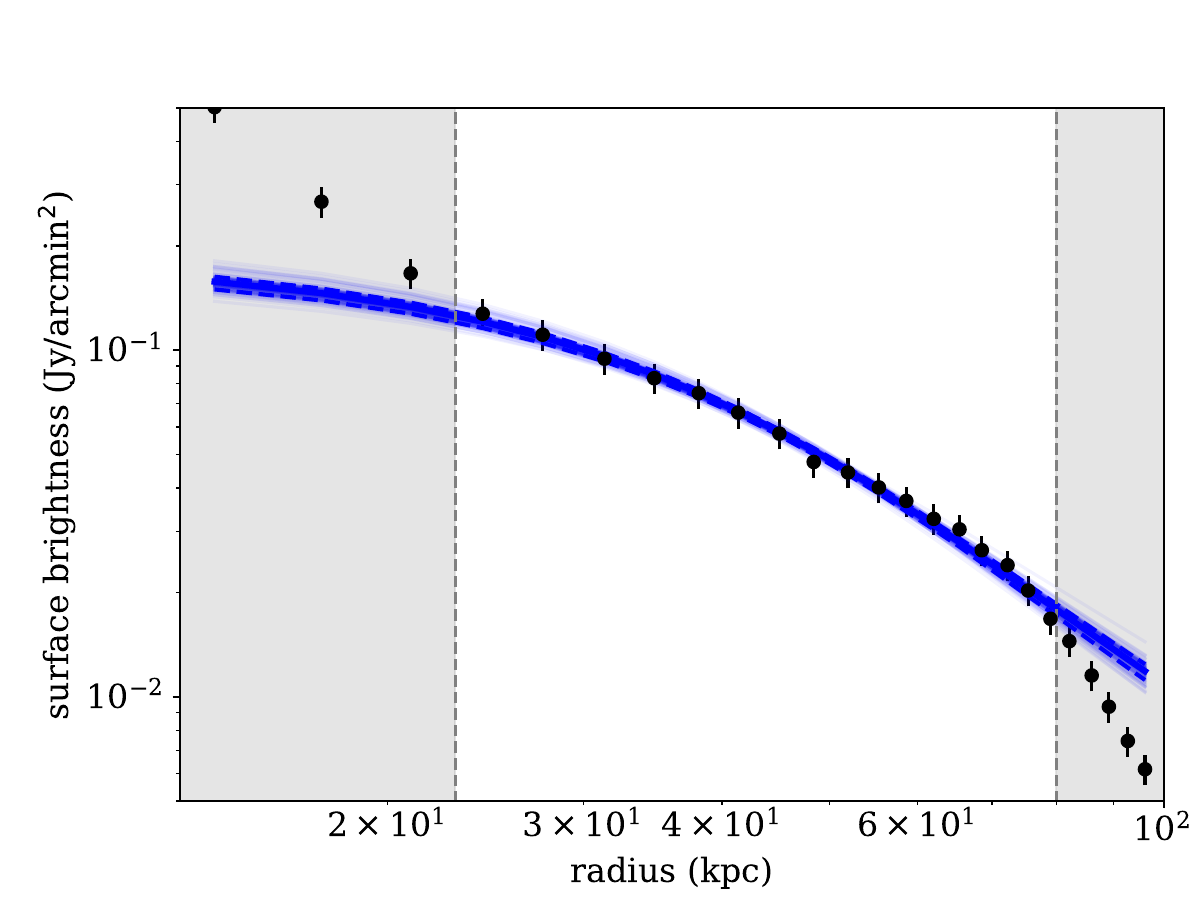}
	\includegraphics[width=0.49\textwidth]{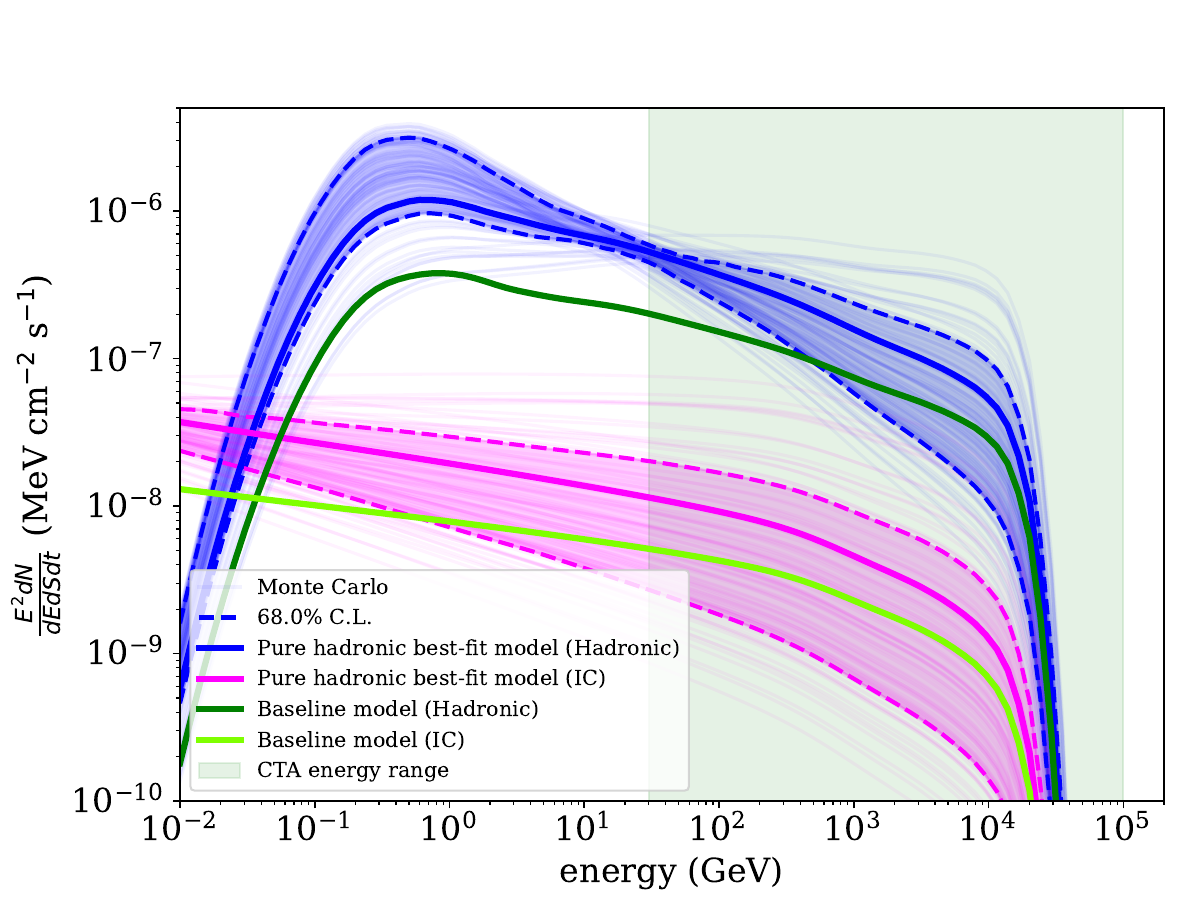}
	\includegraphics[width=0.49\textwidth]{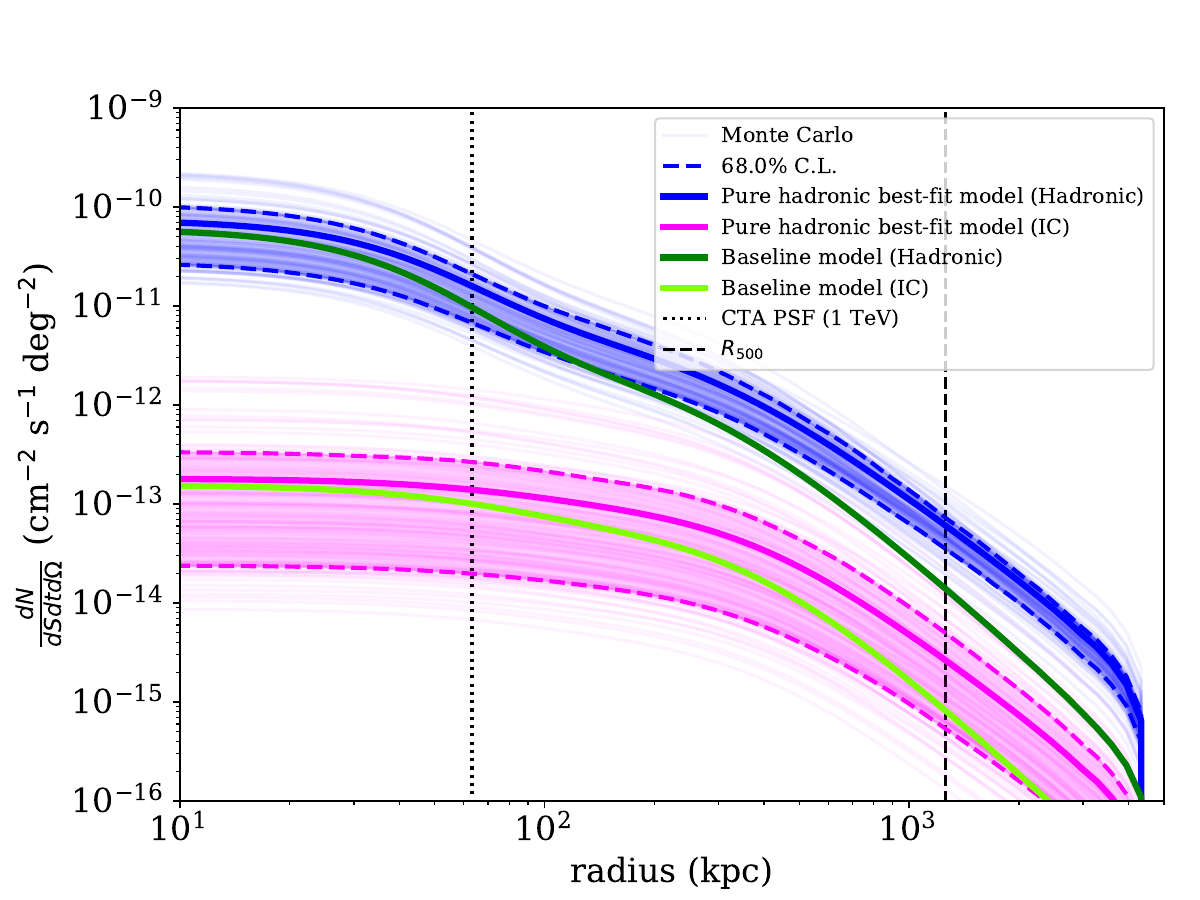}
	\caption{Constraints on the radio and $\gamma$-ray observables in the case of the pure hadronic model, assuming the magnetic field model based on \citep{Taylor2006} and $\eta_B=2/3$. The best-fit models are shown as solid blue lines (pink, in the case of inverse-Compton emission). The dashed lines provide the enclosed 68\% confidence region and 100 model realizations are shown as a light lines. We also report the baseline model for the $\gamma$-ray prediction.
	{\bf Top left panel}: radio spectrum within a 15 arcmin aperture diameter. The data were taken from \citep{Gitti2002}.
	{\bf Top right panel}: radio profile at 1380 MHz. The data were taken from \citep{Pedlar1990}. Following \citep{Zandanel2014b}, we use 10\% uncertainties and reject the points below 23 kpc (contamination from NGC~1275) and above 80 kpc (possible issues with large scale flux recovery). The grey areas correspond to the discarded data.
	{\bf Bottom left panel}: $\gamma$-ray spectrum prediction, computed by cylindrical integration within $\theta_{500}$.
	{\bf Bottom right panel}: $\gamma$-ray projected profile prediction, computed in the range 150 GeV - 50 TeV.}
\label{fig:gamma_ray_observable_hadronic}
\end{figure}

\subsubsection{Pure leptonic contribution}
\label{sec:CR_leptonic_model}
Although it is not physically motivated, as it would require a continuous in situ injection of CRe without protons, we consider the pure leptonic scenario as an exercise to address the inverse-Compton contribution. We use the radio data including CRe only and compute the $\gamma$-ray emission due to inverse-Compton. Since high-energy CRe suffer from major energy losses \citep{Sarazin1999}, the only way to observe inverse-Compton emission at CTA energies would be a scenario in which fresh CRe are injected in the ICM, as for the hadronic model. We consider that all the radio emission arises from a continuous particle injection with a power-law energy spectrum, and we apply energy losses given the magnetic field strength, the CMB at the cluster's redshift and the thermal gas. Given these assumptions, the $\gamma$-ray emission at CTA energies is similar to the inverse-Compton generated in the pure hadronic model, and thus much smaller than typical expectations from hadronic interactions in numerical simulations or in the pure hadronic model. In practice, continuous injection is not guaranteed and the amount of inverse-Compton emission may thus drop dramatically in the CTA energy range. This is why our computation only provides an upper limit to the inverse-Compton signal.

\subsubsection{Turbulent reacceleration model}
The most favored scenario for the acceleration of relativistic electrons generating large-scale emission in the giant radio halos in galaxy clusters is based on turbulent reacceleration \citep{Brunetti2001,Petrosian2001,Brunetti2007,Brunetti2011,Miniati2015,Pinzke2017,Brunetti2020,Nishiwaki2021}. Turbulent reacceleration has been proposed also for mini-halos as an alternative to pure hadronic models \citep{Gitti2002,ZuHone2013}. Turbulence may reaccelerate both primary (CRp and CRe) and secondary particles if they are present in the ICM. In these models, for a given radio luminosity, the level of $\gamma$-rays depends on the ratio between primary and secondary electrons that are reaccelerated. Even in the case where only primary CRp and their secondary products are considered, the $\gamma$-ray emission that is generated in these models is smaller than the one in pure hadronic models \citep{Brunetti2017,Adam2021}. As a consequence, if turbulence plays an important role in the acceleration process in the mini-halo volume, the $\gamma$-ray spectrum calculated in Section \ref{sec:CR_baseline_model} provides an optimistic view of the expected signal. Modeling the expected $\gamma$-ray signal in the CTA energy band from the Perseus cluster in the case that the radio mini-halo originates from turbulent reacceleration is beyond the aim of this paper. Yet as a reference we just mention that modellings in the case of the Coma cluster, which hosts a giant radio halo, predict a $\gamma$-ray signal 4-6 times smaller than that in the pure hadronic scenario \citep{Brunetti2011,Adam2021}. Note that rescaling the CRp normalization of our pure hadronic model ($X_{500} = 4.5 \times 10^{-2}$) by the same amount would give a value that is in line with that of our baseline model ($X_{500} = 10^{-2}$), so that our baseline model is in qualitative agreement with what one can expect from turbulence reacceleration.

In the next sections, we will use the models discussed in Section~\ref{sec:Calibration_of_the_model_parameters} to address the CTA sensitivity to the CR physics of the Perseus cluster in the context of the proposed KSP observations.

\section{Modelling the $\gamma$-ray emission associated with dark matter}
\label{sec:DM_modelling}

In this section, we  build a DM profile for the Perseus cluster following the most up-to-date results both from observations and numerical cosmological simulations. 

Assuming DM is completely composed of WIMPs \citep{Bertone:2004pz, Hooper:2009zm}, we can compute the expected DM-induced $\gamma$-ray emission from any astrophysical source as
\begin{equation}\label{eq:flux-general}
 \frac{d\Phi_{\gamma}}{dE}(\Delta\Omega, l.o.s, E)=\frac{d\phi_{\gamma}}{dE}(E)\times\;J(\Delta\Omega, l.o.s).
\end{equation}
For a given energy $E$, line of sight $l.o.s$, and solid angle $\Delta\Omega$ subtended by the region of interest, $\frac{d\Phi_{\gamma}}{dE}$ is the DM-induced $\gamma$-ray flux, $\frac{d\phi_{\gamma}}{dE}$ contains the spectral information about the expected emission, and $J$, referred as the ``astrophysical factor'', encloses the details of the spatial morphology of the putative signal. In order to perform this DM flux computation, we will assume that WIMPs annihilate into Standard Model (SM) particles in the halo of Perseus (in the following, the annihilation scenario) or that they decay into SM particles (in the following, the decay scenario). According to these two scenarios, the expression in Equation~\ref{eq:flux-general} becomes these two, respectively:
\begin{equation}\label{eqn:flux-scenarios}
\begin{split}
\left. \frac{d\Phi^{ann}_{\gamma}}{dE}(\Delta\Omega, l.o.s, E) = \frac{<\sigma v>}{8\pi m_{\chi}^2}\frac{dN_{\gamma}}{dE_0}\right|_{E_0 = (1+z)E}\times J_{ann}(\Delta\Omega, l.o.s); \\
\left. \frac{d\Phi^{dec}_{\gamma}}{dE}(\Delta\Omega, l.o.s, E) = \frac{1}{4\pi m_{\chi}\tau_{\chi}}\frac{dN_{\gamma}}{dE_0}\right|_{E_0 = (1+z)E}\times J_{dec}(\Delta\Omega, l.o.s),
\end{split}
\end{equation}
where $\frac{dN_{\gamma}}{dE}$ is the number of photons per unit energy we expect from a given annihilation channel and DM particle mass $m_{\chi}$, $<\sigma v>$ is the velocity-averaged  annihilation cross-section in the annihilation scenario and $\tau_{\chi}$ is the mean lifetime of the DM particle in the decay scenario. In deriving these expressions, we have assumed Majorana WIMPs (for Dirac WIMPs it would be necessary to multiply by a factor 1/2 \citep{Bertone:2004pz}. To obtain the corresponding fluxes, we use for $\frac{dN_{\gamma}}{dE}$ the results from \citep{Cirelli2012}, including electro-weak corrections. For the thermally-averaged cross-section, we expect a value around $3\times 10^{-26}$ cm$^{3}$s$^{-1}$ \citep{Steigman:2012nb, Bringmann:2020mgx} to produce the observed DM relic abundance \citep{Planck:2018vyg}, while the main bound for the decay lifetime to account for the observed DM density is the current age of the Universe $\sim10^{17}$s, despite the efforts of trying to obtain tighter constraints from just gravitational probes \citep{Poulin:2016nat}. We recall that for decaying DM only half of the energy budget stored in $m_{\chi}$ is available per SM particle produced in two-body decays. We can also define the spatial morphology of the signal, the so-called J-factor, as:
\begin{equation}\label{eq:j-factors}
\begin{split}
J_{ann}(\Delta\Omega, l.o.s) = \int_{\Delta\Omega}d\Omega\int_{l.o.s}\rho^2_{DM}(r, l)dl;\\
J_{dec}(\Delta\Omega, l.o.s) = \int_{\Delta\Omega}d\Omega\int_{l.o.s}\rho_{DM}(r, l)dl,
\end{split}
\end{equation}
where $\Delta\Omega=2\pi(1-\cos\alpha_{int})$, being $\alpha_{int}$ the integration angle and $\rho_{DM}(r)$ the DM density profile.

From Equation~\ref{eq:j-factors} the main dependencies in both scenarios can be deduced \citep{Sanchez-Conde:2013yxa}:
\begin{equation}\label{eq:J-factors_dependencies}
J_{ann}\propto \frac{M_{200}c^3_{200}}{d_{L}^2};\;\;\;\;
J_{dec}\propto \frac{M_{200}}{d_{L}^2},
\end{equation}
which are the mass $M_{200}$\footnote{Mass enclosed in a radius where an spherical overdensity is 200 times the critical density of the Universe $\rho_{crit}$.}, the luminosity distance to the Earth $d_{L}$ and the concentration $c_{200}$\footnote{Dimensionless quantity that describes how much concentrated halos are and it is dependent on the mass of the object.} for the annihilation case. From these dependencies, one can clearly see why local galaxy clusters are the best targets to consider to test decaying DM \citep{2011ApJ...726L...6C}, since they are the most massive objects in the Universe. For the case of annihilation it may not seem so straightforward. Yet, in the last years some studies where performed \citep{2011JCAP...12..011S} comparing the suitability of galaxy clusters and dwarf spheroidal galaxies (dSphs) for $\gamma$-ray DM searches, concluding that galaxy clusters can also provide competitive results. One key element for this outcome is to consider the distribution of smaller halos that galaxy clusters should host according to the $\Lambda$CDM structure formation theory \citep{Madau:2008fr}, usually called subhalos. The role of these substructures come into play through the $c_{200}$ dependency, since we expect subhalos of a given mass to be much more concentrated than main halos of the same mass. In contrast, in the decay scenario (see Equation~\ref{eq:J-factors_dependencies}) the subhalos do not provide a sizable contribution since their masses are really low compared to the distance. Thus, we will not consider them for this scenario.

The morphology of the expected DM $\gamma$-ray emission is strictly determined by the DM density profile $\rho_{DM}$ and in case of annihilation, particularly by the DM substructure. In the next sections, we present the models we build for the DM distribution in the main halo of Perseus as well as a model for the population of subhalos.

\subsection{Perseus main halo}
\label{sec:DM_main_halo}
The main DM halo of Perseus can be described through a Navarro-Frenk-White (NFW) density profile \citep{Navarro:1995iw, Navarro:1996gj}:
\begin{equation}\label{eqn:NFW}
 \rho_\mathrm{NFW}(r)=\frac{\rho_0}{\left(\frac{r}{r_s}\right)\left(1+\frac{r}{r_s}\right)^{2}},
\end{equation}
where $r_s$ is the scale radius and $\rho_0$ the normalization of the DM density. Introduced as the result of DM-only N-body simulations, the NFW model falls in the family of "cuspy"-like profiles.

To obtain the parameters of the NFW profile, we need to consider a concentration-mass ($c-M$) relation. Consistently, this relation should comprehend the mass scales involved in cluster physics. Taking this into account, we use the parametrization developed in \citep{Sanchez-Conde:2013yxa} for main halos. This concentration-mass relation was found to have an associated $1\sigma$ scatter of 0.14 dex. We build the DM density profile starting from the measured mass of Perseus. We consider the halo to fulfill the spherical collapse model for overdensities $\Delta=200$ times the critical density of the Universe $\rho_{\rm crit}$. This allows us to calculate $M_{200}$ shown in Table~\ref{tab:parameters_nfw} from the measured X-ray mass, $M_{500}$, provided in \citep{Urban2014}. 

We can also obtain the radius $R_{200}$ that contains an enclosed mass $M_{200}$, as
\begin{equation}\label{eqn:r200}
R_{200} = \left( \dfrac{M_{200}}{\frac{4}{3}\pi \Delta \rho_{\rm crit}} \right) ^{1/3}\, ,
\end{equation}
whose corresponding projected angle is
$\theta_{200} = \arctan\left(\frac{R_{200}}{d_{L}}\right)$. Now we can easily compute the scale radius in Equation~\ref{eqn:NFW} as
\begin{equation}
r_{s} \equiv \dfrac{R_{200}}{c_{200}}.
\end{equation}
Finally, the normalization of the density profile $\rho_0$ is obtained by imposing the recovery of $M_{200}$ after the integration over the cluster volume of the profile ({\footnotesize $M_{200}=\int_0^{R_{200}}\int_0^{4\pi}\rho_{\rm NFW}(r)r^2drd\Omega $}) and isolating $\rho_0$:
\begin{equation}\label{eqn:rho_0}
\rho_0 = \frac{2~\Delta_{200}~\rho_{crit}~c_{200}}{3~f(c_{200})}\,,
\end{equation}
where $f(c_{200})=\frac{2}{c_{200}^2}\left(\ln{(1+c_{200})}-\frac{c_{200}}{1+c_{200}}\right)$. Table \ref{tab:parameters_nfw} lists the corresponding parameters of the obtained NFW DM density profile following the above description. 

\begin{table}[h!]
\centering
\caption{NFW density profile parameters for the Perseus galaxy cluster; see Equations~(\ref{eqn:NFW}--\ref{eqn:rho_0}) for the definition of each parameter and accompanying text for details on their derivation. 
}
\begin{tabular}{|c|c|c|c|c|c|}
\hline
$M_{200}$ & $c_{200}$ & $r_s$ & $\log_{10}\rho_0$ & $R_{200}$ & $\theta_{200}$ \\
$[10^{14}$ M$_{\odot}]$ &  & [kpc] & [M$_{\odot}/$kpc$^3$] & [kpc] & [deg] \\ 
\hline
\hline
7.5 & 5.0 & 370.8 & 6.1 & 1865.0 & 1.4 \\ \hline
\end{tabular}
\label{tab:parameters_nfw}
\end{table}

It is interesting to compare the value we obtained for $c_{200}$ and reported in Table \ref{tab:parameters_nfw} with that obtained from observational data. Indeed, from X-ray observations an observational value of $c_{200}=5.0\pm 0.5$ was inferred \citep{Simionescu:2011ii}, which well matches our adopted $c_{200}$.

\subsection{The role of substructures}
\label{sec:DM_substructure_model}

According to the $\Lambda$CDM hierarchical bottom-up structure formation scenario \citep{Kuhlen:2012ft, Zavala:2019gpq}, galaxy clusters should host a significant amount of substructure or subhalos. 
Due to stripping processes, these subhalos are expected to have higher concentrations than main field halos of the same mass (e.g., \citep{Moline:2016pbm}). Because of these high concentrations, we expect them to have a considerable impact in the annihilation fluxes (Equation~\ref{eq:J-factors_dependencies}). This enhancement effect in $J_{ann}$-factor, usually called \textit{boost factor} $B$, was already estimated in \citep{2011JCAP...12..011S} for a selected list of local clusters. These authors found that halo substructure in galaxy clusters could provide boost factors of the order $B\sim40$, in general agreement with current, more refined boost calculations from N-body simulations \citep{Moline:2016pbm, Ando:2019xlm}. In order to account for the contribution of these substructures to the $J_{ann}$-factor, we factorize the computation in the following way:
\begin{equation}\label{eq:j-factor}
\begin{split}
J_{ann}(\Delta\Omega, l.o.s) & = \int_0^{\Delta\Omega}d\Omega\int_{l.o.s}\left(\rho_{main}(r) + \sum_i^{N_{sub}}\rho_{sub}(r)\right)^2dl = \\
& = J^{ann}_{main} + <J^{ann}_{sub}> + <J^{ann}_{cross-prod}>,
\end{split}
\end{equation}
where $N_{sub}$ is the total number of subhalos, $\rho_{main}$ is the DM profile of Perseus' main halo obtained in Section~\ref{sec:DM_main_halo} and $\rho_{sub}$ the DM profile for each subhalo. Due to the complex dynamics that substructures undergo (e.g., tidal stripping, dynamical friction, interaction with baryons, etc. -- see \citep{Green:2005fa, Bringmann:2009vf, Cornell:2013rza, Zavala:2019gpq}), the precise survival probability of the smallest subhalos is not yet known \citep{vandenBosch:2017ynq, vandenBosch:2018tyt, Ogiya:2019del, Aguirre-Santaella:2022kkm}. In any case, when it comes to the subhalo density profile $\rho_{sub}$, deviations from NFW are expected to be pronounced mainly in the outermost subhalo regions, where mass losses are severe. Therefore we keep adopting the NFW profile also to model the subhalo inner structure. We remark that due to their small extension, we do not expect individual subhalos to be spatially resolved by CTA.  Therefore, we decided to average their contribution in $<J^{ann}_{sub}>$. We then model the population of subhalos as:
\begin{equation}\label{eq:subhalos-distribution}
\frac{d^3N}{dVdMdc}= N_{sub}\frac{dP_V}{dV}(R)\frac{dP_M}{dM}(M)\frac{dP_c}{dc}(M,c),
\end{equation}
where $P_i$ with $i = V, M, c$ represents the probability distribution in the volume of the main halo $V$, of the subhalo masses $M$ and the subhalo concentrations $c$. The probability distributions in Equation~\ref{eq:subhalos-distribution} have been modeled based on results of DM-only cosmological simulations, namely from Via Lactea-II (VL-II, \citep{Diemand:2008in, Pieri:2009je}), Aquarius \citep{Springel:2008by, Springel:2008cc} and the work of \citep{Moline:2016pbm}. We describe each of these distributions in the following:
\begin{itemize}
    \item Subhalo distribution within the main halo $\frac{dP_V}{dV}(r)$: As we assume that Perseus main halo is spherically symmetric, the distribution within the volume collapses to the distribution in radius, so in the following we  refer to it simply as the Subhalo Radial Distribution (SRD). Based on N-body simulations, we choose to follow the \textit{antibiased} relation \citep{Pieri:2009je}\footnote{There are other options available for the SRD modelling, e.g., the one based on results from the Aquarius simulations, where they fit the SRD to an Einasto profile \citep{Springel:2008by, Springel:2008cc}. Yet, this provides a subhalo distribution that is compatible compared to the \textit{VL-II} results, which is the one used in this work.}, defined as:
    \begin{equation}
     \frac{dP_V}{dV}(r)=\rho_{main}(r)\frac{r/r_{a}}{1 + \frac{r}{r_a}},
    \end{equation}
    where $r_a$ works as a scale radius and it is defined by the fraction of the total mass that will be in form of subhalos $f_{sub}$. For the distribution of subhalos, we also consider the number of substructure levels to $N_{lvl}=2$ (subhalos inside subhalos)\footnote{In \citep{Sanchez-Conde:2013yxa, Bonnivard:2015pia}, they concluded that adding more levels of substructure only contribute up to a 5\% to the total flux. See, however, the recent \citep{Delos:2022bhp}.
    }.
    
    \item Subhalo mass distribution $\frac{dP_M}{dM}(M)$: Also known as the subhalo mass function (SHMF), is usually modelled using a power-law (e.g., \citep{Lavalle:2007apj}):
    \begin{equation}\label{eq:subhalo-mass-function}
    \frac{dP_{M}}{dM}\propto M^{-\alpha}.
    \end{equation}
    Values range between $\alpha = 1.9$ \citep{Springel:2008by, Springel:2008cc} and $\alpha = 2.0$ \citep{Diemand:2008in}. This slope is key to evaluate the contribution of substructure to the $J_{ann}$-factor \citep{Moline:2016pbm}. For higher values of $\alpha$, i.e., a more numerous population of small subhalos, we obtain proportionally higher boosts. As a way to account for this uncertainty, we will define different benchmark models covering different physical scenarios for the abundance of subhalos. A value of $\alpha = 1.9$, more in line with latest simulations results \citep{Zavala:2019gpq, Moline:2021rza}, will lead to conservative values for the $J_{ann}$-factor,  while $\alpha = 2.0$ will yield an upper bound to subhalo boost values. The mass budget in both cases is a fraction of the total mass, $f_{sub}$. Assuming a value for the minimun subhalo mass of $M_{min}=10^{-6}$M$_{\odot}$ and that the maximum subhalo mass in terms of the host is $M_{max}^{\%}=0.01$\footnote{We note that these values are generally accepted and standard in the community as today, representatives and consistent for the defined models. We could select a lower value for $M_{min}$, which would translate into higher values of the boost or lower boost values if the value was higher. In any case, the variation of these parameters ($M_{min}$, $M_{max}^{\%}$) has been checked to yield $J_{ann}$-factors already encapsulated by our benchmark models.}, we need different values of $f_{sub}$ to conserve the total mass. Integrating the SHMF with the different values of $\alpha$ for the selected $M_{min}$ and $M_{max}^{\%}$, we obtain a value of $f_{sub}=0.182$ for $\alpha=1.9$, and $f_{sub}=0.319$ for $\alpha=2.0$.
    
    \item Subhalo concentrations $\frac{dP_c}{dc}(c, M)$: In the same way we assumed a concentration-mass relation for the main halo, we select a ($c-M$) relation to describe subhalo DM profiles. We adopt the state-of-the-art ($c-M$) subhalo model by \citep{Moline:2016pbm}, which includes a radial dependence of the concentration within the main halo:
    \begin{equation}\label{eq:c-M_subhalos}
    c_{200}(M_{200}, x_{sub}) = c_0 \left[ 
    1 + \sum_{i=0}^{3} \left[ 
    a_{i} \log_{10} 
    \left( \dfrac{M_{200}}{10^8 h^{-1}\, \mathrm{M_{\odot}}} \right)
    \right]^i 
    \right] \times [1 + b\log_{10}(x_{sub})],
    \end{equation}
    where $c_0 = 19.9, \; a_i = [-0.195, 0.089, 0.089], \; b=-0.54$ and $x_{sub}$ refers to subhalo distance with respect to the center of the host halo. The importance of including a ($c-M$) relation specifically derived for subhalos resides, as already stated, in the fact that subhalo concentrations are known to be higher than that of field halos of the same mass \citep{Moline:2016pbm}. Thus, their contribution to the $J_{ann}$-factor is expected to be critical. Note that, in deriving this relation, \citep{Moline:2016pbm} assume NFW profiles for subhalos, keeping our modelling consistent. Every concentration-mass relation is known to exhibit an intrinsic scatter \citep{Bullock:2001jz, Moline:2016pbm}, yet it is highly computationally expensive to take it into account for each subhalo in the field halo. Then, for the sake of this study we decide to neglect it, as its impact on the $J_{ann}$-factor will lie within the spread we have by considering different values for $\alpha$ in the SHMF.
\end{itemize}

After the detailed discussion on the parametrization followed for the subhalo population modelling, we establish three benchmark models for the computation of the J-factors. Each of them represents an expected different level of contribution of the subhalo population to the annihilation flux:
\begin{itemize}
    \item MIN: considers the Perseus main halo and neglects the existence of substructures.
    \item MED: the SRD follows the \textit{antibiased} relation. We adopt $\alpha = 1.9$ for the slope of the SHMF (Equation~ \ref{eq:subhalo-mass-function}), using a coherent subhalo mass fraction of $f_{sub}=0.182$.
    \item MAX: similar to the MED model but we choose $\alpha = 2.0$ for the slope of the SHMF (Equation~\ref{eq:subhalo-mass-function}), with a coherent adjustment of the subhalo mass fraction to $f_{sub}=0.319$.
\end{itemize}

These models and their values are summarized in Table~\ref{tab:benchmark-models}. With the definition of the above benchmark models, we aim to bracket a wide range of possible substructure scenarios. This will translate into a bracketing for the possible values of the $J_{ann}$-factor, being the MED model our best guess and for which we will produce the main results, and the MIN and MAX models as realistic lower and upper limits to the contribution levels of substructures. For decaying DM, we will use a realistic value for $J_{dec}$ assuming the MIN model, since subhalos do not provide a sizeable contribution to it (see Equation~\ref{eq:J-factors_dependencies}).

\begin{table}[h!]
\centering
\caption{Summary of the defined benchmark DM models for the annihilation interaction scenario. From left to right, $\rho_{sub}$ is the main DM profile for both the main halo and the subhalos, $(c-M)_{main}$ is the concentration-mass relation used for Perseus main halo, SRD is the Subhalo Radial Distribution, $(c-M)_{sub}$ is the concentration-mass relation for the subhalos, $\alpha$ is the slope of the SHMF (Equation~\ref{eq:subhalo-mass-function}) and $f_{sub}$ is the fraction of the total mass bound in substructures, see text for more details. Regarding the models, ``SC+14'' refers to the concentration model of \citep{Sanchez-Conde:2013yxa}, $antibiased$ to the SRD from \citep{Pieri:2009je} from Equation~\ref{eq:subhalos-distribution}, ``M+17'' to the ($c-M$) subhalo model of \citep{Moline:2016pbm} from Equation~\ref{eq:c-M_subhalos}.}
\begin{tabular}{ | c | c | c | c | c | c | c |}
\hline
Model & $\rho_{DM}$ & $(c-M)_{main}$ & SRD & $(c-M)_{sub}$ & $\alpha$ & $f_{sub}$\\
\hline
\hline
MIN & NFW & SC+14 & - & - & - & 0 \\
MED & NFW & SC+14 & \textit{Antibiased} & M+17 & 1.9 & 0.182\\
MAX & NFW & SC+14 & \textit{Antibiased} & M+17 & 2.0 & 0.319\\ 
\hline
\end{tabular}
\label{tab:benchmark-models}
\end{table}

\subsection{Dark matter annihilation and decay fluxes}
\label{sec:DM_fluxes}

\begin{figure}[h!]
	\centering
	\includegraphics[width=0.49\textwidth]{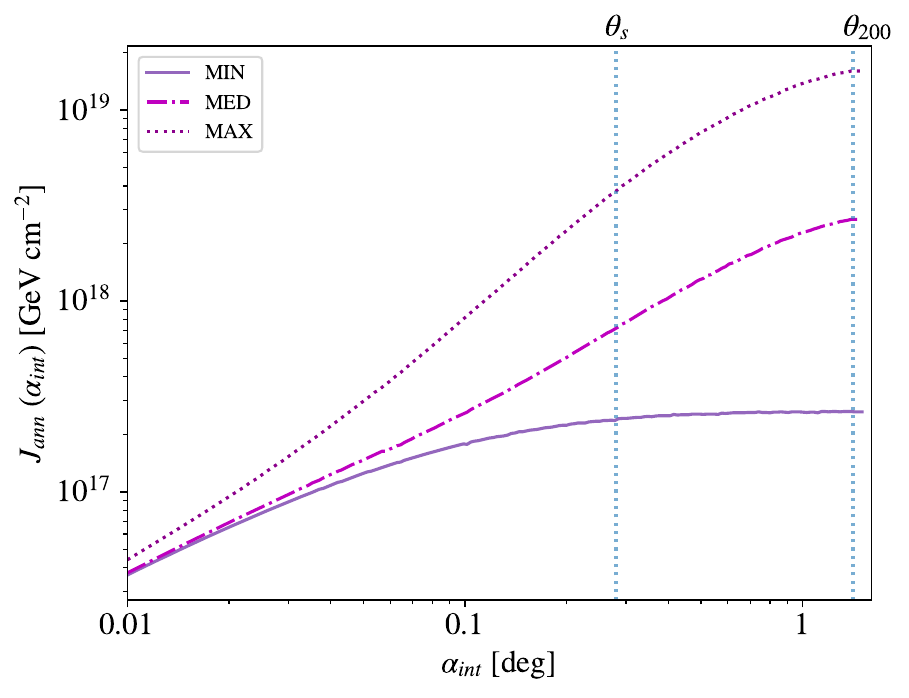}
	\includegraphics[width=0.49\textwidth]{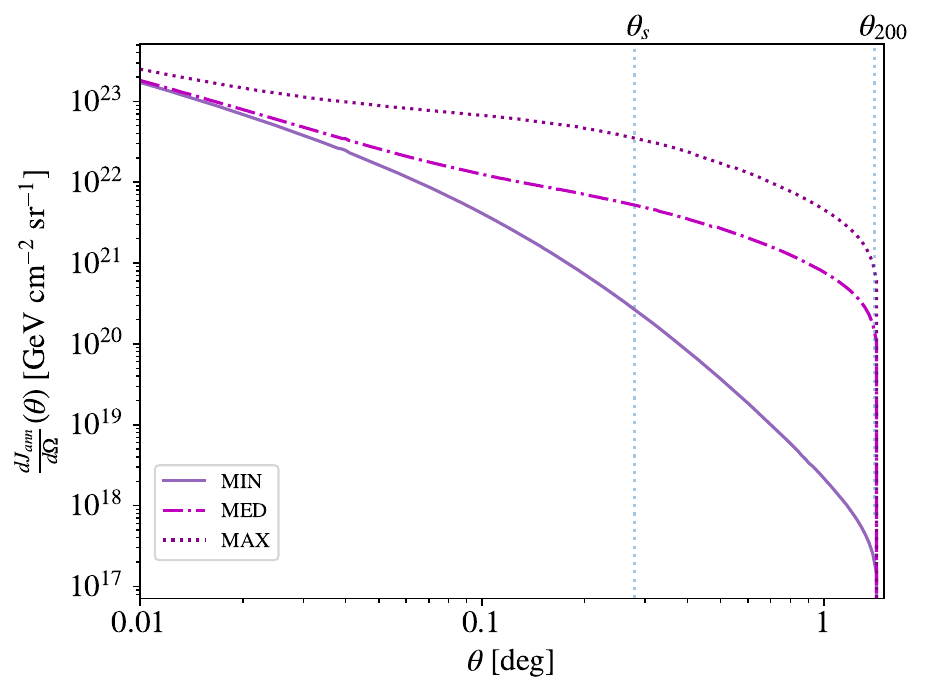}
	\includegraphics[width=0.49\textwidth]{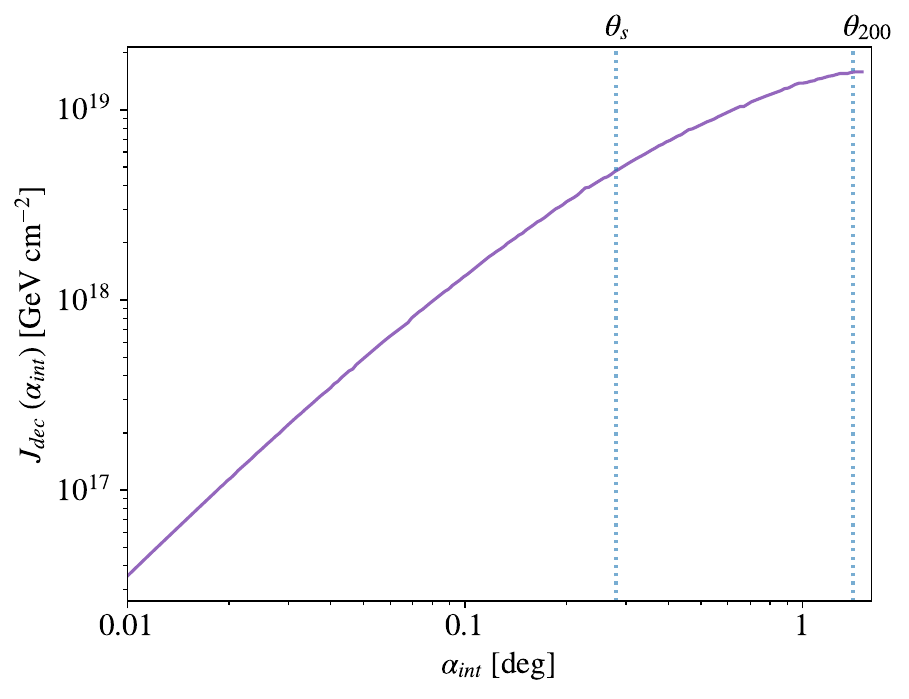}
	\includegraphics[width=0.49\textwidth]{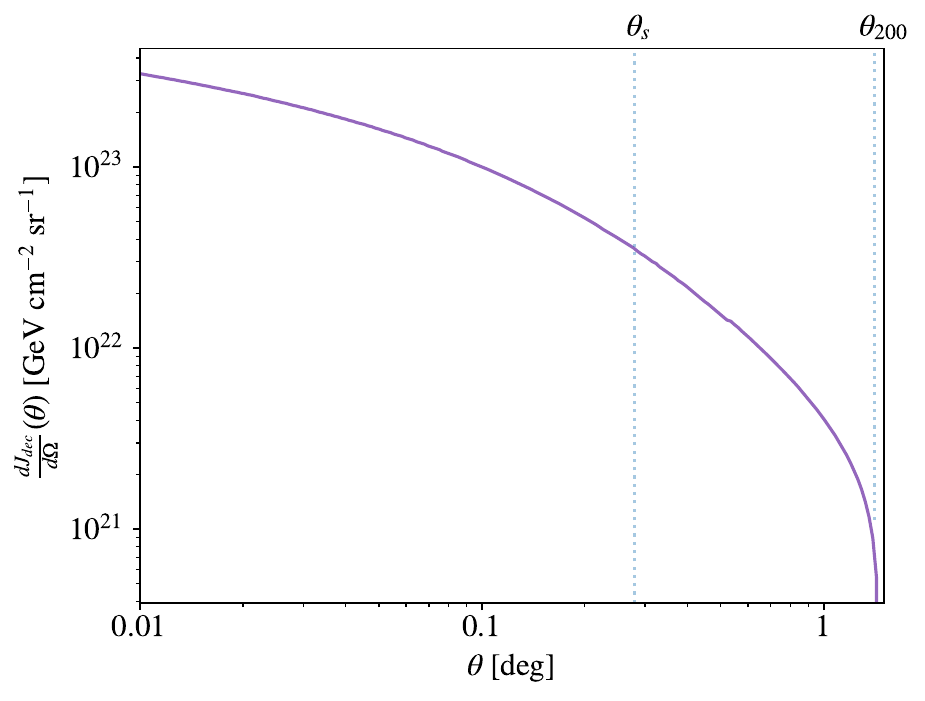}
	\caption{Integrated J-factors (Equation~\ref{eq:j-factors}) versus the integration angle $\alpha_{int}$ (\textbf{left panels}) and differential J-factors versus the radial angle $\theta$ (\textbf{right panels}) for the annihilation scenario (\textbf{top panels}) and decay scenario (\textbf{bottom panels}) for the three substructure benchmark models correspondingly. In all the panels $\theta_{200}$ and the projected angle of the scale radius, $\theta_s$, are shown as a measure of the containment of the emission and the extension of the cluster.}
\label{fig:jtot_perseus}
\end{figure}

Once defined the DM models we use for Perseus cluster and for its expected subhalo population, we are interested in computing their $\gamma$-ray induced fluxes, the J-factors. We use the publicly available \texttt{CLUMPY} code \citep{Charbonnier:2012gf, Bonnivard:2015pia, Hutten:2018aix} to compute them. The obtained integrated J-factors for the cluster and for each benchmark model are summarized in Table \ref{tab:DM-fluxes} and shown in the left panels of Figure~\ref{fig:jtot_perseus}. The right panels in this same figure show the differential J-factors as a function of the angle from the center of Perseus.

\begin{table}[h!]
\centering
\caption{\label{tab:DM-fluxes} Perseus integrated J-factors in the case of the three benchmark annihilation scenarios (MIN, MED, MAX -- see Table \ref{tab:benchmark-models}) and the $J_{dec}$-factor for the decay scenario. In all cases, we integrate the DM signal up to $R_{200}$ ($\alpha_{int}=\theta_{200}$). See text for details.}
\begin{tabular}{| c | c | c |}
\hline
 & Annihilation & Decay \\
\hline
 & $\log_{10} J_{ann}$ (GeV$^2$cm$^{-5}$) & $\log_{10} J_{dec}$ (GeV cm$^{-2}$)\\
\hline
\hline
MIN & 17.42 & 19.20 \\
\hline
MED & 18.43 & - \\
\hline
MAX & 19.20 & - \\
\hline
\end{tabular}
\end{table}
 
From the $J_{ann}$-factor values given in Table \ref{tab:DM-fluxes} for each benchmark model and from Figure~\ref{fig:jtot_perseus} upper panels, we can better understand the impact of including subhalos in our calculations. Indeed, substructures do not only boost the expected annihilation signal but also modify its spatial morphology, subhalos being particularly important in the outskirts of the cluster. The latter is the main reason to also take into account the expected extension of the cluster DM emission in the analysis itself, as it will be described later on. Although the precise properties of the subhalo population within the cluster is the main origin of uncertainty in the computation of the $J_{ann}$-factors, we also wanted to check the impact of the uncertainty in the $M_{200}$ value and the one from the scatter in the concentration-mass relation. The mass uncertainty from the study of \citep{Urban2014} is too small to have an impact itself, thus we compute two different values of $M_{200}$ assuming extreme values for the hydrostatic mass bias ($b_{\rm HSE}= 0, 0.3$) and propagate this to the $J_{ann}$-factor computation. This results into $\sigma_{J_{ann}}= 0.002-0.005$ dex correspondingly, which is two orders of magnitude smaller than the variation introduced from the substructure benchmark models ($\sigma_{J_{ann}}\sim1$ dex). If we also include the 0.14 dex from the scatter of the concentration-mass relation for each of the above extreme mass values, we obtain $\sigma_{J_{ann}}= 0.2$ dex, a $J_{ann}$-factor uncertainty similar to what is typically obtained for dSphs \citep{Fermi-LAT:2015att}. Note though that this $\sigma_{J_{ann}}$ is still well within the uncertainty related to the different considered models for the substructure population. Therefore, from now on we only consider as a theoretical uncertainty the spread among the three substructure benchmark models. As for decay, we compute the uncertainty of $J_{dec}$ from only the use of different mass values and from the scatter in the concentration-mass relation as well, since the substructures do not play a role for this case. We found $\sigma_{J_{dec}}\lessapprox10^{-3}$ dex for all cases, thus we decided to disregard this in the following.

To quantify the enhancement in the total annihilation flux due to the presence of subhalos, we define a subhalo \textit{boost factor} as:
\begin{equation}\label{eq:boost}
B = J_{ann}^X/J_{ann}^{\rm MIN} - 1, 
\end{equation}
where $X$ can be either MED or MAX ($B=0$ means no substructure contribution in our definition\footnote{We recall that other works may use a different definition for the boost, where $B=1$ means no substructure contribution.}). For Perseus and our MED and MAX models, we obtain boost factors of $B=9.2$ and $B=59.3$, respectively. 

We can compare our obtained J-factors for the annihilation scenario with the results in \citep{2011JCAP...12..011S}\footnote{We caution that there exists a factor $\frac{1}{4\pi}$ difference with their J-factor definition.}. For the MIN scenario (no substructure inclusion), these authors obtain $\log_{10}(J_{ann}) = 17.35$ (units of GeV$^2$cm$^{-5}$), very similar to our results (Table~\ref{tab:DM-fluxes}). For the case where they include substructure (which, according to their modelling, would be roughly similar to our MED case), the $J_{ann}$-factors increase up to $\log_{10}(J_{ann}) = 18.23$, also compatible with our MED results (Table~\ref{tab:DM-fluxes}). The \textit{boost factor} that they obtain between their two models is $B=6.6$ (following our definition in Equation~\ref{eq:boost}), a value just around one point lower than our results for the MED model (Table~\ref{tab:DM-fluxes}). This slight mismatch is expected due to some differences in the subhalo modelling in each case. In order to perform a comparison with more recent results, we can examine the boost values we should expect from using the ($c-M$) relation of \citep{Moline:2016pbm}. In this case, for a Perseus-like galaxy cluster halo, they obtain $B\sim9$ for $\alpha=1.9$ and $B\sim72$ for $\alpha=2.0$. The agreement with our MED scenario is absolute, while for MAX we get a relative discrepancy of $\sim 20\%$\footnote{In their calculation the mass in form of subhalos in the cluster is not substracted from the smooth DM distribution, hence this fact leading to an overestimation of the boost, which becomes more important as more mass is modeled as substructures.}. 

A similar comparison can be performed for the decay scenario as well. Given their large masses and relatively close distances, galaxy clusters are the perfect laboratories for probing DM decay and, accordingly, there have been several previous studies in this matter (see refs. in Section~\ref{sec:intro}). One of the latest works on DM decay in the TeV range focused on Perseus and consisted of an observational campaign performed by the MAGIC collaboration \citep{MAGIC:2018tuz}. In this work, authors used the main halo model introduced in \citep{2011JCAP...12..011S} to compute the expected decay flux, obtaining a value of $\log_{10}(J_{dec}) = 19.18$ (units of GeV cm$^{-2}$), compatible within 5$\%$ with our value from Table~\ref{tab:DM-fluxes}.
 
\begin{figure}[h!]
	\centering
	\includegraphics[width=0.8\textwidth]{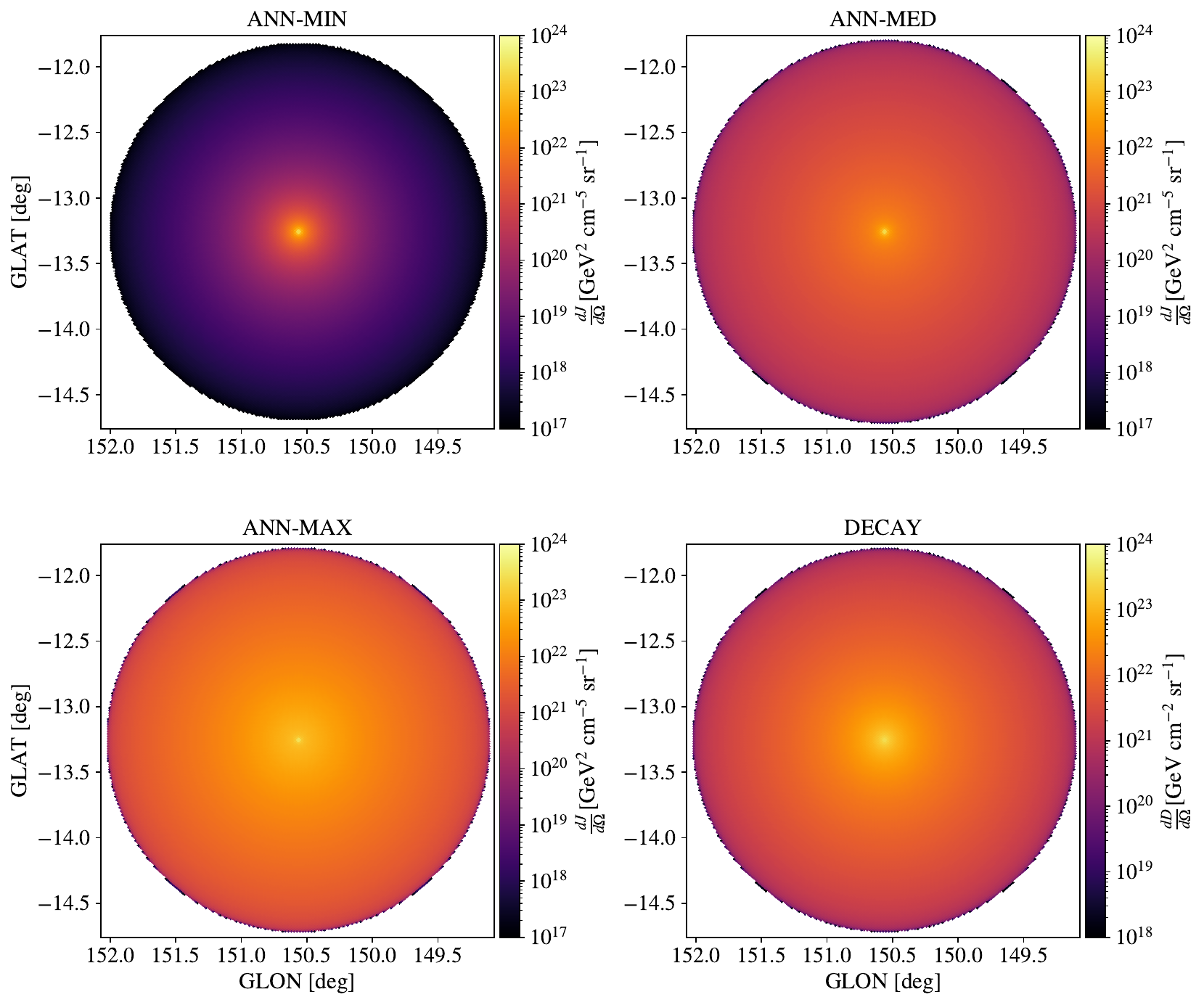}
	\caption{Two-dimensional spatial templates of the expected DM emission in Perseus, quantified in terms of the differential J-factors on the z axis, for the three considered MIN, MED, MAX annihilation scenarios (\textbf{top panels and left bottom panel}) and for decay (\textbf{bottom right panel}). See Tables~\ref{tab:benchmark-models} and \ref{tab:DM-fluxes} for details on each DM model.
	}
\label{fig:DM_J_templates}
\end{figure}

As a last step, we create 2D spatial templates of the expected DM annihilation/decay emissions using \texttt{CLUMPY}. Four maps are created, three for annihilation (each one corresponding to one of our benchmark models in Table \ref{tab:benchmark-models}), and one more for  decay. These maps are all shown in Figure~\ref{fig:DM_J_templates}. As seen already in Figure~\ref{fig:jtot_perseus}, the presence of substructures mainly impacts the morphology of the DM signal in the outer regions of the cluster. 

\section{Observation setup}\label{sec:Observation_setup}
Before considering the CTA data analysis, we discuss the configuration of the observations. In addition to the instrumental background due to CR induced air showers, we present the modelling of the known sources located in the Perseus cluster region which may affect the analysis. We also investigate how the observing strategy will impact the CTA sensitivity to a putative cluster signal.

\subsection{CTA pointing configuration}
\label{sec:CTA_configiuration}
\begin{table}
	\caption{CTA observation configuration summary.}
	\begin{center}
	\begin{tabular}{|c|c|}
	\hline
	Observing time & 300 hours\\
	IRF & prod3b-v2 $^{\dagger}$, North\_z20\_50h \\
	Pointing offset from cluster center & 1.0 deg \\
	Azimuthal position offset & [0,45,90,135,180,225,270,315] deg \\
	\hline
	\end{tabular}
	\end{center}
\footnotesize{Notes. $^{\dagger}$ This IRF corresponds to the full configuration with 15 MST and 4 LST. More recent IRFs are now also available, corresponding to the CTA initial configuration including 9 MST and 4 LST (prod5 \citep{CTA_IRFbrod5}). The corresponding on axis sensitivity is reduced by a factor of about 30\% at high energies ($E \gtrsim 500$ GeV), but the off-axis sensitivity is slightly improved.}
	\label{tab:cta_config}
\end{table}

The CTA observing setup will consist in a set of pointings, with a radial offset with respect to the Perseus cluster center, $\theta_{\rm pointing}$. Such offset is necessary to analyse the data using classical ON-OFF techniques used in Imaging Air Cherenkov Telescopes (IACTs), in whose case the background is estimated using a mirror region equivalent to the region of interest assuming the azimuthal symmetry of the background with respect to the pointing center \citep{Knodlseder2019}. For template-based analysis, in which the full region of interest is modeled using dedicated templates that describe the different sky components and are fit to the data, such offset could be set to zero. Various position angles of the pointing offset can also be considered. This would allow us to cover a larger field around the cluster and thus to increase the chance of serendipitous discovery of new sources. The CTA field of view depends on the energy, given the field of view of the respective telescope class dominating each energy regime: about 4.3 deg diameter for the large-sized telescope (LST) and 7.5 deg diameter for the medium-sized telescope (MST). See \citep{Actis2011} and \citep{CTA2019}.

A total of 300 hours are proposed for observations of the Perseus cluster in the course of the CTA key science program and this is what we assume as our ON-source time for the remaining. Nonetheless, we consider a dead-time fraction of 5\% so that the effective observing time is 285 hours. The observing time will be split into the different pointings. The pointing directions are split between 8 positions, equally spaced from 0 to 315 deg around the cluster reference center, with steps of 45 deg. Nevertheless, we stress that our results do not significantly depend on the number of position directions that we consider.

Because the Perseus cluster field is only observable from the CTA north site (La Palma), we consider the instrument response function (IRF, which connects the sky $\gamma$-ray signal to the actual measurement) associated with long observations \citep{CTA_IRFprod3b} from the north site\footnote{The IRFs are available at \url{https://www.cta-observatory.org/science/ctao-performance/} for ``Alpha configuration'' or prod5 and \url{https://zenodo.org/record/5163273\#.Y5nebbKZNQ0} for prod3b.}. We neglect any variation of the zenith angle of the target source during the observations and assume that the zenith angle is fixed to 20 deg for all pointings. For comparison, the zenith angle ranges from 12 to 36 deg for the MAGIC observations presented in \citep{Aleksic2012}, also performed from La Palma and within which range there is no significant difference. The IRFs are also used to model the instrumental background. Note that IRFs describing the initial CTA array configuration are now available: 9 MST and 4 LST instead of the 15 MST and 4 LST used for the present work. They would imply a reduced sensitivity by about 30\% at high energies, compared to the IRFs used here, although the off-axis sensitivity improved and could make up for some of the deficit in telescope number. 

\begin{figure}
	\centering
	\includegraphics[width=1\textwidth]{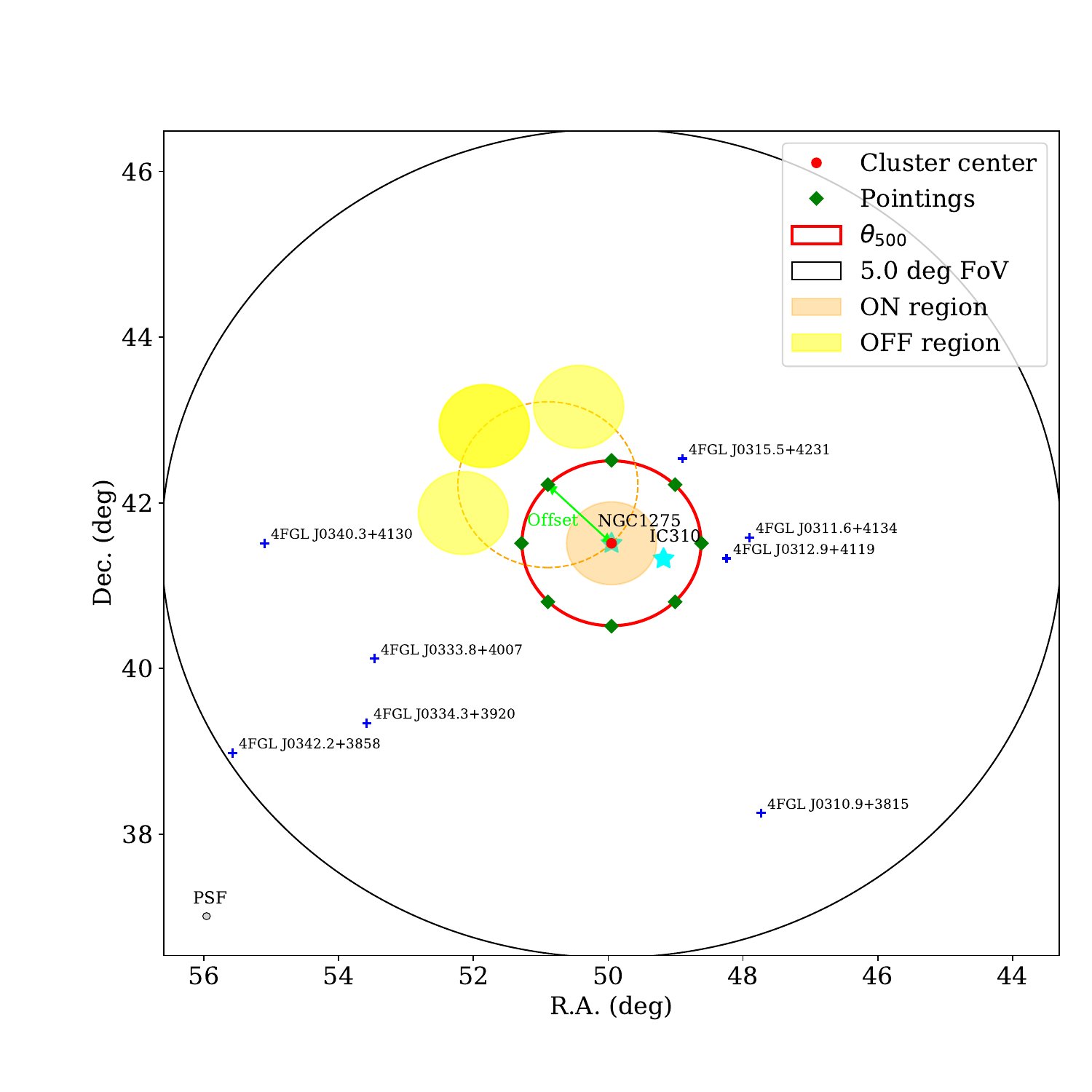}
	\caption{Schematic view of the Perseus cluster region and the assumed observation setup. An illustration of the ON-source and OFF-source regions is shown for one pointing, assuming a 0.5 deg aperture radius. IC~310 and NGC~1275 are observed at very high energy and are shown as cyan stars. 4FGL sources within a 5 deg radius around the cluster center are also depicted as blue crosses. For comparison, the point spread function (PSF) of CTA at 1 TeV is shown in the bottom left corner.
 }
\label{fig:fov_scheme}
\end{figure}

We provide a summary of the CTA observation setup in Table~\ref{tab:cta_config}. A schematic view of the CTA observations configuration is provided in Figure~\ref{fig:fov_scheme}. We include NGC~1275 and IC~310 (see discussions in Section~\ref{sec:background_sky}). The X-ray peak (reference center) is aligned with NGC~1275 and IC~310 is offset toward the southwest by about 0.6 deg but remains within $\theta_{500}$. For a given pointing, the OFF regions are spread over a circle around the pointing for which the radius corresponds to the pointing offset from the source. This is illustrated by the yellow (OFF) and orange (ON) filled regions for which we have used a radius of 0.5 deg. Also shown are the locations of those sources in the \textit{Fermi}-LAT 4FGL catalog \citep{Fermi-LAT:2019yla,Ballet:2020hze} within five degrees from the cluster center.

\subsection{Background sky}
\label{sec:background_sky}
In addition to the diffuse emission from the Perseus cluster, it is necessary to model other sources that are located around the target cluster. At very high energies, two $\gamma$-ray point sources have been detected by the MAGIC telescopes, namely NGC~1275 and IC~310.

NGC~1275, the brightest galaxy in the Perseus cluster, is a variable source presenting flare activities up to fifty times its mean flux at $E>100$ GeV, with observed day-by-day variability \citep{MAGIC2018}. It is not possible to predict what will be the exact state of the source at the time of CTA observations. Yet, we expect that a large majority of the data will be obtained when the source is in a quiescent state and periods with high intensity can be removed from analysis. We therefore assume that the spectrum of NGC~1275 follows the quiescent energy spectrum obtained from MAGIC observations as presented in \citep{Ahnen2016}. This spectrum is well described, as a function of energy $E$, by a simple power-law given by 
\begin{equation}
\frac{dN}{dEdSdt} = 2.1 \times 10^{-11} \left(\frac{E}{200 \ {\rm GeV}}\right)^{-3.6} \ {\rm cm}^{-2} {\rm s}^{-1} {\rm TeV}^{-1}.
\label{eq:spectrum_NGC1275}
\end{equation}

IC~310, a member galaxy of the Perseus cluster, is also a variable source presenting high-amplitude and short duration flares \citep{Aleksic2014}. While its spectral shape was observed to not significantly change, its amplitude does by a factor of up to $\sim 7$. Similarly to NGC~1275, its spectrum is modeled as a power-law,
\begin{equation}
\frac{dN}{dEdSdt} = 0.741 \times 10^{-12} \left(\frac{E}{1 \ {\rm TeV}}\right)^{-1.81} \ {\rm cm}^{-2} {\rm s}^{-1} {\rm TeV}^{-1},
\label{eq:spectrum_IC310}
\end{equation}
corresponding to the quiescent state of the source \citep{Aleksic2014}. This source state corresponds to the large majority of the reported observations. We note that, because we apply EBL absorption a posteriori when performing simulations, we use the \emph{intrinsic} MAGIC best-fit model. The equatorial coordinates of IC~310 are set to (R.A., Dec.) = (49.179,+41.325) deg.

In addition to NGC~1275 and IC~310, several sources have been detected at lower energies with \textit{Fermi}-LAT around the Perseus cluster. In appendix \ref{app:Fermi_source_in_Perseus_field} we list these sources and estimate their fluxes in the CTA energy range. Given their location, none of them should affect the analysis.

Finally, we consider the contribution from the Galactic diffuse emission in the region around the Perseus cluster using the work by \citep{Luque:2022buq}. Within $\theta_{500}$, the emission is expected to be smooth and present a soft gradient. Compared to other contaminant sources, the instrumental background, and the cluster diffuse emission (e.g., baseline CR model), the Galactic $\gamma$-ray foreground should be largely subdominant over all the considered energy range. We refer to Appendix \ref{app:Galactic_foreground_estimate} for more details. Because of the uncertainties in the foreground model, its spatial structure and the fact that it is subdominant, it is not considered in the following.

\subsection{Towards an optimal observing strategy}
In this Section, we explore how the observation setup parameters described in Section~\ref{sec:CTA_configiuration}, together with the background sky discussed in Section~\ref{sec:background_sky}, will affect the expected detection significance. To do so, and as it will be the case in Section~\ref{sec:CTA_CR_sensitivity}, we use the {\tt ctools} software\footnote{\url{http://cta.irap.omp.eu/ctools/}} \citep{Knodlseder2016}. We focus on the CR component since the observation setup selection will be driven by the CR case. We have also checked the influence of different setups for the DM case (Appendix~\ref{app:DM_on_off}) and the chosen one is appropriate for both science cases. Our background model includes both the instrumental background and the two AGNs NGC~1275 and IC~310. We consider the case of ON-OFF analysis, and the case of template fitting.

\paragraph{ON-OFF analysis}
We first consider the case of a classical ON-OFF analysis. A pointing offset, $\theta_{\rm pointing}$, is necessary to define the OFF regions that will be used to monitor the background (see Figure~\ref{fig:fov_scheme}). In addition to the pointing offset, the aperture radius of the ON region will affect the detection significance depending on the shape of the signal. For instance, a small aperture will be favored for compact sources because it will lead to a lower instrumental background while keeping most of the signal photon counts. By contrast, a larger aperture will be preferred for more extended sources. The size of the ON region is also limited by the pointing offset, for which we aim at testing the impact. Here, we request at least 3 independent OFF regions so that the radius of the ON region, $\theta_{\rm ON}$, is constrained to $\theta_{\rm ON} \leq \sqrt{2} \theta_{\rm pointing}$. Such condition will also affect the maximum significance for a given pointing offset. We also consider the possibility of masking the known point sources. We use an energy-dependent aperture radius proportional to the point spread function (PSF) 68\% containment angle $\theta_{\rm PS}(E) = N_{\rm PSF} \times {\rm PSF}(E)$, where $N_{\rm PSF}$ can also be varied and for which the optimal value may depend on the diffuse cluster model.

In order to optimise the observation setup, we compute the expected significance according to \citep{Li1983}:
\begin{equation}
\sigma = \sqrt{2 \times \left(d \times {\rm ln}\left(\frac{d}{m}\right)+m-d\right)}.
\end{equation}
The quantity $d$ corresponds to the expected photon counts (between 150 GeV and 50 TeV) of the diffuse cluster signal plus background in the ON region, after accounting for the point-source mask. The quantity $m$ corresponds to the expected photon counts associated with the background only for the same region. The background includes both the instrumental background and the point sources, assuming that their spectral energy distribution (SED) is known (see Section~\ref{sec:background_sky}) and kept fix. The cluster and point-source models are convolved with the IRFs. We compute $\sigma$ for a set of pointing offset values, for each of which we vary the values of the ON region aperture $\theta_{ON}$ and the size of the point-source mask $\theta_{PS}$. The counts are summed over all the energy bins.

\paragraph{Template fitting}
Beside the classical ON-OFF analysis technique, we consider the case of template fitting, which might be more appropriate for diffuse sources assuming that the background can be properly modeled \citep{CTA:2020qlo}, as it will be further explored in Section~\ref{sec:CTA_CR_sensitivity}. To do so we perform simulations of the observations by varying the pointing offset, and use the maximum likelihood technique to fit the data and recover the test statistic value. We fit for the normalization of the cluster model, for the instrumental background normalization and spectral tilt, and for the normalization and spectral index of the two point sources. We use the square root of the recovered test statistic ($\sqrt{TS}$) as an estimate of the significance, with
\begin{equation}
TS = 2\times\ln\frac{\mathcal{L}(H_1)}{\mathcal{L}(H_0)},
\label{eq:TS_definition}
\end{equation}
where $\mathcal{L}(H_1)$ is the Poisson likelihood associated with the assumed emission model (i.e., including the cluster) and $\mathcal{L}(H_0)$ corresponds to the Poisson likelihood of the null-hypothesis (i.e., without including the cluster) \citep{Rico:2020vlg}. We perform this test for the different diffuse cluster models. We perform several simulations for each case to obtain a measurement of the uncertainty on the recovered significance. 

\begin{figure}
	\centering
	\includegraphics[width=1.0\textwidth]{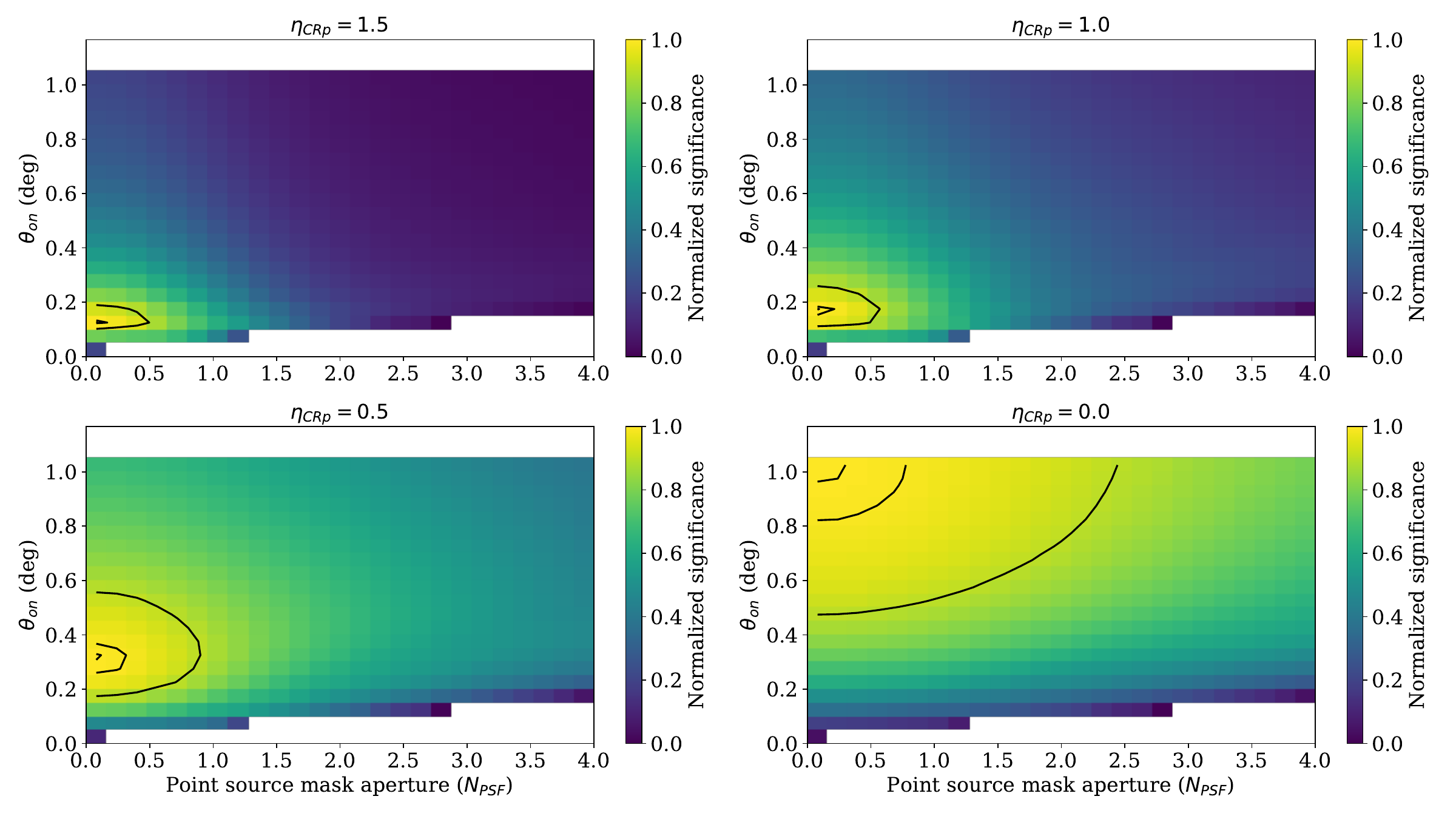}
	\caption{Normalized signal-to-noise ratio as a function of the radius of the ON region and the radius used to define the point-source mask (in units of $N_{\rm PSF}$), for different cluster models defined via the $\eta_{\rm CRp}$ parameter. Black lines correspond to 90\%, 99\% and 99.9\% of the peak significance. The pointing offset was set to 1.5 deg in the present case.}
\label{fig:snr_versus_pointing_offset1}
\end{figure}

\begin{figure}
	\centering
 	\includegraphics[width=0.99\textwidth]{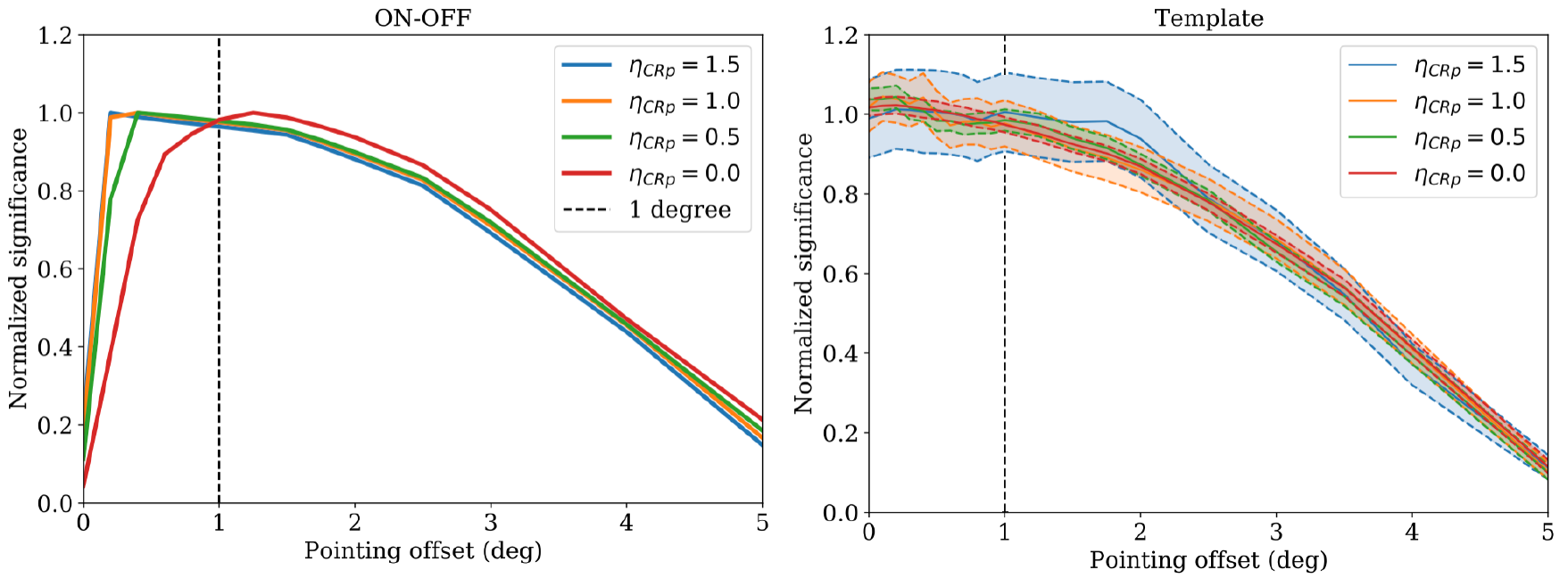}
	\caption{Normalized significance as a function of the pointing offset in the case of the ON-OFF analysis (left) and the template analysis (right).}
\label{fig:snr_versus_pointing_offset2}
\end{figure}

\paragraph{Results and discussions}
In Figure~\ref{fig:snr_versus_pointing_offset1}, we show the expected significance of the ON-OFF analysis as a function of $\theta_{\rm PS}$ and $N_{\rm PSF}$ in the case of $\theta_{\rm pointing} = 1.5$ deg. Note that the point-source mask aperture cannot be larger than the size of the ON region because NGC~1275 sits at the center of the cluster. Although the best significance always corresponds to no point-source mask ($\theta_{\rm PS} = 0$), because the point source was accounted for in the background, we can observe that the recovered significance as a function of the mask and the ON region aperture highly depends on the cluster signal. This implies that the precision in the modeling of NGC~12175 will be more critical as the cluster diffuse emission is more compact (higher value of $\eta_{\rm CRp}$). For instance, when fixing the ON region aperture to its optimal value, masking the point source with $N_{\rm PSF}=1$ will reduce the significance by about 1\% for $\eta_{\rm CRp} = 0$ and about 50\% for $\eta_{\rm CRp} = 1.5$. In practice, the optimization of the mask aperture will also depend on how well the central point source, NGC~1275, can be modeled. For instance, understanding the tails of the PSF could be a major issue when searching for diffuse emission, but we leave these investigations for future work, when the real CTA data will be available. We also observe that the best ON region aperture increases with the extent of the cluster (parameter $\eta_{\rm CRp}$), going from about 0.4 deg to 1 deg for the considered cases and using 1.5 deg as the pointing offset. We also compute the maximal value of the significance as a function of the pointing offset, as shown in Figure~\ref{fig:snr_versus_pointing_offset2}, left panel. As expected, the shape of the curve depends on the diffuse cluster model and we can observe maximal values between about 0.2 deg and 1.5 deg for all the cases.

The results of the template fitting analysis are presented in Figure~\ref{fig:snr_versus_pointing_offset2}, right panel, where we can observe the normalized significance as a function of the pointing offset for all four cluster models considered (fixing $X_{500} = 10^{-2}$, $\alpha_{\rm CRp} = 2.3$, and varying the spatial parameter $\eta_{\rm CRp}=(0, 0.5, 1, 1.5)$). The significance smoothly decreases with the pointing offset, depending on the combination of the effective area and the instrumental background. Unlike for the ON-OFF analysis, no significant significance decrease is observed for small pointing offsets. We note that the background associated with NGC~1275, relative to the cluster signal, does not depend on the offset and acts as an extra background. For this reason, the sensitivity as a function of the radius is flattened by the presence of the central galaxy. We do not observe any significant impact of the cluster diffuse model in this case. The significance remains relatively constant up to a radius of about 1 deg and vanishes beyond.

According to the results presented above and in order to allow for ON-OFF analyses while maximizing the significance of the template analyses, the pointing offset will be set to 1.0 deg in the following. This would enable to choose between either an ON-OFF or a template-based analysis after the actual observations have been made without a significant lose of sensitivity in each case.

\section{CTA sensitivity to CR induced $\gamma$-ray emission from Perseus}\label{sec:CTA_CR_sensitivity}
In this section, we use the cluster models defined in Section~\ref{sec:Modeling_CR} together with the observation setup and background sky defined in Section~\ref{sec:Observation_setup} in order to predict the CTA sensitivity to CR induced $\gamma$-ray emission from the Perseus cluster. This is done via the use of the publicly available package {\tt KESACCO} (Keen Event Simulation and Analysis for CTA Cluster Observations\footnote{\url{https://github.com/remi-adam/kesacco}}). {\tt KESACCO} is a {\tt python} package that is based on the {\tt ctools} software \citep{Knodlseder2016}, and dedicated to perform the simulation and analysis of galaxy clusters observed with CTA. In Appendix~\ref{app:validation_gammapy_cr}, we validate our results using the {\tt gammapy} software \citep{gammapy:2017}. We first discuss the data preparation. We then quantify the degeneracy between the different signal and background components. Finally, we perform the analysis of simulated data using different methods and under different assumption for the signal to address the sensitivity of CTA to the ICM induced $\gamma$-ray emission.

\subsection{Data preparation}
The analysis and results presented in this section are based on event files simulated according to the setup defined above. The events are binned in energy and according to their sky coordinates. We consider 30 bins in energy and 0.02 deg sky pixels to make sure that the PSF and spectrum sampling is sufficient, unless otherwise specified. The region of interest is centered on the Perseus cluster and is 3 deg wide in diameter. Although we consider eight different pointing positions, we use a stacked analysis in which all data from multiple observations are stacked into a single counts cube for each sky and energy bin. Unless otherwise stated, we focus on 3D template analysis as it is expected to be more appropriate in the case of such extended sources and given the presence of the two AGN. Indeed, this method will allow us to constrain the shape of the diffuse Perseus cluster $\gamma$-ray emission in a straightforward way.

The CTA sensitivity is expected to be best around a few TeV, but covers energies from 20 GeV to 300 TeV. Because of the EBL attenuation, the differential flux of all sources which we account for in this paper (all located at the same redshift) is expected to drop drastically around 30 TeV (see Figure~\ref{fig:gamma_ray_observable_sim}, top left panel). Therefore, we consider a maximum energy of 50 TeV for the analysis. We also conservatively increase the low energy threshold to 150 GeV. This choice is driven by the fact that we aim at constraining the spectrum and the shape of the ICM induced emission in the presence of the steep spectrum source, NGC~1275, which strongly dominates at low energy where the PSF is also larger. Such a threshold will allow us to avoid any significant bias that could arise due to mismodeling, but its value could be adapted when the true data will be available. This choice will not apply to the DM search in Section~\ref{sec:CTA_DM_sensitivity}. 

\subsection{Sky components and degeneracy}\label{sec:Sky_components_and_degeneracy}
\begin{figure}
	\centering
    \includegraphics[height=5.7cm]{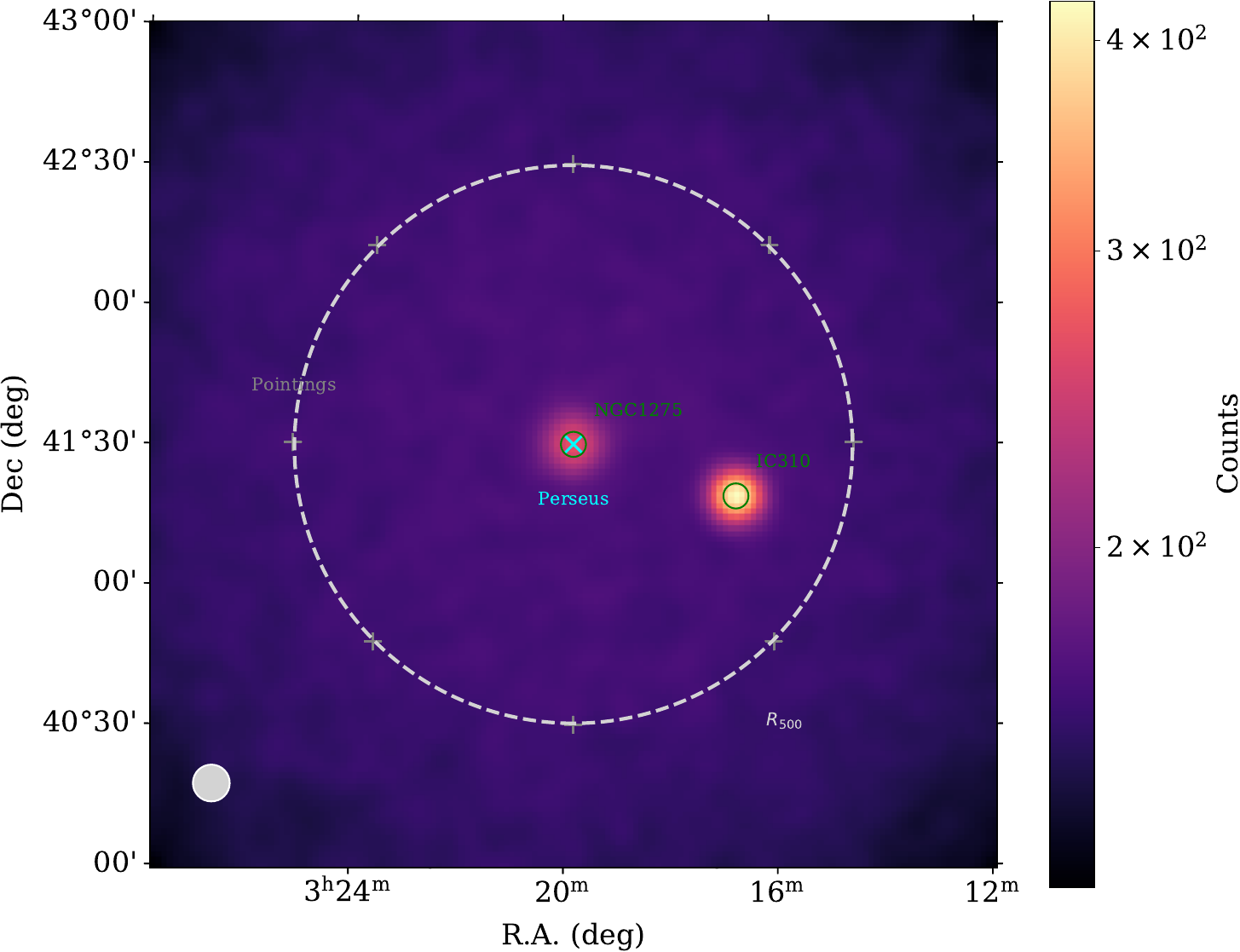}
	\includegraphics[height=5.7cm]{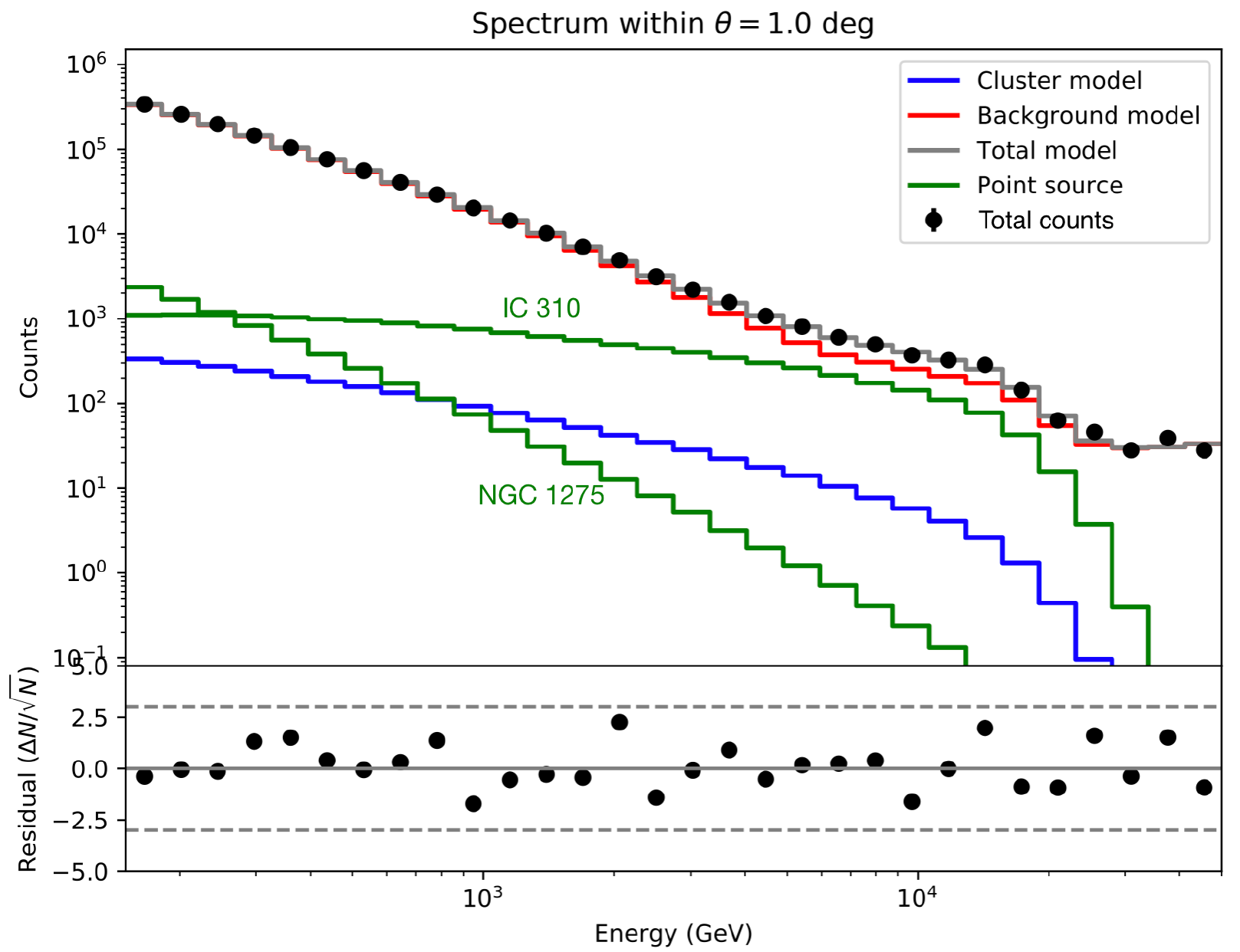}
	\caption{
	{\bf Left panel:} combined counts per pixel map, over the full energy range (150 GeV - 50 TeV). The bottom left grey circle gives the PSF at 1 TeV, and also accounts for an extra 0.1 deg smoothing used for visual purpose. The grey crosses give the pointing coordinates, the green circles show the position of the point sources, the grey dashed lines show the radius $\theta_{500}$ and the location of the cluster center is indicated by the cyan cross.
	{\bf Right panel:} counts per bin measured as a function of energy within one degree from the cluster center. Contributions from the instrumental background, the point sources and the cluster diffuse emission are shown. This is in the case of the baseline model. The normalized residual is also shown at the bottom of the figure.
 }
\label{fig:sky_model_counts}
\end{figure}

We first investigate how the components of the sky model compare to each other and how degenerate they are. We focus on the region enclosed within 1 deg from the cluster center($\sim \theta_{500}$). This region contains the two point sources and a large fraction of the cluster flux. In Figure~\ref{fig:sky_model_counts} (left panel), we present the stacked sky map including both the astrophysical and instrumental background, computed over the full energy range. The location of the different pointing centers is shown, falling very close to the radius $\theta_{500}$. The location of the two point sources are indicated as green circles and they are both clearly visible in the map. The cluster diffuse emission (in the case of our baseline model here) is blended with NGC~1275 and cannot be clearly distinguished at this stage. Figure~\ref{fig:sky_model_counts} (right panel) shows how the different model components compare to each other. The instrumental background is the dominant component, by about an order of magnitude, depending on the energy. NGC~1275 dominates at low energy and IC~310 dominates at high energy. The cluster emission is subdominant in this case (baseline cluster model) given the area that is considered. Note that the signal from IC~310 can be easily masked given its location.

\begin{figure}
	\centering
	\includegraphics[width=0.8\textwidth]{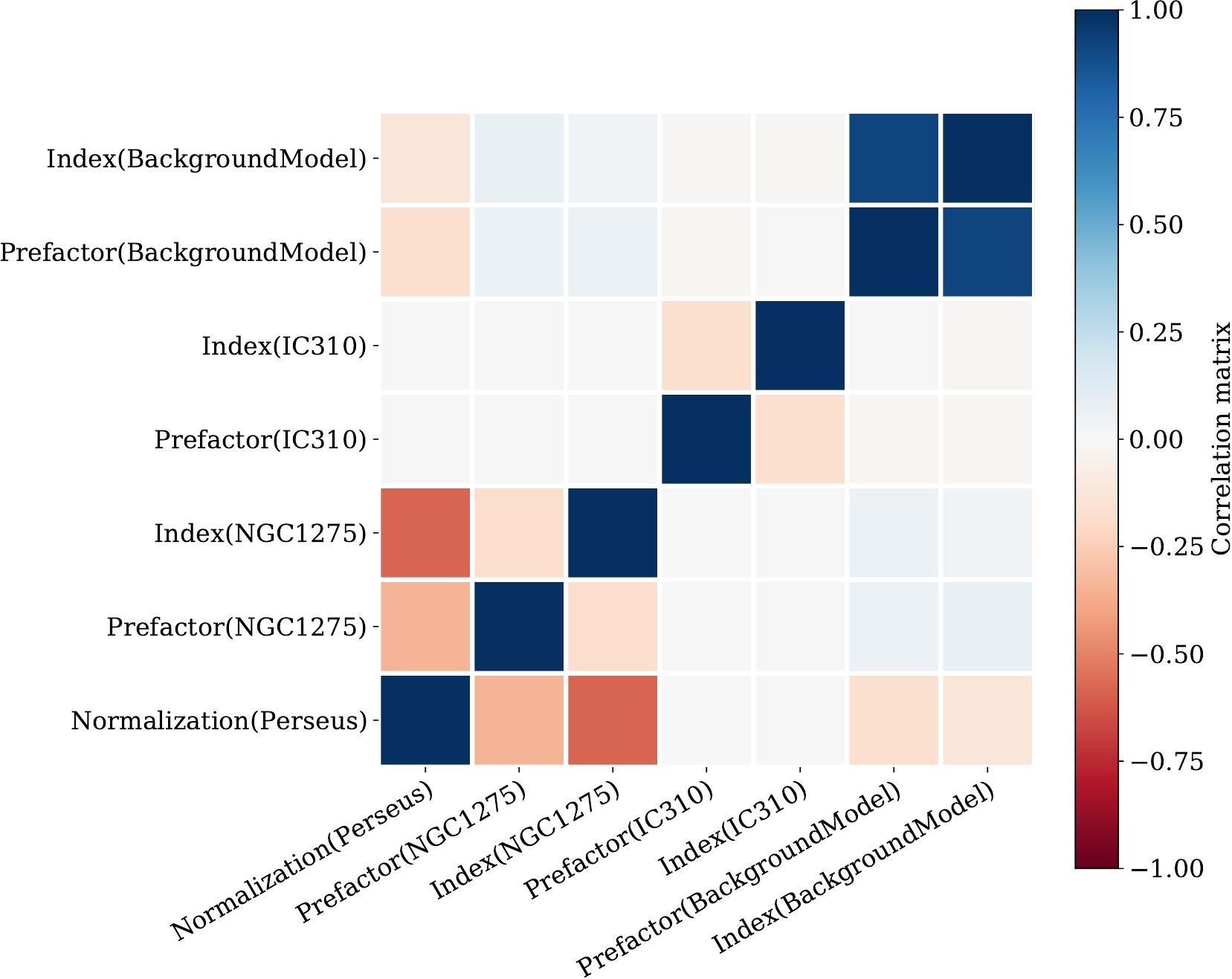}
	\caption{Correlation matrix between the different free parameters included in our model. The cluster model corresponds to the baseline model.}
\label{fig:sky_model_correlation_matrix}
\end{figure}

Secondly, we perform the joint likelihood fit of all the model free parameters: normalization and spectral index of the instrumental background, normalization and spectral index of both NGC~1275 and IC~310, and amplitude of the cluster model. We extract the correlation matrix between these components as reported in Figure~\ref{fig:sky_model_correlation_matrix} for the baseline cluster model. The parameters associated with IC~310 are not significantly degenerate with any other model components given its sky location. On the other hand, small degeneracies are observed between the cluster amplitude and the instrumental background (correlation $\sim 0.10$), as well as the parameters associated with NGC~1275. We note that the degeneracy depends on the cluster model parameters. For instance, a more extended cluster presents less degeneracy with NGC~1275 but is more degenerate with the instrumental background. We also note that increasing the size of the region of interest leads to a slightly lower instrumental background degeneracy because more leverage is available. The degeneracy also depends on the energy range used to select the data.

According to the degeneracy between the model components and the amplitude of these components within the central region of the field of view, we expect that the instrumental background will lead to most of the uncertainties associated with the cluster model constraints that can be extracted. The presence of NGC~1275 is also significantly affecting the properties that can be extracted for the Perseus cluster.

\subsection{Probing the parameter space with CTA}\label{sec:Probing_the_parameter_space_with_CTA}
We first estimate the parameter space constraint that can be obtained with CTA. We perform event simulations without including the cluster signal, but still account for it afterwards.

We use our {\tt ctools} framework to fit for the cluster together with the other sky model components and extract the upper limit on the cluster flux normalization. We also extract the corresponding upper limit on the overall cluster $\gamma$-ray flux. This is reproduced for 50 simulations in order to compute the mean upper limit and its standard deviation (implying an uncertainty on the mean and standard deviation of the upper limit of less than 5\%, see appendix~\ref{app:CR_convergence}), and for the different models that we test: spectral indices $\alpha_{\rm CRp} = \big[ 2.0, 2.2, 2.4, 2.6, 2.8, 3.0 \big] $ and spatial scalings $\eta_{\rm CRp} = \big[ 1.5, 1.0, 0.5, 0.0 \big]$. A higher spatial scaling indicates a more compact cluster, as can be seen by referring to Section~\ref{sec:Modeling_CR} and Figure~\ref{fig:gamma_ray_observable_sim}.

\begin{figure}
	\centering
	\includegraphics[width=0.9\textwidth]{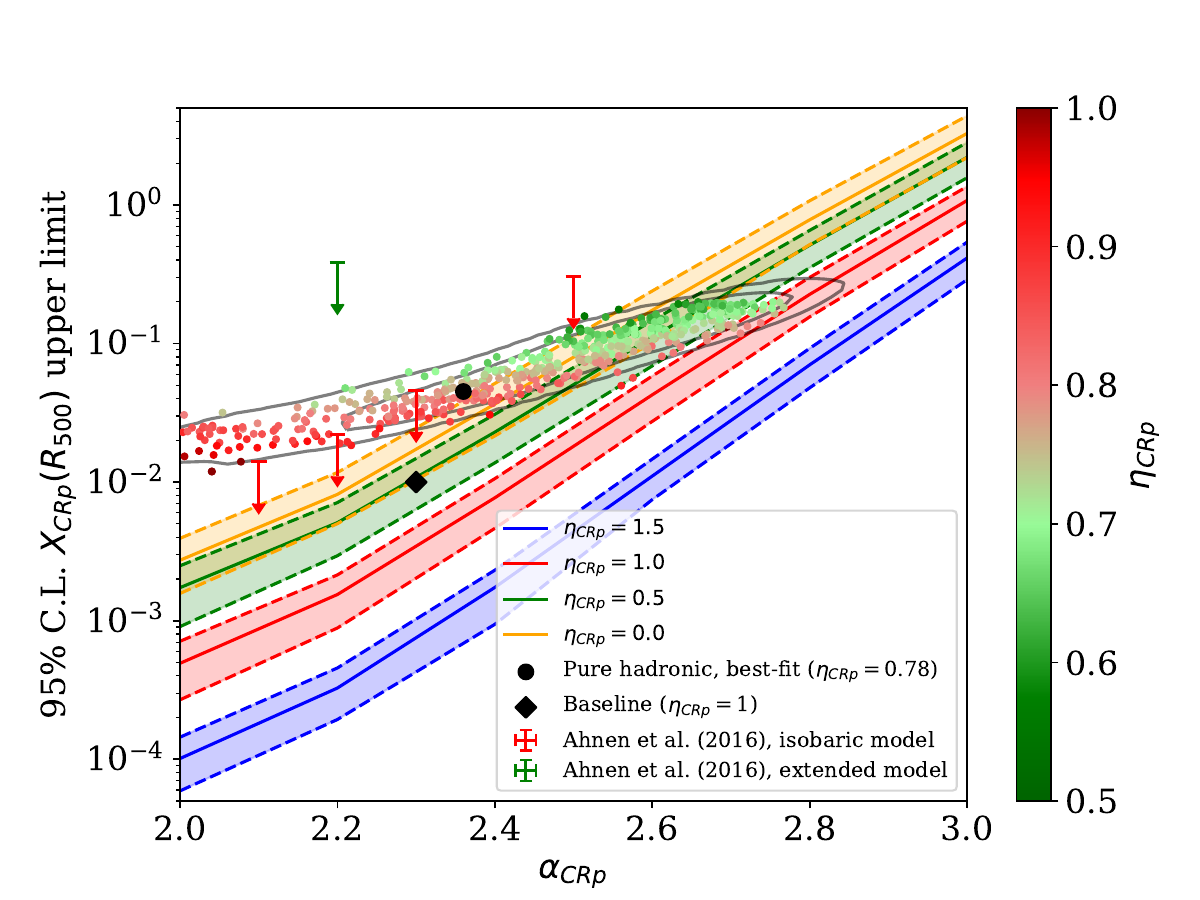}
	\caption{Exclusion limit at 95\% confidence level for the parameter $X_{\rm CRp}$, as a function of $\alpha_{\rm CRp}$ and for different values of $\eta_{\rm CRp}$, assuming non-detection of the signal. The shaded area represent the 68\% confidence interval. The limits obtained by MAGIC \citep{Ahnen2016} are reported for their isobaric model (roughly corresponding to our $\eta_{\rm CRp}=1$ model) and their extended model (roughly corresponding to our $\eta_{\rm CRp}=0.5$ model). We also report the baseline model and the best-fit parameters in the case of the pure hadronic scenario (where $\eta_{\rm CRp} = 0.78$, and given the reference magnetic field model based on \citep{Taylor2006}) in black. To highlight the statistical uncertainties in the pure hadronic model, a set of 500 samples randomly extracted from the MCMC chains is shown, color-coded by the value of $\eta_{\rm CRp}$, together with the 68 and 95\% confidence intervals reported as grey lines.}
\label{fig:CR_95percent_limits_X}
\end{figure}

\begin{figure}
	\centering
	\includegraphics[width=0.69\textwidth]{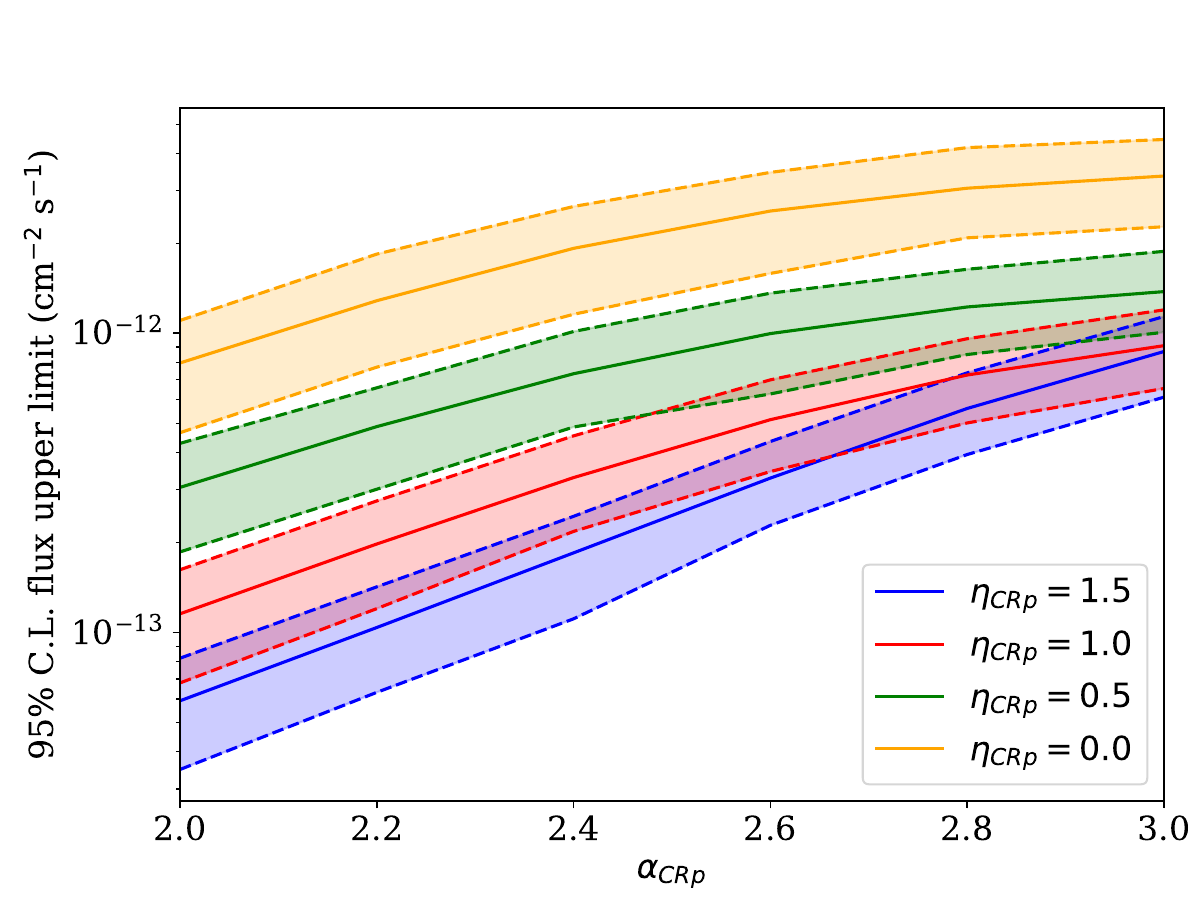}
	\caption{Same as Figure~\ref{fig:CR_95percent_limits_X} for the flux upper limit for energies in the range $[0.15, 50]$ TeV, with the same color code.}
\label{fig:CR_95percent_limits_F}
\end{figure}

Figures~\ref{fig:CR_95percent_limits_X} and \ref{fig:CR_95percent_limits_F} show the expected parameter space constraints in case of non-detection, and the flux upper limit in the range $[0.15-50]$ TeV for 285 hours of observations, respectively. In both figures, the exclusion limit at 95\% confidence interval is shown, with the shaded areas representing the standard deviation. Figure~\ref{fig:CR_95percent_limits_X} shows the upper limit in $\gamma$-ray flux as a function of the spectral index $\alpha_{\rm CRp}$, for different spatial scalings. A flux upper limit down to $10^{-13}$ cm$^{-2}$s$^{-1}$ is expected with CTA. Figure~\ref{fig:CR_95percent_limits_F} shows $X_{\rm CRp}(R_{500})$, the CR to thermal energy density ratio, as a function of spectral index. In the case of non-detection we are able to put a constraint down to $X_{\rm CRp}(R_{500}) \sim 10^{-4}$ for a very compact CRp distribution. The most recent constraints put forth on diffuse $\gamma$-ray emission from Perseus come from the MAGIC telescope observations and analysis by \citep{Ahnen2016}. As mentioned in Section~\ref{sec:CR_leptonic_model}, the $\gamma$-ray and thermal emission modelling used by MAGIC have overestimated the thermal energy, meaning that the MAGIC constraints in the left panel are actually too optimistic. Either way, it is evident that CTA will be able to put a constraint that is about an order of magnitude deeper than MAGIC.

Both figures show that a more compact cluster would give better constraints. From the previous sections we know that in the CTA energy range, a lower spectral index gives more diffuse $\gamma$-ray emission (see Figure~\ref{fig:gamma_ray_observable_sim}), hence the non-detection of diffuse $\gamma$-ray emission puts a better constraint for lower spectral index numbers.

We also show that CTA will be able to exclude the pure hadronic best-fit model, assuming \citep{Taylor2006} magnetic field model with $\eta_B=2/3$. Accounting for statistical uncertainties and assuming the most pessimistic magnetic field model (about 30\% change, see Section~\ref{app:impact_of_magnetic_field}), CTA should still be able to exclude significantly most of the parameters allowed in the pure hadronic scenario, although a small region of the parameter space with high $\alpha_{\rm CRp}$ and low $\eta_{\rm CRp}$ may still be viable in the case of non-detection, despite the higher value of $X_{500}$ that would be implied.

\subsection{Measuring the cosmic-ray properties in the case of detection}
In this section, we consider the baseline and the pure hadronic models, and produce the corresponding event simulations. We then investigate the sensitivity of CTA to constrain the model parameters. We consider two different scenarios. In the first one (Section~\ref{sec:Spectral_constraints_only}), we focus only on the recovered spectral properties by constraining the background model independently from the ICM and assuming that the cluster spatial shape is known. Then, in Section~\ref{sec:Joint_spectral_imaging_constraints}, we consider simultaneously both the spectral and spatial properties together with all background components. The two methods are completely independent from each other. In the following, we analyse and discuss representative simulations of the data corresponding to our two models.

\subsubsection{Spectral constraints only}\label{sec:Spectral_constraints_only}
\begin{figure}
	\centering
	\includegraphics[width=0.45\textwidth]{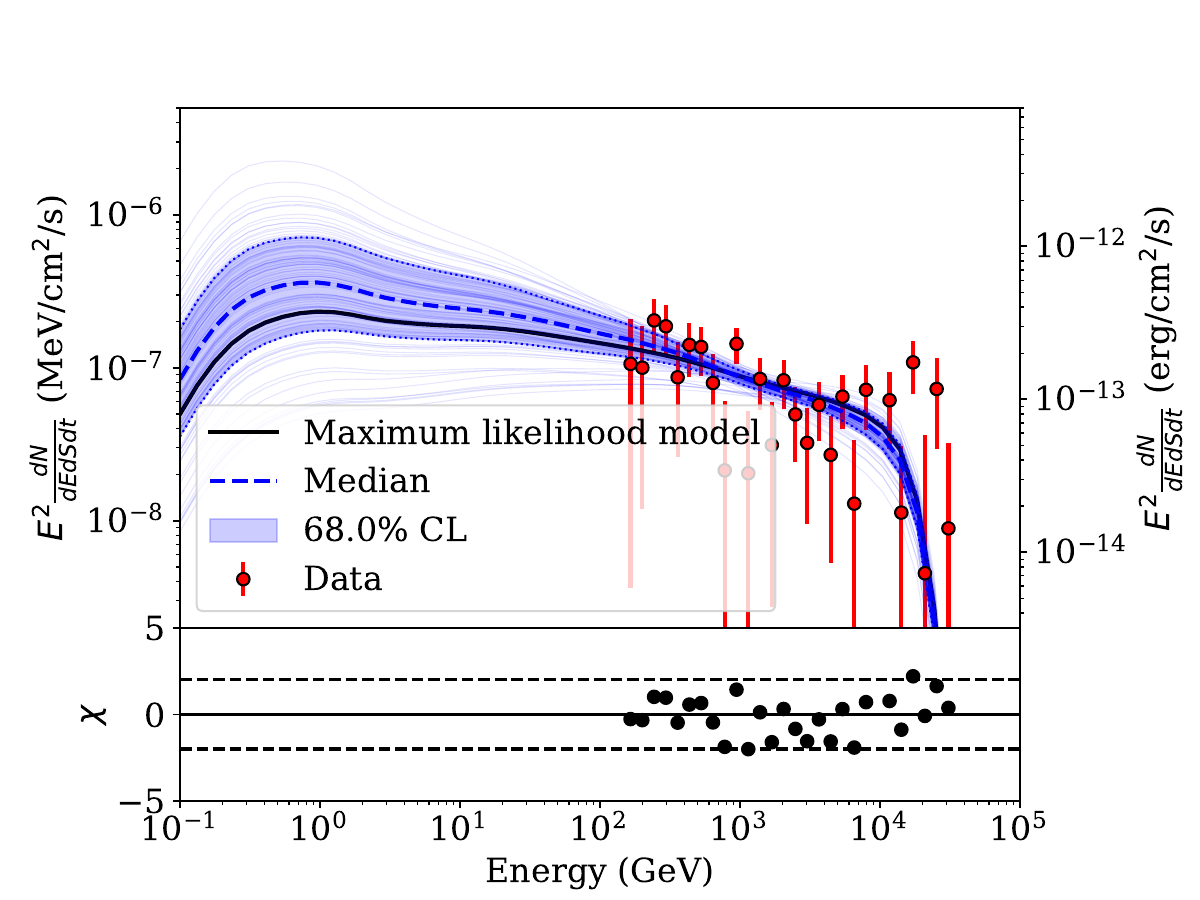}
	\includegraphics[width=0.45\textwidth]{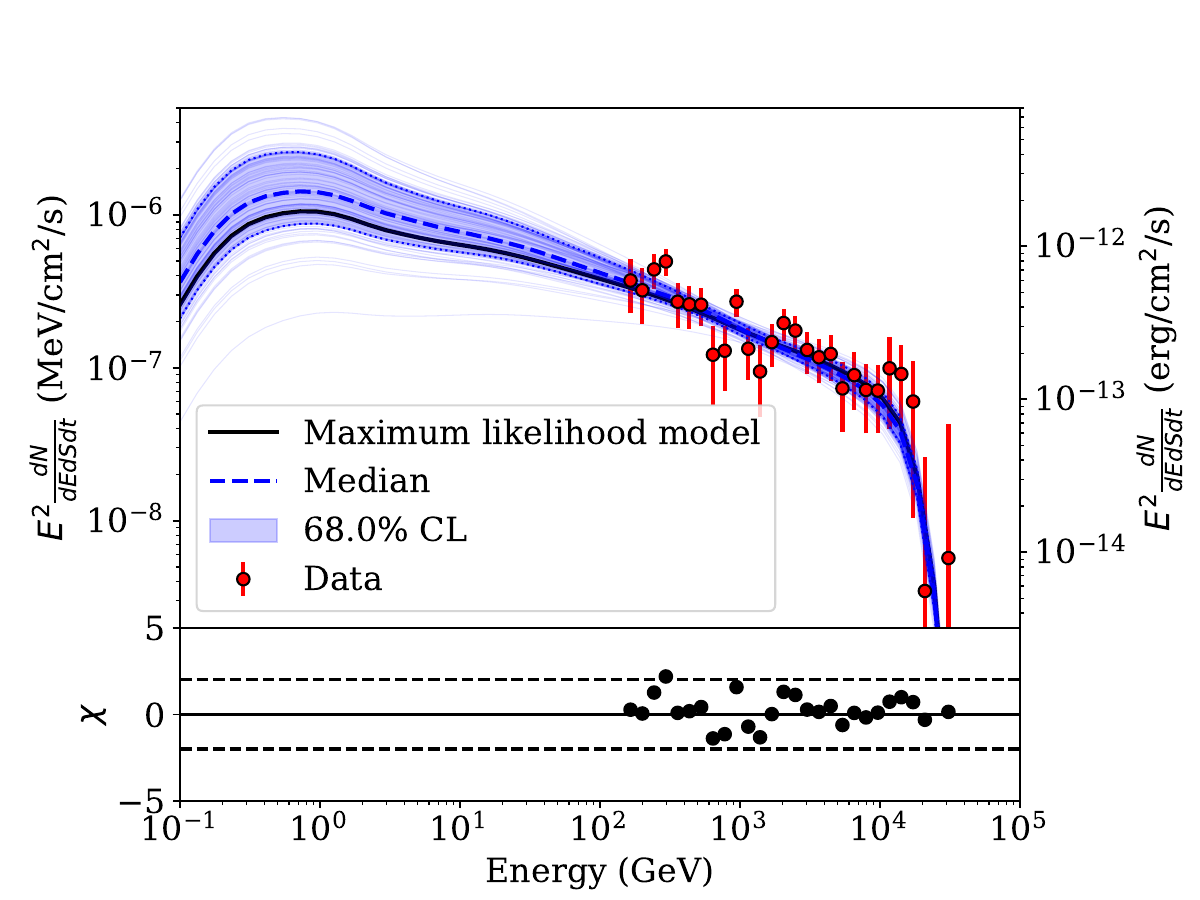}
	\includegraphics[width=0.45\textwidth]{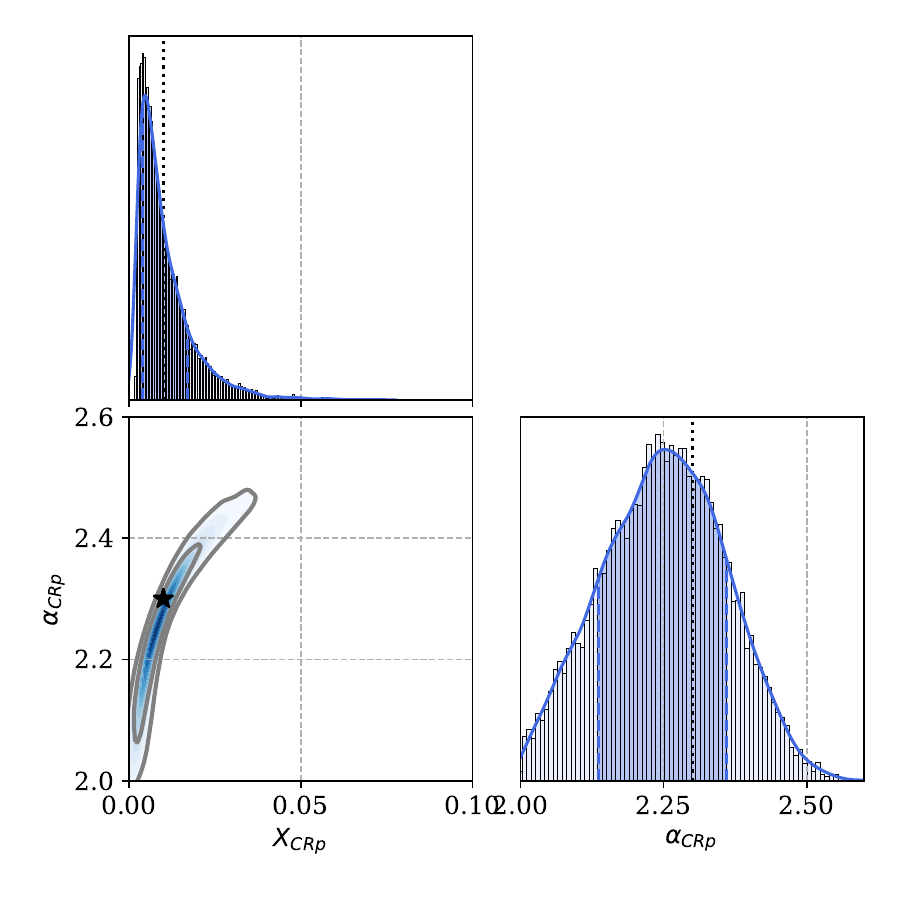}
	\includegraphics[width=0.45\textwidth]{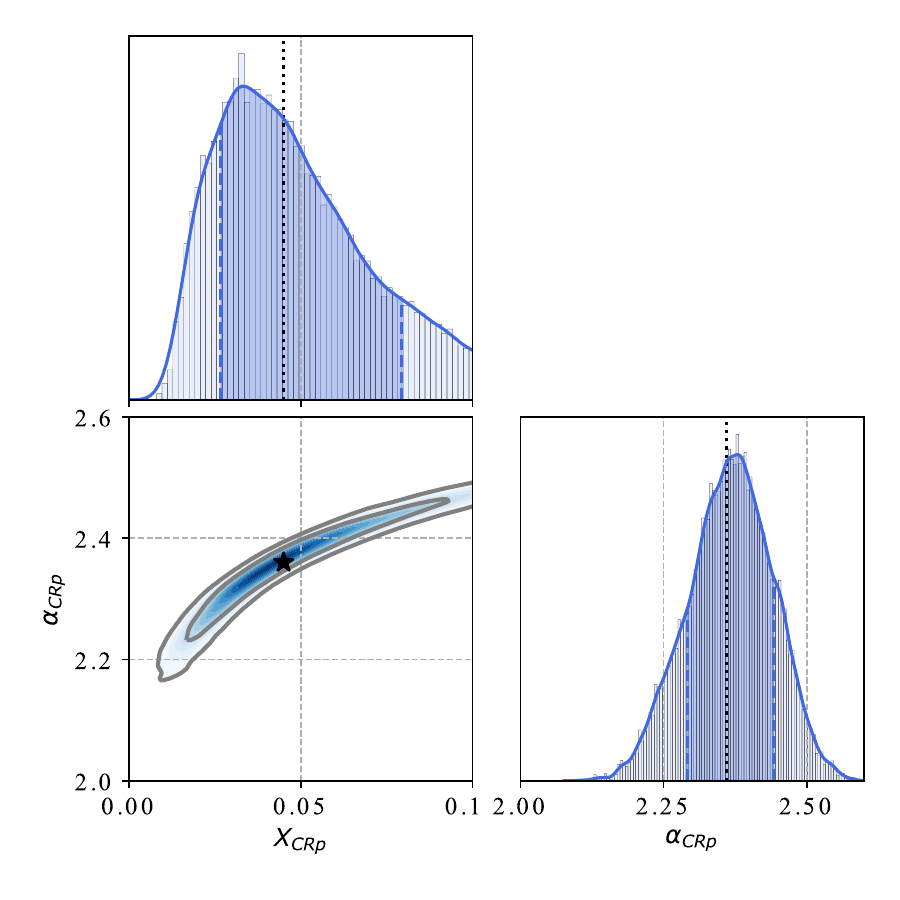}
	\caption{Spectral constraints. 
	The left panels correspond to the baseline model while the right panels correspond to the pure hadronic best-fit model (see Table~\ref{tab:model_summary}).
	{\bf Top panels}: recovered spectrum (red points) and their error bars (according to the likelihood scan curvature), best-fit model (black), 68\% confidence level (blue region) and a set of 100 realizations (light blue). The residual normalized by the error bars, $\chi$, is shown at the bottom. The dashed lines correspond to $\chi = \pm 2$. {\bf Bottom panels}: constraints on the parameter space. Contours provide the 68\% and 95\% confidence region and the black star represents the injected model. The blue shaded region on the one dimensional histograms provides the 68\% confidence region after marginalization.
 }
\label{fig:CR_spectral_constraints}
\end{figure}
We first focus on constraining the spectral energy distribution only. We start by using a likelihood fit to constrain the Perseus cluster normalization and all the other free sky parameters (see Figure~\ref{fig:sky_model_correlation_matrix}). To obtain the detection significance, we apply the Test Statistics defined in Equation~\ref{eq:TS_definition}. As a reference, we obtain a test statistic value of $\sqrt{TS} \sim 6.5$ in the case of the baseline model, and $\sqrt{TS} \sim 11$ for the pure hadronic model. We then extract the spectrum of the diffuse cluster component thanks to the dedicated {\tt csspec} function (from {\tt ctools}) using the true input cluster spatial template to do so, which we keep fixed. The instrumental background is left free but the point sources are kept fixed to their maximum likelihood values. This procedure allows us to extract the Perseus cluster flux normalization in each independent energy bin. Additionally, the {\tt csspec} function is used to recover the full likelihood scan for the normalization in each bin.

In order to measure the posterior likelihood in the parameter space, we employ an MCMC approach using the {\tt emcee} package \citep{Foreman2013}, following the method used by \citep{Adam2021}. The sampling is performed according to the log likelihood function defined as
\begin{equation}
{\rm ln} \ \mathcal{L}(\vec{\theta} | D) = \sum_i {\rm interp} \left({\rm ln} \ \mathcal{L}_i(F_{\rm scan}, F_i(\vec{\theta}))\right).
\end{equation}
The parameters $\vec{\theta} \equiv \left(X_{500}, \alpha_{\rm CRp}\right)$ correspond to the normalization and the CRp spectral index, respectively, and $D$ to the data points. In each energy bin $i$, $\mathcal{L}_i$ is the likelihood scan extracted with {\tt csspec} as a function of the flux $F_{\rm scan}$. This quantity is interpolated at the location of any model to be tested against the data, $F_i(\vec{\theta})$, and summed over all bins. We use flat priors ($X_{500}>0$, $2<\alpha_{\rm CRp}<5$) on the parameters and check a posteriori that the limits do not affect our results.

Once the MCMC has converged, we remove the burn-in phase and the chains provide us also with the posterior probability distribution function. We also recompute 100 SED models using parameters randomly sampled from the chains and measure the median and 68\% confidence limit envelop on the recovered spectrum. The best-fit model is computed using the parameters that correspond to the maximum likelihood point in the chains.

In Figure~\ref{fig:CR_spectral_constraints}, we present the recovered spectrum and the constraints in the parameter space obtained in the case of the baseline and pure hadronic best-fit models. In both cases the best-fit describes the data well, as can be observed in the residual. The constraints in the parameter space show that both input models are recovered with the 68\% confidence interval. The pure hadronic model allows us to better constrain the model given the higher signal-to-noise ratio. We note that the normalization and the spectral index of the CRp distributions are highly degenerate. This is particularly true for CTA because no leverage is available near the spectral bump around 1 GeV. Because of this degeneracy, an amplitude up to four times the input value is still allowed in the case of the baseline model. This reduces to about twice the input value in the case of the pure hadronic model, for which the signal is stronger. The error on the spectral slope is about 0.1 and 0.07 for the baseline and pure hadronic models, respectively. While the details are beyond the scope of this paper, we note that combining lower energy data (e.g., \textit{Fermi}-LAT) to CTA, even in the case of non-detection, could help break the parameter degeneracy.

\subsubsection{Joint spectral-imaging constraints}\label{sec:Joint_spectral_imaging_constraints}
The analysis described in Section~\ref{sec:Spectral_constraints_only} is standard, but intrinsically limited by the fact that the cluster profile is a priori unknown. This might lead to significant systematic effects when recovering the spectral energy distribution if it is improperly modeled. In addition, the uncertainties in the background (diffuse and point sources) are expected to affect the constraints on the cluster diffuse emission, as discussed in Section~\ref{sec:Sky_components_and_degeneracy}. Although the error bars on the extracted cluster spectrum are expected to account for the correlation between the different components, the respective parameters are not co-varied when constraining the model to reduce the computing time. To mitigate these possible biases, we consider the joint-fit of the full sky model simultaneously, as described below. This allows us to sample the full parameter space. While the analysis described in Section~\ref{sec:Spectral_constraints_only} is relatively quick, the one described hereafter is significantly more demanding in term of computing time, and the two are therefore complementary.

We compute the IRF convolved model counts cube, in spatial and energy bin $i$, as
\begin{equation}
M_i(\vec{\theta}) = C_i\left(X_{500},\alpha_{\rm CRp},\eta_{\rm CRp}\right) + 
\sum_{j \in [1,2]} {\rm PS}_i^{(j)}(A_{\rm PS}^{(j)},\alpha_{\rm PS}^{(j)}) + B_i(A_{\rm bkg},\alpha_{\rm bkg}).
\label{eq:CR_model_cube}
\end{equation}
The component $C_i\left(X_{500},\alpha_{\rm CRp},\eta_{\rm CRp}\right)$ corresponds to the cluster model, which depends on the parameters $X_{500}$, $\alpha_{\rm CRp}$, and $\eta_{\rm CRp}$. The components ${\rm PS}_i^{(1,2)}(A_{\rm PS}^{(1,2)},\alpha_{\rm PS}^{(1,2)})$ represent the point-source contribution, NGC~1275 and IC~310, respectively. The component $B_i(A_{\rm bkg},\alpha_{\rm bkg})$ is the diffuse instrumental background, for which we assume a fixed shape with a floating spectral index and normalization. The parameter space includes nine parameters,
\begin{equation}
\vec{\theta} \equiv \left(X_{500},\alpha_{\rm CRp},\eta_{\rm CRp}, A_{\rm PS}^{(1,2)}, \alpha_{\rm PS}^{(1,2)}, A_{\rm bkg},\alpha_{\rm bkg}\right)
\end{equation}

As done in Section~\ref{sec:Spectral_constraints_only}, we use an MCMC method to sample the parameter space. The likelihood function is defined according to Poisson statistics as
\begin{equation}
{\rm ln} \mathcal{L}(\vec{\theta} | D) = \sum_i \Tilde{M}_i(\vec{\theta}) - d_i {\rm ln}(\Tilde{M}_i(\vec{\theta})),
\label{eq:likelihood_def}
\end{equation}
with $d_i$ the data counts cube. Since the convolution of a model with the IRF is computationally expensive, it is not possible to compute $M_i(\vec{\theta})$ for all the MCMC test parameters. Therefore, we compute the models associated to all the components on predefined grids and $\Tilde{M}_i(\vec{\theta})$ corresponds to the model $M_i(\vec{\theta})$ after interpolation at the exact location of the requested parameters. Each model component amplitude only accounts for a linear scaling and does not need to be included into the grid. The cluster model grid is thus bi-dimensional, defined as a function of the non-trivial parameters $\alpha_{\rm CRp}$ and $\eta_{\rm CRp}$ (typically $10 \times 10$ points allows us to recover any model with percent level precision). For the point sources and the instrumental background, the slopes $\alpha_{\rm PS,bkg}$ are the only relevant parameters and the grids are therefore one-dimensional. The range over which the model grids are computed correspond to flat prior and we check a posteriori that it does not affect our result. We note that an extra prior is used for the cluster normalization, as $X_{500} \in [0, 20 \times X_{500}^{(\rm input \ model)}]$ to avoid sampling nonphysical models in the case the signal-to-noise ratio is too low to provide a reliable constraint on this parameter. In order to highlight the impact of the point sources on the cluster parameters constraint, we also perform the fit using a prior on the point-source parameters. The latter is Gaussian, centered on the input parameter values and with a standard deviation corresponding to 5\% of the parameter values.

\begin{table}[h]
	\caption{Recovered parameters of the joint spectral-imaging analysis for the baseline model, baseline model plus prior on NGC~1275, and the pure hadronic best-fit model cases.
 }
	\begin{center}
	\resizebox{\textwidth}{!} {
	\begin{tabular}{ccc|cc|cc|cc}
 	\hline
	$\frac{X_{\rm CRp}}{X_{\rm CRp,0}}$ & $\eta_{\rm CRp}$ & $\alpha_{\rm CRp}$ & $A_{\rm bkg}$ & $\Delta \alpha_{\rm bkg}$ & $A_{\rm ps,1}$ & $\Delta \alpha_{\rm ps,1}$ & $A_{\rm ps,2}$ & $\Delta \alpha_{\rm ps,2}$\\
	\hline
	\hline
 	\multicolumn{9}{c}{Baseline model} \\
    \hline
	$1.69_{-1.21}^{+3.59}$ & $1.17_{-0.24}^{+0.25}$ & $2.45_{-0.20}^{+0.18}$ & $0.9992_{-0.0016}^{+0.0016}$ & $-0.0003_{-0.0011}^{+0.0011}$ & $0.932_{-0.060}^{+0.042}$ & $0.027_{-0.102}^{+0.079}$ & $1.010_{-0.011}^{+0.011}$ & $-0.010_{-0.010}^{+0.010}$ \\ 
	\hline
  	\multicolumn{9}{c}{Baseline model plus NGC~1275 prior} \\
    \hline
	$0.89_{-0.57}^{+1.48}$ & $1.10_{-0.17}^{+0.19}$ & $2.35_{-0.16}^{+0.14}$ & $0.9993_{-0.0016}^{+0.0015}$ & $-0.0002_{-0.0011}^{+0.0010}$ & $0.963_{-0.033}^{+0.030}$ & $0.019_{-0.043}^{+0.043}$ & $1.007_{-0.010}^{+0.010}$ & $-0.010_{-0.010}^{+0.010}$\\
    \hline
  	\multicolumn{9}{c}{Pure hadronic best-fit model} \\
    \hline
	$0.85_{-0.38}^{+0.77}$ & $0.87_{-0.11}^{+0.11}$ & $2.38_{-0.09}^{+0.10}$ & $1.000_{-0.0015}^{+0.0015}$ & $-0.0014_{-0.0010}^{+0.0010}$ & $0.969_{-0.043}^{+0.037}$ & $0.0078_{-0.081}^{+0.071}$ & $1.011_{-0.011}^{+0.011}$ & $-0.015_{-0.010}^{+0.010}$\\
    \hline
	\end{tabular}
	}
	\end{center}
	\label{tab:posterior_fit}
\end{table}

\begin{figure}
	\centering
	\includegraphics[width=1.0\textwidth]{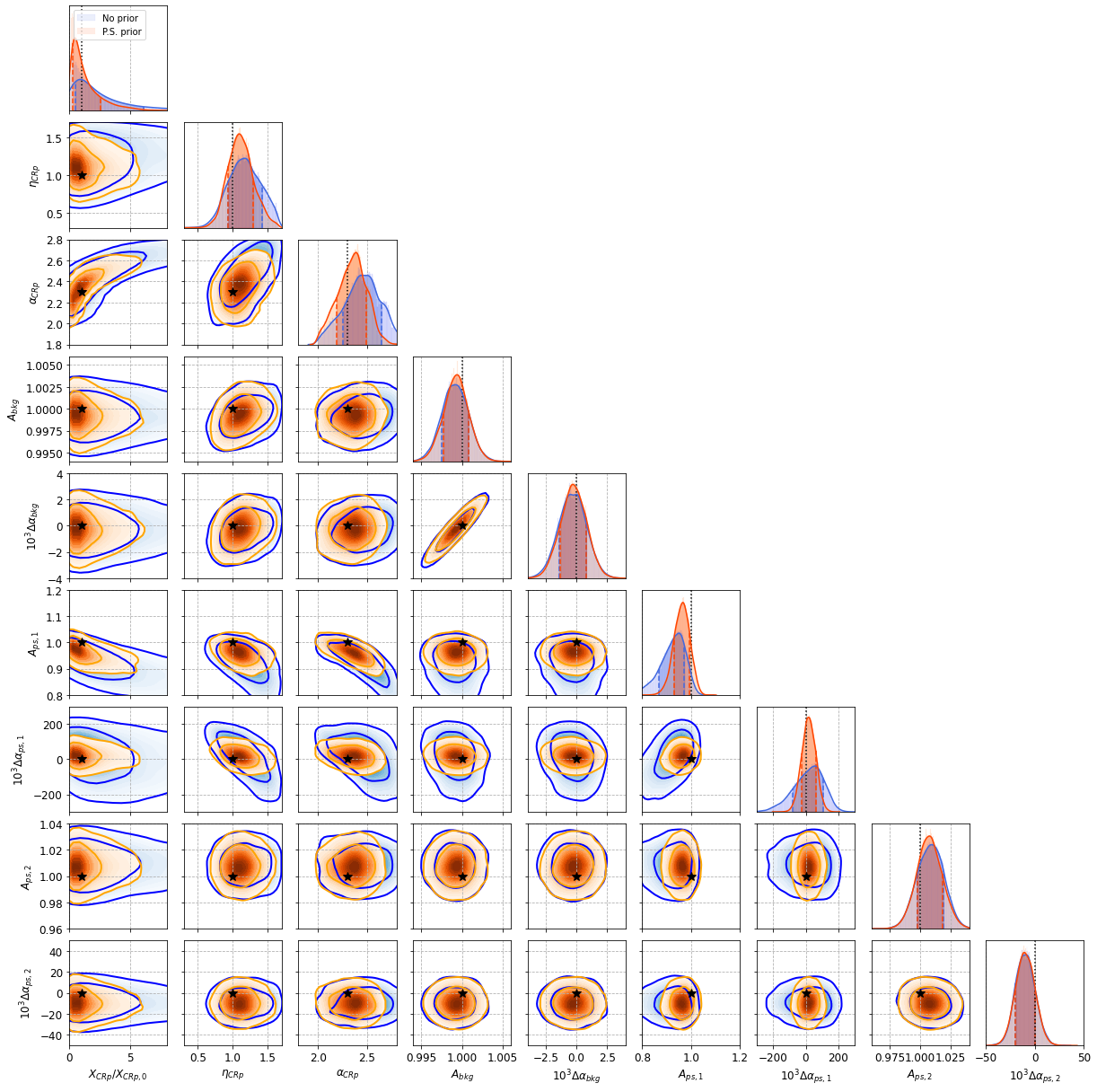}
	\caption{Posterior constraint on the full parameter space in the case of the baseline model. Contours provide the 68\% and 95\% confidence interval. The black star shows the input model parameters. The constraints are reported in the standard case (blue) and in the case where we assume a 5\% prior on the flux and spectral indices of point sources (orange). The CRp normalisation is given relative to the input one, $X_{CRp,0}=10^{-2}$.}
\label{fig:CR_triangle_plot_basline}
\end{figure}
Figure~\ref{fig:CR_triangle_plot_basline} presents the posterior distribution in the full parameter space for the baseline cluster model (the case of the pure hadronic model is available in Appendix~\ref{app:Parameter_constraints_pure_hadronic_model}).Table~\ref{tab:posterior_fit} also reports the recovered parameters and their uncertainties in the different cases. The cluster normalisation, the point sources and the instrumental background parameters are shown relative to their input values. We can see that for this simulated data, the input model is recovered within 95\% confidence interval for all the parameters (and 68\% for most of them). While the observed posterior distribution is non-trivial for most of the parameters, we have tested that in the limit of high S/N, the parameter space is well described by a multivariate Gaussian function. As discussed in the case of the spectral only analysis (Figure~\ref{fig:CR_spectral_constraints}), strong degeneracy is observed between the cluster normalization and the CRp slope. On the other hand, the cluster profile parameter is not strongly degenerate with the other cluster parameters, but is anti-correlated with the parameters associated with NGC~1275. Indeed, as the cluster profile gets more compact, it resembles more to the point source so that increasing $\eta_{\rm CRp}$ leads to a smaller point-source normalization. As the amplitude and slope of NGC~1275 are positively correlated, the same trend is observed between $\eta_{\rm CRp}$ and $\alpha_{\rm PS}^{(1)}$. The cluster normalization is also anti-correlated with NGC~1275 parameters for similar reasons. No significant correlation is observed between the cluster model and IC~310, as expected given its sky coordinates. The instrumental background normalization and slope is slightly correlated with the cluster parameters, in particular with the profile parameter, since lowering the value of $\eta_{\rm CRp}$ leads to a flatter signal that slightly mimics the instrumental background. The correlation is thus positive between $\eta_{\rm CRp}$ and $A_{\rm bkg}$. As expected, the amplitude and spectral slope of the instrumental background and point sources are correlated. Compared to the spectral constraints only, the error bars on the parameter $\alpha_{\rm CRp}$ increase from 0.10 to 0.17, highlighting the importance of constraining simultaneously all the components. The error on the cluster profile parameter, $\eta_{\rm CRp}$, is about 0.26. The normalization is constrained to $X_{500} \lesssim 0.1$ (at 68\% limit), which is significantly larger than in the case of spectral constraints only. 

When using the prior on point source, we observe a strong reduction of the volume allowed by the cluster parameters (especially avoiding too high values of the normalization), in agreement with the degeneracy between components. On the other hand, the instrumental background parameters are nearly unchanged. The posterior constraints on NGC~1275 parameters is strongly improved, as can be expected due to the prior, but the constraints on IC~310 is nearly unchanged because the CTA data already provide a constraint better than 5\% on the corresponding parameters. 

We note that the steep spectrum of NGC~1275 makes it very sensitive to any uncertainty in IRF convolution of the model, which in turns may affect the cluster reconstruction when all parameters are allowed to vary, due to degeneracy. This may cause notable systematic effects if improperly accounted for in the case of a low significance, but we leave its detailed investigation for future work when the true data will be taken.

\begin{figure}
	\centering
	\includegraphics[width=1.0\textwidth]{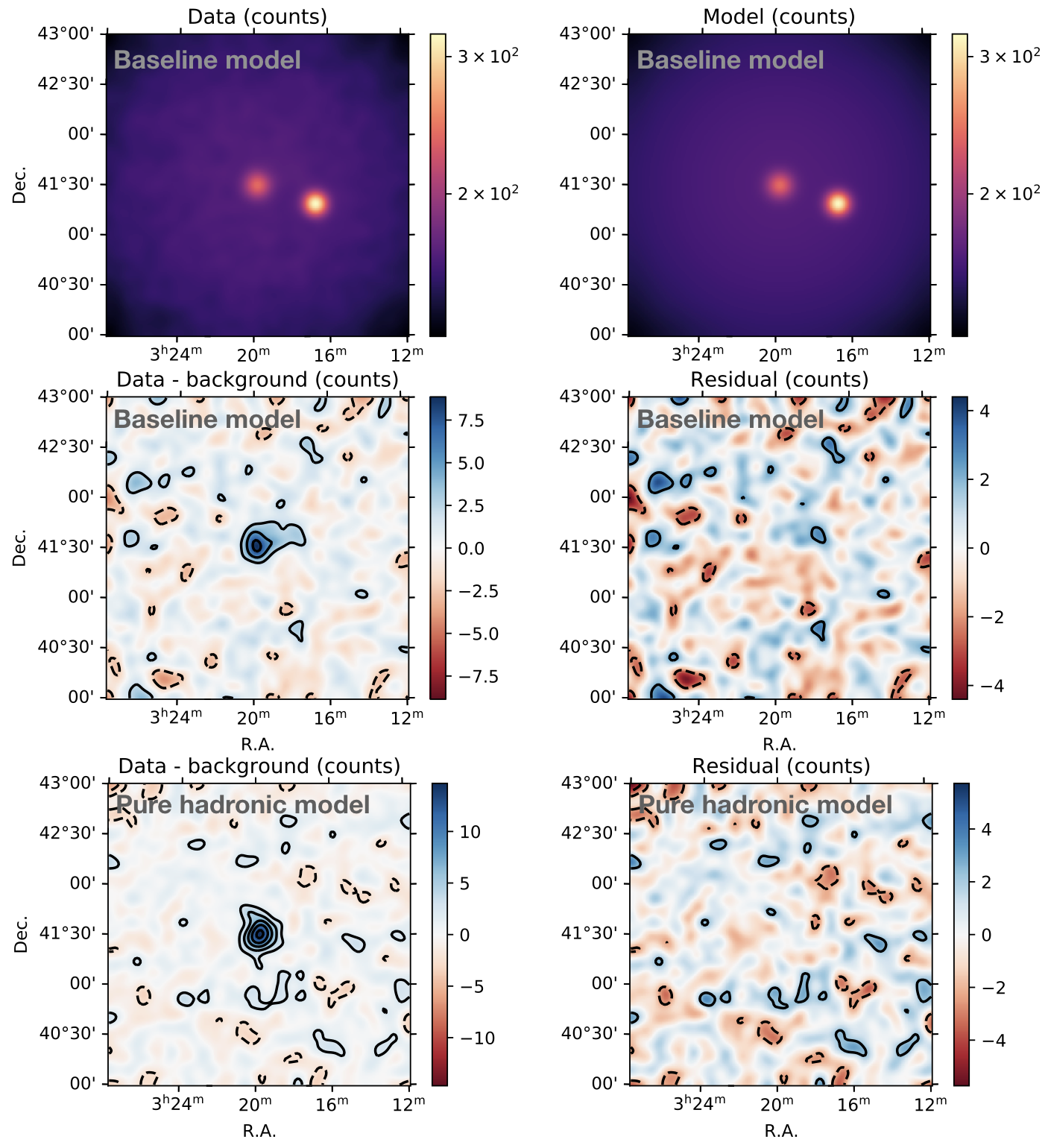}
	\caption{Comparison between the data and the best-fit model, for the full energy range. Units are in counts per pixel.
	{\bf Top left panel:} data (baseline model only).
	{\bf Top right panel:} best-fit model (baseline model only).
    {\bf Bottom and middle left panel:} residual between the data and the background components (including point sources) for the baseline and pure hadronic models, respectively. 
	{\bf Bottom and middle right panel:} total residual between the data and the model for the baseline and the pure hadronic models, respectively.
	Contours provide the signal-to-noise ratio with $2\sigma$ spacing. The maps where smoothed with a Gaussian kernel with 0.15 deg for visualization purpose.}
\label{fig:CR_residual_map_specimg}
\end{figure}
In Figure~\ref{fig:CR_residual_map_specimg}, we show the input data, model and best-fit residual (including the cluster or not) after summing over all the energy bins. We can see that the best-fit model describes the data well. The signal is dominated by the instrumental background on large scales, and by the point sources on small scales. Once these components are subtracted, we can observe the faint signal associated with the cluster emission.

\begin{figure}
	\centering
	\includegraphics[width=0.45\textwidth]{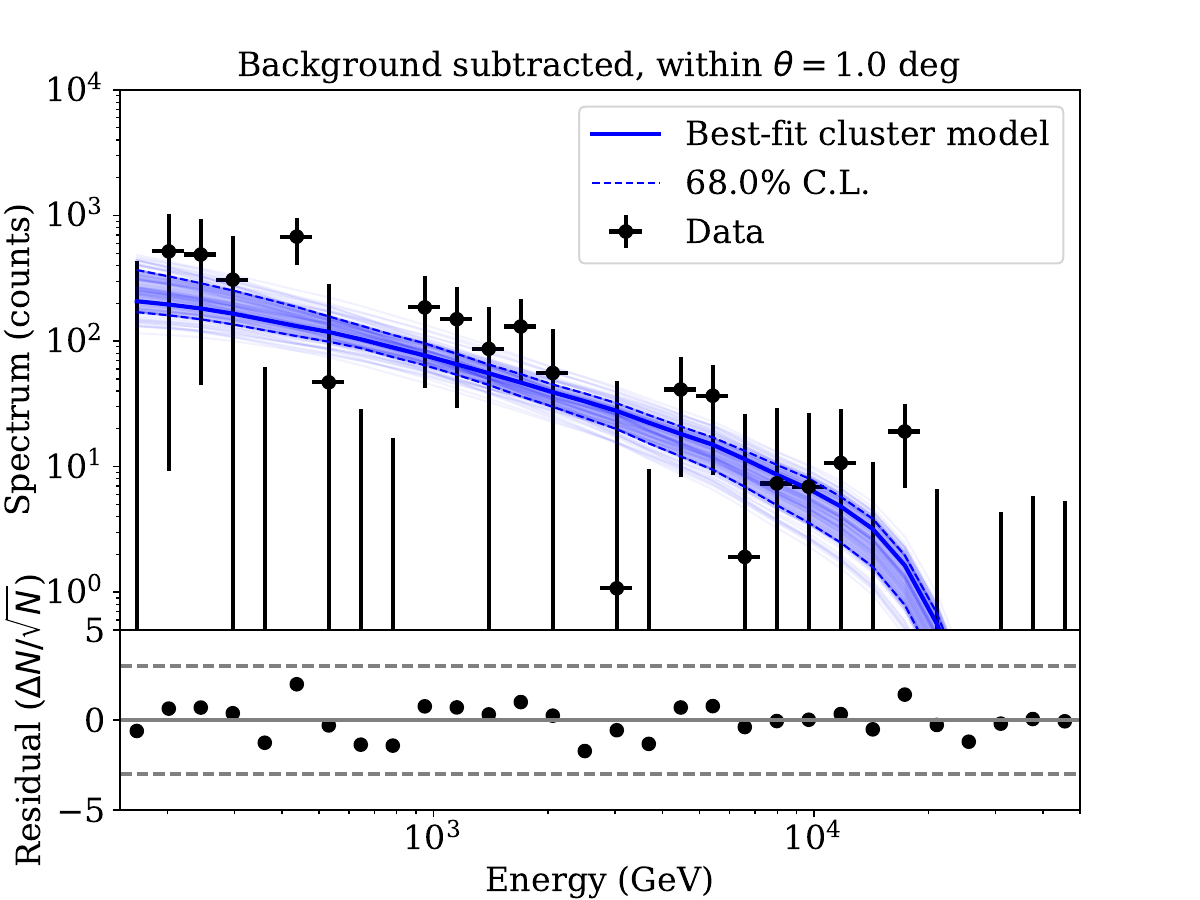}
	\includegraphics[width=0.45\textwidth]{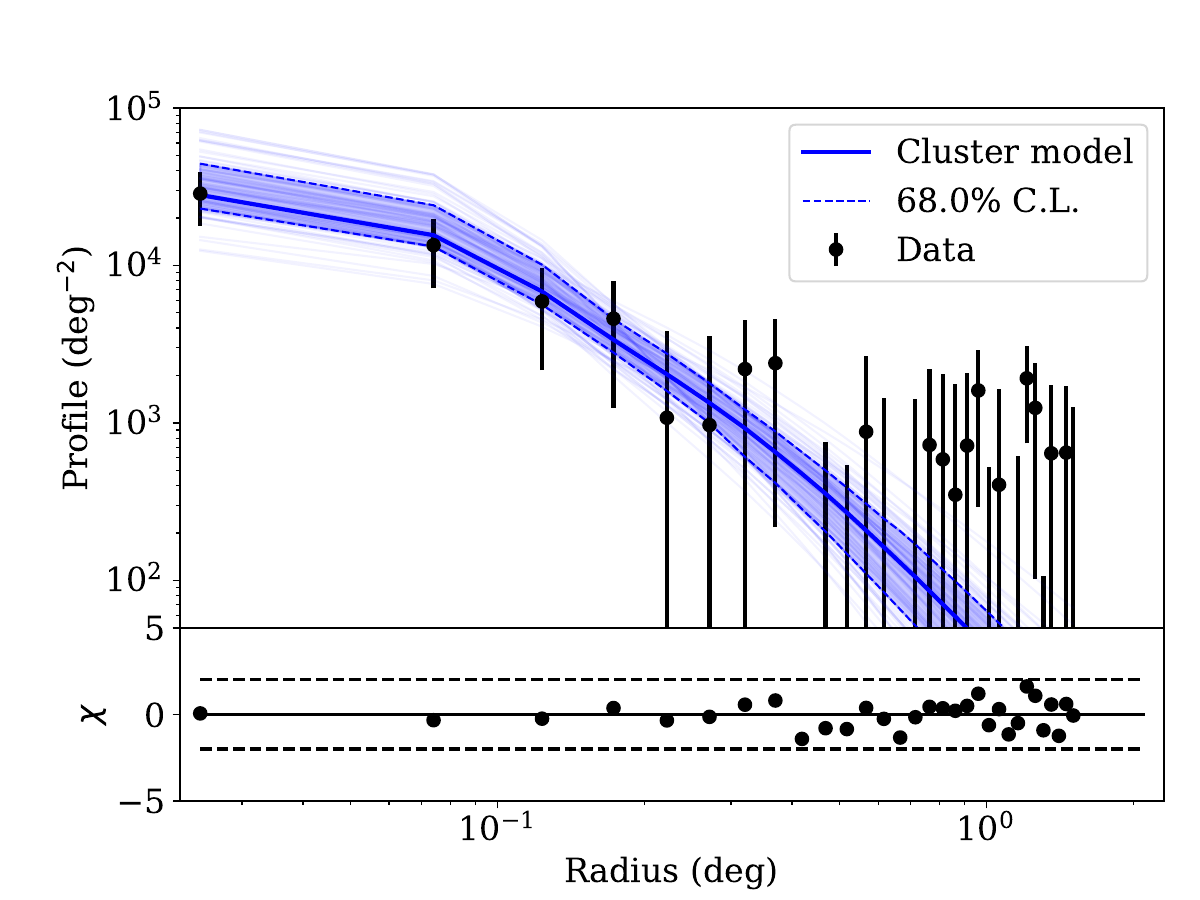}
 	\includegraphics[width=0.45\textwidth]{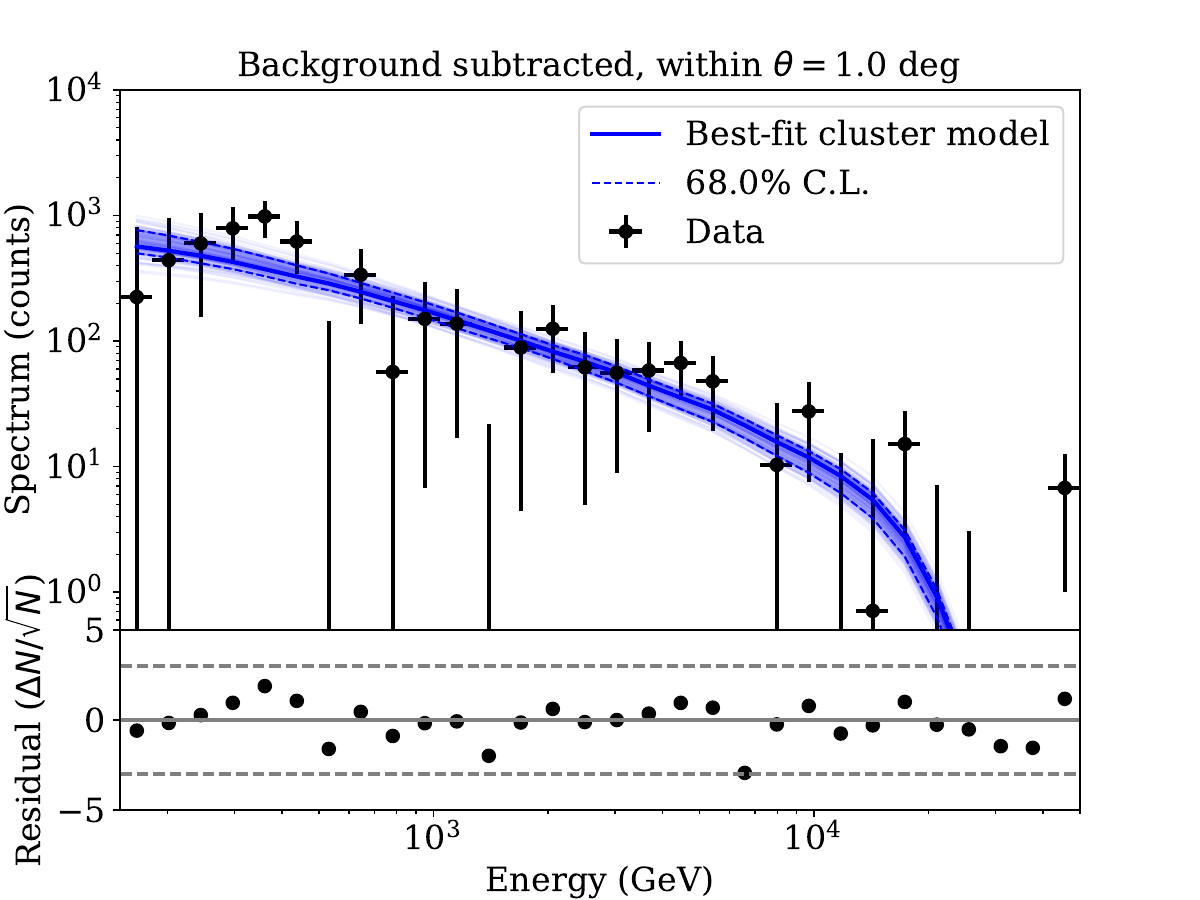}
	\includegraphics[width=0.45\textwidth]{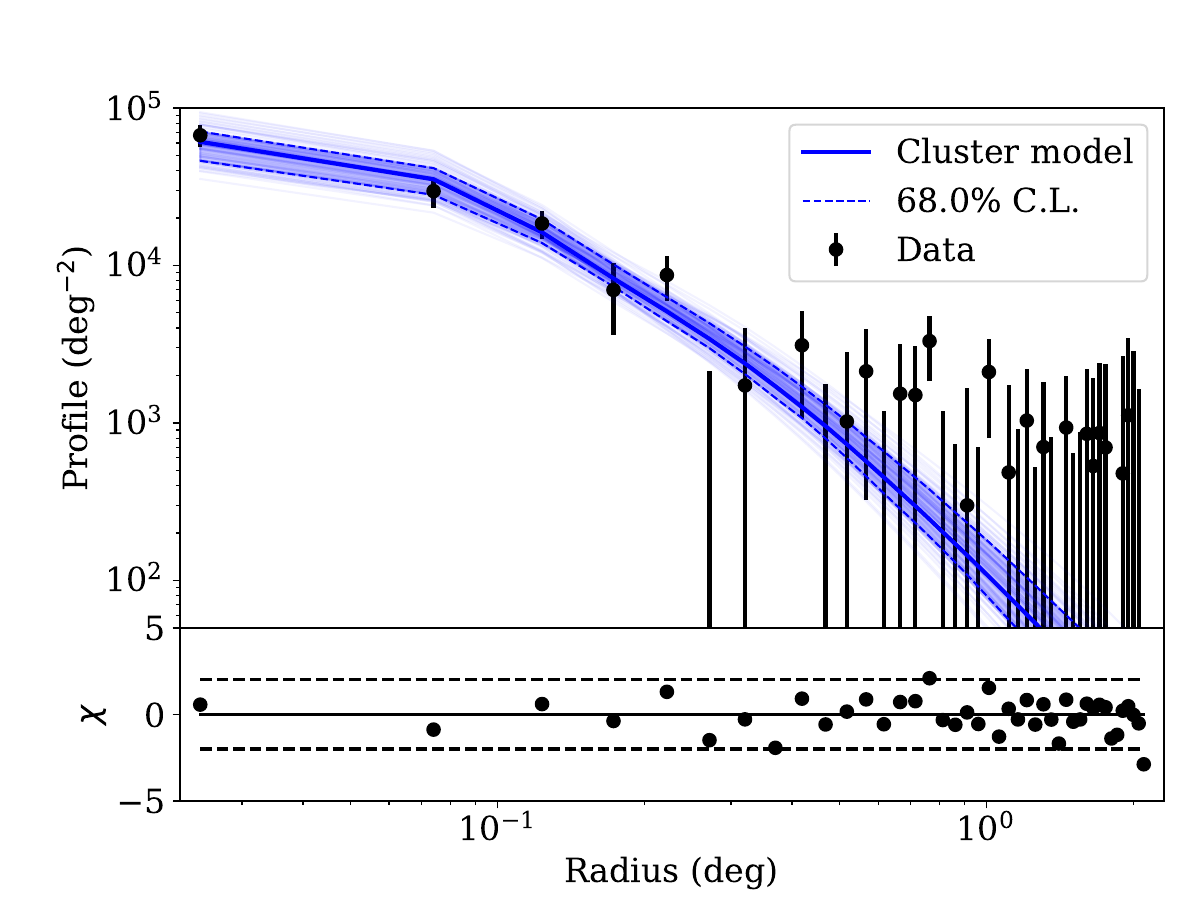}
	\caption{{\bf Left panel:} cluster diffuse emission counts recovered within an aperture radius of 1 deg, after instrumental background and point-source subtraction. {\bf Right panel:} cluster diffuse emission profile computed over all the energy bins, after instrumental background and point-source subtraction. The top panel is in the case of the baseline model and the bottom panel for the best-fit pure hadronic model.}
\label{fig:CR_residual_profile_spectrum_specimg}
\end{figure}
In Figure~\ref{fig:CR_residual_profile_spectrum_specimg}, we show the recovered cluster diffuse emission spectrum (left) and profile (right) after subtracting the point sources and the instrumental background. The spectrum is computed within 1 deg radius, which nearly corresponds to $\theta_{500}$ and the profile is computed by summing over the energy bins. The spectrum is overall well constrained over all the CTA energy range for this specific model, allowing us to recover the spectral distribution of the CRp. The cluster emission is detected on the profile up to about 0.2 deg radius (given the binning of 0.05 deg per radial bin), which allows us to constrain the shape of the signal well.

\subsection{Discussions}\label{sec:Discussions}
The CTA observations are expected to provide unprecedented constraints on the diffuse $\gamma$-ray emission from the Perseus cluster. In this section we review its sensitivity to cluster scale CR physics and discuss the assumptions of the present analysis and caveats.

\begin{figure}
	\centering
	\includegraphics[width=1.0\textwidth]{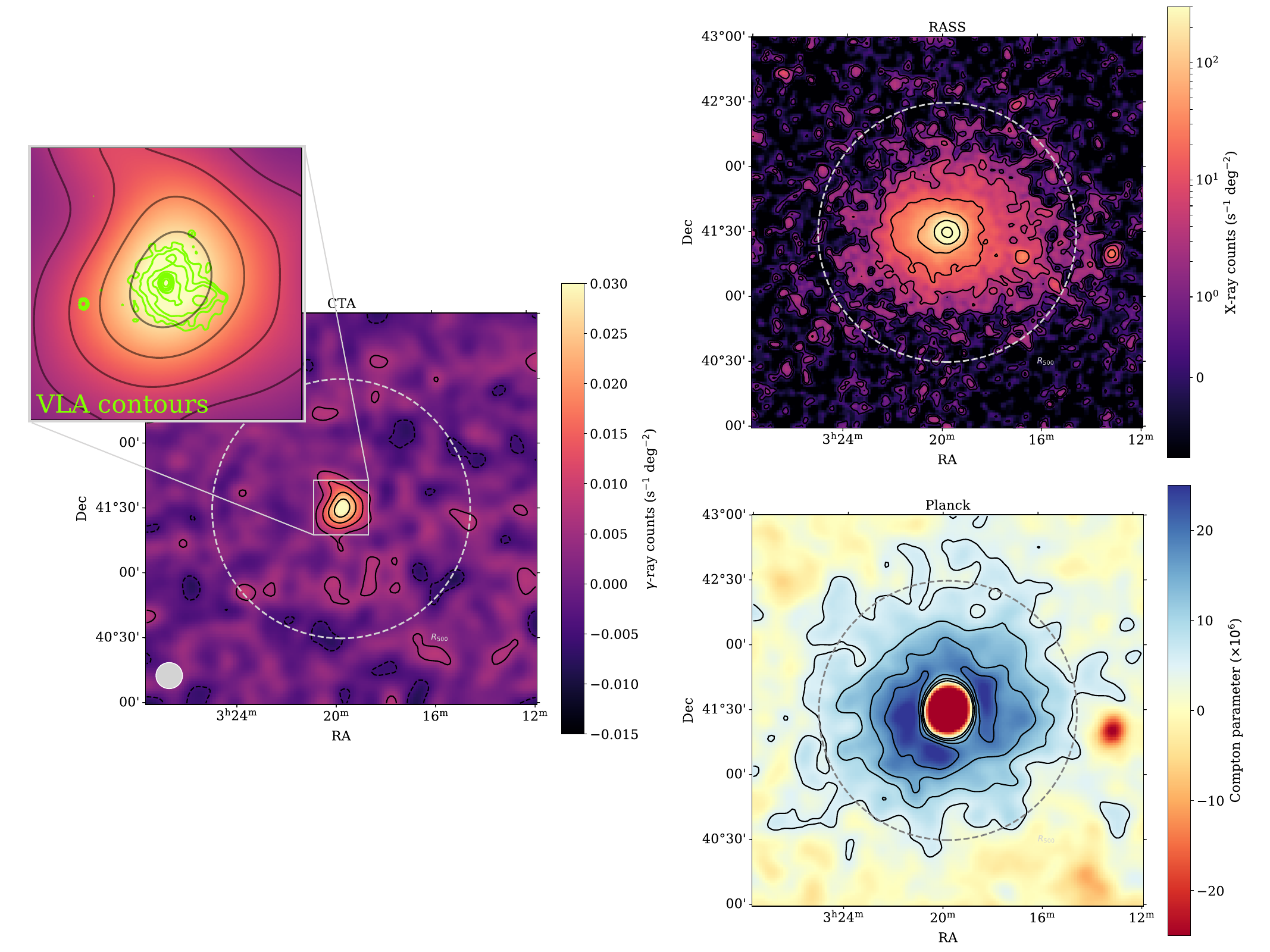}
	\caption{Comparison between the CTA simulated map (background subtracted, pure hadronic best-fit model case, with \citep{Taylor2006} plus $\eta_B=2/3$ magnetic field) and multi-wavelength data.
	{\bf Left panel:} CTA surface brightness together with signal-to-noise ratio (with 2$\sigma$ steps). The map was smoothed with a 0.15 deg Gaussian for visual purposes and the bottom left grey circle accounts for the effective map resolution. The top left subplot shows a zoom comparison between CTA and the radio mini-halo as imaged by the VLA at 1.4 GHz (extracted from the NASA/IPAC extragalactic database). VLA contours are logarithmically spaced.
    {\bf Top right panel:} ROSAT \citep{Truemper1993} X-ray image extracted from the ROSAT All Sky Survey. Contours are logarithmically spaced.
	{\bf Bottom right panel:} Compton parameter from \citep{Planck2016XXII}. The central region appears negative due to the strong contamination from NGC~1275, and another radio source is visible on the West. Contours are linearly spaced.}
\label{fig:CR_multiL}
\end{figure}

\paragraph{Toward a multi-wavelength view of the Perseus cluster from radio to $\gamma$-rays}
We first highlight how CTA observations would compare to data at other wavelengths, in particular in terms of the scales that are probed, in Figure~\ref{fig:CR_multiL}. We consider the most optimistic case that is the pure hadronic model (see Figure~\ref{fig:CR_residual_map_specimg} for the baseline model). We compare a realization of the CTA (background subtracted) image to radio data (VLA, 1.4 GHz), X-ray (ROSAT All Sky Survey), and tSZ (\textit{Planck}). The CTA signal, while slightly resolved, is expected to be relatively compact and is only detected up to about $\theta_{500}/4$. The mini-halo is also very compact (at least at 1.4 GHz), slightly extended when compared to the CTA image angular resolution. In contrast, the X-ray and tSZ images that trace the thermal gas density squared and the pressure, respectively. We note that the tSZ signal is contaminated by two radio galaxies that appear negative on the map, including NGC~1275 at the center. The $\gamma$-ray signal morphology is expected to compare well with X-ray data assuming that the CR follow the thermal gas well. In contrast, if the CR follow a distribution comparable to the temperature it would resemble more the tSZ image ($\propto P_e = n_e k_B T$).

\paragraph{The energy budget of the Perseus cluster}
The cluster ICM is dominated by its thermal component, but non-thermal contributions (turbulence, magnetic fields, cosmic-rays) should be non-negligible and are only poorly known to date. The amplitude of the CTA detection (or upper limit) will provide a direct measurement (or a further constraint) of the energy stored in the CR component, and thus the CR pressure. Noteworthy, this will also provide a measurement of one of the contributions to the hydrostatic mass bias, which is one of the key ingredients for cluster cosmology (see \citep{Pratt2019} for a review). Nonetheless, even in the case of high significance detection, the CTA constraints on this parameter would have large uncertainties due to the degeneracy with other parameters (see Figure~\ref{fig:CR_spectral_constraints} and \ref{fig:CR_triangle_plot_basline}), unless strong priors are available for the shape and the spectrum of CRp via other wavelengths. This could be the case thanks to high quality radio observations, assuming specific scenario (e.g., pure hadronic model). The use of lower energy data, such as \textit{Fermi}-LAT, could also help breaking the degeneracy between the parameters of the cluster components, but also provide a useful prior on the spectrum of NGC~1275. The measurement of the CR to thermal energy will be easier in the case of a harder CRp spectrum and a more compact profile.

\paragraph{Cosmic-ray acceleration mechanism and transport}
If the cluster is detected, the CTA data will allow us constraining the spectral index of the CRp that induce the $\gamma$-ray emission. The CRp spectrum (and amplitude) reflects the acceleration mechanism at play and gives estimates of the CR acceleration efficiency. Such measurement will therefore have important implications to discriminate between the various models that remain poorly constrained to date \citep{Brunetti2014,Bykov2019}. The CTA sensitivity to diffuse emission will, however, loosen as $\alpha_{\rm CRp}$ increases. For instance, assuming that the CR spatial distribution follows that of the thermal gas, it will not be possible to detect the cluster for a CR slope larger than 2.6 if the CR to thermal energy ratio is lower than about 5\% (Figure~\ref{fig:CR_95percent_limits_X} and~\ref{fig:CR_95percent_limits_F}). Assuming that the CR pressure support within $R_{500}$ is less than 20\%, as can be reasonably expected (e.g., \citep{Pratt2019}), CTA will be limited to probe $\alpha_{\rm CRp} < [2.6, 2.9]$ depending on the spatial distribution ($\eta_{\rm CRp} \in [0, 1.5]$).

Thanks to the resolved observations provided by CTA, the data will also be sensitive to the spatial profile of the $\gamma$-ray emission, which is related to the spatial distribution of the CRp in the cluster. In turn, the cluster CR profile will give precious information about the production sites and transport processes (advection and turbulent motion imply a more compact profile, while diffusion and streaming flatten it). In this paper, we have chosen to model the CR profile using a scaling with respect to the thermal gas density, which allows us to test the effect of transport via a single parameter, $\eta_{\rm CRp}$. We stress that this choice is not unique and the present work could be extended to test any model. However, given the current limited knowledge about the CR production and transport in the ICM, simple models might already be sufficient for such investigations. The precision on the CRp spatial distribution will depend on CR to thermal energy ratio, the spectral distribution of CR, and the shape of the CR profile itself since a more compact distribution will lead to a larger signal-to-noise ratio. For hard spectra, CTA should be able to detect the diffuse Perseus emission even in the case of flat CR distributions, provided that the CR to thermal energy ratio is sufficient ($X_{500} \gtrsim 10^{-2}$).

\paragraph{Probing the time integrated AGN and starburst activity in Perseus}
In this paper, we have considered that the cluster ICM was filled with CRp following a given spectral and spatial distribution, regardless of their detailed origin. While they may arise from shocks and turbulence, it is also possible that accumulated injection from compact sources play an important role \citep{Rephaeli2016}. Therefore, the CTA observations will be sensitive to the time integrated AGN cosmic-ray outburst and starburst activity. Since the number of AGNs and starbursts in the Perseus cluster can be estimated, this will provide a unique way to estimate the average energy injection via CRp from such sources in galaxy clusters. The level of $\gamma$-ray emission will depend on how CR from AGN and starbursts are processed, so that acceleration, transport and injection are closely linked to each other.

\paragraph{Constraints on the magnetic field strength in the pure hadronic model}
The cluster non-thermal diffuse emission that we aim at probing with CTA is physically connected with the diffuse synchrotron emission observed at radio wavelengths, the latter being produced by the entangled combination of CRe and magnetic field (see Section~\ref{sec:non-thermal_radiative_processes}). Therefore, it is in principle possible to constrain the magnetic field strength using the joint analysis of radio and $\gamma$-ray data, assuming a given physical scenario such as the pure hadronic model (as done for the Coma cluster \citep{Brunetti2012,Brunetti2017}). In the present analysis, we have assumed a fixed magnetic field model (\citep{Taylor2006} plus $\eta_B=2/3$ as a reference, with $<B(10{\rm kpc})>\sim 25$ $\mu$G) when considering the pure hadronic scenario. For instance, neglecting uncertainties arising from the radio data, our reference magnetic field model implies a test statistic $TS=125$ ($\equiv 11 \sigma$ detection) for CTA in the pure hadronic scenario. Lowering the $\gamma$-ray emission, so that the cluster is not detected, would imply lowering the amount of secondary CRe by the same amount. The magnetic field can then be adjusted so that the radio synchrotron emission remains the same.

In Appendix~\ref{app:impact_of_magnetic_field}, we show how the different models of magnetic field considered in this paper affect our model parameters and translate into $\gamma$-ray emission (Table~\ref{tab:model_summary}). For example, the model giving the lowest magnetic field profile, by a factor of about 2.5 within 100 kpc from the center compared to our reference, implies a normalization that is 60\% larger with nearly unchanged CR slope and spatial distribution, and therefore an increased $\gamma$-ray emission by about 60\% as well. This shows that in turn, constraints can be obtained on the magnetic field strength when fixing the radio and $\gamma$-ray observables. Given the current uncertainties in the radio spatial and spectral distribution of the Perseus cluster, we only provide a qualitative discussion. However, we note that current and future radio observations\footnote{e.g., with the Square Kilometer Array (\url{https://www.skatelescope.org/}) and its precursors.} should provide an unprecedented measurement of the diffuse radio emission from the Perseus cluster, which could then be combined with CTA observations to measure the magnetic field. This could be further combined with independent measurements of the magnetic field (e.g., Faraday rotation measure) to test CR physics and magnetic field in the ICM and particle acceleration mechanisms.

\paragraph{Analysis choices and CTA configuration}
The analysis presented in this paper rely on several choices that are expected to affect our sensitivity results. We discuss the most relevant ones hereafter.

Most of the analysis described in this paper has been performed using a template fitting analysis. Indeed, it was necessary in order to not only test for the detection of the cluster, but also model and constrain the spectral and spatial distribution of the emission. Because it uses more information than classical ON-OFF technique, the template fitting analysis is expected to provide more accurate constraints. On the other hand, it might be more sensitive to systematic effects due to the mismodelling of the data. 

We decided to select events in the energy range between 150 GeV and 50 TeV. While the exact value of the upper bound does not affect the results, because no signal is expected above $\sim 30$ TeV, reducing the minimal energy is expected to increase statistics and improve the CTA constraints. On one hand, we use a conservative value that is relatively large compared to what can be expected with CTA. On the other hand, low energy data may be more difficult to model and in addition to the instrumental background, NGC~1275 strongly dominates at low energies. Therefore, it is likely that the improvement in sensitivity will not be major when pushing to lower energy thresholds.

We conservatively selected events within a $3 \times 3$ deg$^2$ region around the cluster. While no significant cluster diffusion emission is expected beyond this region, it is likely that the corresponding data will help constraining the instrumental background, provided that it can be properly modeled. This might, in turn, help breaking the cluster - instrumental background degeneracy and slightly improve the CTA constraints.

A new version of the CTA IRFs has been released since the analysis presented here was completed. They correspond to the initial array configuration, made of 4 LST plus 9 MST (the so-called CTAO Alpha configuration, prod5-IRFs) instead of the one with 4 LST plus 15 MST (prod3-IRFS) used here. The reduction in sensitivity is mostly effective at high energy ($E \gtrsim 500$ GeV), by a factor of up to 30\%, but improved off-axis sensitivity. Therefore we would expect the results presented in this paper to be affected by such sensitivity reduction. Assuming the initial array configuration would imply a reduction of the upper limits (Figure~\ref{fig:CR_95percent_limits_X} and~\ref{fig:CR_95percent_limits_F}) and a widening of the uncertainties on the cluster recovered parameters (Figures~\ref{fig:CR_spectral_constraints} and \ref{fig:CR_triangle_plot_basline}) by up to 30\%, although the improved off-axis sensitivity should mitigate this effect. Cluster models with harder spectra (small $\alpha_{\rm CRp}$) should be the most affected because a larger part of their flux arises from the high energy part of their spectrum.

\section{CTA sensitivity to DM induced $\gamma$-ray emission from Perseus}\label{sec:CTA_DM_sensitivity}

In this section we investigate the sensitivity of CTA to the DM emission model presented in Section~\ref{sec:DM_modelling} for the cluster. The results shown in the following have been obtained using the \texttt{gammapy} (v.0.18.2) open-source code \citep{gammapy:2017} \footnote{All the notebooks and scripts developed for the DM analysis can be obtained in \url{https://github.com/peroju/dmtools_gammapy}.}. 


Perseus represents a complex environment where multiple $\gamma$-ray sources coexist. Indeed, with the current generations of IACTs, these astrophysical sources have posed large difficulties to the search for a putative DM signal in the area, as it is hard to avoid `contamination' from the astrophysical sources and to disentangle between these and DM, both spectrally and spatially. In contrast, due to its expected improvement in sensitivity, angular resolution and larger FoV, CTA will allow for a much better discrimination by means of analysis techniques that were traditionally discarded for this kind of telescopes and used only for all-sky $\gamma$-ray facilities such as \textit{Fermi}-LAT. In particular, in our DM search we perform a template-based analysis, which allows to take into account the rest of expected $\gamma$-ray emissions as backgrounds, and is in line with the analysis previously performed in Section~\ref{sec:CTA_CR_sensitivity} for CRs and also with recent studies in the Galactic center region by the CTA Consortium \citep{CTA:2020qlo}. This type of analysis, although computationally expensive, is expected to provide not only the highest DM sensitivity but also to yield the most realistic one\footnote{There is an on-going debate on the possible introduction of bias due to the use of theoretical models in the template-based methods.}. Yet, mainly for the sake of comparison with previous IACT works and in order to provide a first-order sensitivity study in our work as well, we also carry out the so-called ON-OFF analysis, i.e. the most standard analysis strategy adopted by some current IACTs. Full details of such analysis are left for Appendix~\ref{app:DM_on_off} with the corresponding comparison with template-based results.

To perform the simulation of the DM-induced $\gamma$-ray emission from Perseus, we adopt the observation setup described in Section~\ref{sec:Observation_setup}, i.e. a pointing with one degree offset from the cluster center. Similarly to the CR analysis, we stack the data from the different pointings in a single 3D cube, containing the proposed 300h observation time. We adopt ten energy bins, starting from 50 GeV up to 100 TeV. This allows us to exploit the whole range of CTA's sensitivity as well as to overlap with the DM parameter space previously explored by the \textit{Fermi}-LAT (at the lowest considered energies) and by the H.E.S.S and MAGIC telescopes (in the TeV regime). We use a spatial binning of 0.02 deg per pixel and our field of view diameter is set to 5 deg.

\begin{figure}[h!]
	\centering
	\includegraphics[width=0.49\textwidth]{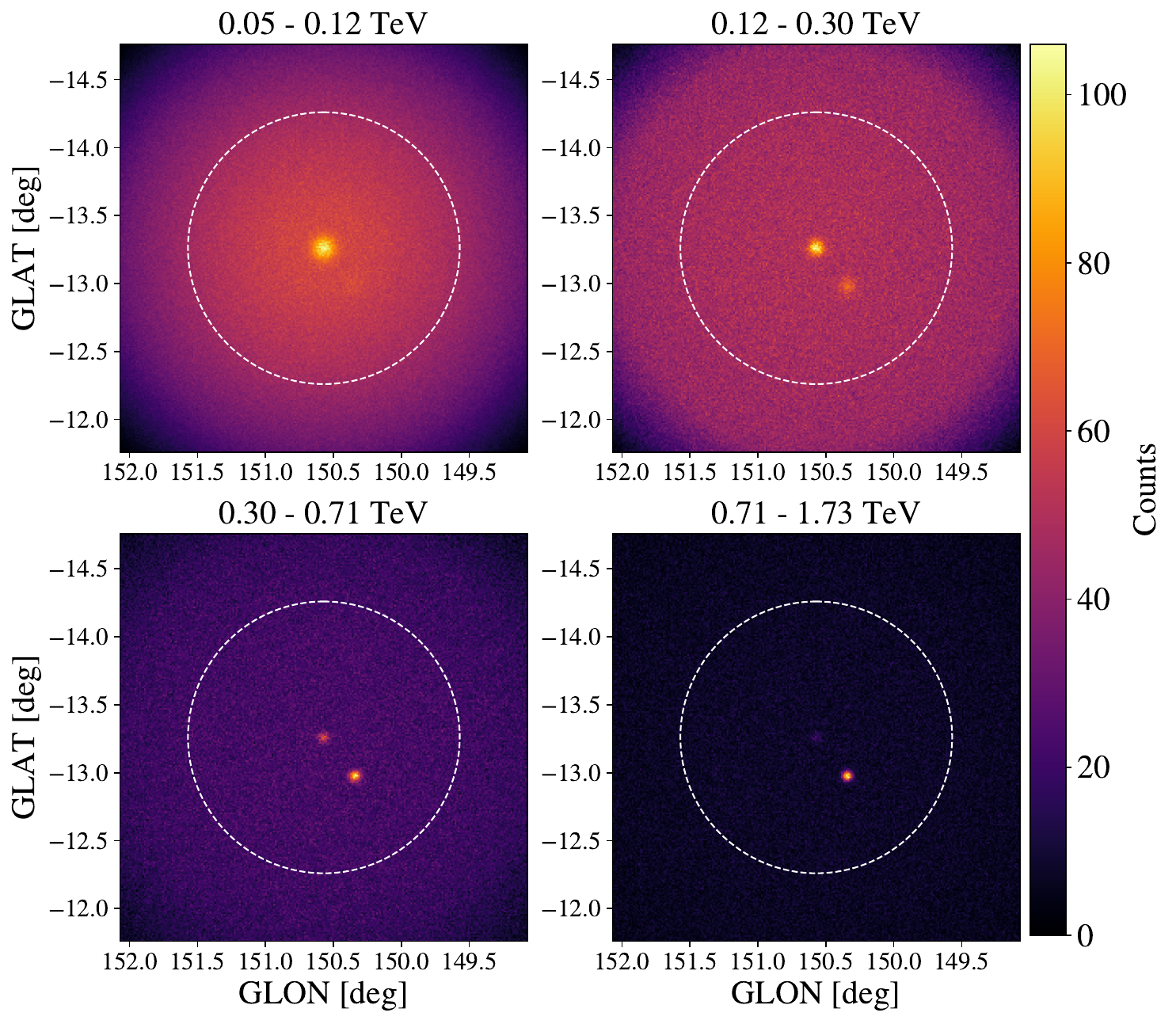}
	\includegraphics[width=0.49\textwidth]{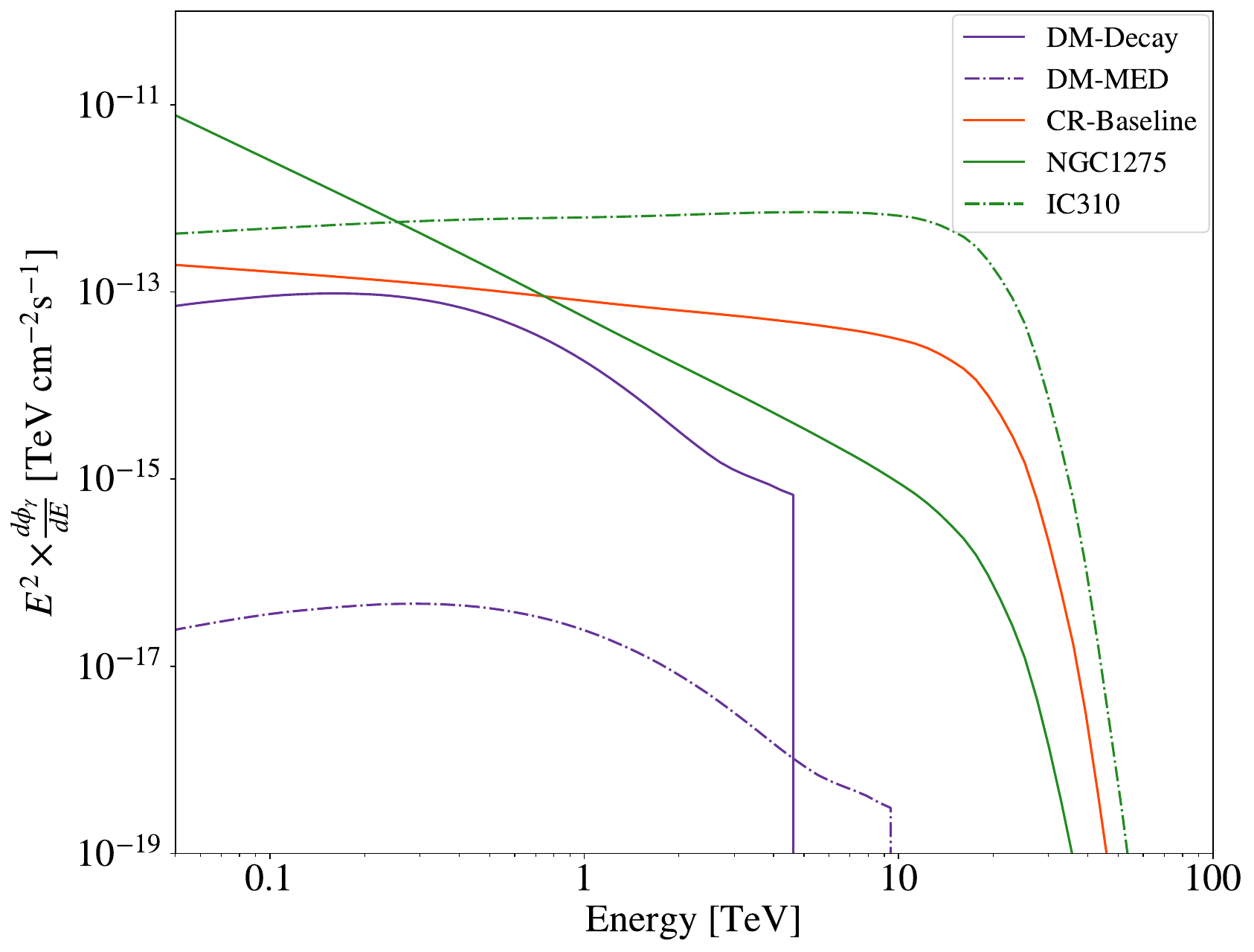}
	\caption{\textbf{Left panel:} Simulation of an observation of the Perseus region with CTA. The color informs on the expected counts. This example is for the MED annihilation model, with a DM particle of $m_{\chi}=10$ TeV,  $b\bar{b}$ channel and $<\sigma v>=3\times10^{-26}$ cm$^{2}$ s$^{-1}$. We adopted the `Baseline' model for the CR-induced emission (Section~\ref{sec:CR_baseline_model}) and included the two AGNs in the region via the descriptions in Equations~\ref{eq:spectrum_NGC1275} and \ref{eq:spectrum_IC310}. We only show the first four energy bins of our analysis, each of them corresponding to one panel. The white dashed circle goes over the different proposed pointings (all of them at 1 deg from the cluster's center). \textbf{Right panel:} Comparison of the spectra of the different simulated $\gamma$-ray components integrated up to $R_{200}$. We also include the spectrum of a $m_{\chi}=10$ TeV DM particle decaying via the $b\bar{b}$ channel. We recall that all these emissions do not happen through the entire CTA FoV.
    }
\label{fig:DM_simulations_spectrum}
\end{figure}

To model the DM distribution in the cluster, we use the 2D templates described in Section~\ref{sec:DM_fluxes} for each of the three benchmark models in the case of annihilation and for the one of decay. For annihilation, we simulate a putative $\gamma$-ray signal assuming a branching ratio of 100\% to the $b\bar{b}$ annihilation channel, a DM mass $m_{\chi}=10$ TeV\footnote{The effect of the EBL on the DM simulated spectrum will be much more significant if we consider $m_{\chi}>30$ TeV, in which case the characteristic DM cutoff would be substituted by the EBL cutoff due to $\gamma$-ray attenuation.} and a velocity-averaged cross-section matching the thermal value of $<\sigma v>=3\times 10^{-26}$ cm$^{3}$s$^{-1}$. For decay, we adopt $\tau_{\chi}=10^{27}$ s as the DM particle lifetime. The expected $\gamma$-ray contribution from the existing AGNs in the area is accounted for via the models described in Section~\ref{sec:background_sky}, while the CR-induced $\gamma$-ray emission is described using the spectral and spatial template corresponding to the `Baseline' benchmark model (Section~\ref{sec:CR_baseline_model}). We remark that those $\gamma$-rays originated from both the AGNs and CRs are considered as backgrounds in our DM search, together with the CTA instrumental background modelled by the IRFs. In total, we create four sets of simulations, one per each DM template (MIN, MED, MAX and DEC - see Table~\ref{tab:benchmark-models}), and always adopting the same combination of backgrounds. An example of a simulation can be seen in Figure~\ref{fig:DM_simulations_spectrum} for the MED annihilation model. We also compare in the same figure the spectra of the different $\gamma$-ray sources in Perseus to the ones for annihilation and decay. This comparison shows that in the annihilation scenario the $\gamma$-ray flux is expected to be orders of magnitude below the astrophysical backgrounds for realistic cross-section values. Yet, in the case of decay, we can have comparable fluxes (adopting $\tau_{\chi}=10^{27}$ s, which is indeed very close to current constraints, \citep{2012ApJ...761...91A}). 

Finally, in order to obtain statistically meaningful results, we create 100 different simulations with the same observation setup, and use the corresponding mean for the data analysis itself. A study of the stability of our simulation results with respect to the number of simulations is included in Appendix~\ref{app:DM_gammapy_convergence}, which shows that a reasonable compromise in terms of convergence and computation time is indeed achieved after a few tens of simulations in almost all cases. Thus, in total, we produce 400 simulations, 100 per each benchmark DM model (MIN, MED, MAX and DEC), which we now proceed to analyze.

\subsection{Template-fitting analysis and DM sensitivity}\label{sec:DM_template_fitting}

To search for DM emission, we first define for convenience the normalization of the DM-induced $\gamma$-ray flux $A_{\chi}$, as $<\sigma v> = A_{\chi}\times <\sigma v>_{thermal}$, for the three annihilation models, and $\tau_{\chi} = \tau_{ref}/A_{\chi}$ (Equation~\ref{eqn:flux-scenarios}), where $\tau_{ref}=10^{27}$ s, for the decay scenario. We fit this normalization for each simulation using the \texttt{iminuit} \citep{iminuit} backend of \texttt{gammapy} to the theoretical emission of different DM particle candidates. To cover the available mass range for WIMPs, we perform the fit to fourteen DM masses across the defined energy range and for two representative annihilation/decay channels, i.e. $b\bar{b}$ and $\tau^+\tau^-$\footnote{Initially we also included the $W^+W^-$ channel and noticed that it produced similar results to $b\bar{b}$, as expected, so we decided not to show it in the following for the sake of clarity.}.

Together with the DM normalization, we apply a joint-likelihood fitting including the relevant parameters of the considered astrophysical backgrounds. For example, since the CR-related flux is considered as a background for this analysis, we only allow for an overall normalization on the CR-induced $\gamma$-ray flux for the `Baseline' model. Then, the complete flux model to fit the simulations is defined, in spatial and energy bin $i$, as:
\begin{equation}
M_i(\vec{\theta}) = {\rm DM}_i\left(A_{\chi}\right) + {\rm CR}_i\left(A_{\rm CR}\right) + 
\sum_{j \in [1,2]} {\rm PS}_i^{(j)}(A_{\rm PS}^{(j)},\alpha_{\rm PS}^{(j)}) + B_i(A_{\rm bkg},\alpha_{\rm bkg}).
\label{eq:DM_model_cube}
\end{equation}
In total, we fit eight different parameters:
\begin{equation}
\vec{\theta} \equiv \left(A_{\chi}, A_{\rm CR}, A_{\rm PS}^{(1,2)}, \alpha_{\rm PS}^{(1,2)}, A_{\rm bkg},\alpha_{\rm bkg}\right),
\label{eq:DM_parameters}
\end{equation}
where all the parameters have been previously defined (see Section~\ref{sec:Joint_spectral_imaging_constraints} and Equation~\ref{eq:CR_model_cube}). The likelihood function is the one corresponding to Poissonian data (Equation~\ref{eq:likelihood_def}). We also adopt a flat prior on $A_{\chi}$, to allow only for zero or positive values\footnote{This method is usually known as the bounded likelihood method \citep{Rolke:2004mj}.}. This joint-likelihood fit will enable to check for intrinsic correlations and to account for them when estimating the size of the uncertainties. To obtain the detection significance of the DM-induced $\gamma$-ray emission, we apply the Test Statistics as defined in Equation~(\ref{eq:TS_definition}). We define a detection when $TS=25$, which roughly corresponds to $5\sigma$.

In neither of the MIN, MED, MAX annihilation or decay scenarios, and for none of the channels considered, a statistically significant detection is predicted. We proceed to compute the 95\% confidence level (C.L.) upper limits (lower limits for decay) of the DM normalization parameter $A_{\chi}$ using the likelihood profile method \citep{Rolke:2004mj, Rico:2020vlg}. For the DM normalization, we assume a one-sided distribution (we recall the flat prior $A_{\chi}\geq0$), thus the 95\% C.L. corresponds to a change of $\Delta TS = 2.71$, with respect to the best fit.

\subsection{Projected sensitivity to annihilating DM}
\label{sec:DM_result_annihil}
Assuming the templates and spectra for DM annihilation (Section~\ref{sec:DM_modelling}), we compute the 95\% C.L. upper limits for $A_{\chi}$ for all considered cases, convert this value to $<\sigma v>$ (the standard parameter to be constrained for annihilating DM) and average the results for 100 simulations. 

\begin{figure}
	\centering
	\includegraphics[width=0.49\textwidth]{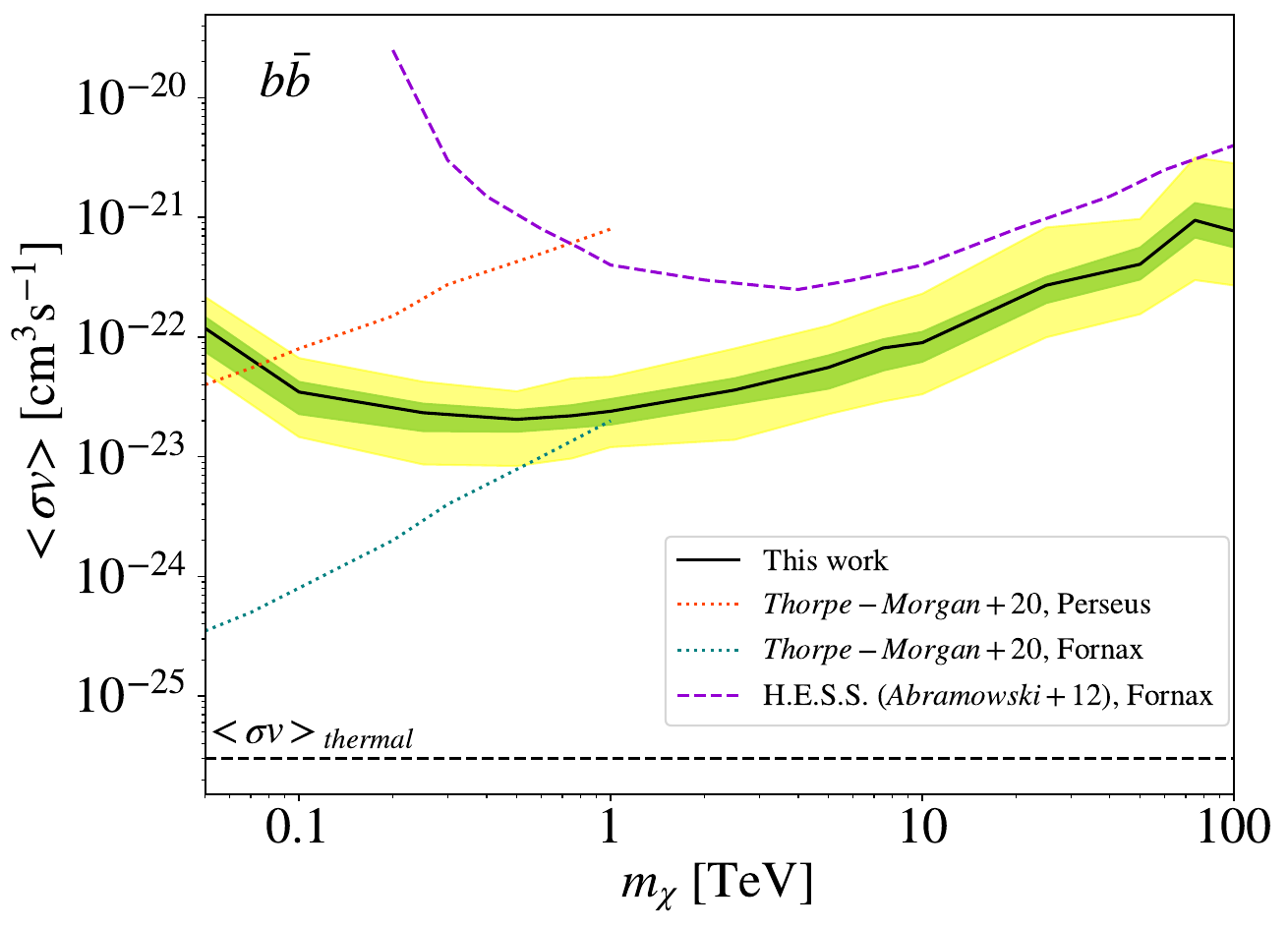}
	\includegraphics[width=0.49\textwidth]{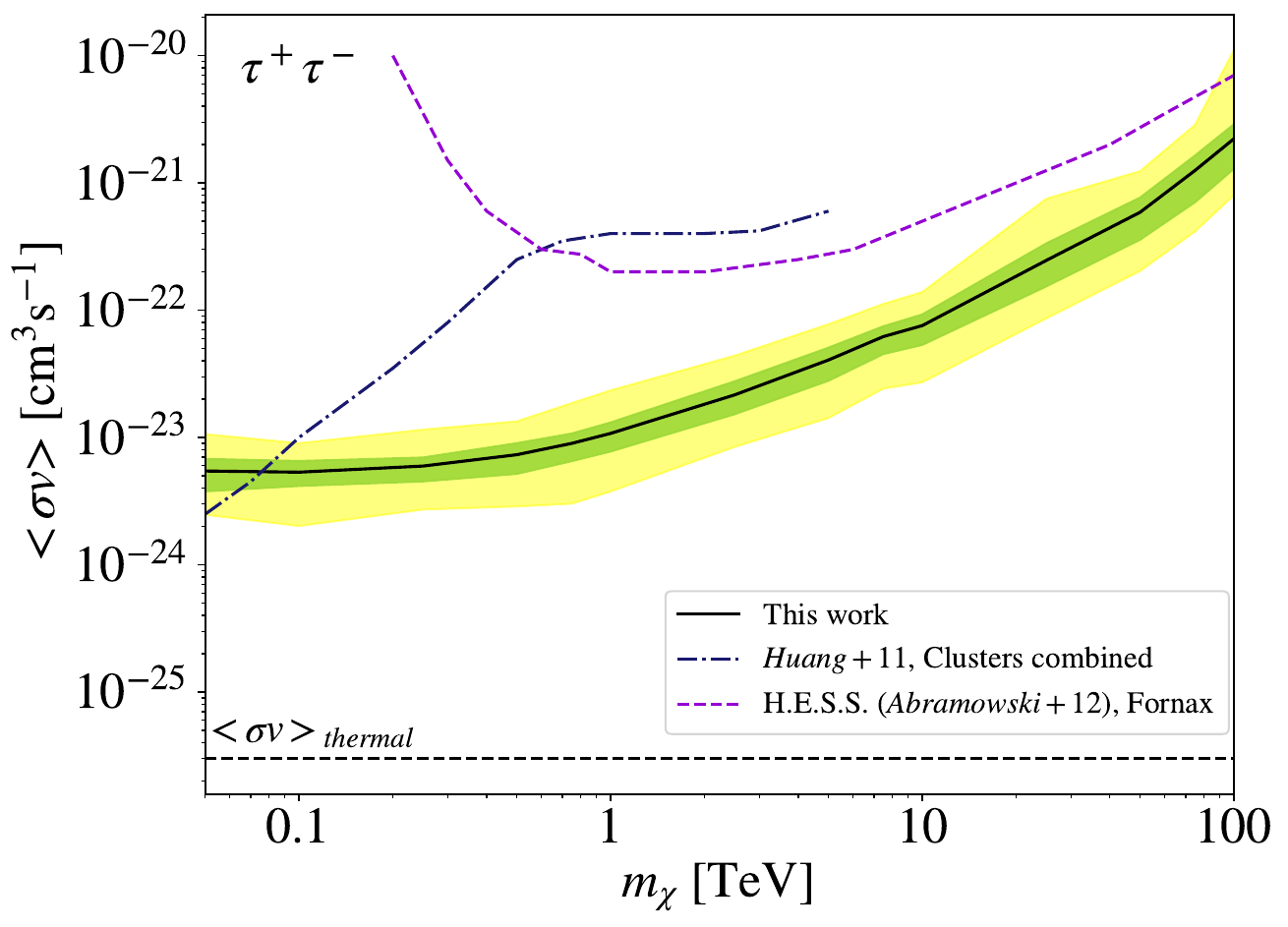}
	\caption{Sensitivity of CTA to a DM annihilation signal from the Perseus cluster. Curves represent the 95\% C.L. upper limits on the velocity-averaged cross-section versus the DM mass for the MED annihilation model. The green (yellow) band shows the $1\sigma$ ($2\sigma$) scatter of the projected limits. The black dashed line is the thermal relic cross-section ($<\sigma v>_{thermal} = 3\times10^{-26}$ cm$^{3}$ s$^{-1}$). \textbf{Left panel:} Cross-section upper limits for the $b\bar{b}$ channel (\textbf{right panel} for $\tau^+\tau^-$ channel) in comparison with the most recent results on DM-annihilation searches in galaxy clusters using $Fermi$-LAT (\citep{Thorpe-Morgan:2020czg} with orange dotted lines; \citep{Huang:2011xr} with blue dot-dashed lines) and H.E.S.S (\citep{2012ApJ...750..123A}, with purple dashed lines). All of our results are shown for the template fitting analysis, unless stated otherwise.
	}
\label{fig:DM_annihil_results}
\end{figure}

In Figure~\ref{fig:DM_annihil_results}, we show the CTA projected sensitivity as 95\% C.L. upper limits on the velocity-averaged cross-section versus the DM mass, for the MED annihilation model and the two considered representative annihilation channels. As it can be seen, the most constraining limits are obtained for the $\tau^+\tau^-$ channel, although also the $b\bar{b}$ channel yields similar results for DM masses above $\sim$10 TeV. In all cases, our exclusion limits are more than $\sim\mathcal{O}(10^2)$ above the thermal relic cross-section value. Yet, we note that they will be the most constraining ones (in the TeV energy range) considering galaxy clusters as DM targets. This can also be seen in both panels of Figure~\ref{fig:DM_annihil_results}, where we show a comparison to recent works. 
The latest cluster DM limits obtained by currently operating IACTs come from the observation of the Fornax cluster (14.5 hours) by the H.E.S.S Collaboration \citep{2012ApJ...750..123A}, and are around one order of magnitude weaker than our CTA predictions at a few TeV (i.e. at the peak of H.E.S.S. sensitivity). 
We also compare our CTA predictions with limits from \textit{Fermi}-LAT in the sub-TeV WIMP mass range. In \citep{Thorpe-Morgan:2020czg}, authors analyze 12 years of \textit{Fermi}-LAT data for a sample of five clusters and, in the absence of a signal, set constraints only for the $b\bar{b}$ channel. Perseus is the cluster yielding their weakest limits while Fornax gives the most constraining ones. The CTA sensitivity predictions for Perseus are always better than the obtained in \citep{Thorpe-Morgan:2020czg} in the whole range but below 100 GeV, where both results are still comparable. But even for Fornax, CTA's improvement in sensitivity will allow us to set the tightest DM constraints for DM masses above 1 TeV. 
In the case of the $\tau^+\tau^-$ annihilation channel, we can compare our results with those in \citep{Huang:2011xr}, where authors combine 3 years of \textit{Fermi}-LAT data from 8 galaxy clusters to derive the corresponding DM limits. As a consequence of using a considerable reduced amount of data compared to \citep{Thorpe-Morgan:2020czg}, their limits are considerably weaker in comparison and cannot compete to those from CTA. In conclusion, CTA DM limits from Perseus will be the most constraining ones from these class of targets in the TeV energy range.

\begin{figure}[h!]
	\centering
	\includegraphics[width=0.7\textwidth]{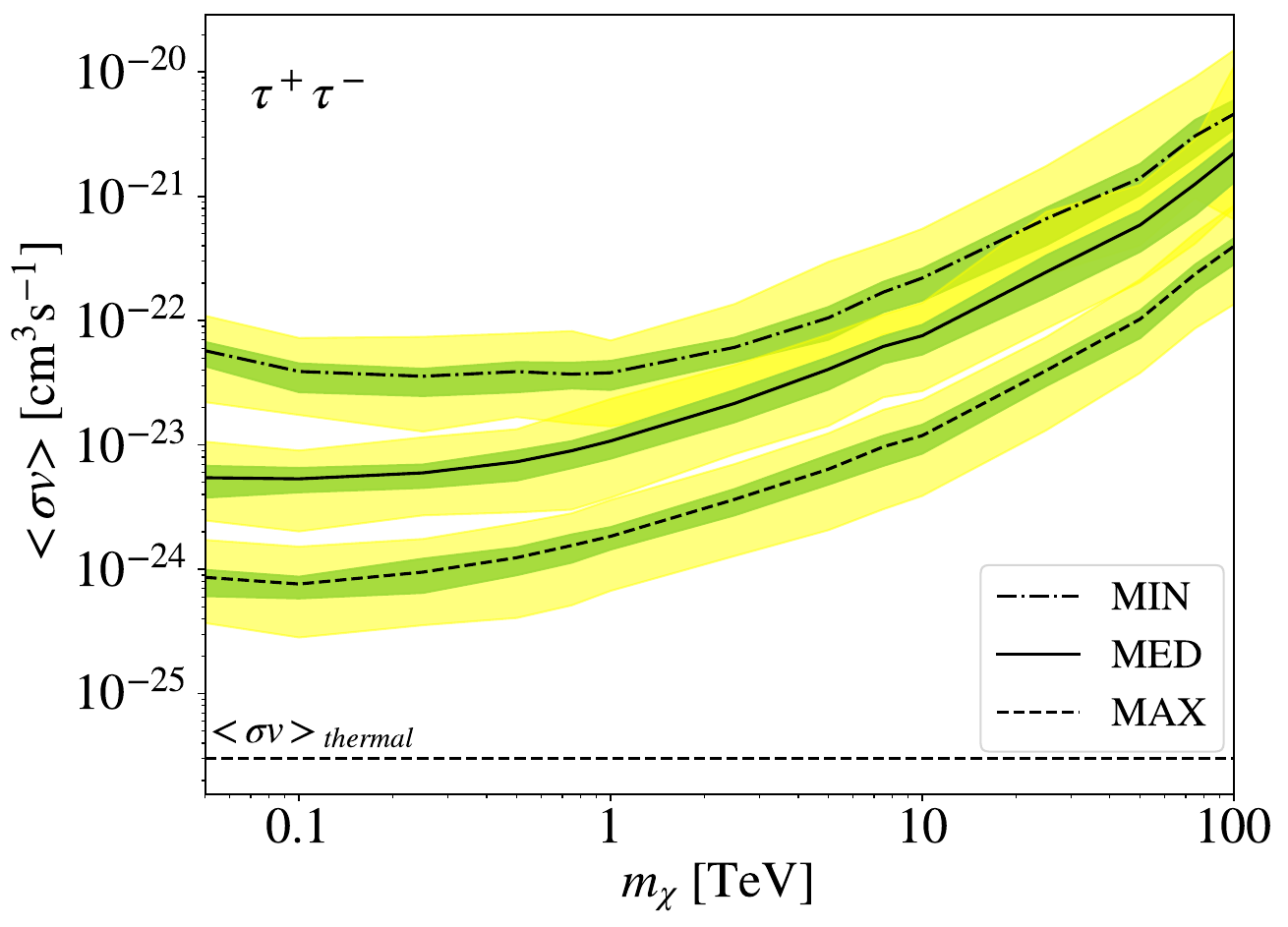}
	\caption{95\% C.L. upper limits of the velocity-averaged cross-section versus the DM mass for the MIN (dot-dashed line), MED (solid) and MAX (dashed) (see Table~\ref{tab:DM-fluxes}) subhalo models and the $\tau^+\tau^-$ annihilation channel. The green (yellow) band represents the $1\sigma$ ($2\sigma$) scatter of the projected limits. The black dashed line represents the thermal relic cross-section ($<\sigma v>_{thermal} = 3\times10^{-26}$ cm$^{3}$ s$^{-1}$). All of our results are shown for the template fitting analysis, unless stated otherwise.}
\label{fig:DM_annihil_results_subs}
\end{figure}

By far, the largest uncertainty on the obtained cross-section upper limits is the modelling and contribution of cluster DM subhalos to the total $J_{ann}$-factor. This was the reason to build three different annihilation benchmark models in the first place (MIN, MED and MAX -- Section~\ref{sec:DM_substructure_model}), each of them representative of very diverse levels of substructure and their contribution to the annihilation flux. To properly quantify the impact of this uncertainty in our limits, in addition to the 95\% C.L mean upper limit for the MED case presented above, we also compute limits for the MIN and MAX benchmark models. These results are shown in Figure~\ref{fig:DM_annihil_results_subs} for the $\tau^+\tau^-$ annihilation channel, and reveal that our limits can be modified substantially depending on the considered subhalo scenario. In particular, the \textit{boost factor} associated to our MED model ($B_{MED}=9.2$) translates into an improvement of a factor $\sim7$ with respect to the MIN benchmark model. In the case of the MAX model ($B_{MAX}=59.3$), the limits improve up to $\mathcal{O}$(10) times the MED model constraints.\footnote{Note that the improvement in the limits takes into account the precise impact of the subhalo boost on the spatial morphology of the DM signal as well; thus the relation of this improvement with the boost value is not linear, as naively expected.} Previous studies have also accounted for the subhalo contribution in their limits. In \citep{Thorpe-Morgan:2020czg}, for instance, authors show that their boosted DM model improves their limits only by a factor $\sim1.5$, indeed a natural consequence of the low contribution of their substructure model to the annihilation flux. In earlier work by \citep{Huang:2011xr}, authors gained up to  $\sim\mathcal{O}(10^{2})$ of improvement in their constraints due to subhalos, however they assumed boosts $B=500-1200$, which from recent studies can be considered as overly optimistic (e.g., \citep{Ando:2019xlm} and references therein). As for the constraints from the H.E.S.S Collaboration \citep{2012ApJ...750..123A}, their boosted MED ($B=10$) and MAX ($B=100$) models improved their limits by a factor $\sim7$ and $\sim75$, respectively, in reasonable agreement with our results (we note though that, even with these improvements in their limits, our CTA projected constraints for the MED model are more restrictive). 
All in all, our results in Figure~\ref{fig:DM_annihil_results_subs} illustrate the key role that halo substructure can have to discover/rule-out WIMP DM.  

Other uncertainties affect the results in this section as well, namely the one coming from the scatter in the concentration-mass relation and the uncertainty in the estimation of Perseus mass. Yet, the level of these uncertainties ($\sigma_{J_{ann}}=0.2$ dex; see Section~\ref{sec:DM_fluxes}) is well below the one induced by the modelling of the subhalo population ($\sigma_{B_{MAX}}=2$ dex). As a result, we note that the variation of upper limits shown in Figure~\ref{fig:DM_annihil_results_subs} among our three different subhalo models encompasses by far these other, second-order uncertainties. Finally, we remark that these results are for the use of the ``Baseline'' CR model, since different models for the CR densities may also impact our results (see Appendix~\ref{app:DM_interplay}).

\subsection{Projected sensitivity to decaying DM}
\label{sec:DM_result_decay}

Similarly to the annihilating DM case, we compute 95\% C.L. upper limits for $A_{\chi}$, convert this value to $\tau_{\chi}$ (the standard parameter to be constrained for decaying DM) and average the results of 100 simulations. We note that as $A_{\chi}$ and $\tau_{\chi}$ are inversely proportional, the upper limits on $A_{\chi}$ translating into lower limits for $\tau_{\chi}$.

\begin{figure}[h!]
	\centering
	\includegraphics[width=0.49\textwidth]{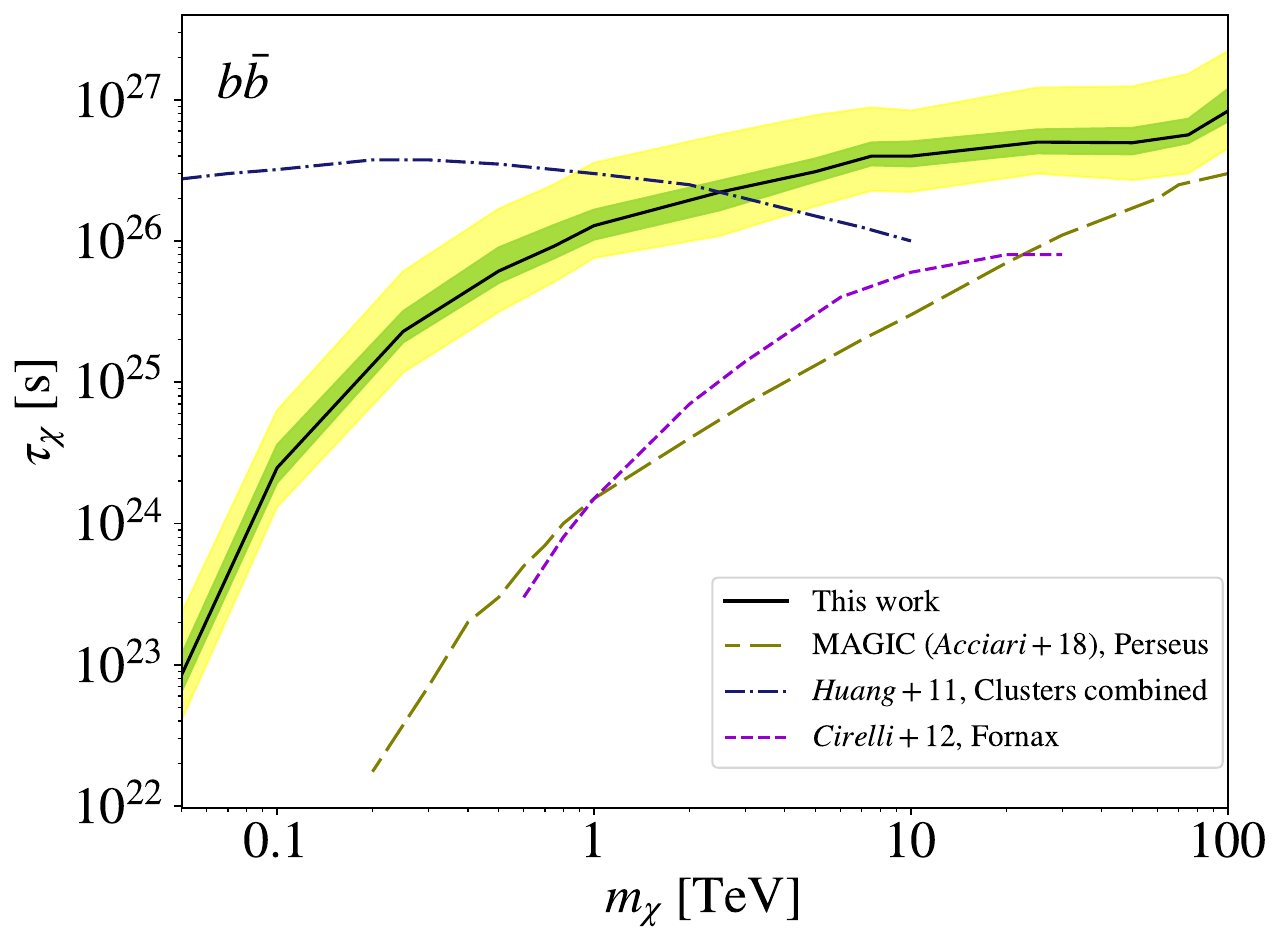}
	\includegraphics[width=0.49\textwidth]{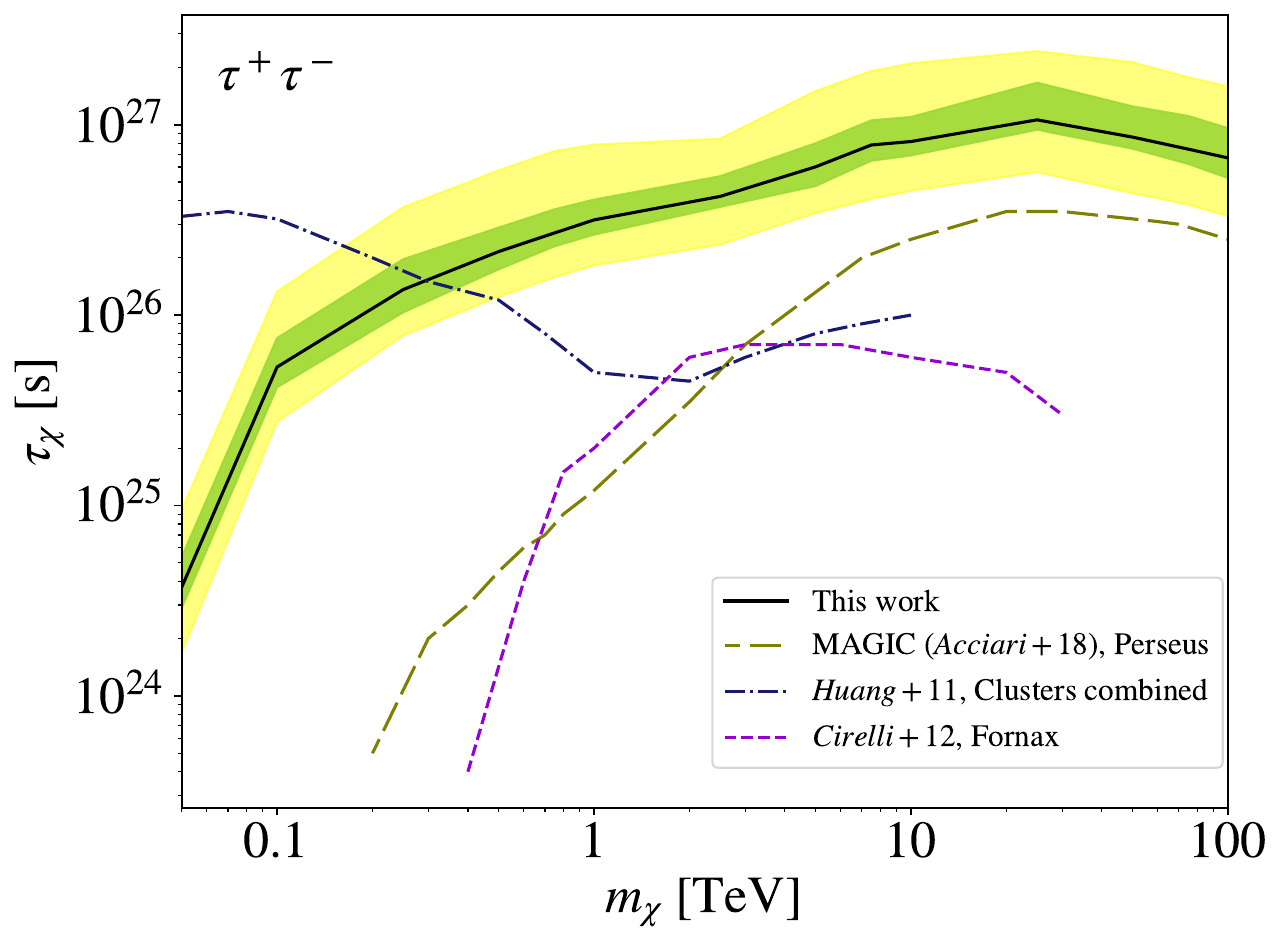}
	\caption{Sensitivity of CTA to a DM decay signal from the Perseus cluster, at 95\% C.L., in terms of the mean lower limits of the lifetime of the DM particle versus the DM mass. The green (yellow) band shows the $1\sigma$ ($2\sigma$) scatter of the projected limits. \textbf{Left panel:} Mean lifetime lower limits for the $b\bar{b}$ channel (\textbf{right panel} for $\tau^+\tau^-$ channel) in comparison with the most recent results on DM decay in galaxy clusters using MAGIC data (olive long-dashed lines; \citep{MAGIC:2018tuz}), $Fermi$-LAT data (blue dot-dashed lines; \citep{Huang:2011xr}) and H.E.S.S data (purple dashed lines; \citep{Cirelli:2012ut}). All of our results are shown for the template fitting analysis, unless stated otherwise.
	}
\label{fig:DM_decay_results}
\end{figure}

We show, in Figure~\ref{fig:DM_decay_results}, the projected sensitivity of CTA as the 95\% lower limits on the DM particle lifetime versus the DM mass for our DEC model. As for the annihilation constraints, the most constraining limits are for the $\tau^+\tau^-$ channel, reaching similar values than the $b\bar{b}$ channel for masses above $\sim$10 TeV. We also compare our CTA projections to the most up-to-date constraints on DM decay in clusters. The most recent work comes from the MAGIC Collaboration \citep{MAGIC:2018tuz}, where authors analyze 202h of data targeting the Perseus cluster. Their limits yield the tightest constraints for masses above $\sim1$ TeV for the $\tau^+\tau^-$ channel, yet CTA will improve these limits by more than two orders of magnitude at lower masses and up to a factor 7 for masses higher than 5 TeV. This large improvement in the low mass range is probably twofold: i) the improvement of up to one order of magnitude in the sensitivity of CTA with respect to MAGIC\footnote{\url{https://www.cta-observatory.org/science/ctao-performance/}} in the lower energy range; ii) the adoption of a mask of 0.1 deg in the MAGIC analysis to avoid contamination from NGC~1275, thus loosing a considerable amount of data. Another comparable work from existing IACTs comes from \citep{Cirelli:2012ut}, where  authors analyze 14.5h of H.E.S.S. observations of the Fornax cluster. Though covering a smaller range of DM masses, their DM decay limits, both for the $b\bar{b}$ and $\tau^+\tau^-$ channels, are of the same order of magnitude than the MAGIC limits for Perseus \citep{MAGIC:2018tuz}. Lastly, we also compare in Figure~\ref{fig:DM_decay_results} our CTA predictions to the constraints obtained using 3 years of \textit{Fermi}-LAT data from 8 galaxy clusters \citep{Huang:2011xr}. These limits prevail up to DM masses of a few TeV, where CTA starts to yield first comparable and then, quickly, stronger constraints for both decay channels. From all these comparisons, we conclude that CTA will not only be able to test an unexplored region in the DM decay parameter space, but should yield the most stringent constraints from $\gamma$-ray DM decay searches above 1 TeV\footnote{The IceCube Collaboration provides more constraining limits above 100 TeV \citep{IceCube:2018tkk}.}.

\begin{figure}[h!]
	\centering
	\includegraphics[width=0.49\textwidth]{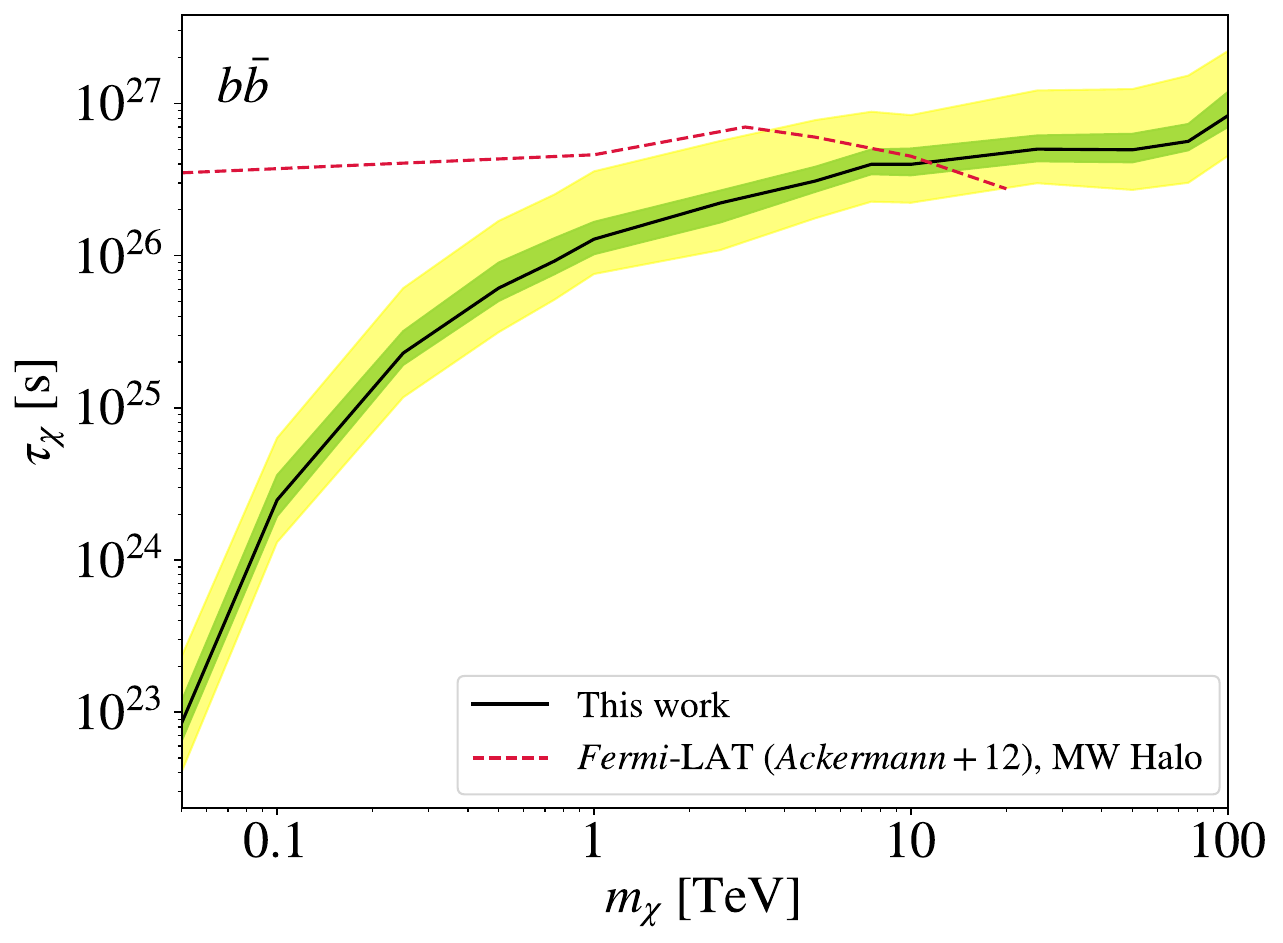}
	\includegraphics[width=0.49\textwidth]{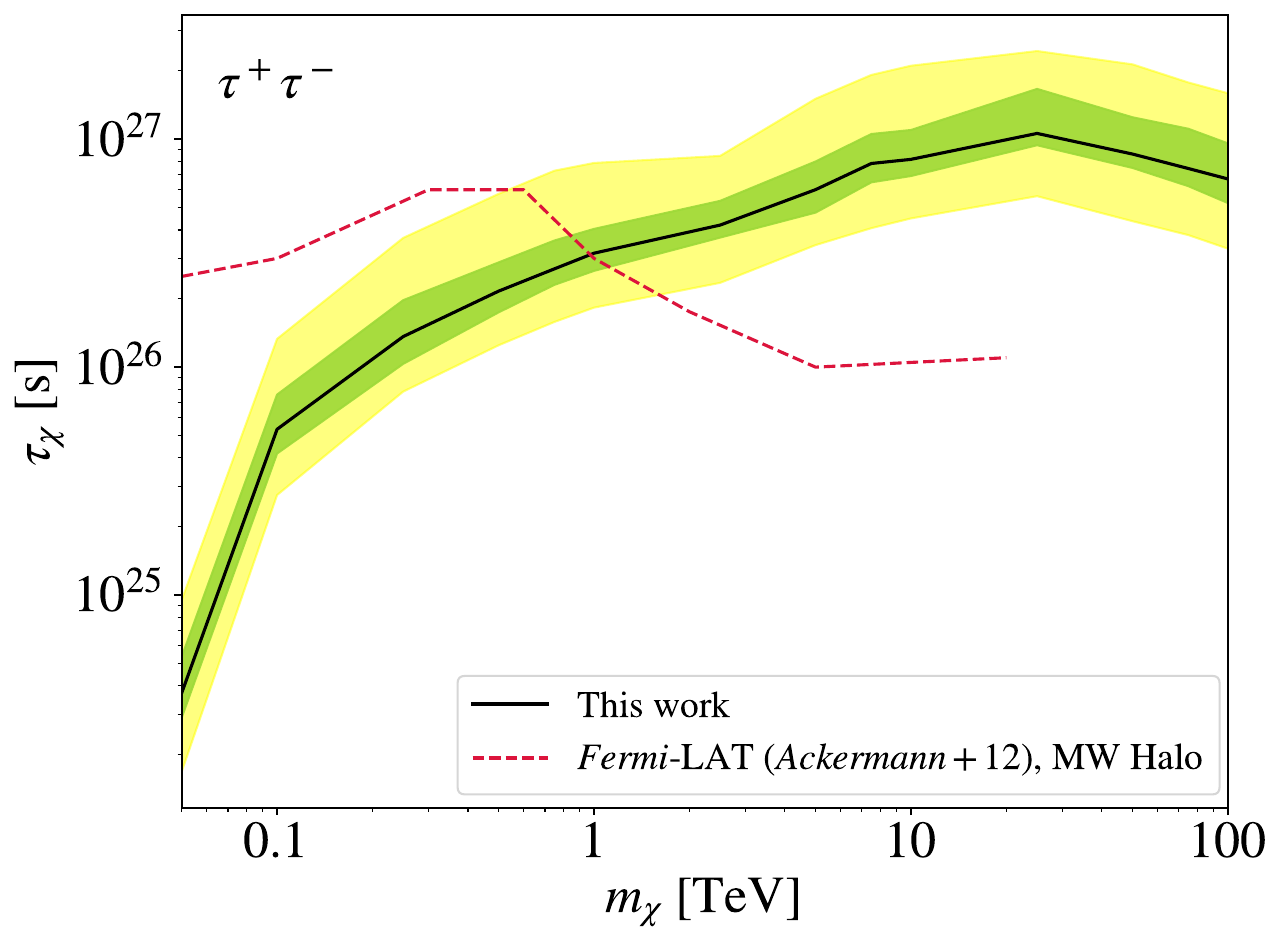}
	\caption{95\% C.L. mean lower limits of the lifetime of the DM particle versus the DM mass (black lines) for $b\bar{b}$ (\textbf{left panel}) and $\tau^+\tau^-$ (\textbf{right panel}) decay channels, in comparison with the limits obtained by \citep{2012ApJ...761...91A} (corresponding to NFW 5$\sigma$ free source fits, in dashed red lines). The green (yellow) band shows the $1\sigma$ ($2\sigma$) scatter of the projected limits. All of our results are shown for the template fitting analysis, unless stated otherwise.
 }
\label{fig:DM_decay_gc}
\end{figure}

Finally, we compare in Figure~\ref{fig:DM_decay_gc} our CTA projections to the most constraining DM decay results from $\gamma$-ray observations so far \citep{2012ApJ...761...91A}. In the latter work, the \textit{Fermi}-LAT Collaboration analyzes 1 year of LAT data from the Milky Way halo. The CTA exclusion curves, as already seen in previous comparisons with \textit{Fermi}-LAT limits, become more constraining above DM masses around a few TeV (few hundreds of GeV for the case of the $\tau^+\tau^-$ channel)\footnote{$Fermi$-LAT observations of the Extragalactic $\gamma$-Ray Background \citep{Huang:2011xr} also yield DM decay limits that are comparable to the ones from the Milky Way Halo.}. From these comparisons, we conclude that the null-detection of a DM decay signal in Perseus with CTA would still provide unprecedented constraints on DM decay models at the TeV mass scale.

\section{Summary and conclusions}\label{sec:Conclusions}
In this work, we have analyzed the CTA sensitivity to detect diffuse $\gamma$-ray emission from the Perseus galaxy cluster. We have considered the possible contribution from CR and from DM annihilation and decay. We have assumed 300 hours of observations as proposed by the CTA consortium as a key science project \citep{CTA2019}.

We built a CR-induced $\gamma$-ray emission model using the {\tt MINOT} software \citep{Adam2020} (Section~\ref{sec:Modeling_CR}). Our model relies on a description of the thermal gas pressure and density, the magnetic field strength, and a parameterization of the spatial and spectral distribution of the CRp. We calibrated the model components according to available data in the literature (Figure~\ref{fig:thermal_and_mag_model}). The thermal part is expected to be accurate from the core to the outskirt and was kept fixed. The magnetic field was modeled by considering available measurements and different assumptions and scaling to estimate the associated systematic uncertainty. The CRp parameters were calibrated according to several scenarios including a baseline model and the pure hadronic model (Figure~\ref{fig:thermal_and_mag_model} and Table~\ref{tab:model_summary}). The description of the background sky was made according to the literature, noting that at least two point sources will affect the CTA observations (Section~\ref{sec:background_sky}). The observation setup was investigated focusing on the offset between the cluster center and the pointing (Figure~\ref{fig:snr_versus_pointing_offset2}). Finally, the CTA sensitivity to diffuse $\gamma$-ray emission was investigated (Section~\ref{sec:CTA_CR_sensitivity}). We considered both the case of non-detection in which we computed the expected exclusion limit that CTA will obtain (Figure~\ref{fig:CR_95percent_limits_X}), and the case of specific models for which a detection is expected. In the case of detection, we investigated the constraints that CTA will be able to provide regarding the spatial and spectral properties of the $\gamma$-ray diffuse emission (Figures~\ref{fig:CR_spectral_constraints} and \ref{fig:CR_triangle_plot_basline}).

Our main findings with regard to $\gamma$-ray emission are as follows.
\begin{itemize}
\item The pure hadronic model, as calibrated using existing radio data, implies about 5\% of CRp energy relative to the thermal energy within $R_{500}$, a CRp spectral index of $\sim 2.3$ and a CRp profile slightly shallower than the thermal gas density, assuming a magnetic field strength model based on existing data, although we note that parameters are strongly degenerate. Even when fixing the magnetic field strength, the uncertainty on the $\gamma$-ray flux corresponds to a factor of about two. This large uncertainty is mainly due to the limited spectral coverage of the available radio data that we used. The systematic uncertainty associated with the magnetic field implies an additional factor of a few in the uncertainty, depending on the radius.
\item According to our modelling and the CTA IRF, we found that a pointing offset of about 1 deg was an optimal choice. This is obtained by considering both ON-OFF analysis and template-fitting techniques and for both CR and DM related analysis.
\item In the case of non detection, we find that CTA should improve the current limits on the CRp content of the Perseus cluster by about an order of magnitude. Assuming a standard scenario ($\alpha_{\rm CRp} = 2.3$ and $\eta_{\rm CRp} = 1$), CTA should be able to constrain the CRp to thermal energy ratio within $R_{500}$, $X_{500}$, down to about a $3 \times 10^{-3}$.
\item Assuming the pure hadronic model, CTA should allow us to detect the ICM induced diffuse emission with a high significance. The spectral index of the CRp should be constrained to an uncertainty of about $\pm 0.1$ and the spatial distribution down to about 10\% precision.
\end{itemize}

CTA observations of Perseus will allow us to address fundamental questions related to the underlying mechanism that accelerates particles in the ICM (such as shocks, turbulent reacceleration, the direct injection of CR from AGN), the direct injection of CR from AGN. CTA will also allow us to test the physics associated with CR transport in the ICM. According to the results presented in this paper, CTA will provide unprecedented constraints on the physics associated with particle acceleration in galaxy clusters.

In our work, we also investigated the potential of CTA to search for DM in Perseus. To model its DM content, we assume the DM to be entirely composed of WIMPs. We consider two different DM scenarios, where the WIMP annihilates or decays into SM particles. We build the DM density profile of the cluster accounting for the smooth DM distribution in the main halo plus the abundant DM in the form of substructures, or subhalos, that these massive objects are expected to host. The main halo is modelled starting from X-ray data, that allows for an estimate of the cluster's mass. For that mass, we then build a tailored NFW profile for Perseus following results from numerical cosmological simulations at these scales. The contribution of the subhalo population to the expected DM flux is the main uncertainty in our DM annihilation model (subhalos do not play a role for decay), and simply reflects the existing ongoing debates in the literature on the precise properties of these objects. To bracket this uncertainty, we define three benchmark models (Table~\ref{tab:benchmark-models}), each of them representative of the different contribution that subhalos may have in the computation of the DM annihilation flux: the MIN model, where substructures are completely neglected; the MED model, our best guess according to the most recent results from numerical simulations; and the MAX model, defined to provide an upper limit to the role that substructures could play for the DM flux. From these, we compute the corresponding $J_{ann}$ factors for the MIN, MED and MAX model and $J_{dec}$ for the MIN case (summarized in Table~\ref{tab:DM-fluxes}) and produce 2D spatial templates of the expected emissions (Figure~\ref{fig:DM_J_templates}). The effect of substructures in the $J_{ann}$-factor is quantified in terms of the so-called subhalo \textit{boost factor}. More precisely, we find $B_{MED}\sim9$ and $B_{MAX}\sim59$ for the MED and MAX models, respectively. As expected, substructure affects mainly the morphology and strength of the DM signal in the outskirts of the cluster (Figure~\ref{fig:jtot_perseus}).

Adopting these DM templates and a particular DM spectrum (Figure~\ref{fig:DM_simulations_spectrum}), we then create simulations of CTA observations of Perseus, that take also into account the $\gamma$-ray emission from the CR (baseline model), the two AGNs in the area (NGC~1275 and IC~310) and the instrumental background. All of these components act as background in our DM search analysis. In a next step, we fit these observations to $b\bar{b}$ and $\tau^+\tau^-$ annihilation/decay channels and several DM masses using a template-fitting approach. In the absence of a DM signal in all the considered DM scenarios, we find the following results:
\begin{itemize}
    \item DM annihilation: We obtain 95\% C.L. upper limits on the annihilation cross-section for different DM masses (Figure~\ref{fig:DM_annihil_results}). The best results are for the $\tau^+\tau^-$ annihilation channel, reaching values of $<\sigma v>\sim5\times10^{-24}$ cm$^3$s$^{-1}$ for DM masses up to $\sim$1 TeV. For masses above few TeV, our limits become less constraining as the sensitivity of the CTA weakens, reaching values up to $10^{-21}$ cm$^3$s$^{-1}$. These limits are between 2-4 orders of magnitude above the value of the thermal-relic cross-section. Different prescriptions for the subhalo population in Perseus weaken (MIN) or strengthen (MAX) our results by a factor $\sim\mathcal{O}(10)$ (Figure~\ref{fig:DM_annihil_results_subs}).
    
    \item DM decay: We obtain  95\% C.L. lower limits on the DM particle lifetime for different DM masses (Figure~\ref{fig:DM_decay_results}). The best results are again for the $\tau^+\tau^-$ decay channel, reaching values of $\tau_{\chi}\sim 10^{27}$s for DM masses in the range of 10-30 TeV.
    
\end{itemize}

To put these DM results into the more general context: for WIMP annihilation, our prospects show that CTA will be able to provide the best constraints from $\gamma$-ray DM searches in galaxy clusters above 1 TeV (Figure~\ref{fig:DM_annihil_results}). Note, though, that CTA is not expected to reach the thermal relic cross-section value from these observations\footnote{Yet, we did not include possible flux enhancements due to the Sommerfeld effect \citep{Sommerfeld1931, Lacroix:2022cjm}, that could drastically impact the constraints especially in the TeV range. This will be done elsewhere.}. On the other hand, the sensitivity of CTA to DM decay in Perseus will be particularly significant. Indeed, comparing our predictions with other existing constraints, CTA will test an unexplored region of the DM decay parameter space for TeV WIMPs. This will allow CTA to set unprecedented constraints on the DM particle lifetime at these masses (Figures~\ref{fig:DM_decay_results} and~\ref{fig:DM_decay_gc}). All these DM results not only demonstrate the superb capabilities of CTA to search for DM and to test the preferred DM models, especially for heavy WIMPs, but also the excellent potential of galaxy clusters as excellent targets for $\gamma$-ray DM searches.

\acknowledgments
This work was conducted in the context of the CTA Consortium (mainly the CTA DMEP and CR Working Groups and the CTA Galaxy Clusters Task Force). We gratefully acknowledge financial support from the following agencies and organizations:

\bigskip

State Committee of Science of Armenia, Armenia;
The Australian Research Council, Astronomy Australia Ltd, The University of Adelaide, Australian National University, Monash University, The University of New South Wales, The University of Sydney, Western Sydney University, Australia;
Federal Ministry of Education, Science and Research, and Innsbruck University, Austria;
Conselho Nacional de Desenvolvimento Cient\'{\i}fico e Tecnol\'{o}gico (CNPq), Funda\c{c}\~{a}o de Amparo \`{a} Pesquisa do Estado do Rio de Janeiro (FAPERJ), Funda\c{c}\~{a}o de Amparo \`{a} Pesquisa do Estado de S\~{a}o Paulo (FAPESP), Funda\c{c}\~{a}o de Apoio \`{a} Ci\^encia, Tecnologia e Inova\c{c}\~{a}o do Paran\'a - Funda\c{c}\~{a}o Arauc\'aria, Ministry of Science, Technology, Innovations and Communications (MCTIC), Brasil;
Ministry of Education and Science, National RI Roadmap Project DO1-153/28.08.2018, Bulgaria;
The Natural Sciences and Engineering Research Council of Canada and the Canadian Space Agency, Canada;
CONICYT-Chile grants CATA AFB 170002, ANID PIA/APOYO AFB 180002, ACT 1406, FONDECYT-Chile grants, 1161463, 1170171, 1190886, 1171421, 1170345, 1201582, Gemini-ANID 32180007, Chile, W.M. gratefully acknowledges support by the ANID BASAL projects ACE210002 and FB210003, and FONDECYT 11190853;
Croatian Science Foundation, Rudjer Boskovic Institute, University of Osijek, University of Rijeka, University of Split, Faculty of Electrical Engineering, Mechanical Engineering and Naval Architecture, University of Zagreb, Faculty of Electrical Engineering and Computing, Croatia;
Ministry of Education, Youth and Sports, MEYS LM2015046, LM2018105, LTT17006, EU/MEYS CZ.02.1.01/0.0/0.0/16\_013/0001403, CZ.02.\\1.01/0.0/0.0/18\_046/0016007 and CZ.02.1.01/0.0/0.0/16\_019/0000754, Czech Republic; \\Academy of Finland (grant nr.317636 and 320045), Finland;
Ministry of Higher Education and Research, CNRS-INSU and CNRS-IN2P3, CEA-Irfu, ANR, Regional Council Ile de France, Labex ENIGMASS, OCEVU, OSUG2020 and P2IO, France;
Max Planck Society, BMBF, DESY, Helmholtz Association, TU Dortmunt University grant DFG SFB 1491, Germany;
Department of Atomic Energy, Department of Science and Technology, India;
Istituto Nazionale di Astrofisica (INAF), Istituto Nazionale di Fisica Nucleare (INFN), MIUR, Istituto Nazionale di Astrofisica (INAF-OABRERA) Grant Fondazione Cariplo/Regione Lombardia ID 2014-1980/RST\_ERC, Italy;
ICRR, University of Tokyo, JSPS, MEXT, Japan;
Netherlands Research School for Astronomy (NOVA), Netherlands Organization for Scientific Research (NWO), Netherlands;
University of Oslo, Norway;
Ministry of Science and Higher Education, DIR/WK/2017/12, the National Centre for Research and Development and the National Science Centre, UMO-2016/22/M/ST9/00583, Poland;
Slovenian Research Agency, grants P1-0031, P1-0385, I0-0033, J1-9146, J1-1700, N1-0111, and the Young Researcher program, Slovenia; 
South African Department of Science and Technology and National Research Foundation through the South African Gamma-Ray Astronomy Programme, South Africa;
The Spanish groups acknowledge the Spanish Ministry of Science and Innovation and the Spanish Research State Agency (AEI) through the government budget lines PGE2021/28.06.000X.411.01, PGE2022/28.06.000X.411.01 and PGE2022/28.06.000X.711.04, and grants PGC2018-095161-B-I00, PGC2018-095512-B-I00, PID2019-104114RB-C31,  PID\\2019-107847RB-C44, PID2019-104114RB-C32, PID2019-105510GB-C31, PID2019-104114RB-C33, PID2019-107847RB-C41, PID2019-107847RB-C43, PID2019-107847RB-C42, PID2019-107988GB-C22; the ``Centro de Excelencia Severo Ochoa'' program through grants no. SEV-2017-0709, CEX2019-000920-S, CEX2020-001007-S; the ``Unidad de Excelencia Mar\'ia de Maeztu'' program through grants no. CEX2019-000918-M, CEX2020-001058-M; the ``Ram\'on y Cajal'' program through grant RYC-2017-22665; the ``Juan de la Cierva-Incorporaci\'on'' program through grants no. IJC2018-037195-I, IJC2019-040315-I. They also acknowledge the La Caixa Banking Foundation, grant no. LCF/BQ/PI21/11830030; the ``Programa Operativo'' FEDER 2014-2020, Consejer\'ia de Econom\'ia y Conocimiento de la Junta de Andaluc\'ia (Ref. 1257737), PAIDI 2020 (Ref. P18-FR-1580) and Universidad de Ja\'en; ``Programa Operativo de Crecimiento Inteligente'' FEDER 2014-2020 (Ref. ESFRI-2017-IAC-12), Ministerio de Ciencia e Innovaci\'on, 15\% co-financed by Consejer\'ia de Econom\'ia, Industria, Comercio y Conocimiento del Gobierno de Canarias; the ``CERCA'' program of the Generalitat de Catalunya; and the European Union’s ``Horizon 2020'' GA:824064 and NextGenerationEU.
Swedish Research Council, Royal Physiographic Society of Lund, Royal Swedish Academy of Sciences, The Swedish National Infrastructure for Computing (SNIC) at Lunarc (Lund), Sweden;
State Secretariat for Education, Research and Innovation (SERI) and Swiss National Science Foundation (SNSF), Switzerland;
Durham University, Leverhulme Trust, Liverpool University, University of Leicester, University of Oxford, Royal Society, Science and Technology Facilities Council, UK;
U.S. National Science Foundation, U.S. Department of Energy, Argonne National Laboratory, Barnard College, University of California, University of Chicago, Columbia University, Georgia Institute of Technology, Institute for Nuclear and Particle Astrophysics (INPAC-MRPI program), Iowa State University, the Smithsonian Institution, V.V.D. is funded by NSF grant AST-1911061, Washington University McDonnell Center for the Space Sciences, The University of Wisconsin and the Wisconsin Alumni Research Foundation, USA.

The research leading to these results has received funding from the European Union's Seventh Framework Programme (FP7/2007-2013) under grant agreements No~262053 and No~317446.
This project is receiving funding from the European Union's Horizon 2020 research and innovation programs under agreement No~676134. This research has made use of the CTA instrument response functions provided by the CTA Consortium and Observatory, see \url{http://www.cta-observatory.org/science/ctao-performance/} for more details.

\bigskip

JPR work was supported by grant SEV-2016-0597-17-2 funded by MCIN/AEI/10.13039/\\501100011033 and ``ESF Investing in your future''. MASC was also supported by the {\it Atracci\'on de Talento} contracts no. 2016-T1/TIC-1542 and 2020-5A/TIC-19725 granted by the Comunidad de Madrid in Spain. The work of JPR and MASC was additionally supported by the grants PGC2018-095161-B-I00 and CEX2020-001007-S, both funded by MCIN/AEI/10.13039/\\501100011033 and by ``ERDF A way of making Europe''. MH acknowledges funding from the Max Planck Society, the University of Tokyo, and the ICRR Inter-University Research Program in the Fiscal Years 2021 and 2022. PTL is supported by the Swedish Research Council under contract 2019-05135. SHC work is supported by UNAM-PAPIIT IG101323.

This research made use of \texttt{gammapy},\footnote{\url{https://www.gammapy.org}} a community-developed core Python package for TeV $\gamma$-ray astronomy \citep{gammapy:2017}. This research made use of Astropy, a community-developed core Python package for Astronomy \citep{Astropy2013}, in addition to NumPy \citep{VanDerWalt2011}, SciPy \citep{Jones2001}, Healpy \citep{Zonca2019} and Ipython \citep{Perez2007}. Figures were generated using Matplotlib \citep{Hunter2007}.

We acknowledge support from the CNRS/IN2P3 Computing Center (Lyon - France) for providing computing and data-processing resources needed for this work. We thank the support of the Hydra HPC cluster in Instituto de Física Teórica (IFT UAM-CSIC) for the computing time and resources.

\appendix
\clearpage
\newpage
\section{Validation of the thermal model using \textit{Planck} data}\label{app:validation_thermal_model_planck}
The thermal Sunyaev-Zel'dovich (tSZ) effect provides a direct measurement of the electron thermal pressure in galaxy clusters. In order to validate our thermal gas pressure model calibration, which is based on the indirect inference of the pressure from X-ray observations, we compare the tSZ prediction of our model to \textit{Planck} data \citep{Planck2016XXII} obtained with MILCA \citep{Hurier2013}.

\begin{figure}[h!]
	\centering
	\includegraphics[width=0.79\textwidth]{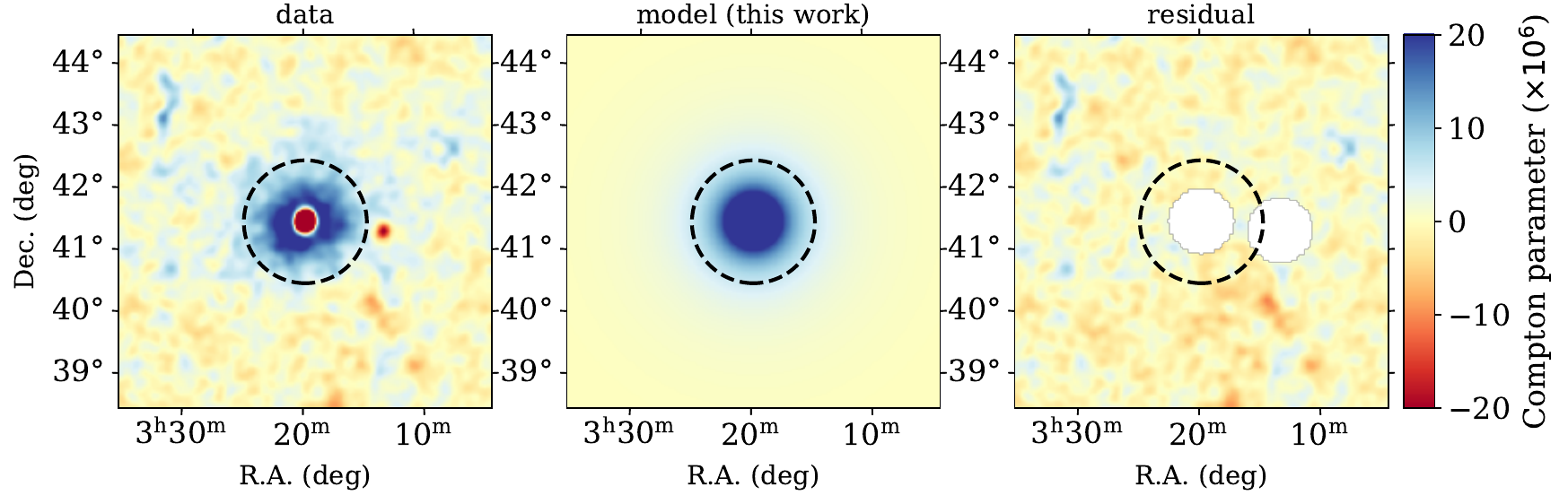}
	\caption{Comparison between the Compton parameter map measured by \textit{Planck} and that inferred from our thermal model. {\bf Left panel}: raw data. {\bf Middle}: model. {\bf Right panel}: residual. The dashed circle indicates $\theta_{500}$. Radio sources appear as negative point sources on the Compton parameter map. The two white circles are masks for NGC~1275 and another radio source.}
\label{fig:ymap_data_versus_model}
\end{figure}

\begin{figure}[h!]
	\centering
	\includegraphics[width=0.50\textwidth]{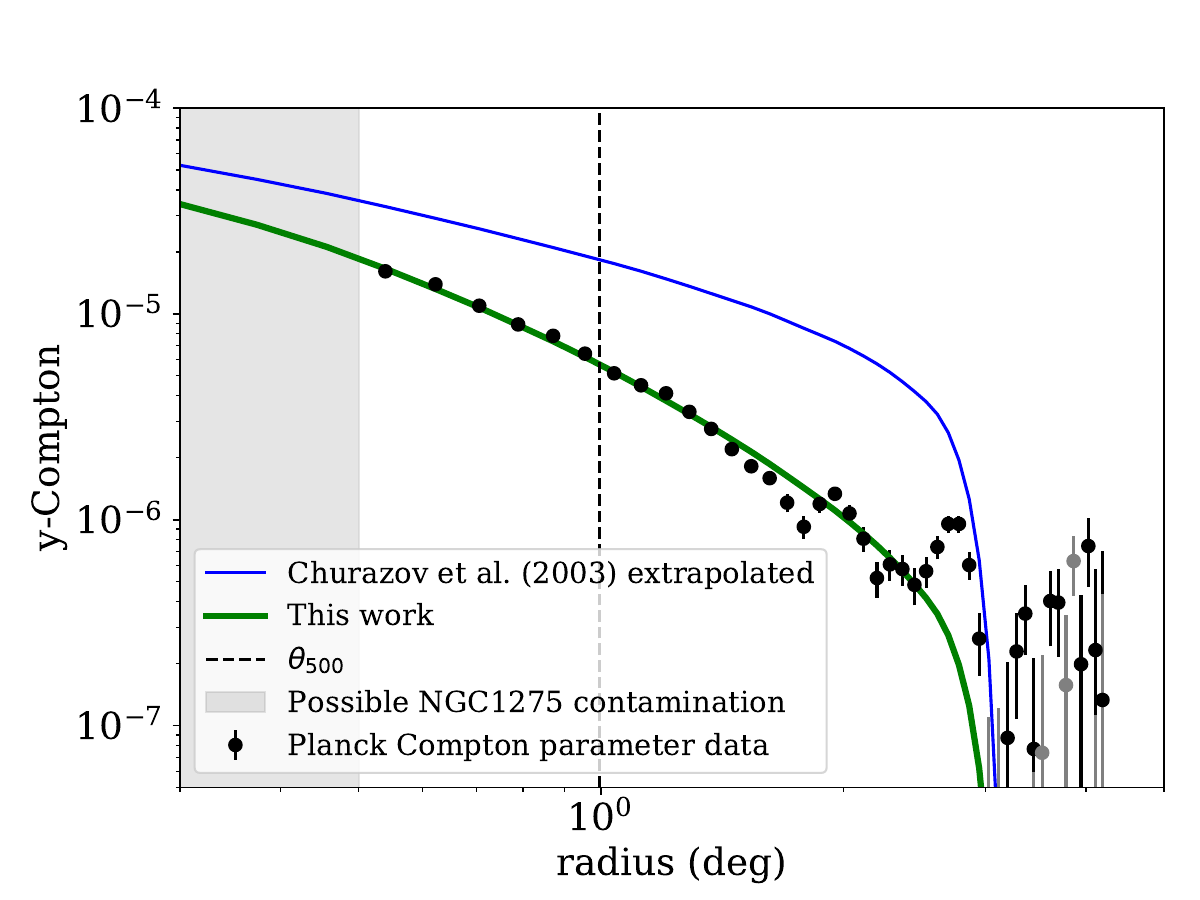}
	\caption{Comparison between the Compton parameter profile measured by \textit{Planck} and that inferred from our thermal model. We also show the result of the extrapolation of the model by \citep{Churazov2003}.}
\label{fig:yprofile_data_versus_model}
\end{figure}

We extract a MILCA sky patch centered on the Perseus cluster and compute its profile. Uncertainties are computed according to the standard deviation of the map in regions free from emission. We project our tSZ model on the same sky patch, convolve the map to the 10 arcmin \textit{Planck} beam and compare it to the data in Figures~\ref{fig:ymap_data_versus_model} (map) and \ref{fig:yprofile_data_versus_model} (profile). Radio sources (including NGC~1275) are masked within 30 arcmin.

Our model, as calibrated using X-ray information, provides an excellent match to the \textit{Planck} Compton parameter map and profile, and thus to the cluster thermal pressure and thermal energy distribution. The model is validated up to $\gtrsim 2 R_{500}$. In the core the bright radio source associated with NGC~1275 prohibits the validation below 30 arcmin, but this is where high quality X-ray data are available. We verify that extrapolation of the model by \citep{Churazov2003} leads to an overestimation of the tSZ signal by a factor of a few.

\clearpage
\newpage
\section{Impact of the magnetic field on the model calibration}\label{app:impact_of_magnetic_field}
\begin{figure}[h!]
	\centering
	\includegraphics[width=1\textwidth]{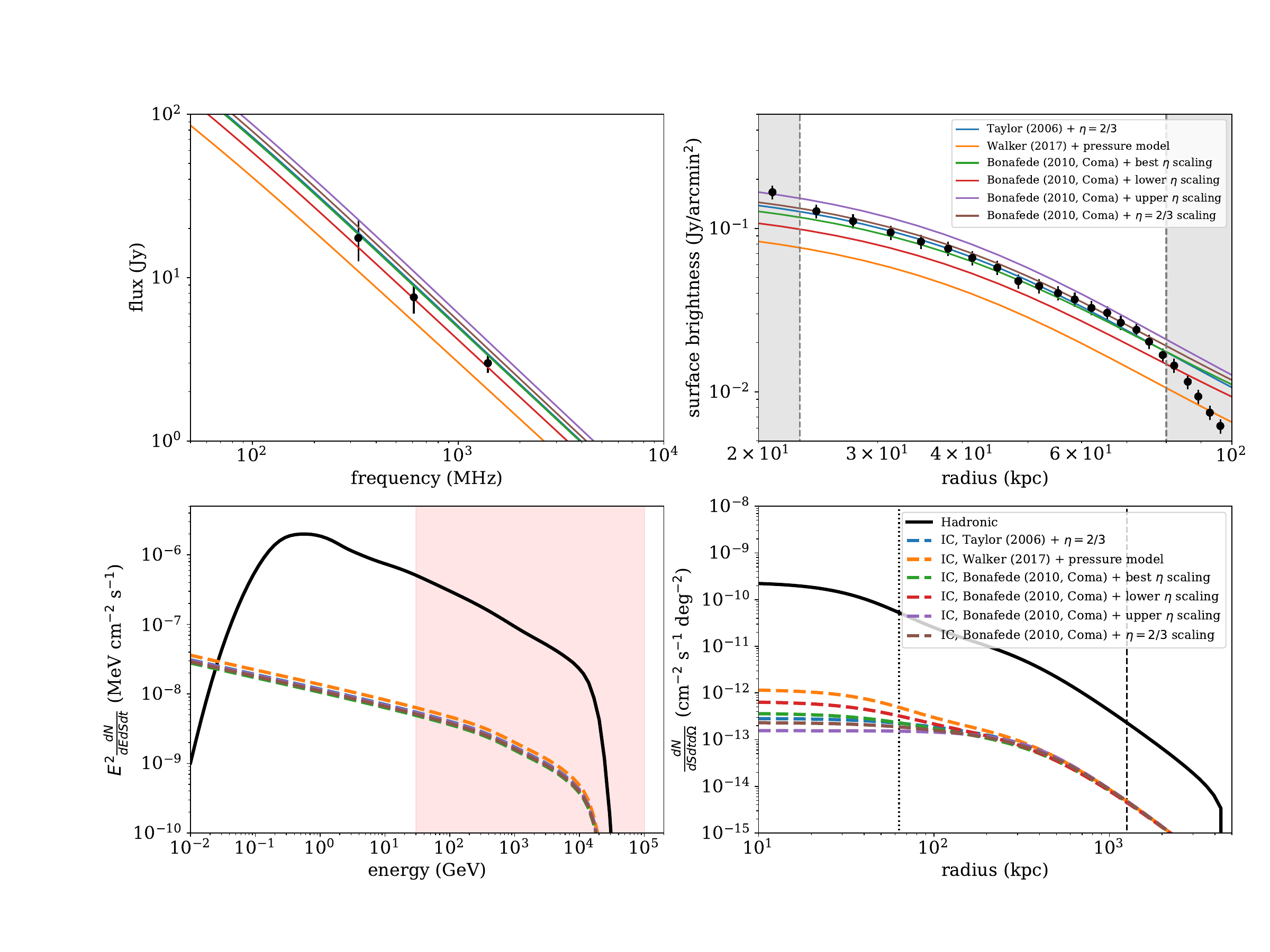}
	\caption{Same as Figure~\ref{fig:gamma_ray_observable_hadronic}, but showing the dependence of the pure hadronic model observables on the magnetic field model (see Section~\ref{sec:magnetic_field_model} and Figure~\ref{fig:thermal_and_mag_model}). The color code is the same for all panels.
	{\bf Top left panel}: radio synchrotron spectrum.
	{\bf Top right panel}: radio synchrotron profile.
	{\bf Bottom left panel}: $\gamma$-ray spectrum. The hadronic component, which does not depend on the magnetic field, is shown in black and the inverse-Compton in color, for the different magnetic field models. The shaded pink area corresponds to the CTA energy range.
	{\bf Bottom right panel}: $\gamma$-ray profile. The dashed lines correspond to the CTA PSF and $\theta_{500}$, respectively. 
 	}
\label{fig:Bfield_effect_on_model_calib}
\end{figure}

In Figure~\ref{fig:Bfield_effect_on_model_calib}, we show how the choice of the magnetic field strength model affects the observable in terms of radio synchrotron emission and $\gamma$-ray emission in the case of the pure hadronic scenario. The model parameters were fixed to $\{X_{\rm CRp,500}, \alpha_{\rm CRp},\eta_{\rm CRp}\} = \{0.075, 2.5, 0.75\}$. Since the radio data are limited to the cluster core, the observable profile is essentially sensitive to the core magnetic field. While the hadronic emission does not depend on the magnetic field, the inverse-Compton signal is affected by the magnetic field via the synchrotron losses that modify the spectral and spatial distribution of the secondary CRe. However, matching the CRp model normalization so that a given magnetic field model would fit the radio data would also imply changing the normalisation of the hadronic and inverse-Compton model, accordingly.

We also note that in the CTA energy range, the radio data implies that the inverse-Compton signal is too small to be detected, and much smaller than the expected hadronic signal. The predicted level of emission would be similar in a pure leptonic model assuming continuous injection (as it is the case in the hadronic model, via secondary particle production). However, in the case of discontinuous particle injection, energy losses are expected to lead to a cutoff in the CR spectrum, which will drastically reduce the amount of inverse-Compton emission at CTA energies. The level of emission seen in Figure~\ref{fig:Bfield_effect_on_model_calib} can thus be considered as an upper limit for the inverse-Compton emission, and it is therefore neglected in the present paper.

\begin{figure}[h!]
	\centering
	\includegraphics[width=0.95\textwidth]{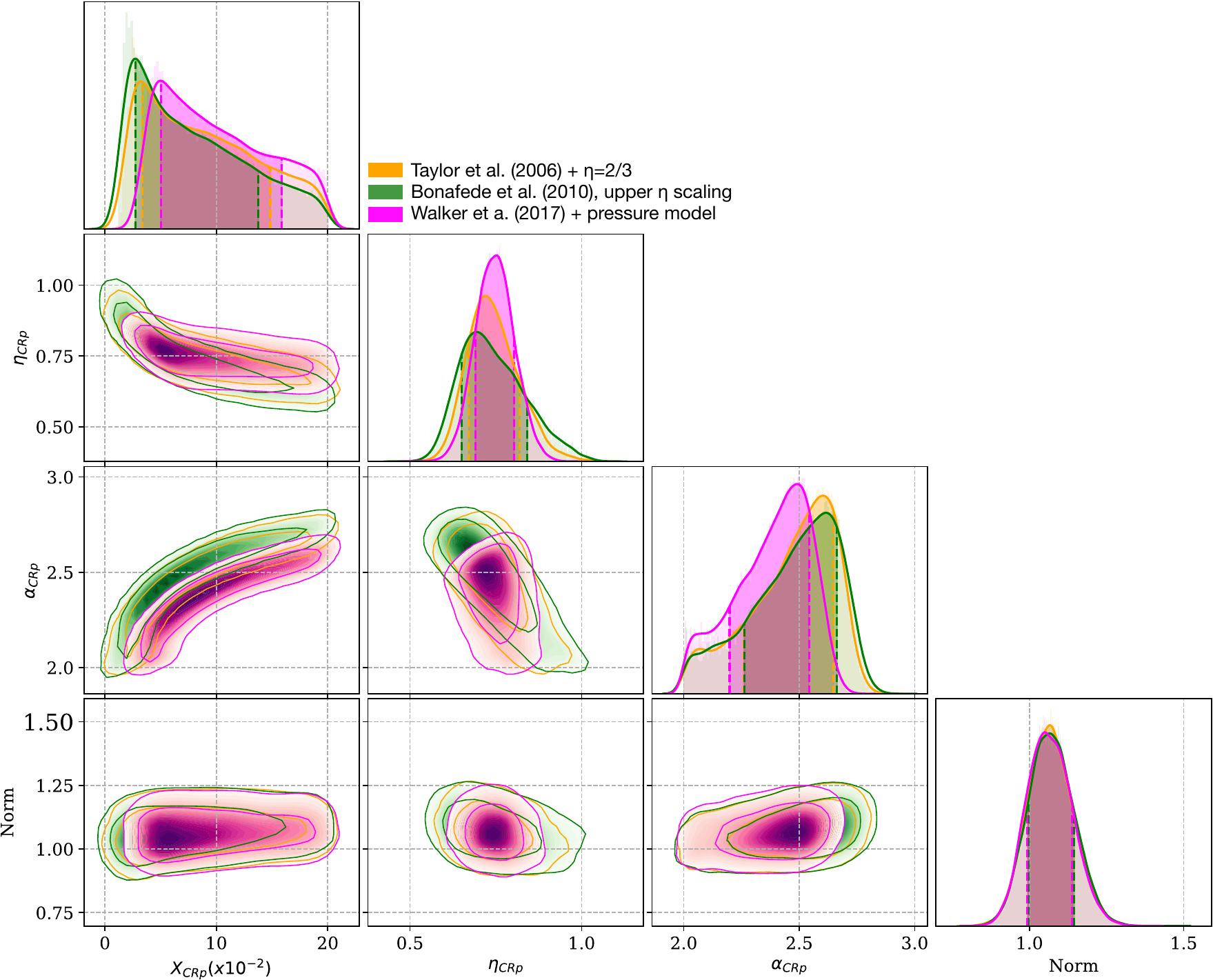}
	\caption{Constraints on the parameter space of the pure hadronic model, for three extreme magnetic field models. See Section~\ref{sec:magnetic_field_model} and Figure~\ref{fig:thermal_and_mag_model}.}
\label{fig:pure_hadronic_parameter_space}
\end{figure}

In Figure~\ref{fig:pure_hadronic_parameter_space}, instead we fit for the CRp parameters $\{X_{\rm CRp,500}$, $\alpha_{\rm CRp}$, and $\eta_{\rm CRp}\}$ by matching the radio synchrotron emission to the data in the pure hadronic scenario (as in Section~\ref{sec:CR_hadronic_model}). We show the constraints on the recovered pure hadronic parameter space for three extreme magnetic field models. As we can see, changing the magnetic field model essentially affects the normalization, and only minor variations are observed on the other parameters. This also mainly leads to a change in the normalization of the $\gamma$-ray observables.

\clearpage
\newpage
\section{\textit{Fermi}-LAT sources around the Perseus cluster}\label{app:Fermi_source_in_Perseus_field}
No sources other than NGC~1275 and IC~310 has been detected within the CTA field of view in the direction of Perseus at very high $\gamma$-ray energies yet. Thus, we do not consider any other object in the modelling of the sky. Yet, the improved CTA sensitivity and increased field of view may allow us to detect other $\gamma$-ray emitters. To highlight such opportunities, we use the \textit{Fermi}-LAT catalog, 4FGL-DR2 \citep{Fermi-LAT:2019yla,Ballet:2020hze}, and select all sources detected within 5 deg from the Perseus reference center. These sources are listed in Table~\ref{tab:4FGL_sources} and their locations are reported in Figure~\ref{fig:fov_scheme}. NGC~1275 and IC~310 are included in the 4FGL source list. The spectra of all of these sources, at \textit{Fermi}-LAT energies, are best described by power-laws, except for the pulsar PSR J0340+4130 and NGC~1275. For reference, and comparison to detected VHE sources NGC~1275 and IC~310, we extrapolate their fluxes up to the CTA energy range at 100 GeV. Most of the obtained values correspond to $\lesssim 10^{-11}$ cm$^{-2}$ TeV$^{-1}$ s$^{-1}$, which is smaller but comparable to the extrapolation of the fluxes of NGC~1275 and IC~310 at the same energy (about $25$ and $5 \times 10^{-11}$ cm$^{-2}$ TeV$^{-1}$ s$^{-1}$) according to equations~\ref{eq:spectrum_NGC1275} and~\ref{eq:spectrum_IC310}. However, we stress that a cutoff in the spectra is possible, especially for distant sources affected by EBL absorption, and these numbers provide optimistic fluxes. Additionally, some of the sources may be variable so that these values only provide rough estimates. Even in the case of a detection, none of these sources is expected to affect the results presented in this paper because of the large angular separation from the cluster center.
\begin{table}[h]
	\caption{4FGL-DR2 sources located within 5 deg from the Perseus cluster center.}
	\begin{center}
	\resizebox{\textwidth}{!} {
	\begin{tabular}{|cccccccc|}
	\hline
	4FGL name & Alt. name & $z^{\dagger}$ & Offset & Type & Spectrum model & PL index & PL extrapolation at 100 GeV \\
	 & & & (deg) & & & & (cm$^{-2}$ TeV$^{-1}$ s$^{-1}$) \\
	\hline
	\hline
    J0319.8+4130 & NGC~1275$^{\star}$ & 0.01756 & 0.0 & RDG & LP & -- & -- \\
    J0316.8+4120 & IC~310$^{\star}$ & 0.01894 & 0.6 & RDG & PL & 1.85 & $4.5 \times 10^{-11}$\\
    J0315.5+4231 & NVSS~J031527+423249 & N/A & 1.3 & BCU & PL & 1.95 & $1.3 \times 10^{-11}$ \\
    J0312.9+4119 & B3~0309+411B & 0.134 & 1.3 & RDG & PL & 2.69 & $1.5 \times 10^{-12}$ \\
    J0311.6+4134 & B3~0308+413 & N/A & 1.5 & BCU & PL & 1.95 & $1.2 \times 10^{-11}$ \\
    J0333.8+4007 & B3~0330+399 & N/A & 3.0 & BCU & PL & 2.17 & $1.3 \times 10^{-11}$ \\
    J0334.3+3920 & 4C~+39.12 & 0.02059 & 3.5 & RDG & PL & 1.81 & $4.0 \times 10^{-11}$ \\
	J0310.9+3815 & B3~0307+380 & 0.816 & 3.7 & FSRQ & PL & 2.40 & $8.3 \times 10^{-12}$ \\
    J0340.3+4130 & PSR~J0340+4130 & 0 & 3.9 & PSR & PLSEC & -- & -- \\
    J0342.2+3858 & GB6~J0342+3858 & 0.945 & 5.0 & FSRQ & PL & 2.26 & $1.2 \times 10^{-11}$ \\
	\hline
	\end{tabular}
	}
	\end{center}
    {\small {\bf Notes.} 
    $^{\star}$ Detected at very high $\gamma$-ray energies ;
    $^{\dagger}$ Taken from the NASA/IPAC extragalactic database ; PL: power-law ; LP: LogParabola ; PLSEC : PLSuperExpCutoff ; RDG: radio galaxies; BCU: blazar candidates of uncertain type ; FSRQ: flat-spectrum radio quasars type of blazar; PSR: pulsar.}    
	\label{tab:4FGL_sources}
\end{table}

\clearpage
\newpage
\section{Galactic foreground estimate}\label{app:Galactic_foreground_estimate}
\begin{figure}[h!]
	\centering
	\includegraphics[height=5.5cm]{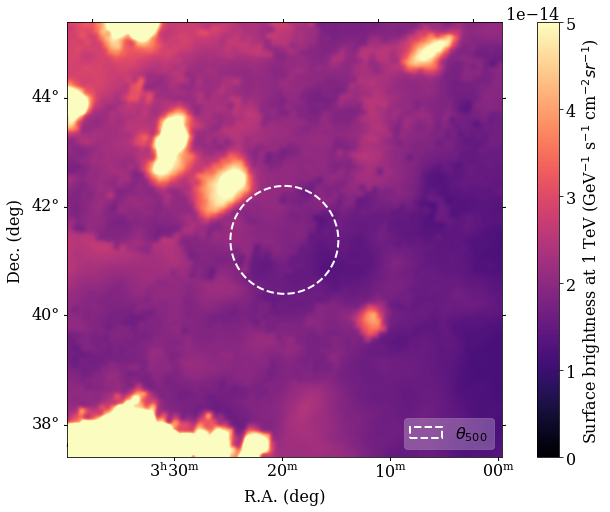}
	\includegraphics[height=5.5cm]{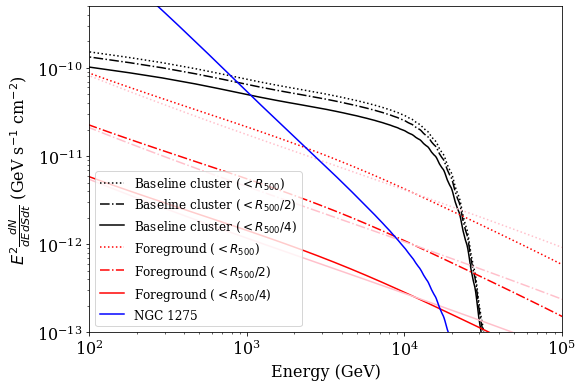}
	\caption{{\bf Left panel:} Surface brightness image of the Galactic foreground contribution at 1 TeV (\emph{min} model). {\bf Right panel:} spectral energy distribution of the Galactic foreground estimates and comparison to the baseline cluster model and the contribution from NGC~1275. Three different aperture radius are used, as indicated in the legend. The \emph{min} model is given in red and the \emph{max} model is given in pink. See \citep{Luque:2022buq, DelaTorreLuque:2022ats} for more details about the modelling.}
\label{fig:galactic_foreground}
\end{figure}

The Galactic foreground is estimated using the models developed by \citep{Luque:2022buq, DelaTorreLuque:2022ats}. We consider the \emph{min} and \emph{max} models as two different estimates. In Figure~\ref{fig:galactic_foreground}, we present the spatial distribution of the diffuse foreground signal at 1 TeV, and the spectrum integrated in different apertures around the cluster center. We can observe that the signal presents a soft gradient over the considered region, plus a few clumps. The latter are not correlated with the cluster and none of them is located within $\theta_{500}$. Even in the case of a large aperture of about 1 deg ($\simeq \theta_{500}$), the integrated diffuse foreground is expected to be subdominant compared to the cluster CR induced signal within the same region (baseline CR model). When reducing the aperture, the Galactic foreground becomes completely negligible compared to the cluster emission since the latter is much more compact.

\clearpage
\newpage
\section{Convergence of the CR parameter constraints versus the number of simulations}\label{app:CR_convergence}
In Section~\ref{sec:Probing_the_parameter_space_with_CTA}, we derived exclusion limits on the CR normalization as a function of the CR spectral and spatial distribution parameters. This was done by averaging the upper limits obtained over a given number of simulations in order to reduce the Poissonian noise.

\begin{figure}[h!]
	\centering
	\includegraphics[width=0.49\textwidth]{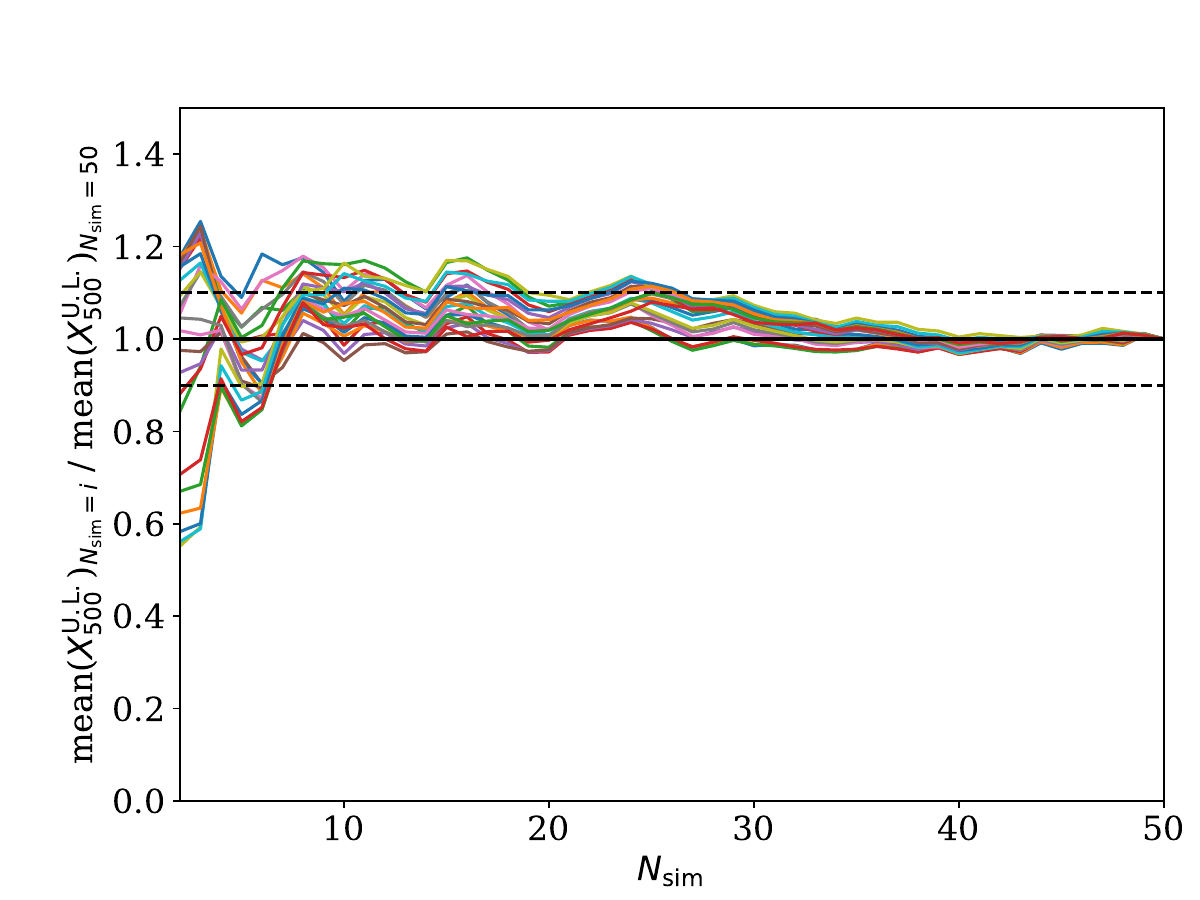}
	\includegraphics[width=0.49\textwidth]{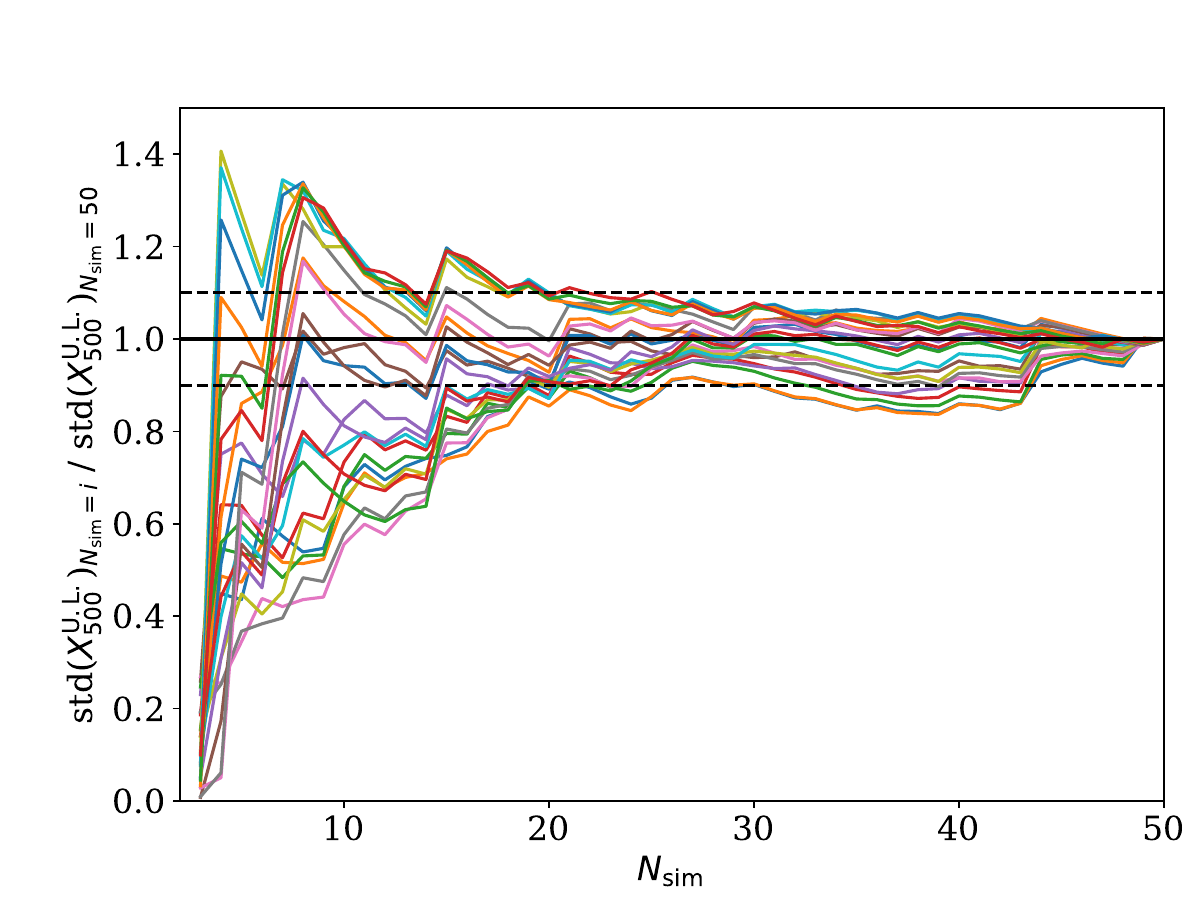}
	\caption{Evolution of the upper limit on the parameters $X_{500} \equiv X_{\rm CRp}(R_{500})$ (95\% confidence level), as a function of the number of simulations performed and for the different model parameters ($\alpha_{\rm CRp}$, $\eta_{\rm CRp}$) presented in Figure~\ref{fig:CR_95percent_limits_X}. 
 {\bf Left}: normalized mean upper limit.
 {\bf Right}: normalized mean standard deviation.
 Note that the results obtained for the different models are correlated because the same underlying Poissonnian realization is used for the different models.
 }
\label{fig:CR_stability_constraint}
\end{figure}

Figures~\ref{fig:CR_stability_constraint} shows the evolution of the constraint on the CR normalisation upper limit (mean and standard deviation), as a function of the number of simulations performed. We can see that averaging the results over 50 simulations is enough to obtain uncertainties on the mean and the standard deviation of the exclusion limit lower than 10\%, which is negligible compared to other sources of uncertainties discussed in the paper. Therefore, we use a set of 50 simulations in the analysis.

\clearpage
\newpage
\section{Validation of the cosmic-ray analysis with {\tt gammapy}}\label{app:validation_gammapy_cr}
\begin{figure}[h!]
	\centering
	\includegraphics[width=0.8\textwidth]{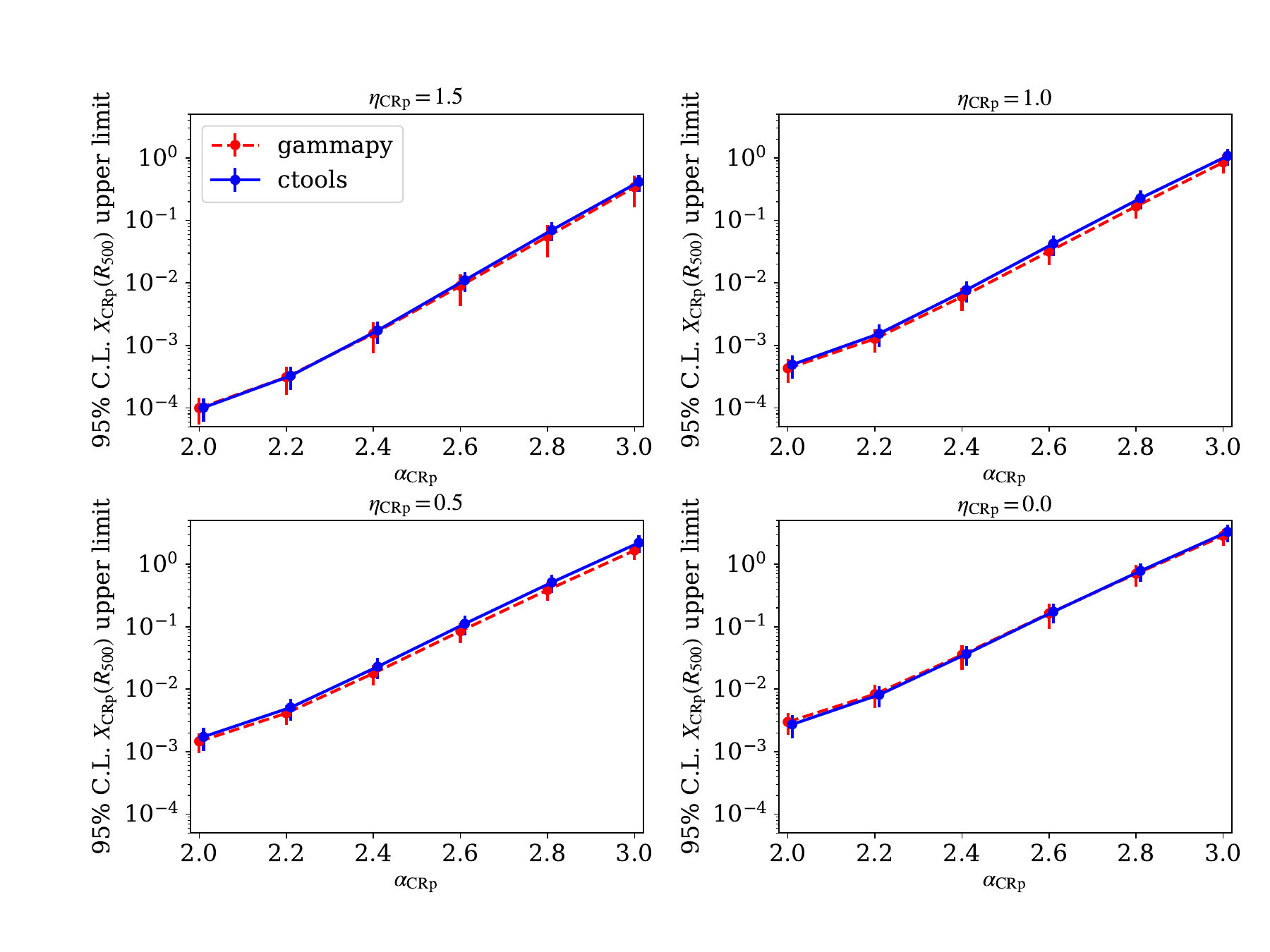}
	\caption{Comparison of the exclusion limit on the parameter $X_{\rm CRp}$ obtained with {\tt ctools} and {\tt gammapy}. The four panels give the results for the different values of $\eta_{\rm CRp}$. The error bars represent the standard deviation of the upper limits and the data points give the means. A set of 50 simulations is used both for {\tt ctools} and {\tt gammapy}.}
\label{fig:CR_gammapy_vs_ctools}
\end{figure}

In this appendix, we validate our baseline cosmic-ray analysis, which relies on {\tt ctools} \citep{Knodlseder2016}, using the alternative software {\tt gammapy} \citep{gammapy:2017}. First, we compute the number count prediction given the sky model discussed in Section~\ref{sec:Observation_setup} using the {\tt gammapy} framework and a similar setup as the one used for {\tt ctools}, but we use only one fixed pointing direction here. We verify that for this setup the two count cubes agree within a few percent, so that the residual difference is negligible compared to other sources of uncertainties. We then reproduce the results presented in Section~\ref{sec:Probing_the_parameter_space_with_CTA} and Figure~\ref{fig:CR_95percent_limits_X} with {\tt gammapy}. In this case, the 95\% upper limits are computed using the {\tt iminuit} software \citep{iminuit}. To reduce the computing time, we free only the normalization of each sky component in the fit and compute how the upper limit changes when considering the full parameter space (normalization plus spectral parameters) from test cases. The corresponding correction, of about 50\% depending on the model, are applied to the results. As shown in Figure~\ref{fig:CR_gammapy_vs_ctools}, {\tt gammapy} and {\tt ctools} give consistent results on the constraints obtained in the cosmic-ray parameter space within error bars. The residual differences are smaller than other sources of uncertainties presented in the paper (e.g., the cluster modeling) and are likely due to the fact that only the normalization parameters are fitted, or the slight differences in the setup. We conclude that using either {\tt ctools} or {\tt gammapy} for the cosmic-ray analysis will not affect the results presented in the paper.

\clearpage
\newpage
\section{Cosmic-ray parameter constraints in the pure hadronic model}\label{app:Parameter_constraints_pure_hadronic_model}
\begin{figure}[h!]
	\centering
	\includegraphics[width=1.0\textwidth]{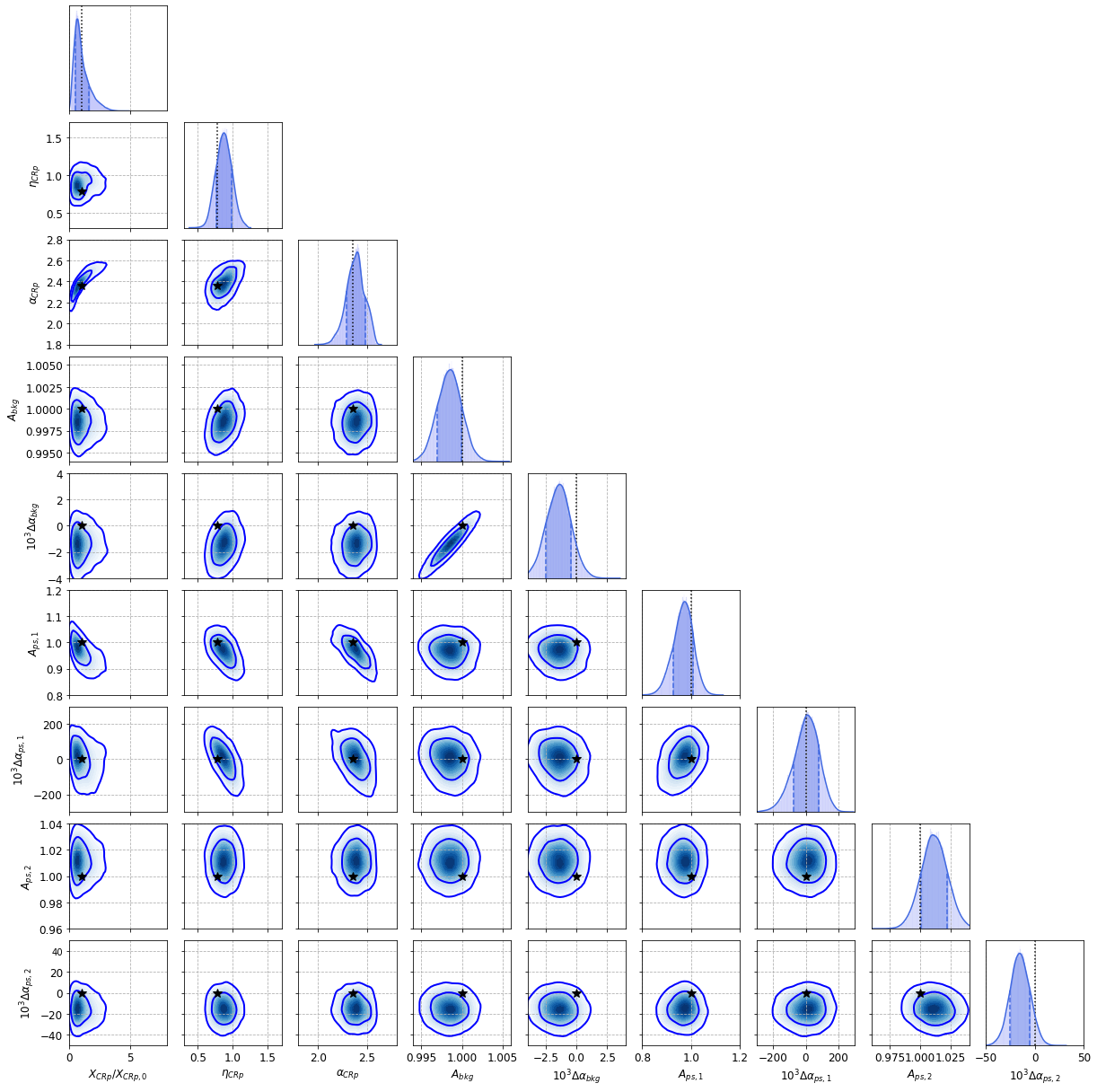}
	\caption{Example of posterior constraint on the full parameter space, as in Figure~\ref{fig:CR_triangle_plot_basline}, but in the case of the pure hadronic model. Contours provide the 68\% and 95\% confidence interval. The black star shows the input model parameters.}
\label{fig:CR_triangle_plot_pure_hadronic}
\end{figure}

In this appendix, we report the constraints in the parameter space obtained as discussed in Section~\ref{sec:Joint_spectral_imaging_constraints}, in Figure~\ref{fig:CR_triangle_plot_pure_hadronic}. In the case of this data realization, the input parameters are recovered within 68\% confidence, or right at the limit, for all parameters.

\clearpage
\newpage
\section{Interplay between the DM components and the astrophysical $\gamma$-ray sources}
\label{app:DM_interplay}

In Sections~\ref{sec:DM_result_annihil} and \ref{sec:DM_result_decay}, we have discussed the results of the template-fitting analysis in terms of the DM parameters. Now we will investigate the results obtained for the rest of free parameters (Equation~\ref{eq:DM_parameters}) and scrutinize the correlations between them. 

Appendix~\ref{app:DM_astro_params} (Table~\ref{tab:recovered_params}) shows the recovered mean values for the parameters regarding the astrophysical backgrounds (CRs, NGC~1275, IC~310, instrumental). In all of the cases we recover, within $1\sigma$, the input values that were used to produce the simulations in the first place, independently of the DM channel or mass that we fit. These results reinforce the idea that the DM component seems to be, at most, mildly correlated with the parameters describing the astrophysical sources in the area. 

\begin{figure}[h!]
	\centering
	\includegraphics[width=0.75\textwidth]{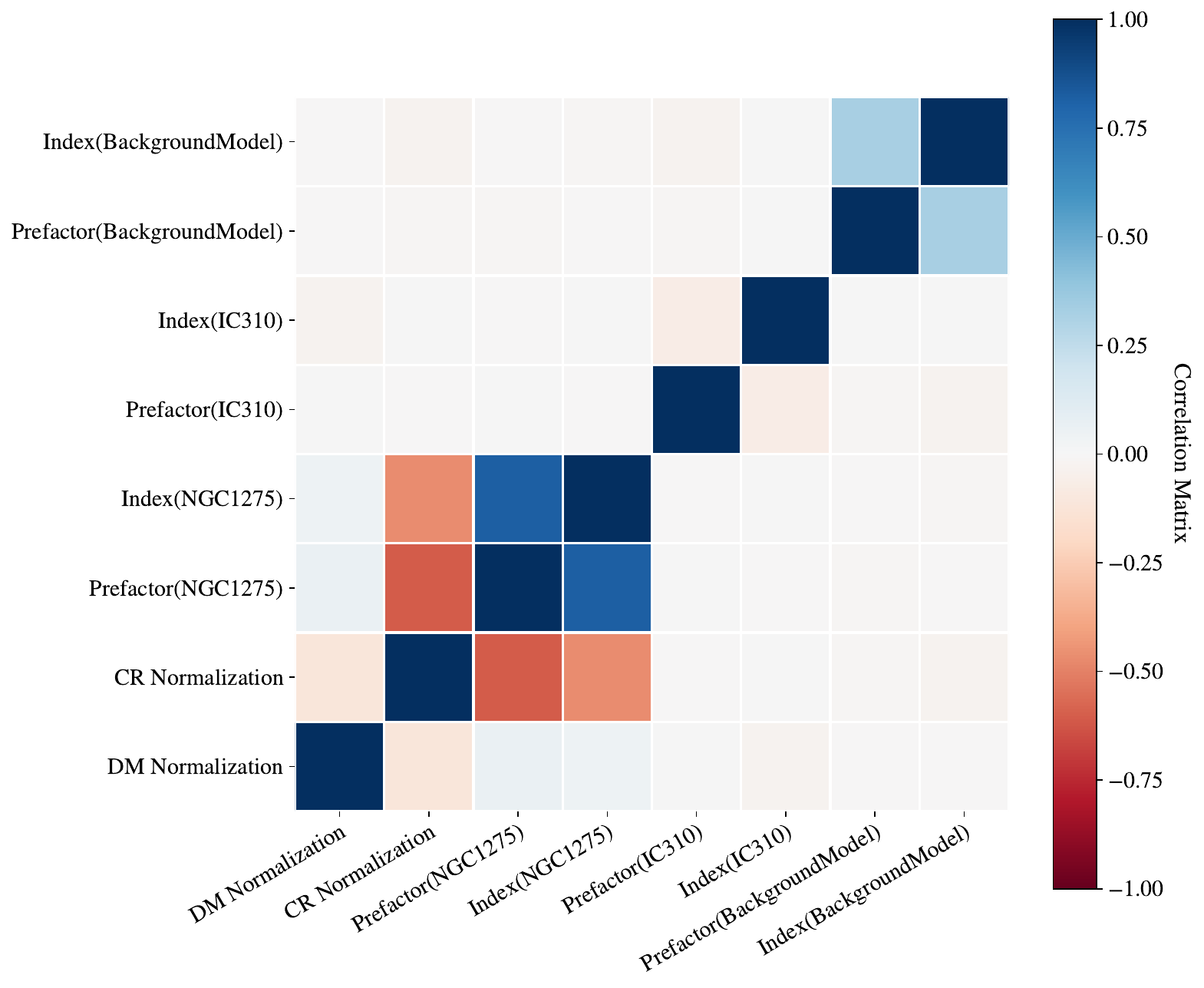}
	\caption{Correlation matrix obtained for a fit to the MED annihilation model, a DM mass of $m_{\chi}=10$ TeV and the $\tau^+\tau^-$ channel. The color scale ranges from -1 for parameters completely degenerated and anticorrelated (in red), to 1 for those degenerated and correlated (in blue). See text for discussion on the found correlations.}
\label{fig:DM_correlation_matrix}
\end{figure}

To properly quantify the correlation between the different parameters we compute the correlation matrix. For this, we use an MCMC method based on the \texttt{emcee} python package \citep{Foreman2013}. We follow the same methodology explained in Section~\ref{sec:Joint_spectral_imaging_constraints} for the astrophysical parameters ($A_{\rm CR}, A_{\rm PS1}, \alpha_{\rm PS1}, A_{\rm PS2}, \alpha_{\rm PS2}, A_{\rm bkg}, \alpha_{\rm bkg}$). We simplify our analysis only including the CR normalization due large amount of computation time that it is needed. For the DM normalization ($A_{\chi}$), we use a flat prior ($A_{\chi}\geq0$) to avoid sampling non-physical values. Since this approach is computationally very expensive, we only perform the fit for a representative DM mass of $m_{\chi}=10$ TeV, the $\tau^+\tau^-$ channel and the MED annihilation model, and use a total of 30 simulated observations. The obtained correlation matrix is shown in Figure~\ref{fig:DM_correlation_matrix}. Indeed, the DM normalization is only mildly anticorrelated with the CR normalization, showing little correlation with the parameters describing NGC~1275. This implies that the results obtained for the limits on DM cross-section and DM lifetime should be robust. We caution that this correlation matrix may be dependent on the considered DM scenario (annihilation/decay), channel or $m_{\chi}$, that were not explored in their totality given the constraints on the computation time. In Figure~\ref{fig:DM_correlation_matrix}, we can also appreciate that, with respect to the correlation matrix shown in Figure~\ref{fig:sky_model_correlation_matrix}, which did not include DM fluxes, the correlation of the CR normalization with the normalization of NGC~1275 is strengthened. Also, NGC~1275 parameters change their correlation sign. In the case of a detection of $\gamma$-rays from Perseus, these correlations should be investigated in depth and properly taken into account in the analysis. 

\section{CTA sensitivity to DM under the ON-OFF observational setup}\label{app:DM_on_off}

The ON-OFF method is one of the most standard observational and analysis approaches adopted by the existing IACTs to obtain DM constraints (e.g., \citep{2012ApJ...750..123A, MAGIC:2018tuz, CTA2019}). This choice is mainly motivated by their limited FoV and angular resolution, which makes it very difficult a (optimal) use of spatial information in large regions of the sky, as needed for a template-fitting analysis. In fact, a large number of IACT DM analyses for clusters even consider the DM-induced $\gamma$-ray fluxes as a point-like source in the target center containing the integrated DM flux of the whole object, this way neglecting its extension as a first approximation to the DM search. 

For the sake of comparison, in this appendix we first compute the CTA sensitivity to DM in Perseus assuming the DM-induced emission as a point-like source. We will only include the instrumental background and will neglect the rest of $\gamma$-ray sources in the area. This methodology, although over-simplistic, will provide a first-order evaluation of the CTA sensitivity to DM in the object and is expected to yield the most stringent (unrealizable) constraints. Yet, as described in Section~\ref{sec:background_sky}, the Perseus cluster hosts a very bright AGN, NGC~1275, in its center, which in a real observation may be difficult to neglect. Thus, in a second step, and in order to perform a more refined, realistic ON-OFF analysis, we use the DM templates to account for the spatial extension of the DM emission and, as adopted in \citep{MAGIC:2018tuz}, we place a circular mask of 0.1 deg. radius in the center of Perseus to avoid AGN contamination. In the observational setup described in Section~\ref{sec:CTA_configiuration}, IC~310 is neither in the ON nor in the OFF regions, thus it is not included in this analysis either. Since in this section we focus on the sensitivity to the DM emission, we neglect the CR component in the following, which indeed has already been properly discussed and considered for obtaining the main DM results in this paper, i.e. those derived via the template-fitting analysis in Section~\ref{sec:CTA_DM_sensitivity}. In total, we create 4 simulations for the ON-OFF analysis, two assuming the MED annihilation and DEC decay scenarios, both integrated to look like a point-like source in the center of Perseus; and two more simulations adopting the MED and DEC DM templates, this time including a central mask. 

We follow the likelihood maximization method and the likelihood ratio test ($TS$), as done in the template-fitting analysis, to search for a signal. The likelihood for the ON-OFF method corresponds to a product of Poissonian likelihoods for the ON and OFF regions, described for each energy bin (i-th) and ON-OFF regions (j-th) as:
\begin{equation}
\mathcal{L}(A_{\chi} | D) = \prod_{ij} \frac{(N^S_{ij} + \kappa_{ij}N^B_{ij})^{N^{ON}_{ij}}}{N^{ON}_{ij}!}e^{-(N^S_{ij}+\kappa_{ij}N^B_{ij})}\times\frac{(N^B_{ij})^{N^{OFF}_{ij}}}{N^{OFF}_{ij}!}e^{-N^B_{ij}},
\label{eq:likelihood_onoff_def}
\end{equation}
where $N^S_{ij}$ is the number of expected signal events in the ON region, $N^B_{ij}$ is the number of expected background events, $\kappa$ is the normalization factor to account for potentially different background acceptance in the ON and OFF regions, $N^{ON}_{ij}$ is the number of observed photons in the ON region and $N^{OFF}_{ij}$ the same but for the OFF regions\footnote{This likelihood is known to result in biased estimates for observations with very low counts, typical scenario for the high energy behaviour of faint sources.}. In this case, we only fit one parameter, the DM normalization $A_{\chi}$, resulting in profiling a one-dimensional likelihood, with the corresponding flat prior of positive normalization values. To perform this analysis, we use the \texttt{gammapy} software package for $\gamma$-ray data analysis, especially the function \texttt{FluxPointsEstimator}, with the \texttt{iminuit} backend. 

\begin{figure}[h!]
	\centering
	\includegraphics[width=0.6\textwidth]{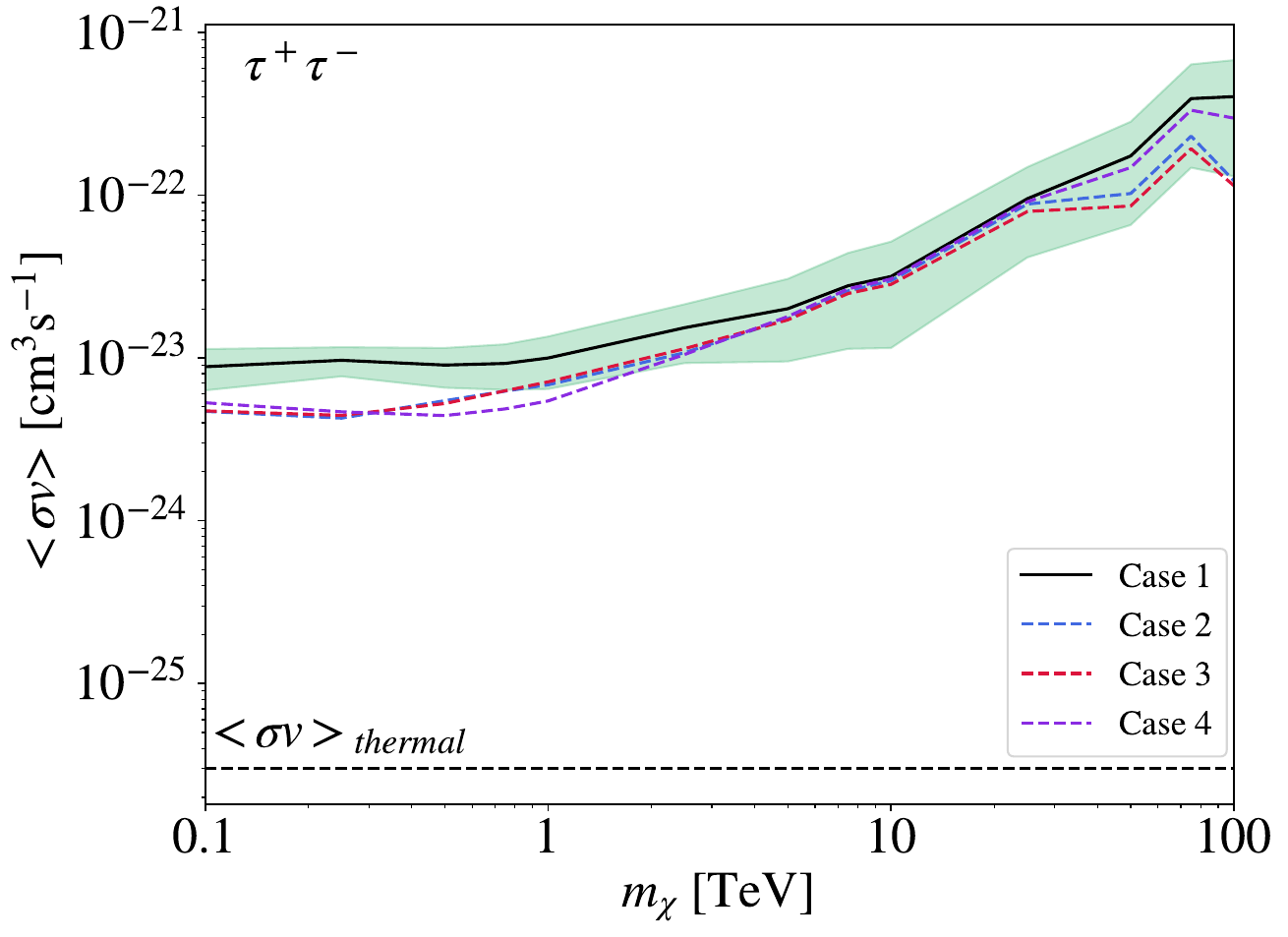}
	\caption{95\% C.L. mean upper limits to the annihilation cross-section under the different ON-OFF configurations shown in Table~\ref{tab:DM_on_off}. In all cases we adopt the MED model and the $\tau^+\tau^-$ channel, and consider the spatial extension of the DM-induced emission in Perseus. The green band represents the $1\sigma$ scatter of the projected limits around ``Case 1''. The mean is computed over 50 Poisson realizations for computational reasons. Note that no mask is included for cases 2, 3 and 4, which explains some of the observed differences among curves. See text for discussion.
	}
\label{fig:DM_onoff_configs}
\end{figure}

No hint of detection neither for annihilation nor for decay is obtained in our simulations, thus we proceed to obtain the corresponding upper (lower for decay) DM limits. For this analysis, we use a number of Poissonian realizations as suggested by our studies in Appendix~\ref{app:DM_gammapy_convergence}. In particular, we average 100 realizations in order to obtain statistically meaningful and stable results. We also check several variations of this final configuration (referred as ``Case 1'' in the following) to test the impact of the details of the observational strategy in the DM limits. In Table~\ref{tab:DM_on_off} we show the different setups here evaluated and Figure~\ref{fig:DM_onoff_configs} the corresponding upper limits in each case. For the latter, we adopt the MED annihilation case and use the 2D spatial template for the DM emission, yet we neglect the mask for computational reasons in cases 2, 3, and 4.

\begin{table}[h!]
\centering
\begin{tabular}{ | c | c | c | c | c | c |}
\hline
Case & $\theta_{\rm pointing}$ [deg] & $\theta_{\rm ON}$ [deg] & N$_{\rm OFF}$ & $\kappa$ \\
\hline
\hline
\textbf{1} & \textbf{1} & \textbf{0.5} & \textbf{3} & \textbf{3} \\
\hline
 2 & 0 & 1 & 3 & 3 \\
\hline
 3 & 0.5 & 0.5 & 3 & 3 \\
\hline
 4 & 1 & 0.5 & 5 & 5 \\
\hline
\end{tabular}
\caption{Different observational setups tested in the ON-OFF analysis of this section. $\theta_{pointing}$ is the pointing offset with respect to the center of Perseus, where the ON region is centered; $\theta_{\rm ON}$ is the aperture radius of the ON region (which for all the considered setups coincide with the aperture radius of each of the OFF regions), N$_{\rm OFF}$ is the number of OFF regions considered, and $\kappa$ is the normalization parameter introduced in Equation~\ref{eq:likelihood_onoff_def}. Our default configuration is ``Case 1'' (highlighted in bold in the table), i.e. the same introduced in Section~\ref{sec:CTA_configiuration}.}
\label{tab:DM_on_off}
\end{table}

Figure~\ref{fig:DM_onoff_configs} shows that all the tested cases lie within the $1\sigma$ scatter band of the ``Case 1'' configuration. Indeed, the discrepancy that is observed at the lowest considered energies is due to the inclusion of the mask for the ``Case 1'' configuration while no mask is present for the rest of cases. We conclude that the configuration selected in  Section~\ref{sec:CTA_configiuration} is also optimal for DM searches in Perseus based on the ON-OFF method. 

\begin{figure}[h!]
	\centering
	\includegraphics[width=0.49\textwidth]{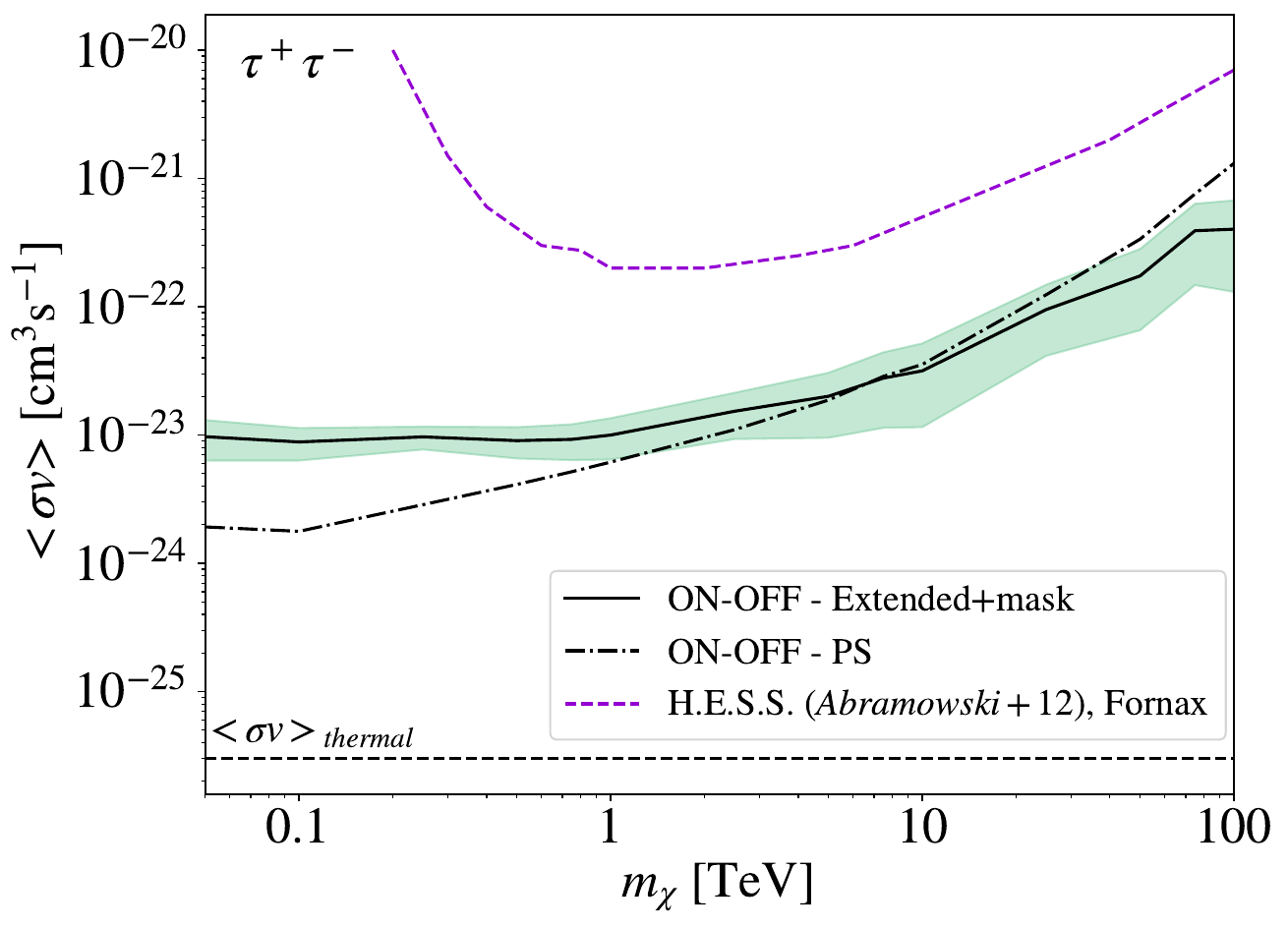}
	\includegraphics[width=0.49\textwidth]{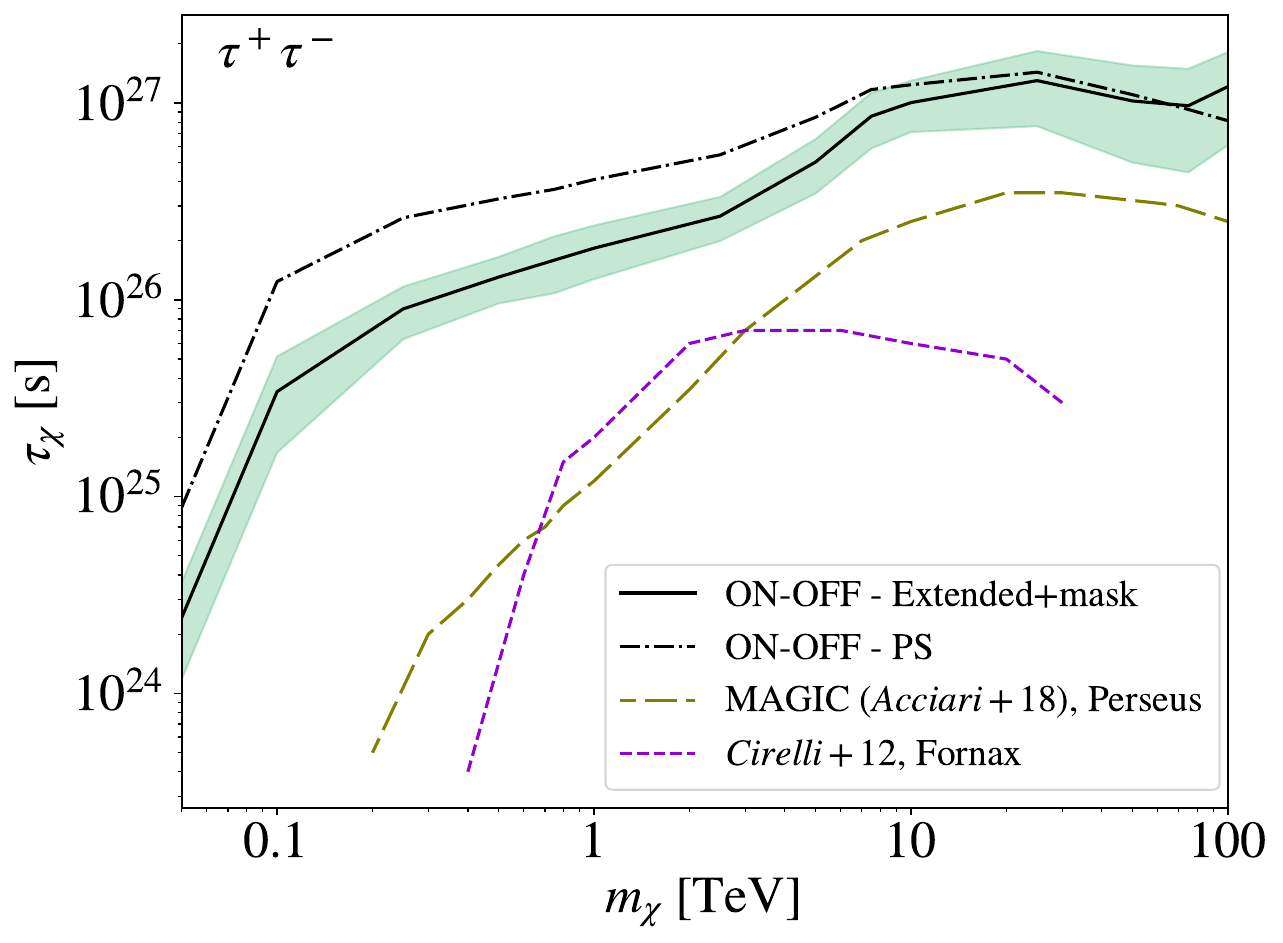}
	\includegraphics[width=0.49\textwidth]{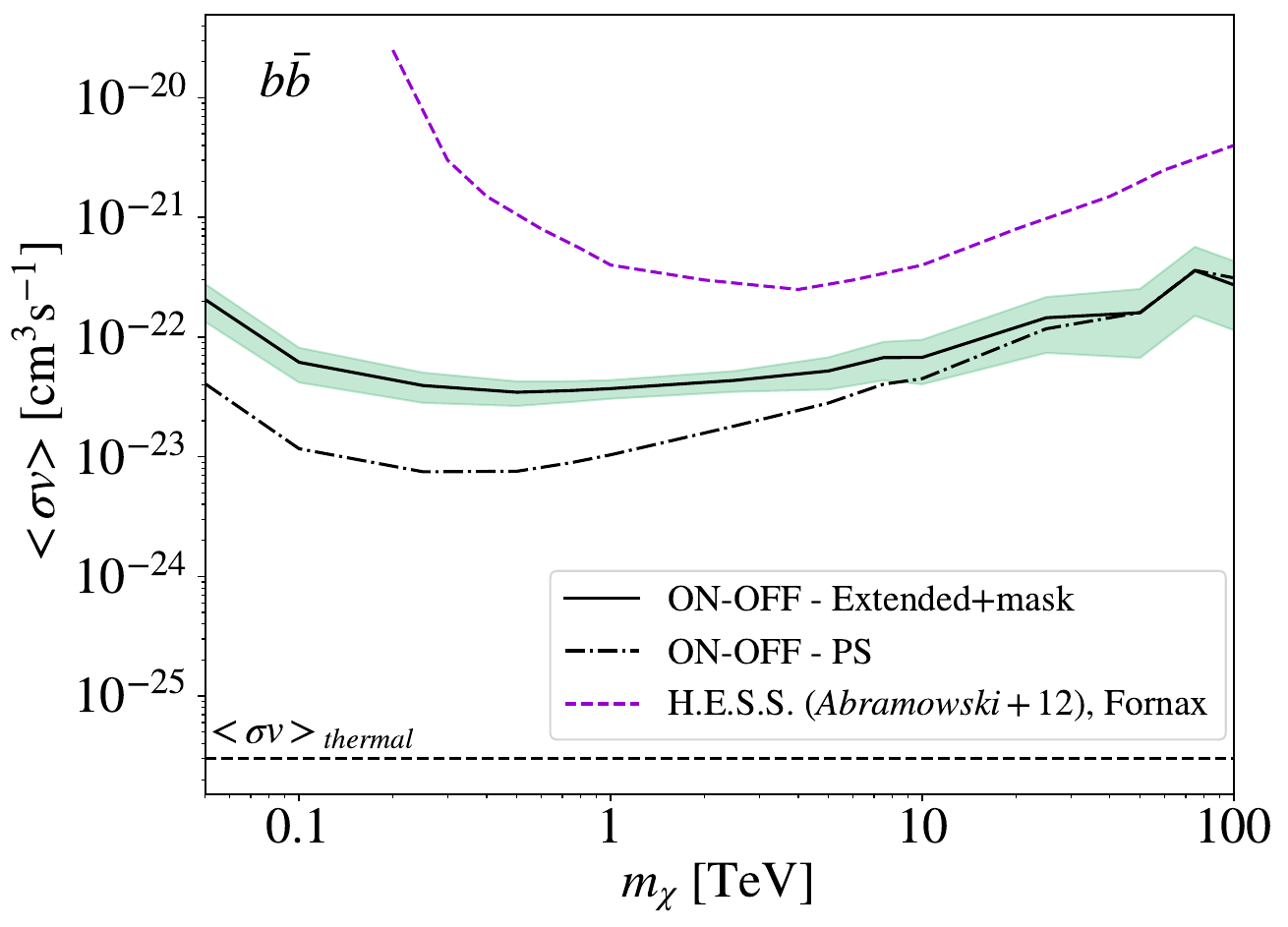}
	\includegraphics[width=0.49\textwidth]{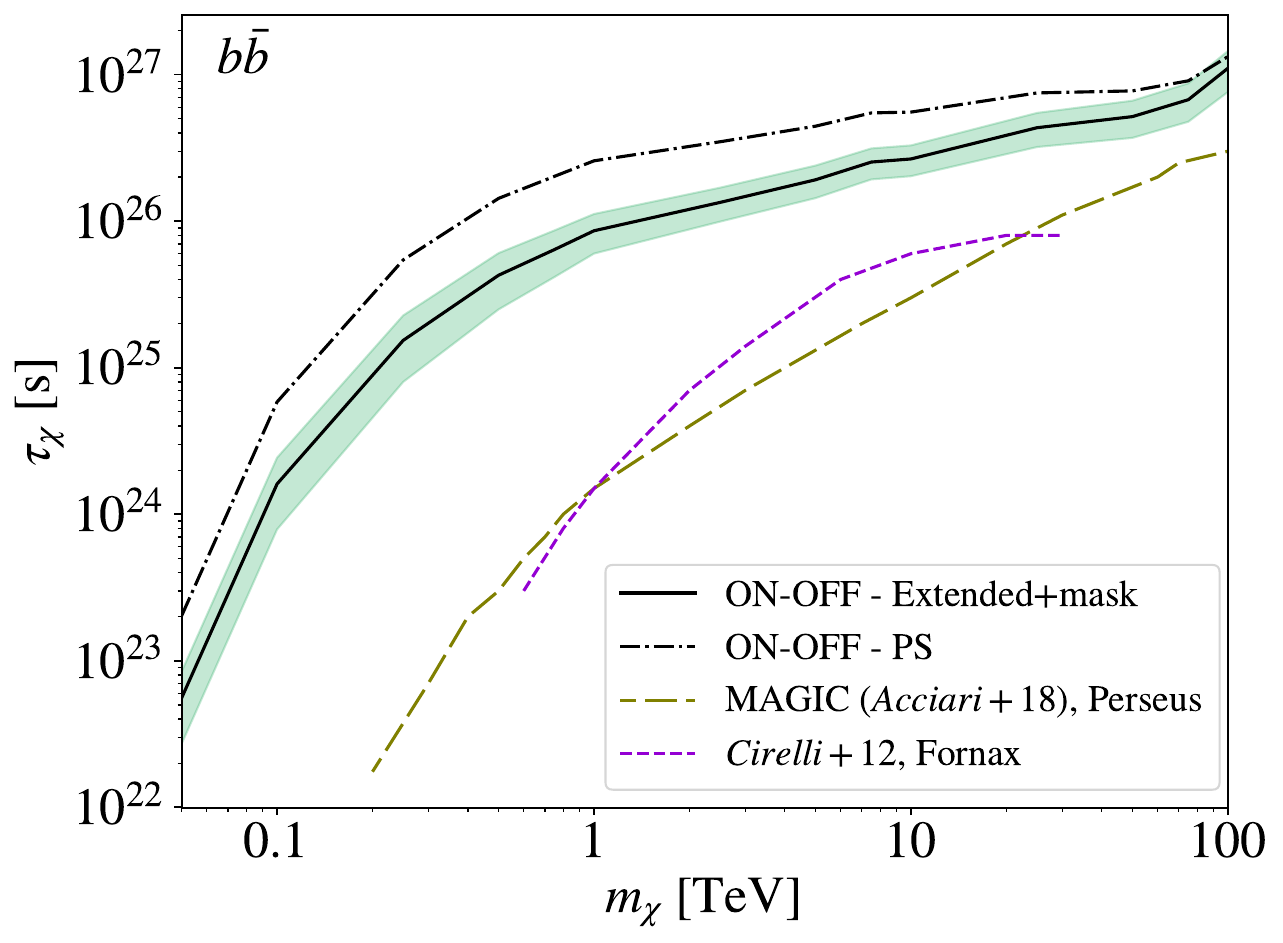}
	\caption{\textbf{Left panels (right panels):} 95\% C.L. mean upper (lower) limits for the annihilation MED model (DEC decay model) for $\tau^+\tau^-$ (\textbf{top panels}) and $b\bar{b}$ (\textbf{bottom panels}) for the ON-OFF ``Case 1'' configuration (see Table~\ref{tab:DM_on_off}). The solid black line shows the results considering the spatial extension of the DM emission plus a mask of 0.1 deg in the center of Perseus, while the dot-dashed black line corresponds to the results for the over-simplistic point-like DM source assumption; see text for details. The green band represents the $1\sigma$ scatter of the projected limits. We also show for comparison the results from the MAGIC observations of Perseus (olive long-dashed lines; \citep{MAGIC:2018tuz}) and the results from the H.E.S.S observations of Fornax (purple dashed lines; \citep{2012ApJ...750..123A, Cirelli:2012ut}).
	}
\label{fig:DM_onoff_baseline}
\end{figure}

\begin{figure}[h!]
	\centering
    \includegraphics[width=0.49\textwidth]{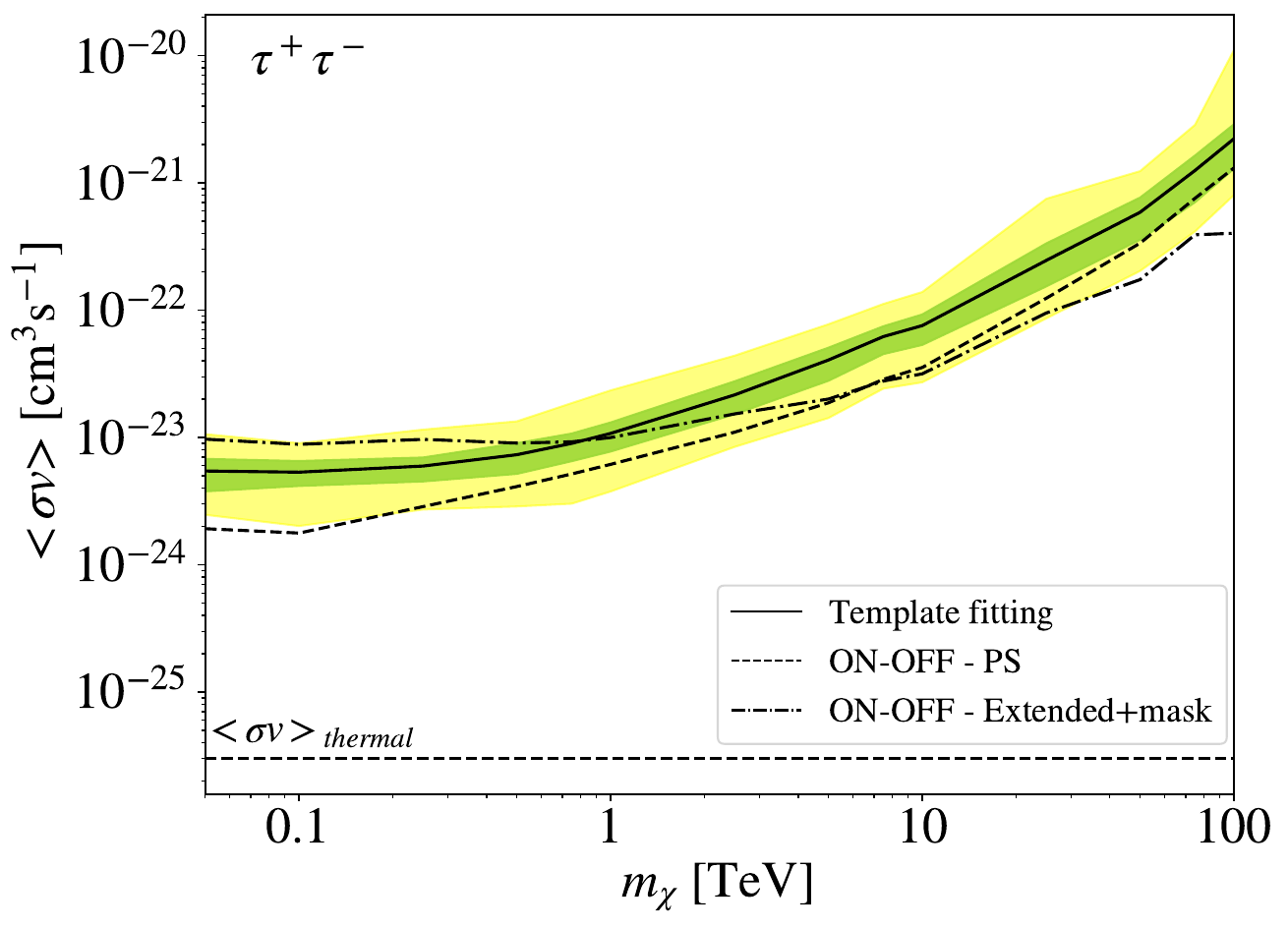}
	\includegraphics[width=0.49\textwidth]{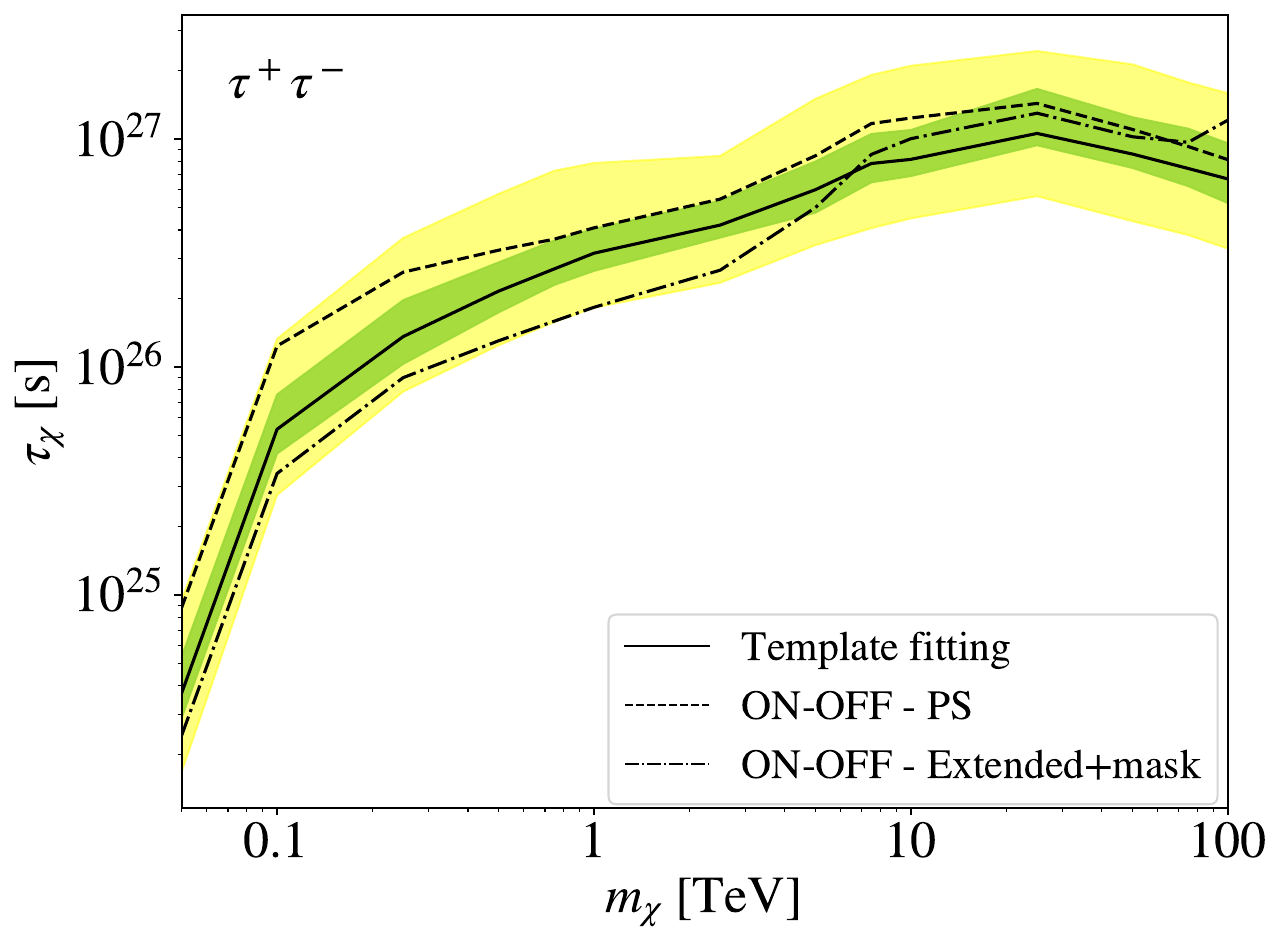}
	\caption{
    Limits for the $\tau^+\tau^-$ channel for three different analysis methods for annihilation (\textbf{left panel}) and decay (\textbf{right panel}). The solid line refers to the template-fitting approach (MED model in case of annihilation, see Sections~\ref{sec:DM_result_annihil} and \ref{sec:DM_result_decay}); the dot-dashed line is for the ON-OFF ``Case 1'' configuration, and the dotted line indicates our most simplistic analysis, i.e. ON-OFF assuming a point source for all the DM in the cluster. The green (yellow) band shows the $1\sigma$ ($2\sigma$) scatter of the projected limits using the template-fitting approach.
	}
\label{fig:DM_results_comps}
\end{figure}

We show in Figure~\ref{fig:DM_onoff_baseline} the 95\% C.L. limits for the canonical ``Case 1'' ON-OFF setup, for the different assumptions on the spatial extension of the DM-induced emission (i.e., point-like and extended), for both the MED annihilation and DEC decay models. At a first glance, we can see that the best limits are reached for the point-like source assumption in all of the cases, as expected. This is even more pronounced at the lowest DM masses considered, since in the case of ``Extended+mask'' the mask removes a comparatively larger fraction of photons, this way weakening the limits.  The comparison with other IACT results on DM searches in galaxy clusters show that CTA prospects are up to more than $\mathcal{O}$(10) constraining, depending on the DM mass range and annihilation/decay channel\footnote{Each individual analysis required of a different -- in many cases very sophisticated -- definition and treatment of the ON-OFF regions, analysis cuts, spatial morphology of the underlying DM signal, etc. Yet, we remind that none of these IACT results were computed using a template-fitting analysis. Thus, in this sense, they represent a fair comparison with our ON-OFF results.}.

Finally, we also explore the impact of the different analysis pipeline on our limits. In the right panel of Figure~\ref{fig:DM_results_comps}, together with our template-fitting analysis results for annihilation (solid line, MED model, see Section~\ref{sec:DM_result_annihil}), we show the results obtained for the canonical ``Case 1'' ON-OFF setup (`ON-OFF - Extended+mask', dot-dashed line). We can appreciate the loss in sensitivity in the lower mass range, mainly due to the central mask. In the high mass range (above $\sim1$ TeV), the limits become a factor up to $\sim4$ times more constraining than the template-fitting ones, thanks to an over-simplistic modelling of the rest of astrophysical sources.  In the same right panel of Figure~\ref{fig:DM_results_comps}, we also include the annihilation results for the point-like source assumption. We can notice that this approach (`ON-OFF - PS'; dashed line) is the most optimistic scenario, as expected, improving the limits of the template-fitting analysis by a factor$\sim2-2.5$, yet being within the $2\sigma$ scatter of the template-fitting results in all the explored mass range.  Although useful and relevant to understand the absolute sensitivity reach for CTA in an idealistic scenario, we recall that these `ON-OFF - PS' limits do not describe a realistic science case and correspond to an overly simplistic setup. An important conclusion from these comparisons among different analyses is that, despite the very different methods and assumptions on the modelling of $\gamma$-ray sources in the area, we obtain upper limits that lie within the $2\sigma$ scatter of the template-fitting results. This demonstrates not only the robustness of the found results but also points towards a low correlation of the DM parameters with respect to those corresponding to the rest of $\gamma$-ray sources in the cluster, including CRs (further investigated in Appendix~\ref{app:DM_interplay}). This effect is also quantified for the decay scenario in the left panel of Figure~\ref{fig:DM_results_comps}. We show our canonical results for the template-fitting analysis (solid line, see Section~\ref{sec:DM_result_decay}), together with the limits resulting from the ``Case 1'' ON-OFF setup (ON-OFF - Extended+mask, dot-dashed line) and the simplistic ON-OFF analysis for which we assume the DM emission to be a point-like source (ON-OFF - PS, dashed line). As expected, the most simplistic approach (ON-OFF - PS) is the one providing the most optimistic constraints, in this case being a factor$\sim2-2.5$ better than the template-fitting results, yet lying within the $2\sigma$ scatter of the latter. We remark that the aim of this overly simplistic approach is simply to understand the maximum sensitivity reach of CTA in an idealistic, unrealizable scenario. As for the results of the ``ON-OFF - Extended+mask'' method, the effect of the mask is clearly visible in the lower mass range as a worsening with respect to the canonical limits, while there is a light improvement for masses above a few TeV, still being withing the $1\sigma$ scatter band. Thus, the use of the different analysis methods produce mean lower limits within the $2\sigma$ scatter of our canonical, template-fitting results, in agreement with that found for annihilation as well.

\section{Convergence of the DM fits versus the number of simulations}\label{app:DM_gammapy_convergence}

\begin{figure}[h!]
	\centering
	\includegraphics[width=0.45\textwidth]{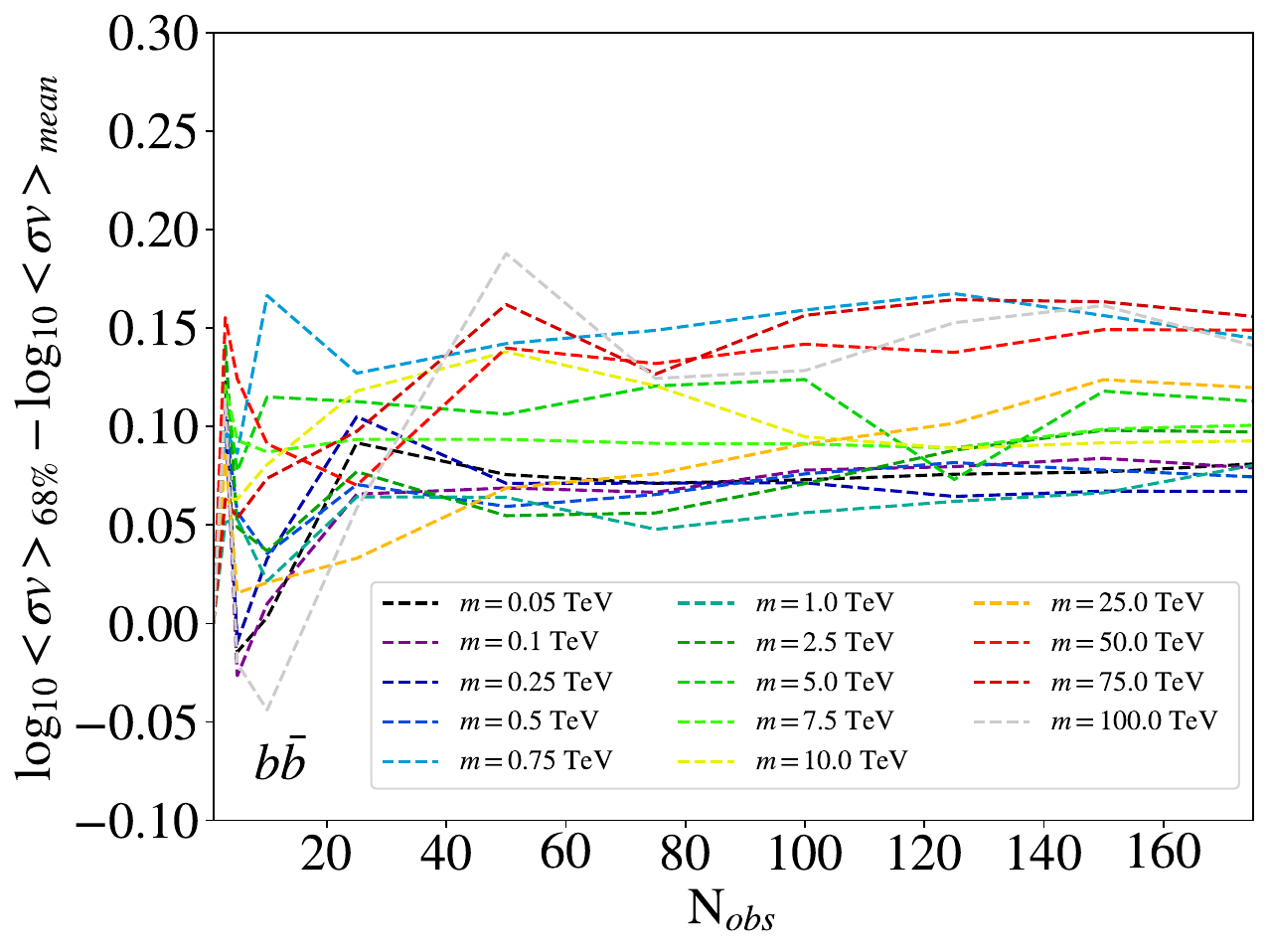}
	\includegraphics[width=0.45\textwidth]{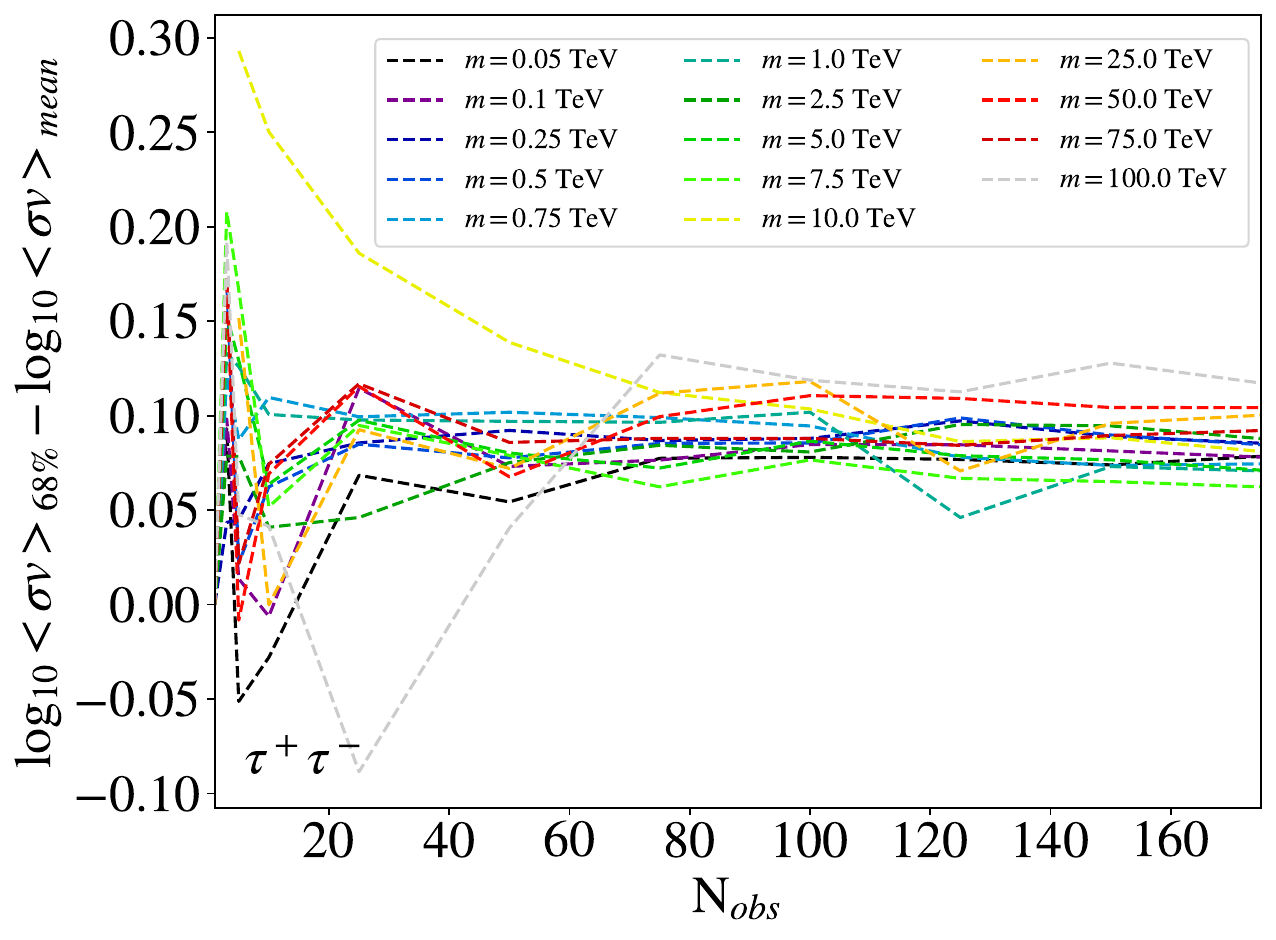}
	\includegraphics[width=0.45\textwidth]{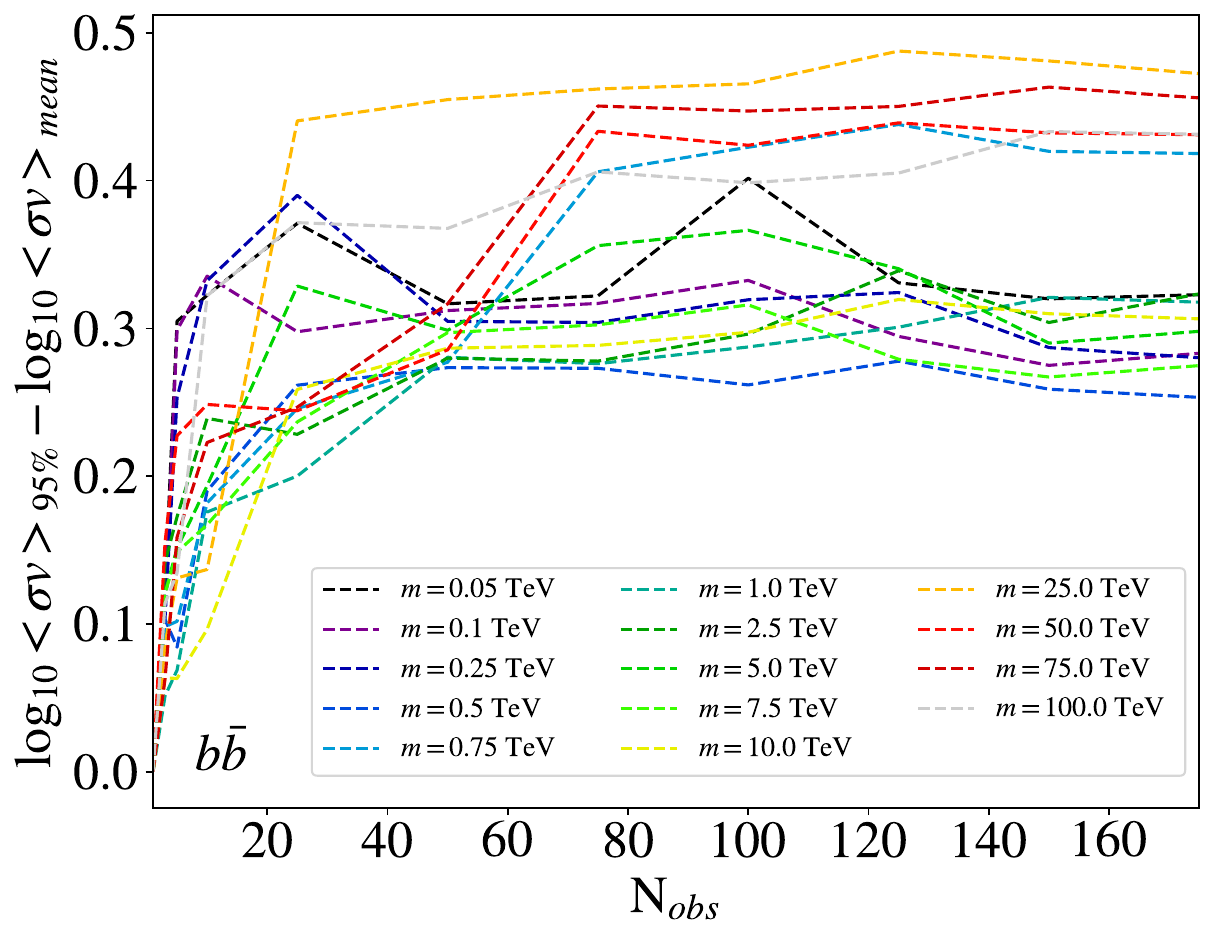}
	\includegraphics[width=0.45\textwidth]{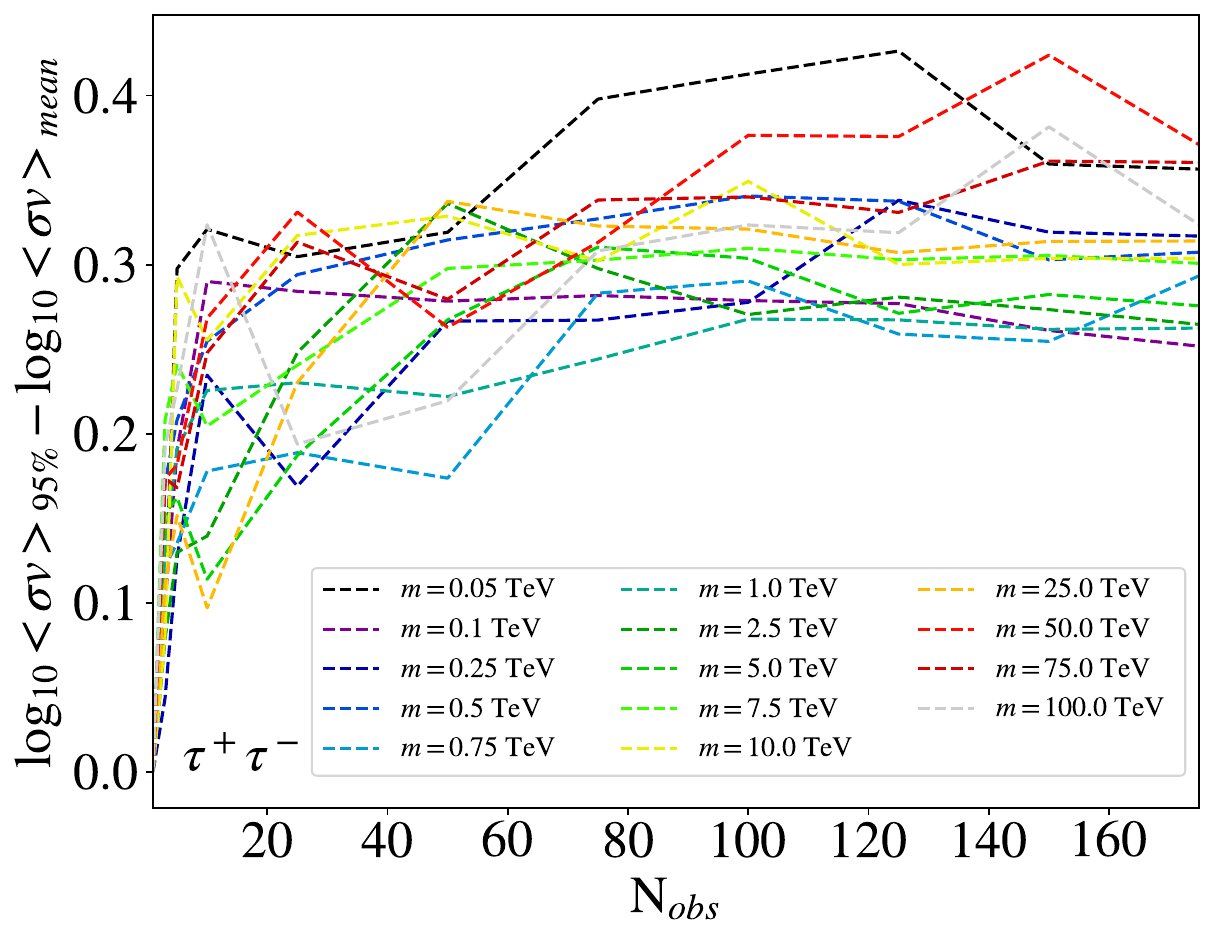}
	\caption{Convergence of the $1\sigma$ (\textbf{top panels}) and $2\sigma$ (\textbf{bottom panels}) bands in the case of the template-fitting analysis technique, for $b\bar{b}$ (\textbf{left panels}) and $\tau^+\tau^-$ (\textbf{right panels}) channels and for different values of the fitted DM mass.
	}
\label{fig:DM_convergence_fitter_t}
\end{figure}

\begin{figure}[h!]
	\centering
	\includegraphics[width=0.46\textwidth]{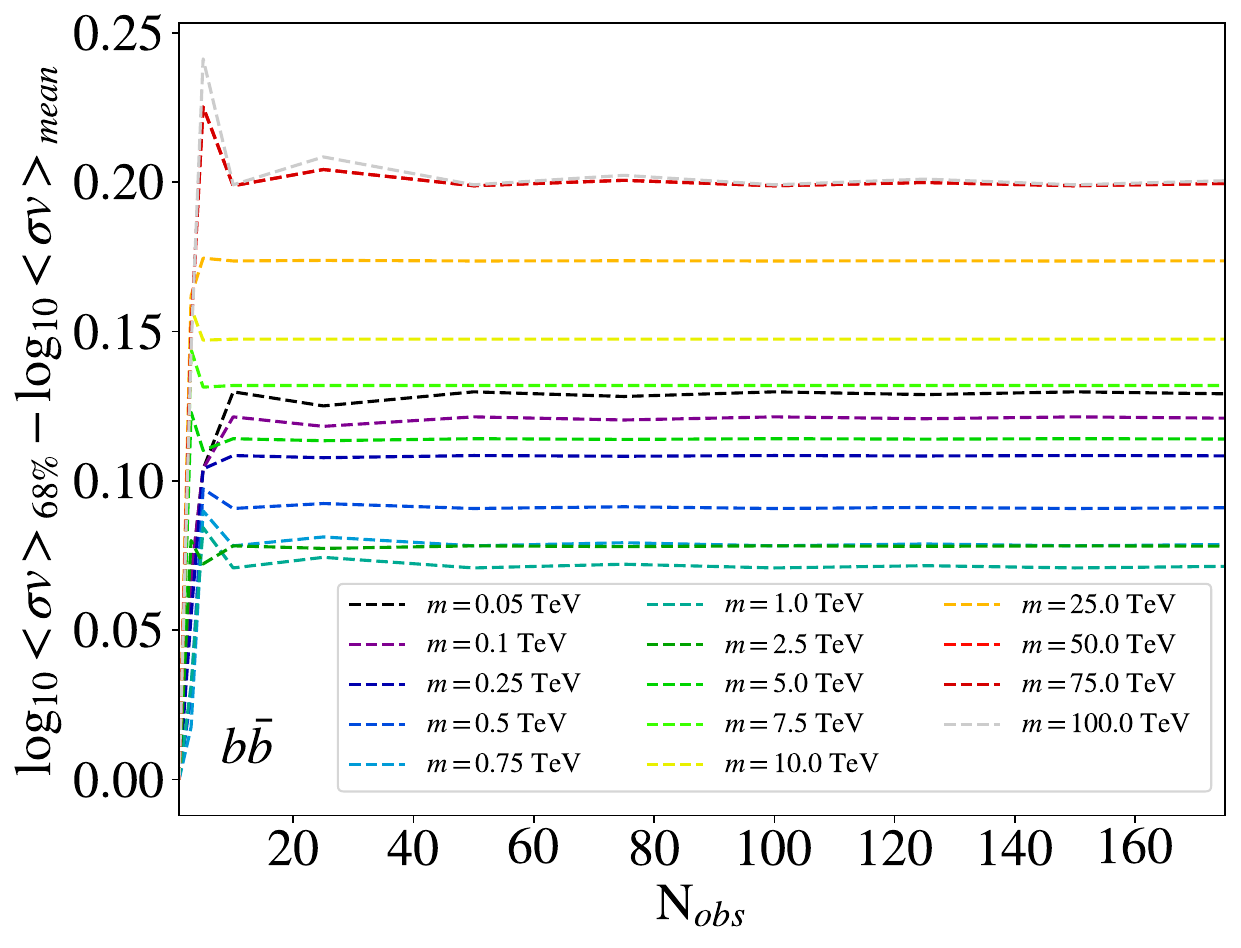}
	\includegraphics[width=0.46\textwidth]{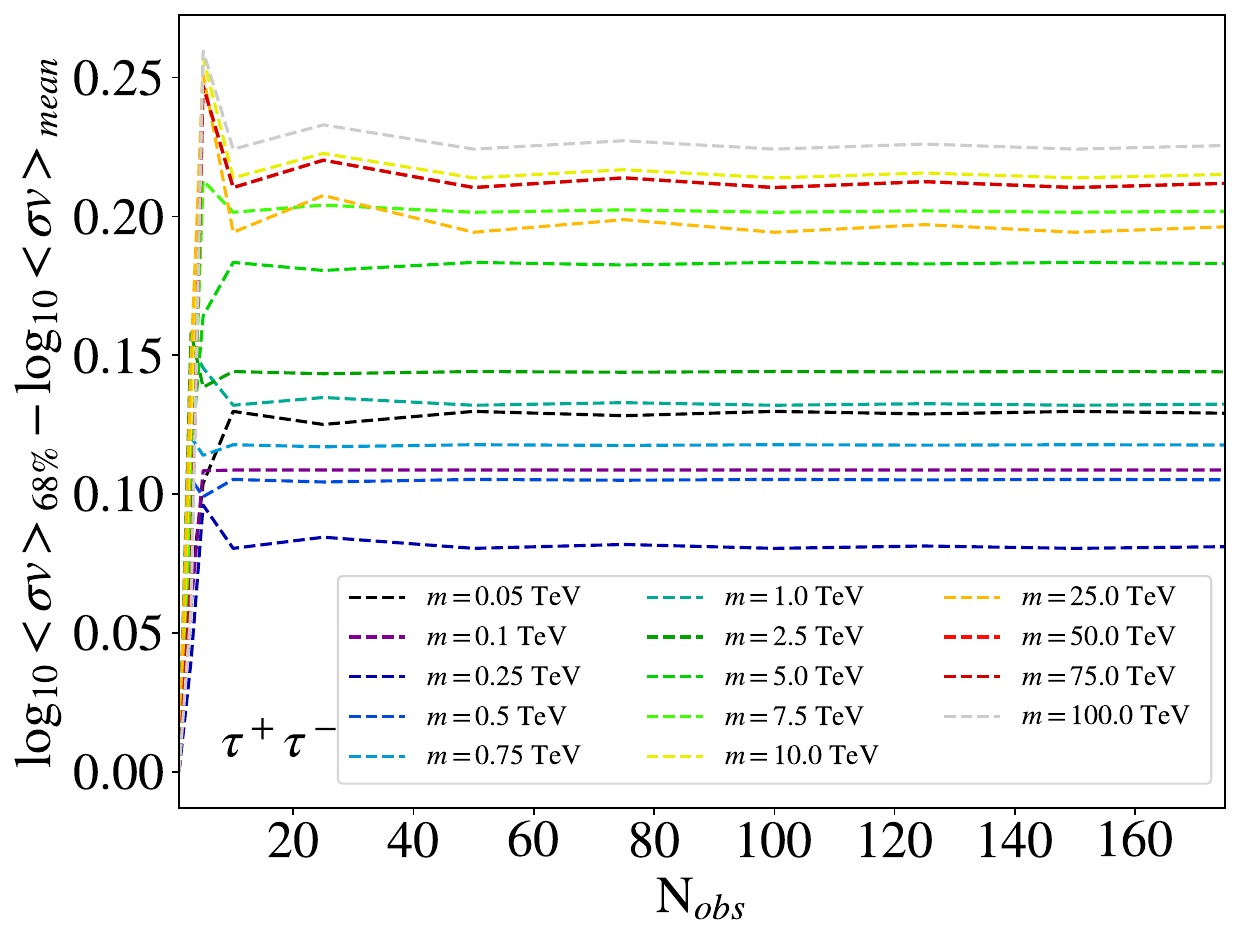}
	\caption{Convergence of the $1\sigma$ band for the case of the ON-OFF analysis technique, for $b\bar{b}$ (\textbf{left panel}) and $\tau^+\tau^-$ (\textbf{right panel}) channels, and for different values of the fitted DM mass.
	}
\label{fig:DM_convergence_fitter_onoff}
\end{figure}

The simulated observations that we use as CTA data (for more details see Section~\ref{sec:CTA_DM_sensitivity}) produce the corresponding photons and events assuming a Poisson distribution. To obtain stable and representative prospects, we need to average the results over a certain number of different outcomes of the Poisson randomization process. To guarantee a number of realizations high enough to have trustful results, at the same time representing a good compromise in terms of computational time, we analyze the evolution of the $1\sigma$ and $2\sigma$ scatter bands corresponding to the DM upper limits (for the MED annihilation scenario) with respect to the number of simulations that are considered in the computation of the mean upper limits. This is shown in Figure~\ref{fig:DM_convergence_fitter_t} and Figure~\ref{fig:DM_convergence_fitter_onoff} for the template-fitting and ON-OFF analysis methods, respectively (note though that in the latter case we only study the $1\sigma$ band).

In Figure~\ref{fig:DM_convergence_fitter_t} we can see that the $1\sigma$ bands converge around 80 realizations to a value of $\sim$0.1 dex, mostly independently of the DM mass or annihilation channel. The $2\sigma$ bands seem to converge instead after around one hundred realizations, to a value between 0.2-0.4 dex depending on the mass (no clear correlation is observed). With these results at hand, we decide to average our DM limits over 100 Poissonian realizations for the template-fitting analysis.

In Figure~\ref{fig:DM_convergence_fitter_onoff} the $1\sigma$ band converges much more quickly, around 50 realizations for any channel or DM mass. In contrast to the case of the template fitting, we appreciate a clear correlation of the convergence value with DM mass: larger masses lead to higher $1\sigma$ values. This is surely related to the likelihood we used for the ON-OFF method (Equation~\ref{eq:likelihood_onoff_def}), which is known to result in biased estimates in case of observations with very low counts, as it is the case of large WIMP masses. Indeed, our simulations including a DM annihilation flux plus the CTA instrumental background have an extremely low number of photons, resulting in a higher error of the estimated upper limit. We decide to present the mean DM limits from the ON-OFF results after averaging over 100 Poissonian realizations. 

\newpage
\clearpage

\section{Recovered astrophysical parameters from the DM template-fitting}
\label{app:DM_astro_params}

In Table~\ref{tab:recovered_params} we show the best-fit values (averaged over 100 simulations) obtained for all the parameters involved in the modelling of the $\gamma$-ray sources considered in the analysis (see Equation~\ref{eq:DM_parameters}). The shown results are for the case of including a DM template corresponding to the MED annihilation model. The recovered best-fit values are all compatible with the input value used to create the simulations within the $1\sigma$ error. Also, we tested that  these values do not seem to be correlated with either the considered DM mass or annihilation channel of the fit, indeed recovering the same value within 10\% of the error in all cases. 

\begin{table}[h!]
\label{tab:recovered_params}
\centering
\caption{Recovered values, with their corresponding errors, of the parameters describing the astrophysical sources in Perseus (see Equation~\ref{eq:DM_parameters}). In all cases, we adopt the MED annihilation model for the DM component considered in the template-fitting. The error corresponds to the symmetrical $1\sigma$ statistical error.}
\begin{tabular}{| c | c | c | c | c | c | }
\hline
Source & Parameter & Units & Input value & Recovered value & Error \\
\hline
\hline
\multirow{2}{*}{NGC~1275} & Amplitude & TeV$^{-1}$ cm$^{-2}$ s$^{-1}$ & $2.1\times 10^{-11}$ & $2.14\times 10^{-11}$ & $7\times 10^{-13}$ \\
 & Index & - &3.6 & 3.59 & 0.02 \\
\hline
\multirow{2}{*}{IC~310} & Amplitude & TeV$^{-1}$ cm$^{-2}$ s$^{-1}$ & $2.1\times 10^{-11}$ & $7.6\times 10^{-13}$ & $2\times 10^{-14}$ \\
 & Index & - & 1.81 & 1.80 & 0.02 \\
\hline
\multirow{2}{*}{IRF-BKG} & Prefactor & - & 1.00 & 0.9999 & 0.0017 \\
 & Index & - & 0.00 & $0.3\times 10^{-4}$ & $6.3\times10^{-4}$ \\
\hline
CR & Prefactor & - & 1.00 & 1.11 & 0.35 \\
\hline
\end{tabular}
\end{table}


\section{CTA sensitivity to DM in Perseus with \texttt{ctools}}\label{app:DM_ctools}

In this appendix, we describe the analysis performed with \texttt{ctools} software to search for $\gamma$-ray DM emission in the simulated CTA observations (see Section~\ref{sec:CTA_configiuration}) of the Perseus cluster. The main goal is to perform a comparison with the results already obtained with the \texttt{gammapy} software, presented in Section~\ref{sec:CTA_DM_sensitivity}, and to also quantify the compatibility among them.

\subsection{Data preparation}
\label{app:ctools_data_prep}

We simulate three different sets of observations of the Perseus cluster using the \texttt{ctools} public code \citep{Knodlseder2016}. \texttt{ctools} is a software to simulate and analyze data for $\gamma$-ray observatories. It is based on the \texttt{gammalib} C-library. To generate the observations, we use the \texttt{ctobssim} tool to simulate $\gamma$-ray events in ten energy bins starting from 30 GeV to 100 TeV, same as used for the \texttt{gammapy} analysis (check Section~\ref{sec:CTA_DM_sensitivity}). The total duration of the observations is 300 h, obtained after stacking 300 individual observations of 1 h duration each (Section~\ref{sec:CTA_configiuration}).

The first two data sets correspond to classical ON-OFF analyses, used in current IACTs. The first ON-OFF setup, as it is explained in Section~\ref{app:DM_on_off} for \texttt{gammapy}, we assume that the DM-induced $\gamma$-ray emission is described by a point source (PS). We also neglect the contributions of the other $\gamma$-ray sources in Perseus, and only consider the instrumental background contribution that is modeled via the IRFs. Also following Section~\ref{app:DM_on_off}, the second ON-OFF data set uses a more realistic modeling of the Perseus cluster by including the emission of NGC~1275. We use the DM templates to model the spatial morphology of the DM signal (see Section~\ref{sec:DM_fluxes}). Additionally, we place a circular mask of 0.1 deg radius, as in \citep{MAGIC:2018tuz}, to block out the bright emission from NGC~1275 (Equation~\ref{eq:spectrum_NGC1275}). The instrumental background is modeled by the IRFs. In the end, we also create four sets of simulations, as done with \texttt{gammapy}, for the ON-OFF observations: two for the MED model annihilation scenario, i.e., one assuming the point-like source (`PS') approximation and another one for the extended source plus the mask (`ES+Mask'); and two for DEC decay scenario, `PS' and `ES+Mask' setups. 

The last set of simulations refers to only one circular ON region of 3 deg radius, considering the contribution of all the $\gamma$-ray sources in the Perseus cluster. NGC~1275 and IC~310 emissions are described by Equations~\ref{eq:spectrum_NGC1275} and \ref{eq:spectrum_IC310}, respectively. The CR-induced $\gamma$-ray emission is described using the `Baseline' model (Section~\ref{sec:CR_baseline_model}). We consider a total of four scenarios for the DM-induced $\gamma$-ray emission, i.e. (MIN, MED, MAX) for DM annihilation, and (DEC) for DM decay.

Finally, for every set of observations we simulate a total of 100 realizations, to consider the statistical background fluctuations and compute mean parameter values, and as $1\sigma$ and $2\sigma$ bands.

\subsection{DM analysis pipeline with \texttt{ctools}}\label{app:ctools_analysis}

We follow the same analysis strategy described in Section~\ref{sec:DM_template_fitting} for \texttt{gammapy}. In the case of \texttt{ctools}, the DM analysis pipeline is available in the \texttt{ctadmtool} public code\footnote{\texttt{ctadmtool} is partially based in \texttt{ctools} and \texttt{cscripts}, and allows the use of different observation setups and analysis strategies. It can be found in \url{https://github.com/sergiohcdna/ctadmtool/tree/development}}. \texttt{ctadmtool} integrates three different steps in the calculation of exclusion limits. In the first step, the $\gamma$-ray flux induced by annihilation/decay of DM is estimated, given the parameters of the DM candidate and the spatial emission template. We use PPPC4DMID \citep{Cirelli:2010xx} to interpolate to the desired values of DM mass. \texttt{ctadmtool} computes this $\gamma$-ray flux to the number of mass points that the user wants to explore. The second step, we use the \texttt{ctlike} tool to estimate the parameters (Equation~\ref{eq:DM_parameters}) that fit the observation, get the correlation matrix and $TS$ (Equation~\ref{eq:TS_definition}) for every component, and compute the $TS$ profiles as a function of the DM normalization. The $TS$ profile is computed by letting free the parameters of all other components in the cluster. In the absence of a signal, the final step is to estimate the upper limits (95\% C.L., $\Delta TS=2.71$ with respect to the best fit) to the flux and convert to exclusion limits of the DM parameters. 

We assume that the DM normalization, $A_{\chi}$, can only take physical values ($A_{\chi}\geq0$), and that a signal detection occurs when $TS\geq25$. We adopt 10 logarithmically-spaced values of the DM mass in the range from 50 GeV to 100 TeV, and assume two representative DM channels, $b\bar{b}$ and $\tau^{+}\tau^{-}$.

\subsection{CTA sensitivity to DM under the ON-OFF observational setup with \texttt{ctools}}\label{app:ctools_DM_sens_onoff}

\begin{figure}[htb!]
    \centering
	\includegraphics[width=0.49\textwidth]{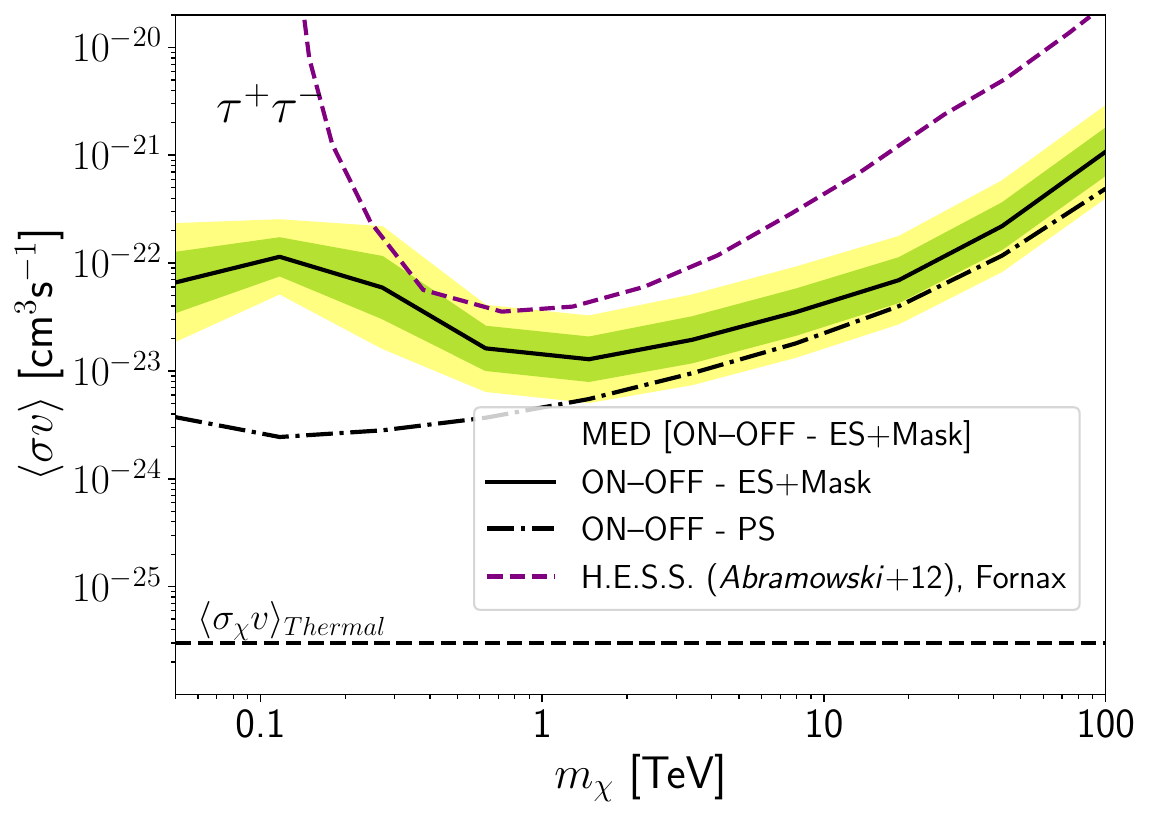}
	\includegraphics[width=0.49\textwidth]{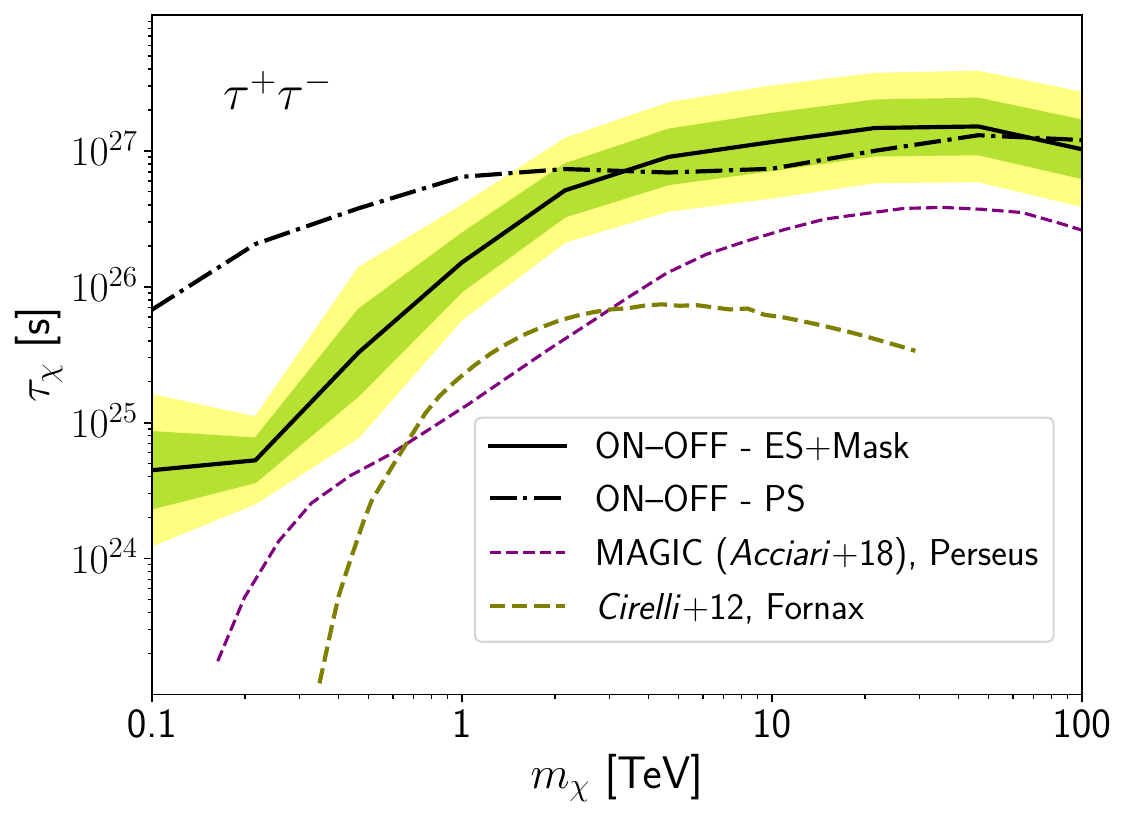}
	\includegraphics[width=0.49\textwidth]{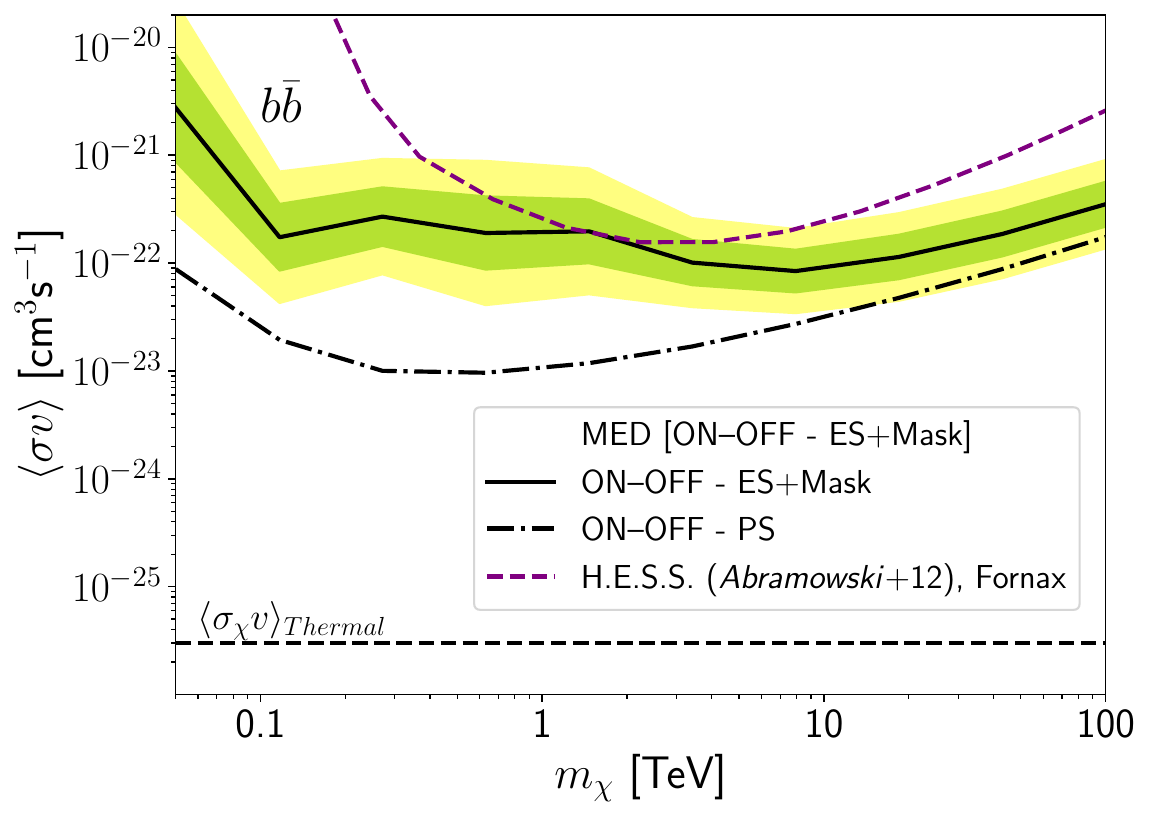}
	\includegraphics[width=0.49\textwidth]{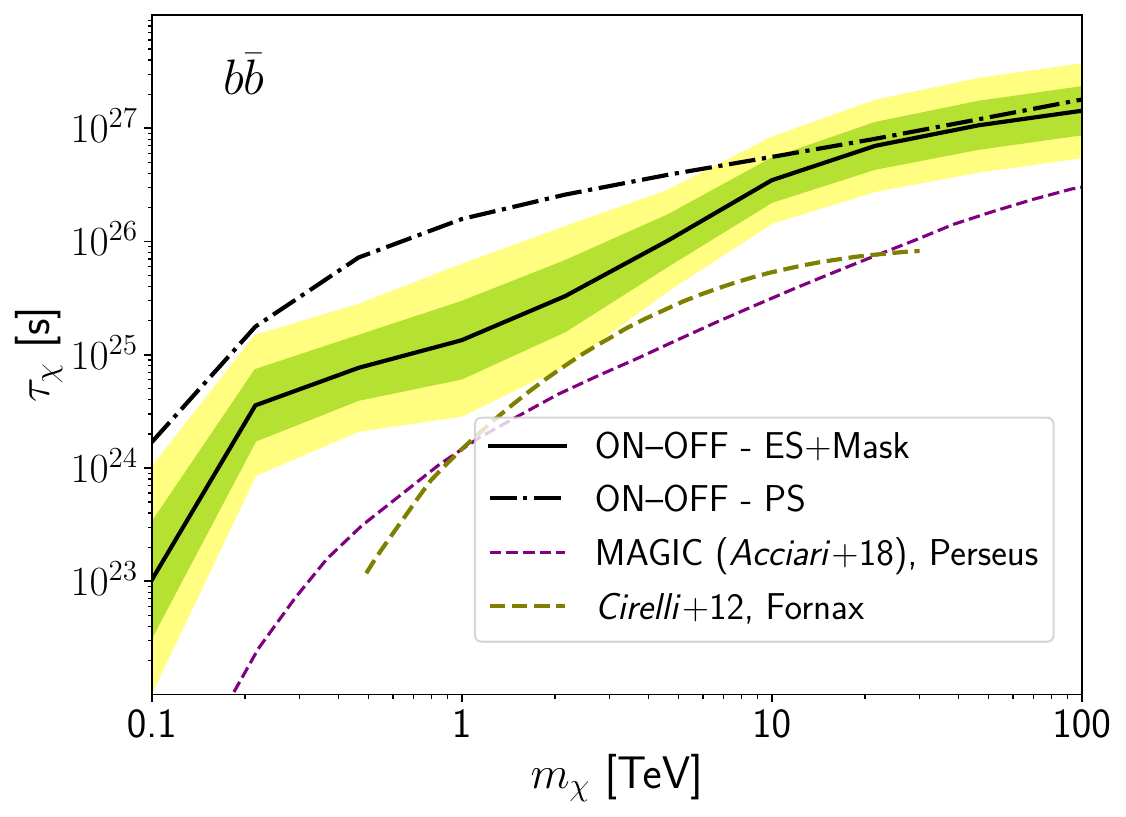}
	\caption{\textbf{Left panels (right panels):} 95\% C.L. mean upper (lower) limits for the annihilation MED model (DEC decay model) for $\tau^+\tau^-$ (\textbf{top panels}) and $b\bar{b}$ (\textbf{bottom panels}), for the ON-OFF configurations. The solid black line shows the results considering the spatial extension of the DM emission plus a mask of 0.1 deg in the center of Perseus ('ES+Mask'), while the dot-dashed black line corresponds to the results for the point-like ('PS') DM source assumption. The green (yellow) band represents the $1\sigma$ ($2\sigma$) scatter of the projected limits. We also show for comparison the results from the MAGIC observations of Perseus (purple dashed lines; \citep{MAGIC:2018tuz}) and from H.E.S.S observations of Fornax (purple dashed lines for annihilation \citep{2012ApJ...750..123A}; olive dashed lines for decay \citep{Cirelli:2012ut}).}
    \label{fig:DM_onoff_results_ctools}
\end{figure}

We do not find a DM signal neither in the annihilation nor in the decay scenarios for the different sets of ON-OFF observations. Thus, we proceed to compute the 95\% C.L. ULs to the DM-induced $\gamma$-ray flux and, from there, calculate exclusion limits of annihilation cross-section and decay lifetime as a function of the DM mass. We show in Figure~\ref{fig:DM_onoff_results_ctools} the 95\% C.L. exclusion limits for the `PS' and `ES+Mask' ON-OFF observational setups, and for the four sets of observations (Section~\ref{app:ctools_data_prep}). We observe that the best limits are obtained for the (unrealistic) `PS' case, in agreement with the results obtained with \texttt{gammapy} (Section~\ref{app:DM_on_off}). Placing a mask on NGC~1275 (`ES+Mask') weakens the limits up to $\mathcal{O}(10)$ for DM mass below 1 TeV (10 TeV) for annihilation/decay channels to $\tau^{+}\tau^{-}$ ($b\bar{b}$). 

\subsection{CTA sensitivity to DM based on template fitting with \texttt{ctools}}\label{app:ctools_DM_sens_tempfitting}

\begin{figure}[htb!]
    \centering
    \includegraphics[width=0.8\textwidth]{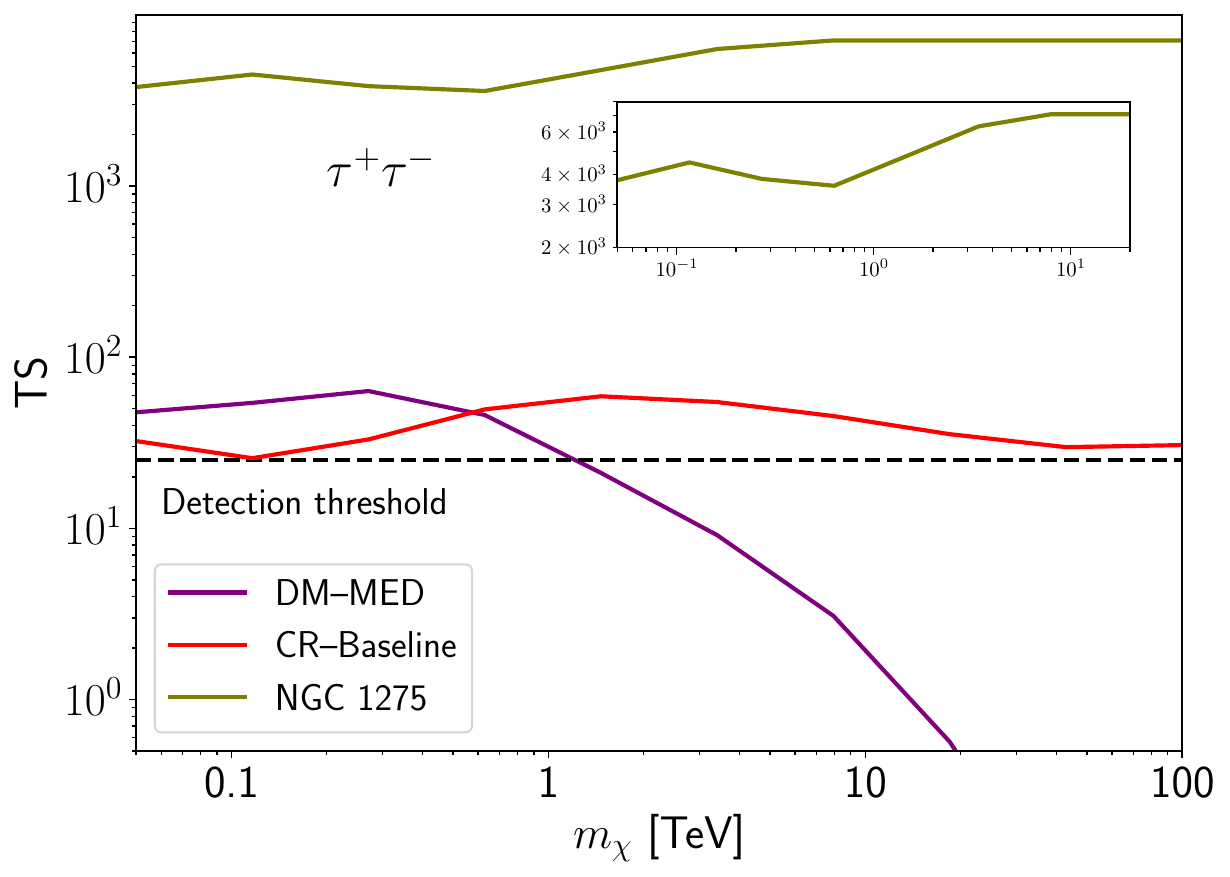}
    \caption{Mean $TS$ values associated to three emission components in the Perseus cluster. The purple, red and olive lines correspond, respectively, to the MED DM emission model with annihilation to channel $\tau^{+}\tau^{-}$, the CR-Baseline model, and NGC~1275. The black dashed line is the detection threshold ($TS\geq25$). We observe that for masses below 1 TeV, the $TS$ for NGC~1275 decreases to the half of its value for DM masses above $\thicksim$10 TeV. This change seems to be associated with a detection of the DM-induced $\gamma$-ray emission that it is not observed neither for the ON-OFF setups nor in the \texttt{gammapy}-based analyses. See text for more details.}
    \label{fig:tscomponents}
\end{figure}

In this case, as done with \texttt{gammapy} in Section~\ref{sec:DM_template_fitting}, we first check with \texttt{ctadmtool} if NGC~1275 can potentially contribute to the DM- and CR-induced $\gamma$-ray emission components in the Perseus cluster, while ideally it should not. To estimate the effect of this ``contamination'' we extract the $TS$ for every component as a function of the DM mass, for 100 realizations of the observations, and compute the mean value of the $TS$ for the (MIN, MED, MAX) annihilation and (DEC) decay scenarios.

Figure~\ref{fig:tscomponents} shows the mean value of the $TS$ for the MED DM-induced $\gamma$-ray emission, CR-Baseline and NGC~1275 as a function of the DM mass used in the fit. For clarity, we only show the results for MED DM scenario, but same results are obtained for the rest of cases. We observe that for DM masses below $\thicksim$1 TeV the $TS$ of NGC~1275 has a decrement of almost a factor 2 with respect to that obtained above $\thicksim$10 TeV. This decrement in NGC~1275 $TS$ is possibly associated with the apparent positive detection of DM-induced $\gamma$-ray signal, not observed neither in the previous ON-OFF analysis with \texttt{ctadmtool} nor in the full \texttt{gammapy} analysis. Moreover, we notice that the CR component has also significant variations in the $TS$ that are possibly correlated with the decrement in NGC~1275 $TS$, starting for DM masses below 10 TeV. These variations though do not change the fact that we always have a detected CR-induced $\gamma$-ray signal (Baseline model). We do not show the $TS$ of IC~310 because it is constant for all the DM masses considered in the fit.

\begin{figure}[htb!]
    \centering
    \includegraphics[width=0.49\textwidth]{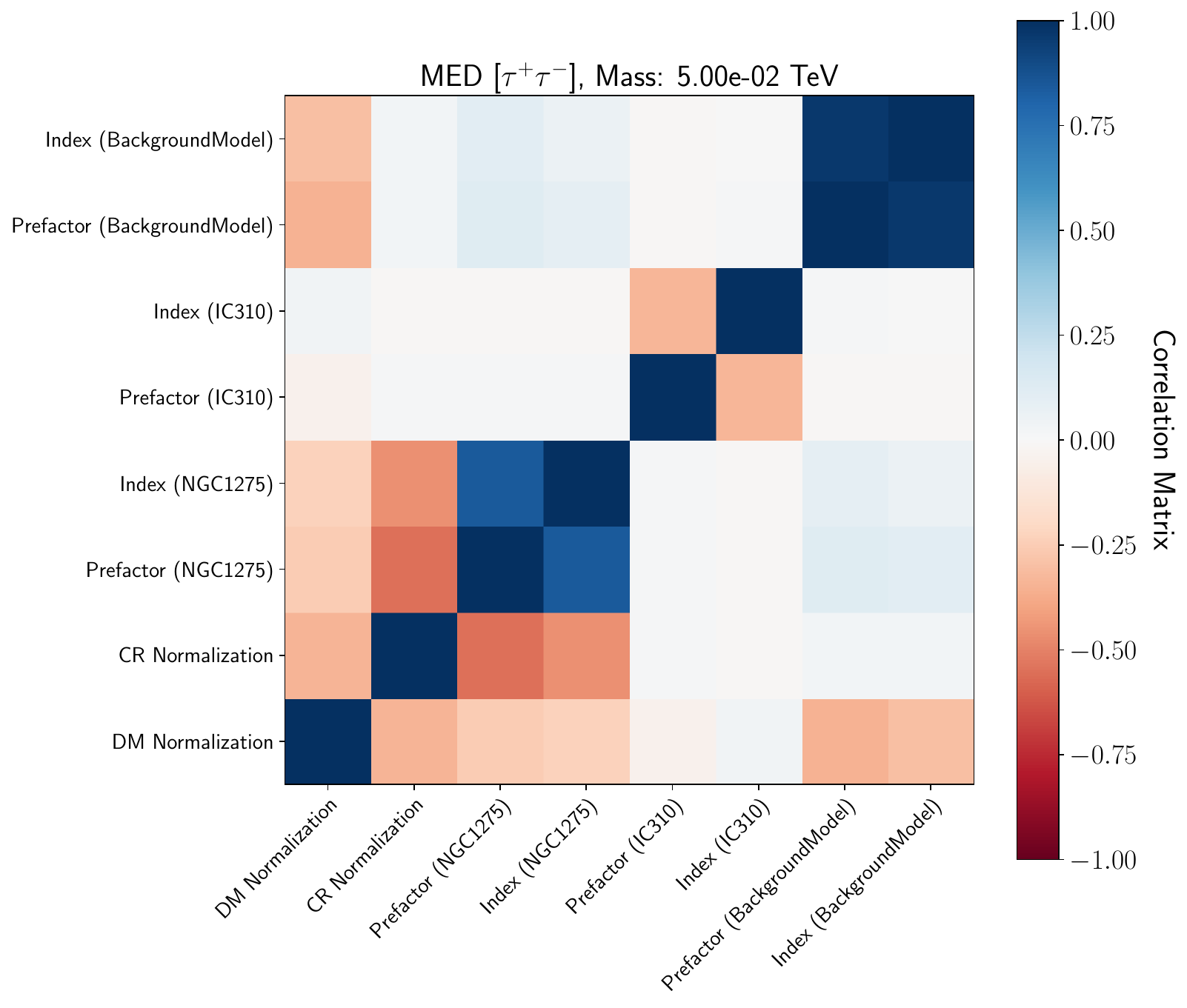}
    \includegraphics[width=0.49\textwidth]{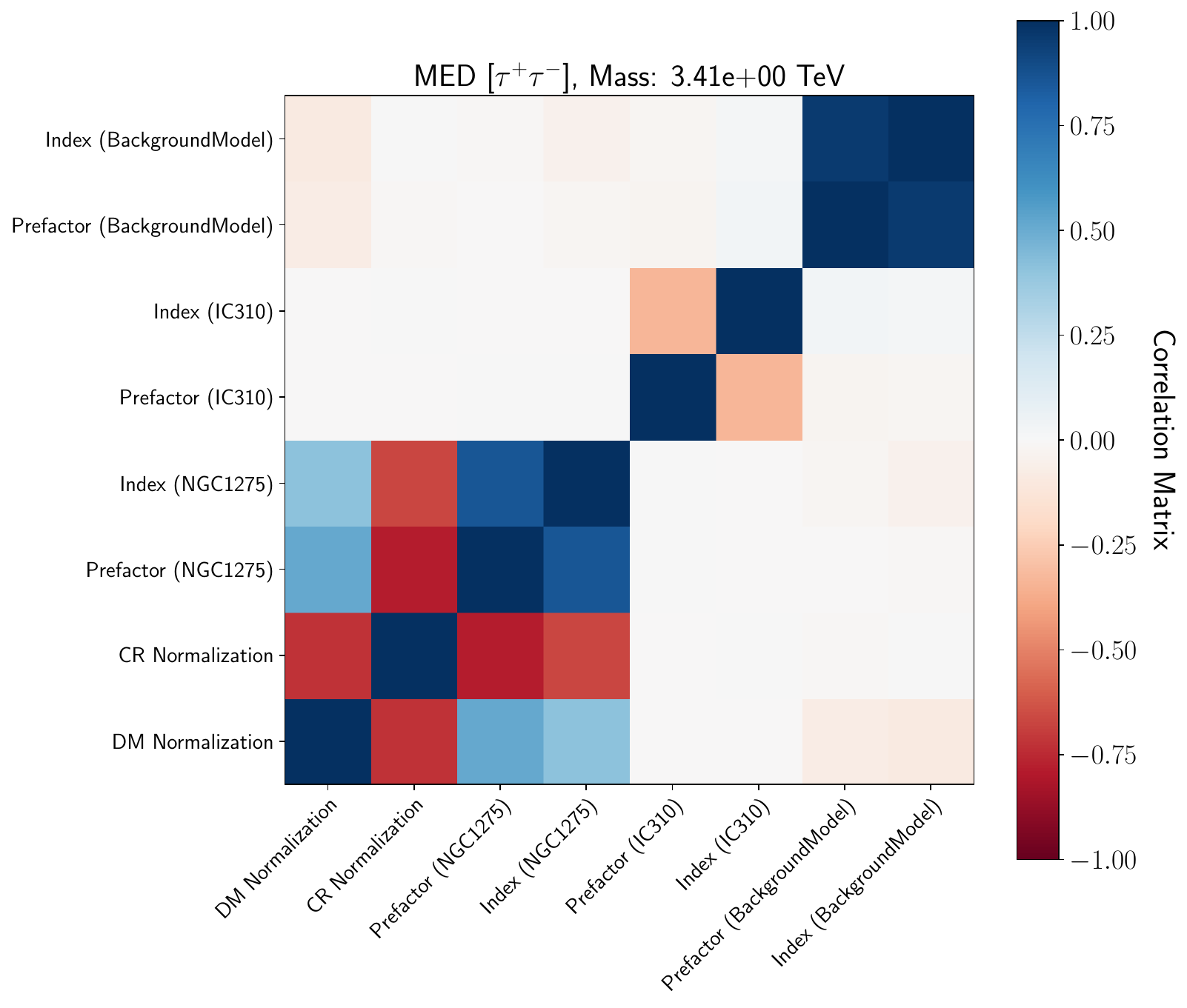}
    \includegraphics[width=0.49\textwidth]{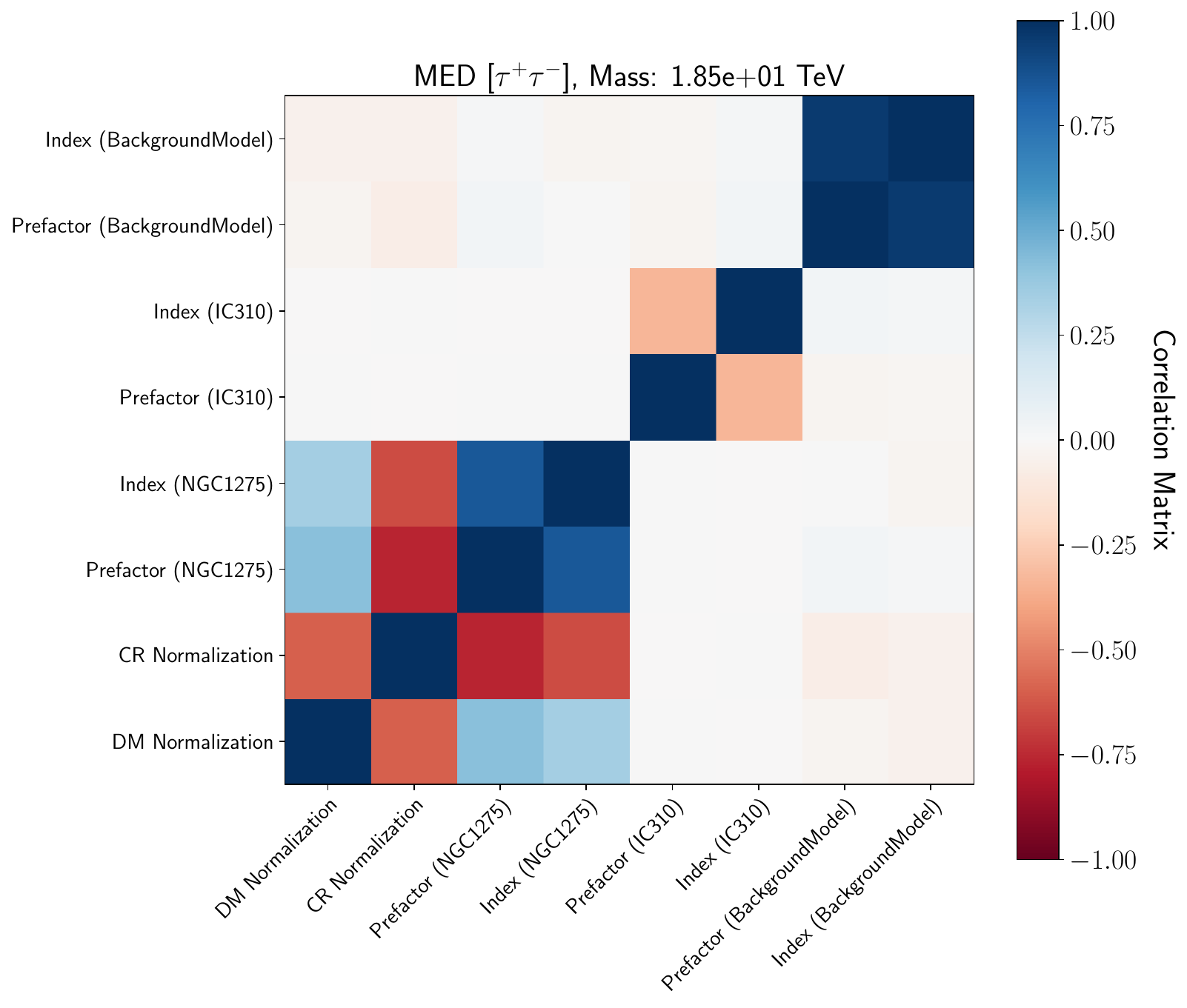}
    \includegraphics[width=0.49\textwidth]{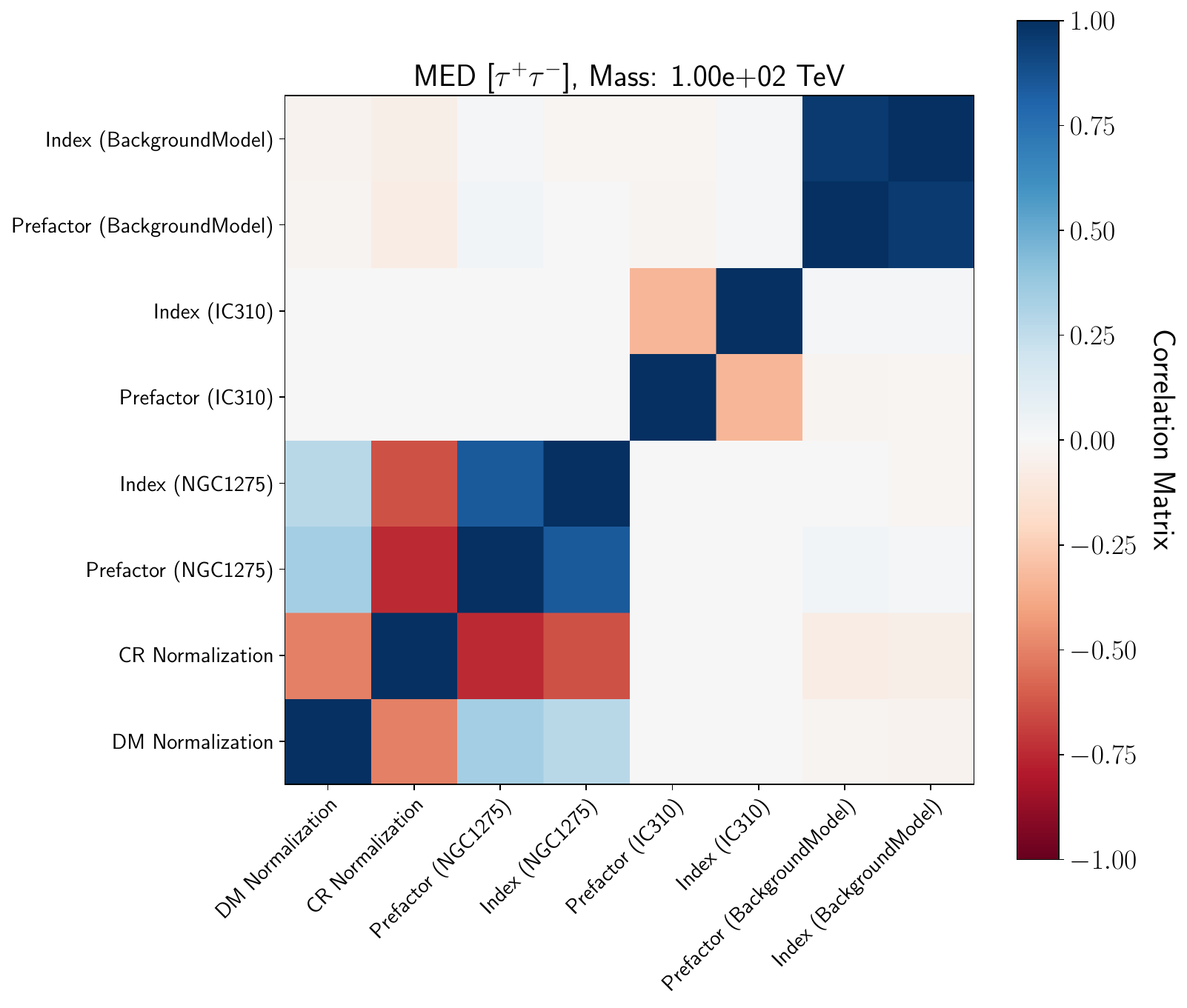}
    \caption{Correlation matrices of the free parameters of the emission models in the Perseus cluster (Equation~\ref{eq:DM_parameters}), for four different DM masses. Shown are the mean values of the correlation matrix obtained for 100 realization of observations assuming the MED annihilation scenario.}
    \label{fig:MEDCorrelation}
\end{figure}

The behaviour and our interpretation of these $TS$ curves as a function of the DM mass is supported by the correlation matrix. Figure~\ref{fig:MEDCorrelation} shows the average correlation matrix obtained for 100 realizations of the $300~\text{h}$ observations considering all the $\gamma$-ray emission components in the cluster. We select four different DM masses according to the different behaviours of the $TS$ in Figure~\ref{fig:tscomponents}. For DM masses below 1 TeV, we see that the DM normalization is anti-correlated to the free parameters of CRs and NGC~1275. We can also observe a mildly anti-correlation with the instrumental background parameters. It is interesting that the anti-correlation between DM and the instrumental background disappears as the DM mass increases. We can also observe that the sign of correlation between DM and NGC~1275 parameters changes around DM masses of $\thicksim$TeV, and then the value of the correlation decreases as the DM mass increases. This is different to the correlation matrix obtained with \texttt{gammapy} (Figure~\ref{fig:DM_correlation_matrix}), where there is not obvious correlation between NGC~1275 model-parameters and DM Normalization. We also find a strong anti-correlation for DM masses above $\thicksim$1 TeV between CR and NGC~1275, also differing with respect to the behaviour obtained with \texttt{gammapy}. IC~310 model parameters do not show any correlation with other parameters, which supports the fact that its $TS$ is constant in the whole range of considered DM masses, as mentioned before. 

From these results, we conclude that the template fitting analysis with the setup described in this section is not sufficient when using \texttt{ctadmtool}, since the strong NGC~1275 emission makes the analysis particularly tricky, and will possible lead to a fictitious detection of a DM-induced $\gamma$-ray signal.

In light of the previous results, in the following we propose and describe a specific analysis strategy to avoid the leaking of NGC~1275 emission into the other emission templates when using \texttt{ctools}. This alternative strategy will enable us to properly compute projected limits to velocity-averaged cross-section and lifetime of DM particles. The strategy is based on placing a mask in the center of the Perseus cluster so as to block out a significant fraction of the emission from NGC~1275. The main drawback of this approach is that, as the CRs and DM emission models also peak in this central region, we will be less sensitive to a putative emission from both contributions. 

\begin{table}[htb!]
    \centering
    \begin{tabular}{|c|c|}
    \hline
    Energy Range & $\theta_{\text{mask}}$ \\
    (TeV) & (deg) \\
    \hline
    \hline
    0.03 - 0.06 & 0.50\\
    0.06 - 0.15 & 0.30\\
    0.15 - 1.00 & 0.20\\
    1.00 - 10.0 & 0.12\\
    10.0 - 100.0 & 0.08\\
    \hline
    \end{tabular}
    \caption{Angular sizes (radii) of the mask applied to the simulation in the center of the Perseus cluster. The size is set to 2 times the value of the angular resolution of the CTA North Array at the energy corresponding to the lower extreme of each energy interval \citep{CTA2019}.}
    \label{tab:sizemask}
\end{table}

We select five different sizes for the mask, each of them applied to a particular energy range between 30 GeV and 100 TeV. More precisely, the angular size of the mask is set to 2 times the value of the CTA angular resolution at the energy corresponding to the lower extreme of each energy interval. Table~\ref{tab:sizemask} provides the corresponding mask sizes and energy ranges. With this definition of the mask, we guarantee that we remove $\thicksim 95\%$ of the photons coming from NGC~1275. More importantly, with this analysis configuration we do not detect a DM signal in the simulation data and confirm that the $TS$ values are not stable across all considered energies. In the same way, we do not detect a signal associated with CR-induced $\gamma$-ray emission, either.

\begin{figure}[htb!]
    \centering
	\includegraphics[width=0.49\textwidth]{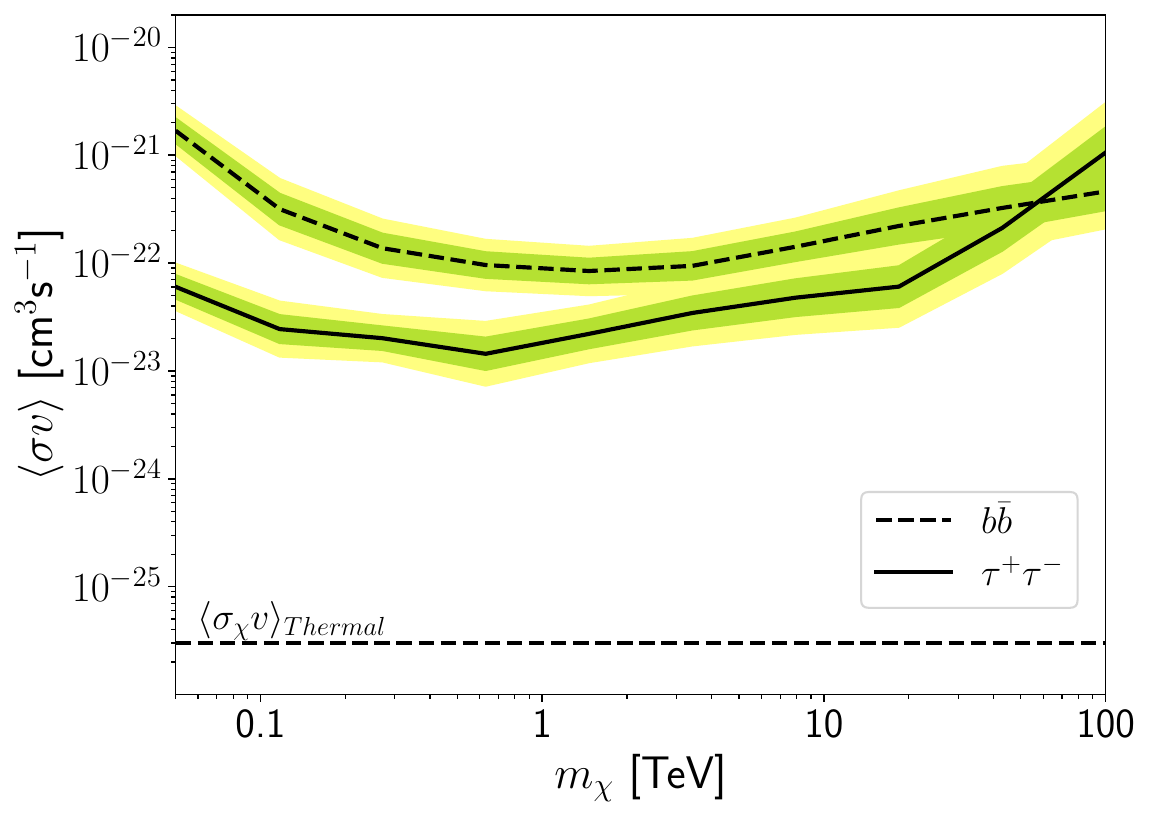}
	\includegraphics[width=0.49\textwidth]{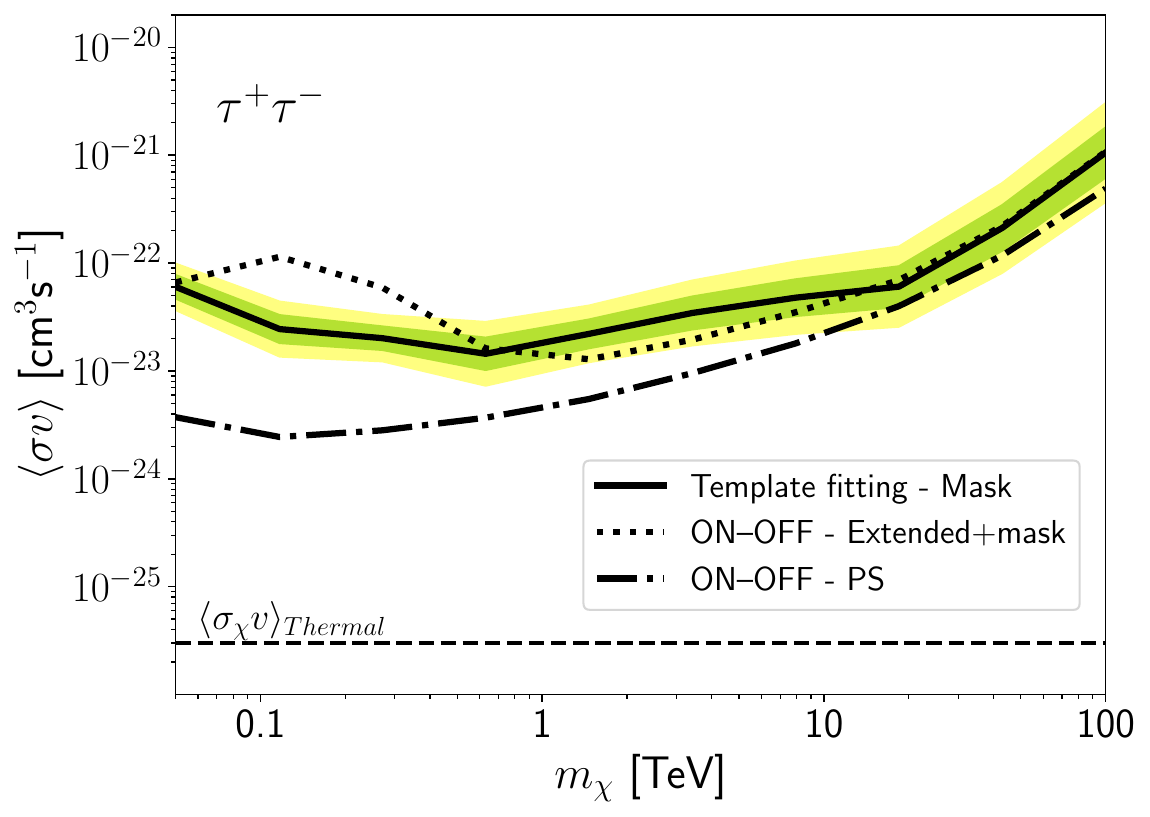}
	\includegraphics[width=0.49\textwidth]{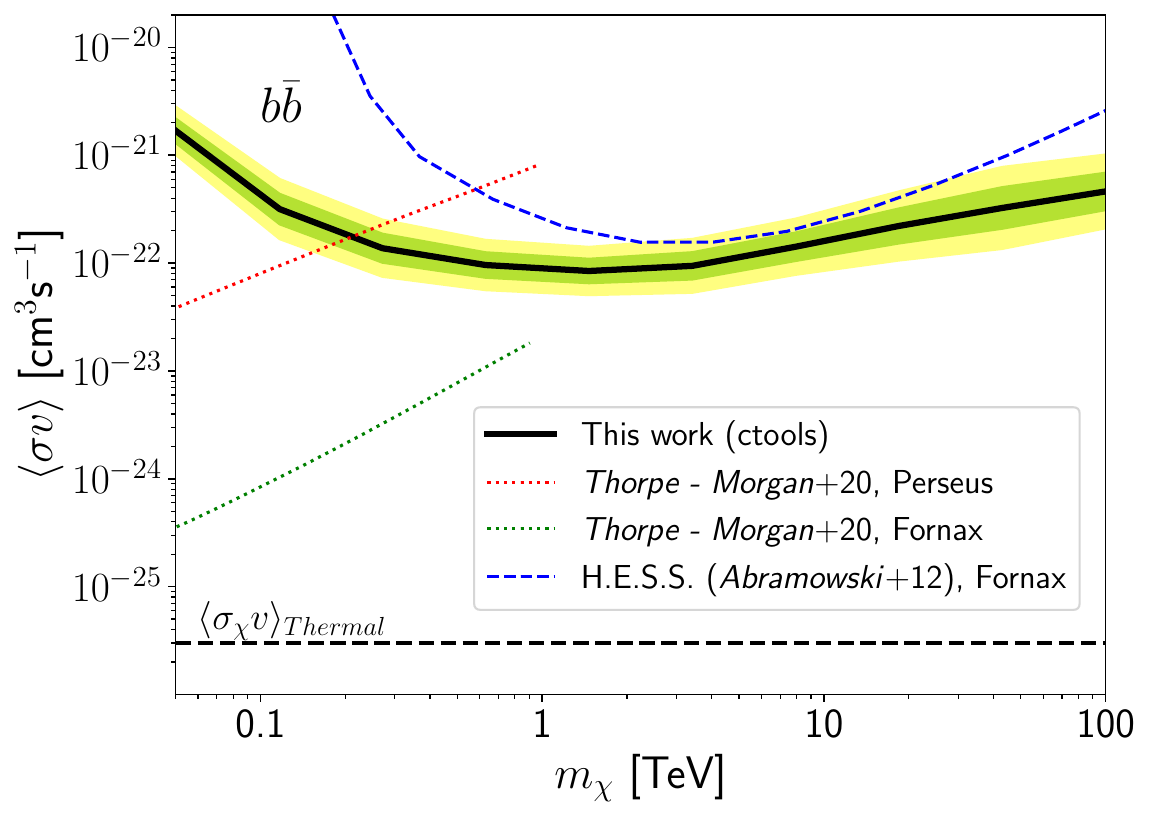}
	\includegraphics[width=0.49\textwidth]{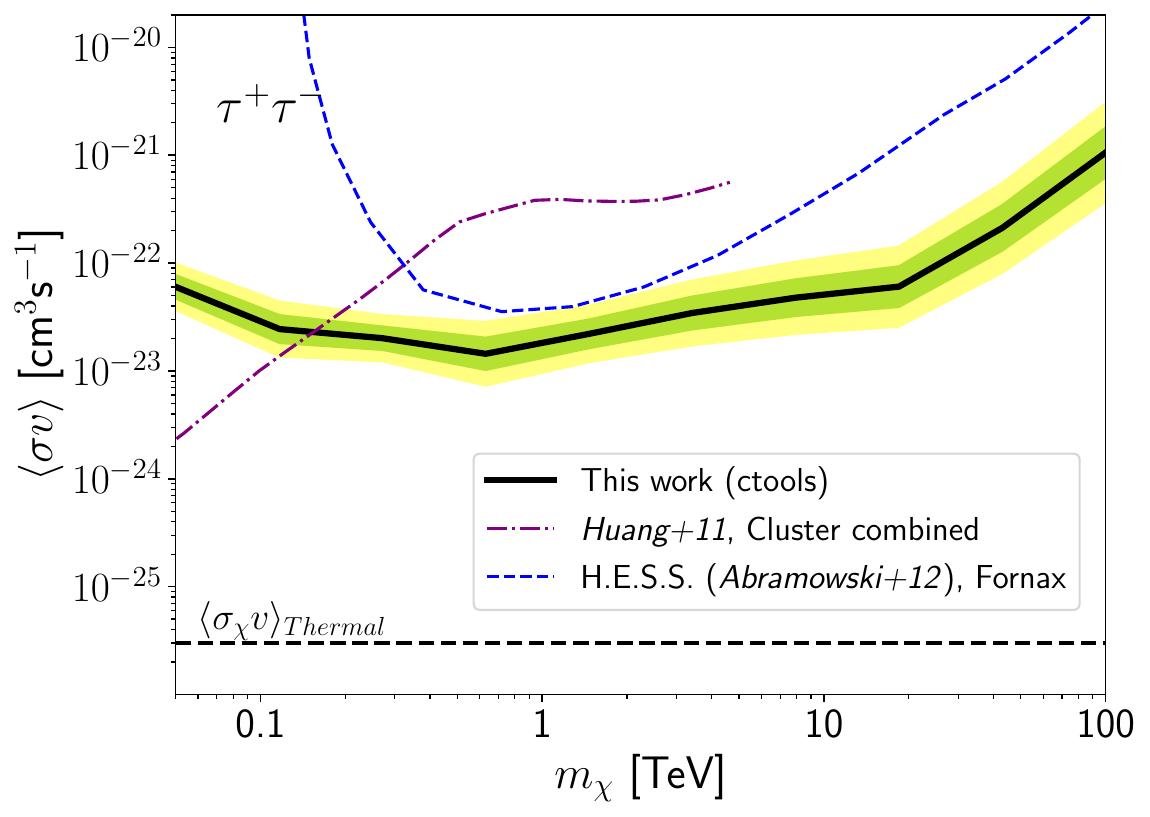}
	\caption{Sensitivity of CTA to a DM annihilation signal from the Perseus cluster, obtained via a template-fitting analysis with \texttt{ctools} and adopting a mask with those properties in Table~\ref{tab:sizemask}; see text for details. Curves represent the 95\% C.L. upper limits on the velocity-averaged cross-section for the MED annihilation model. The green (yellow) band shows the $1\sigma$ ($2\sigma$) scatter of the projected limits. The black dashed line is the thermal relic cross-section ($<\sigma v>_{thermal} = 3\times10^{-26}$ cm$^{3}$ s$^{-1}$). \textbf{Top left panel:} Upper limits for the two considered annihilation channels, $b\bar{b}$ channel (dashed) and $\tau^+\tau^-$ (solid). \textbf{Top right panel:} Limits for the $\tau^+\tau^-$ channel for three different analysis methods. The solid line refers to the template-fitting placing a mask in the center of the cluster; the dot-dashed line is for the ON-OFF analysis assuming a point-like source (PS) for the DM-induced $\gamma$-ray emission, and the dotted line refers to the ON-OFF analysis when we consider a DM spatially-extended emission and place a mask in the center of the Perseus cluster (Section~\ref{app:ctools_data_prep}). \textbf{Bottom panels:} Cross-section upper limits for the $b\bar{b}$ (\textbf{left panel}) and $\tau^+\tau^-$ (\textbf{right panel}) channels in comparison with the most recent results on DM-annihilation searches in galaxy clusters using $Fermi$-LAT (\citep{Thorpe-Morgan:2020czg}, dotted, and \citep{Huang:2011xr}, purple dot-dashed lines) and H.E.S.S (\citep{2012ApJ...750..123A}; blue dashed lines).
    }
\label{fig:DM_anna_results_ctools}
\end{figure}

\begin{figure}[htb!]
    \centering
	\includegraphics[width=0.49\textwidth]{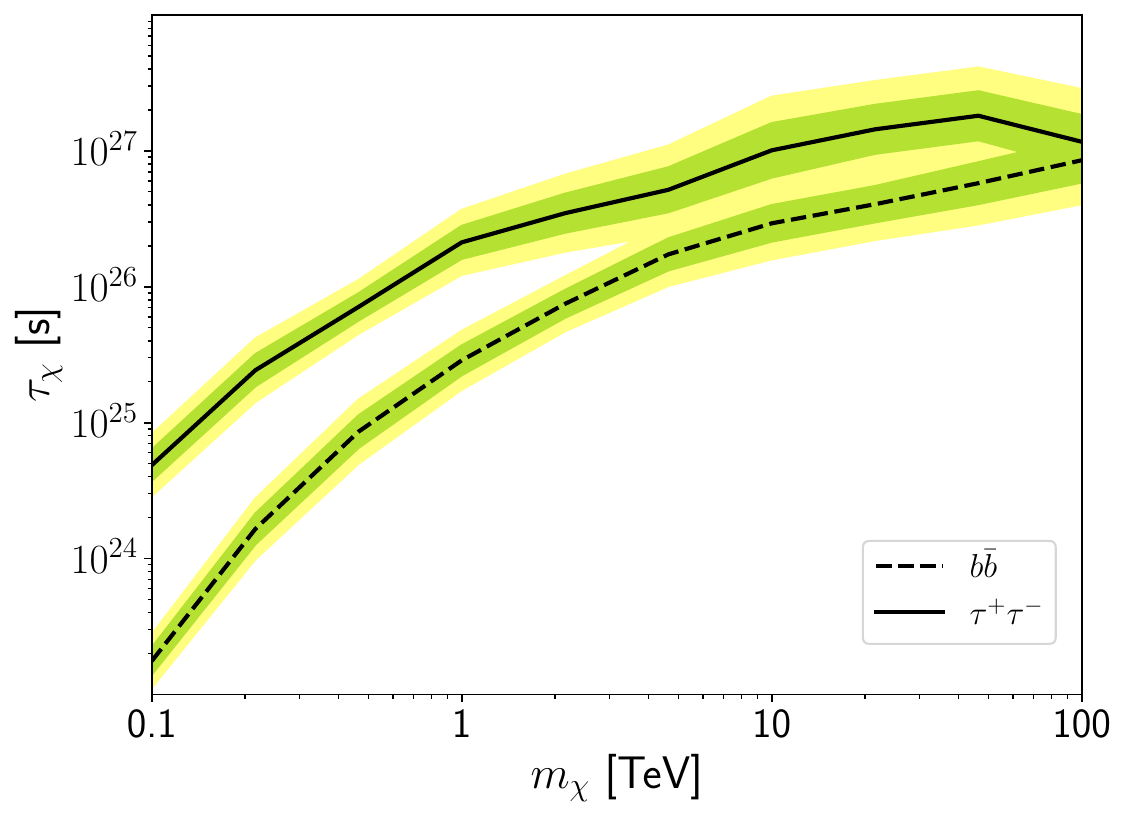}
	\includegraphics[width=0.49\textwidth]{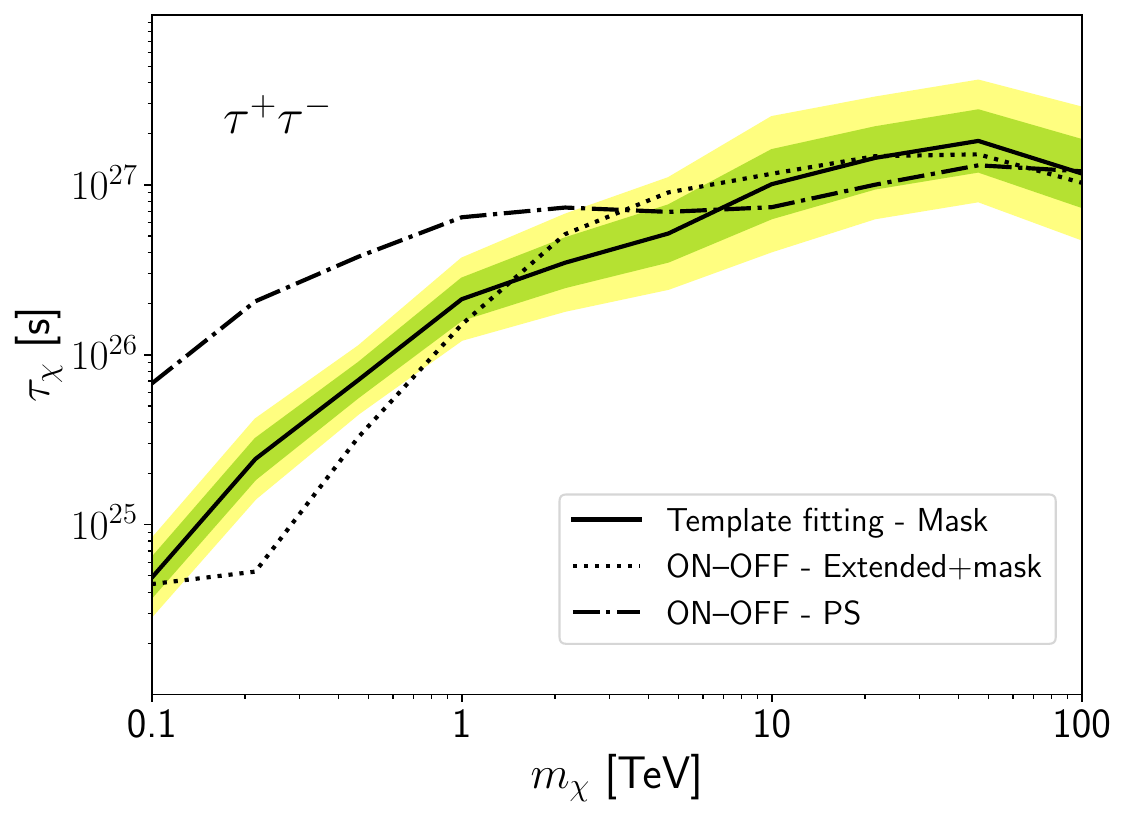}
	\includegraphics[width=0.49\textwidth]{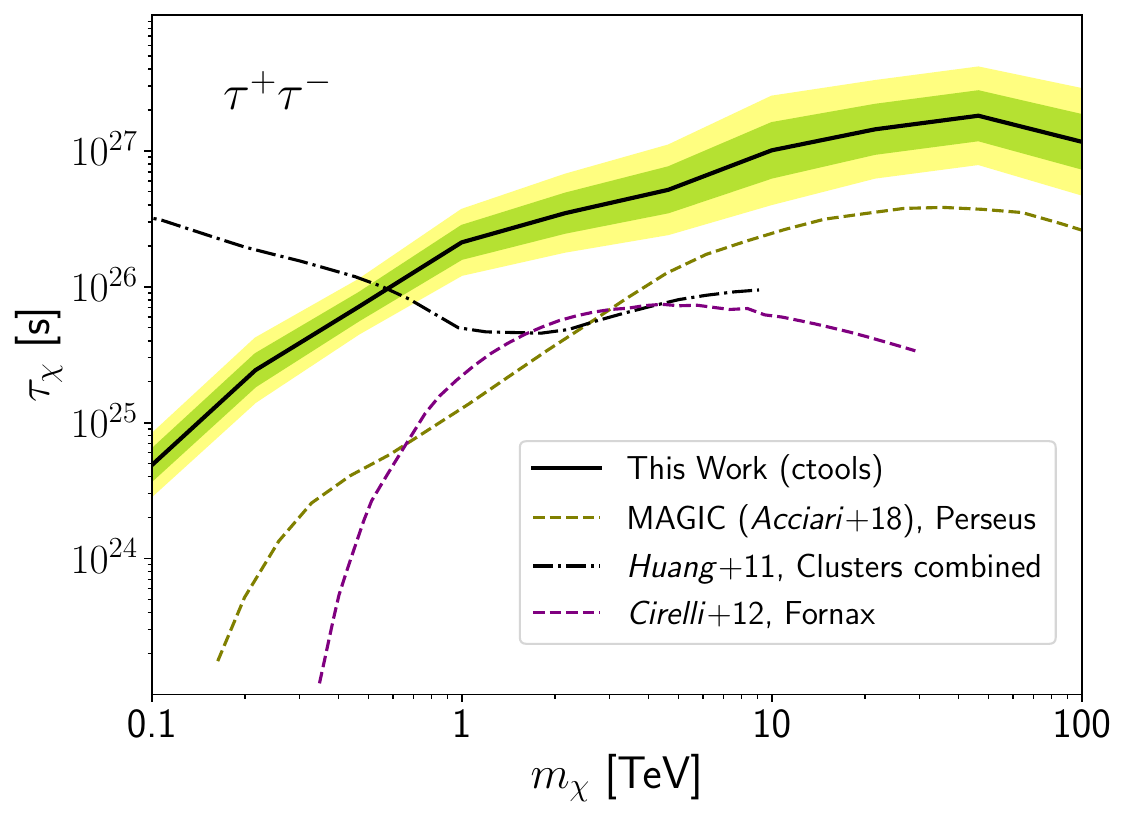}
	\includegraphics[width=0.49\textwidth]{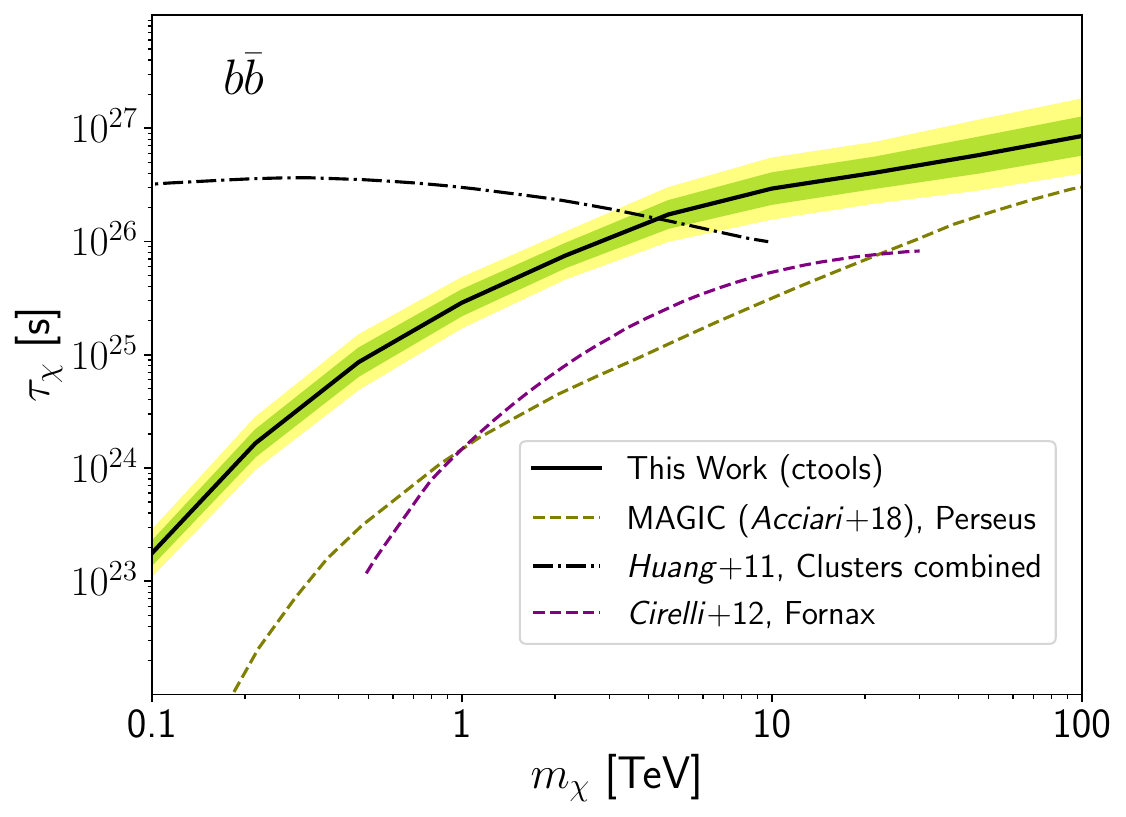}
	\caption{Sensitivity of CTA to a DM decay signal from the Perseus cluster, at 95\% C.L., in terms of the mean lower limits of the lifetime of the DM particle versus DM mass. The green (yellow) band shows the $1\sigma$ ($2\sigma$) scatter of the projected limits. \textbf{Top left panel:} Mean lifetime lower limits for the two considered decay channels, $b\bar{b}$ (dashed line) and $\tau^+\tau^-$ (solid). \textbf{Top right panel:} Mean lifetime lower limits for the $\tau^+\tau^-$ channel for three different analysis methods: template-fitting with a mask on NGC~1275 (solid line), ON-OFF analysis assuming a point-like source (PS) for the DM-induced $\gamma$-ray emission (dot-dashed), and ON-OFF analysis considering a spatially-extended DM emission with a mask in the center of the Perseus cluster (dotted line). See Section~\ref{app:ctools_data_prep} for details. \textbf{Bottom panels:} Mean lifetime lower limits for the $\tau^+\tau^-$ (\textbf{left panel}) and $b\bar{b}$ (\textbf{right panel}) channels in comparison with the most recent results on DM decay in galaxy clusters using MAGIC data (olive dashed lines; \citep{MAGIC:2018tuz}), $Fermi$-LAT data (black dot-dashed lines; \citep{Huang:2011xr}) and H.E.S.S data (purple dashed lines; \citep{Cirelli:2012ut}).
	}
\label{fig:DM_decay_results_ctools}
\end{figure}

In the absence of a clear DM signal, we compute 95\% C.L. exclusion limits to the annihilation cross-section (decay lifetime) versus the DM mass. Figure~\ref{fig:DM_anna_results_ctools} shows the exclusion limits for the MED annihilation scenario for both $b\bar{b}$ and $\tau^{+}\tau^{-}$ channels, while Figure~\ref{fig:DM_decay_results_ctools} shows the results obtained for the DEC decay scenario. In both figures, we follow the same convention to show the results as in Sections~\ref{sec:DM_result_annihil} and \ref{sec:DM_result_decay}, and compare to recent results from other experiments as well. Note that results from both the template-fitting and ON-OFF extended analyses, both for annihilation and decay, are in good agreement for DM masses above 1 TeV. This is expected, as the masks adopted in both types of analyses at these energies is comparable.

\subsection{Comparison with \texttt{gammapy}}

\begin{figure}[htb!]
    \centering
	\includegraphics[width=0.49\textwidth]{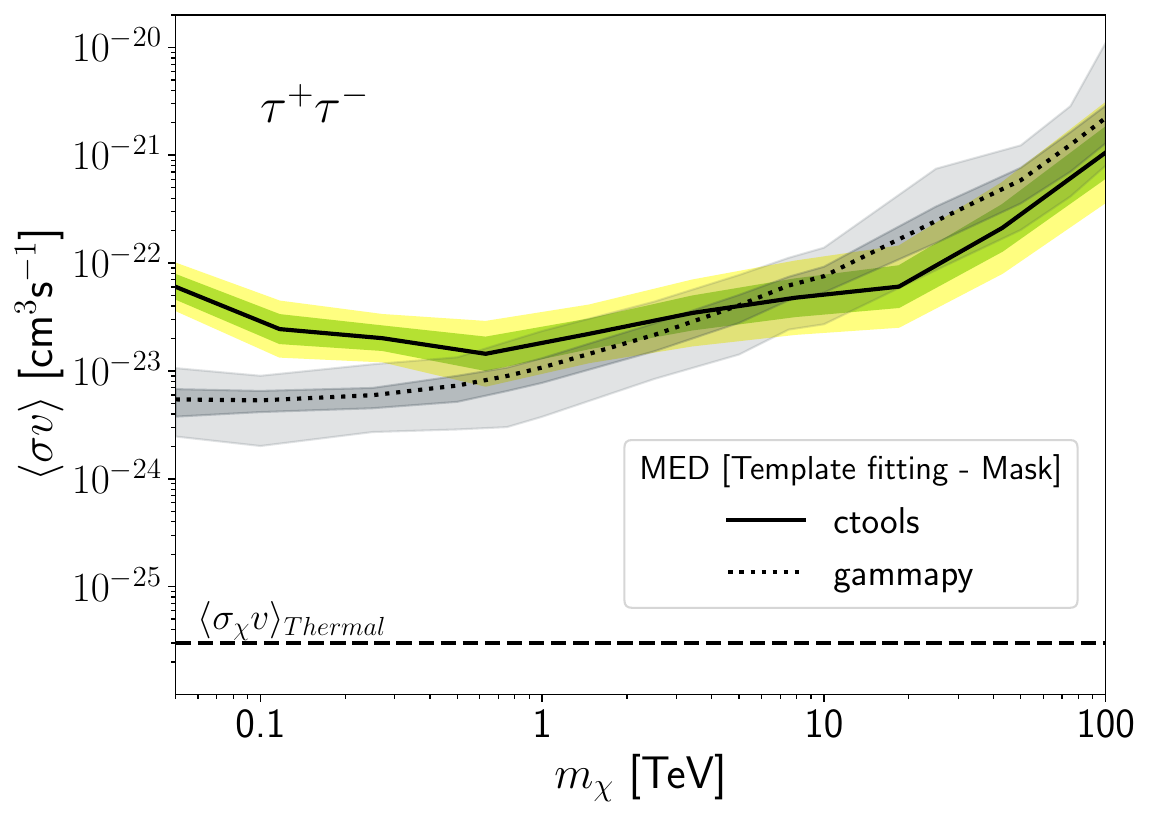}
	\includegraphics[width=0.49\textwidth]{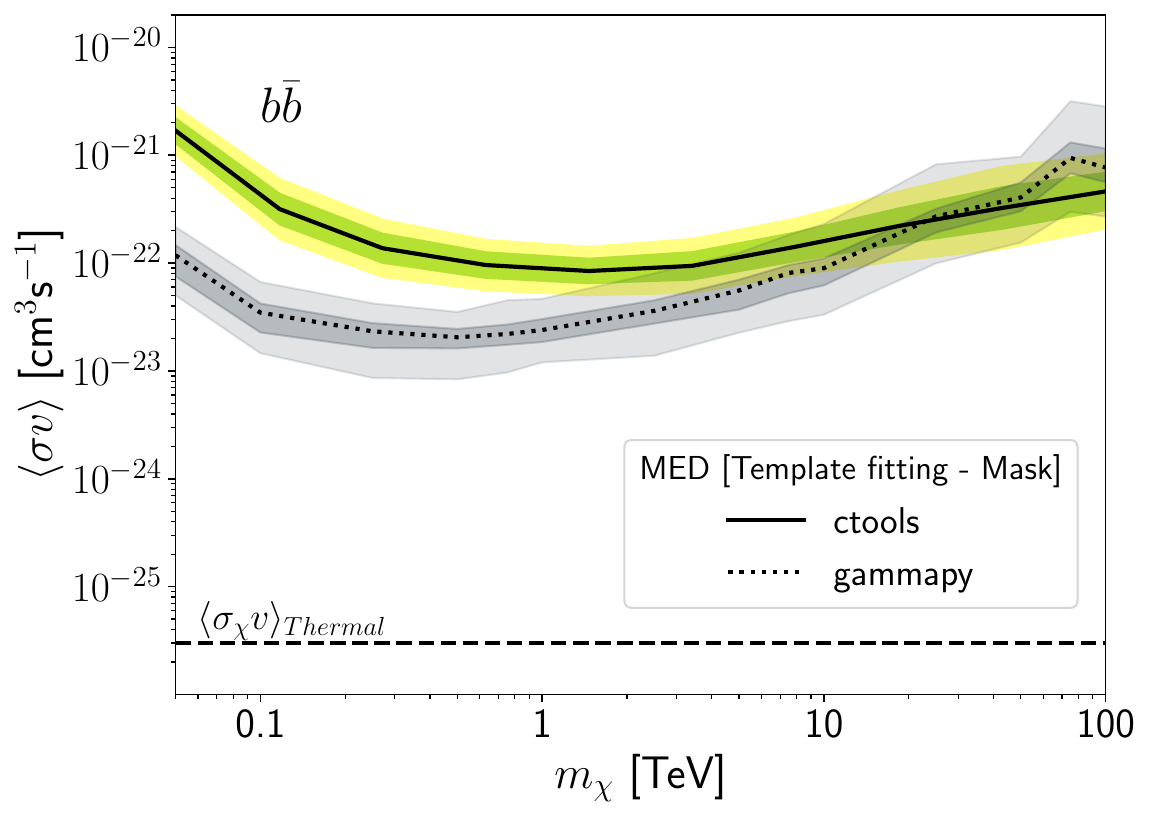}
	\includegraphics[width=0.49\textwidth]{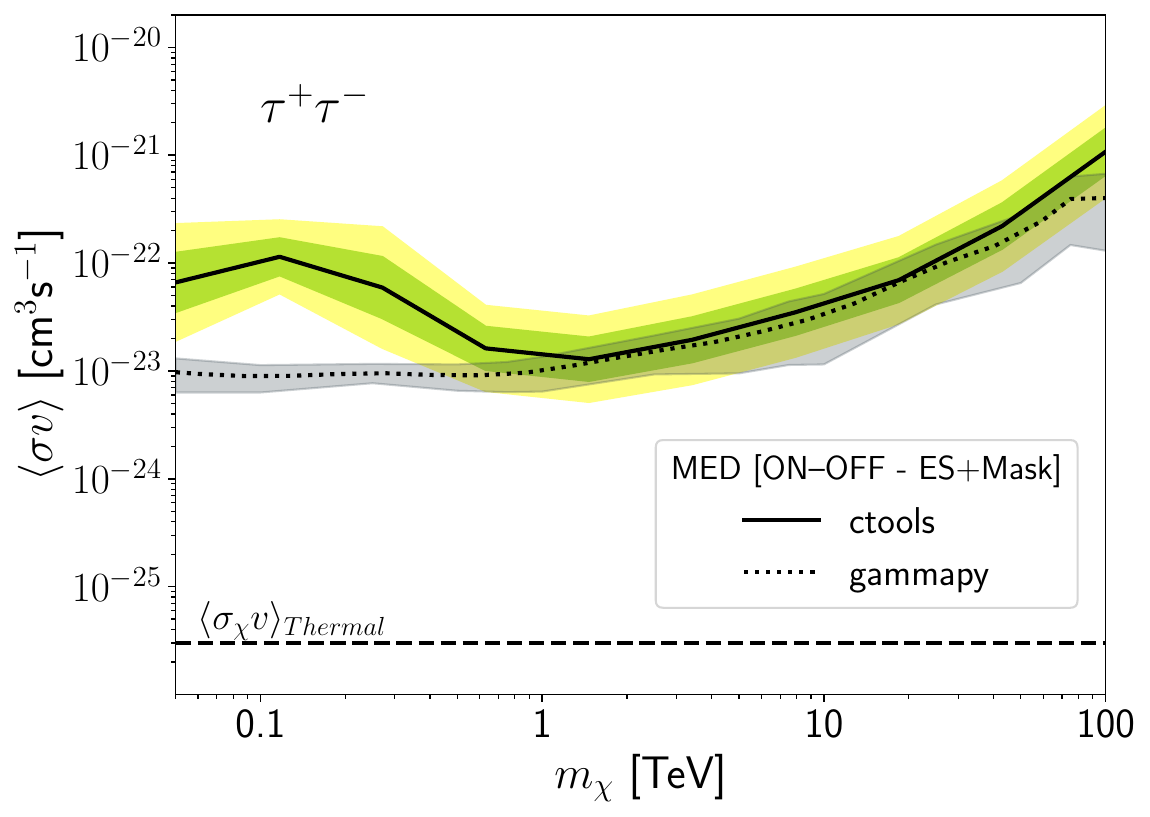}
	\includegraphics[width=0.49\textwidth]{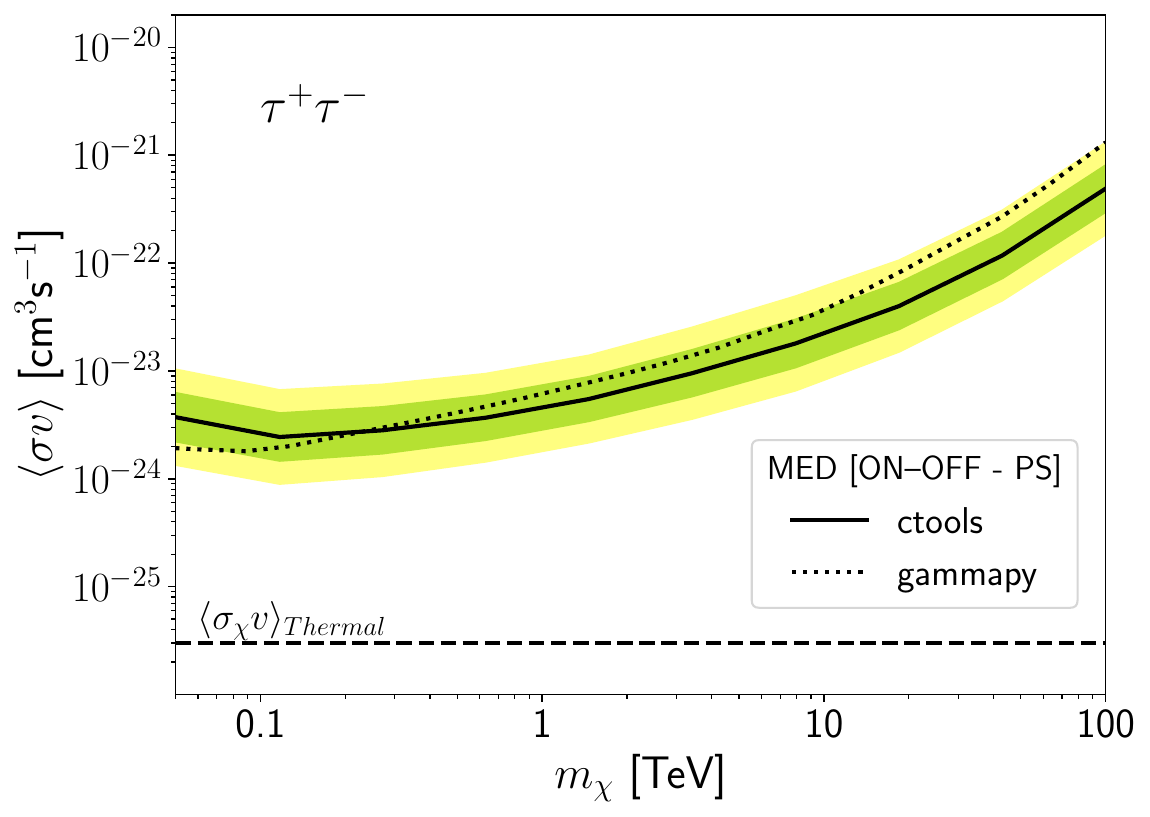}
	\caption{Sensitivity of CTA to a DM annihilation signal from the Perseus cluster, at 95\% C.L., in terms of the mean upper limits of the velocity-averaged cross-section of the DM particle versus the DM mass, obtained via the two CTA analysis softwares used in this work, \texttt{ctools} (solid line) and \texttt{gammapy} (dotted line). The green (yellow) band shows the $1\sigma$ ($2\sigma$) scatter of the projected limits obtained with \texttt{ctools}, and the gray bands (when visible) represent the $1\sigma$ and $2\sigma$ bands for \texttt{gammapy}. The black dashed line is the thermal relic cross-section ($<\sigma v>_{thermal} = 3\times10^{-26}$ cm$^{3}$ s$^{-1}$). \textbf{Top left panel:} Mean cross-section upper limits for the $\tau^+\tau^-$ annihilation channel obtained with \texttt{ctools} (solid) and \texttt{gammapy} (dotted), in both cases adopting the template-fitting analysis approach (Sections~\ref{sec:DM_template_fitting} and \ref{app:ctools_DM_sens_tempfitting}). \textbf{Top right panel:} Mean cross-section upper limits for the $b\bar{b}$ channel for \texttt{ctools} (solid) and \texttt{gammapy} (dotted) for the template fitting (Sections~\ref{sec:DM_template_fitting} and \ref{app:ctools_DM_sens_tempfitting}). \textbf{Bottom panels:} Mean cross-section upper limits for the $\tau^{+}\tau^{-}$ channel for the MED annihilation scenario, and ON-OFF 'ES + Mask' (\textbf{left panel}) and ON-OFF 'PS' analyses (\textbf{right panel}), as well as their comparison with the \texttt{gammapy} results (Section~\ref{app:DM_on_off}).}
\label{fig:DM_anna_comparison}
\end{figure}

\begin{figure}[htb!]
    \centering
	\includegraphics[width=0.49\textwidth]{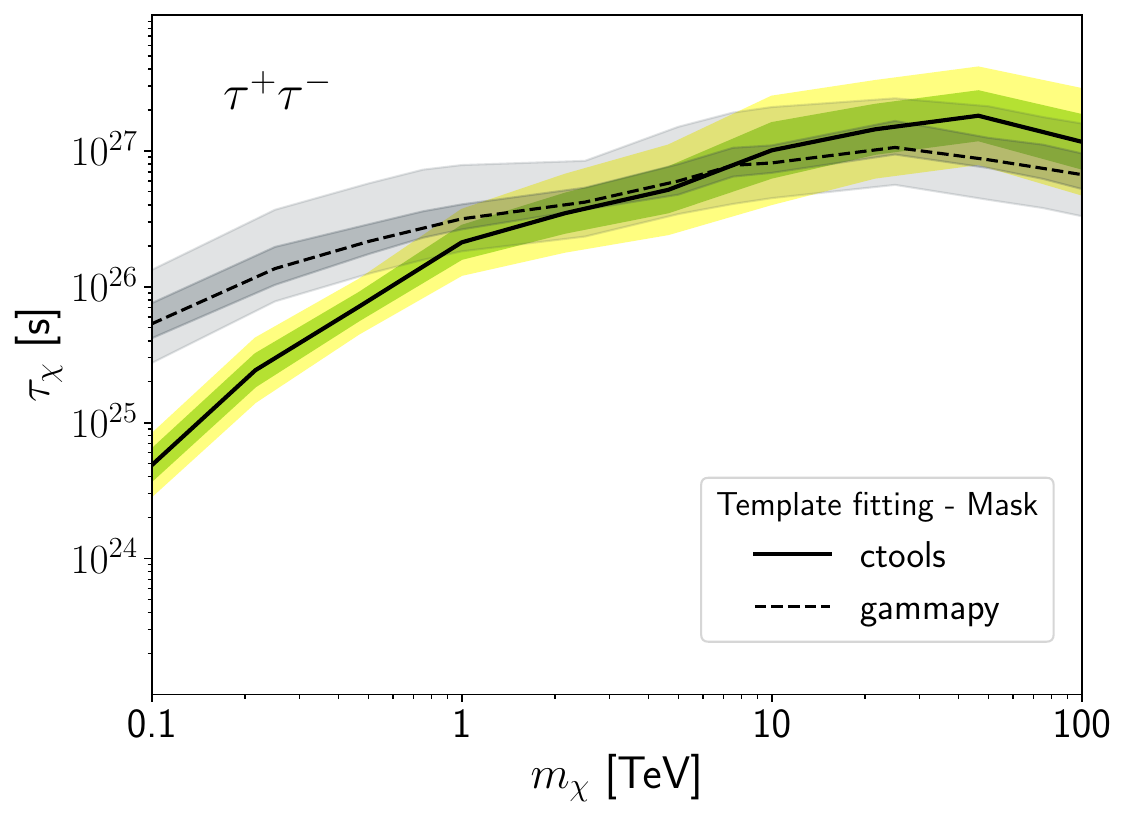}
	\includegraphics[width=0.49\textwidth]{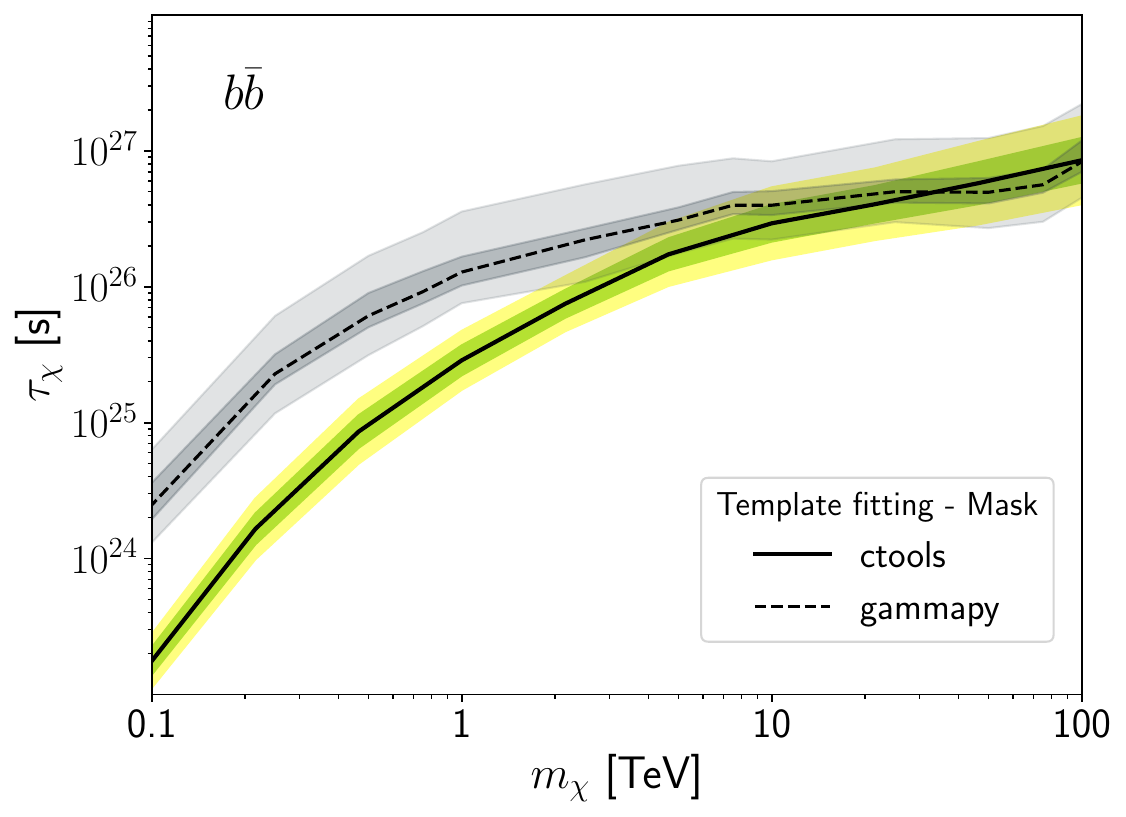}
	\includegraphics[width=0.49\textwidth]{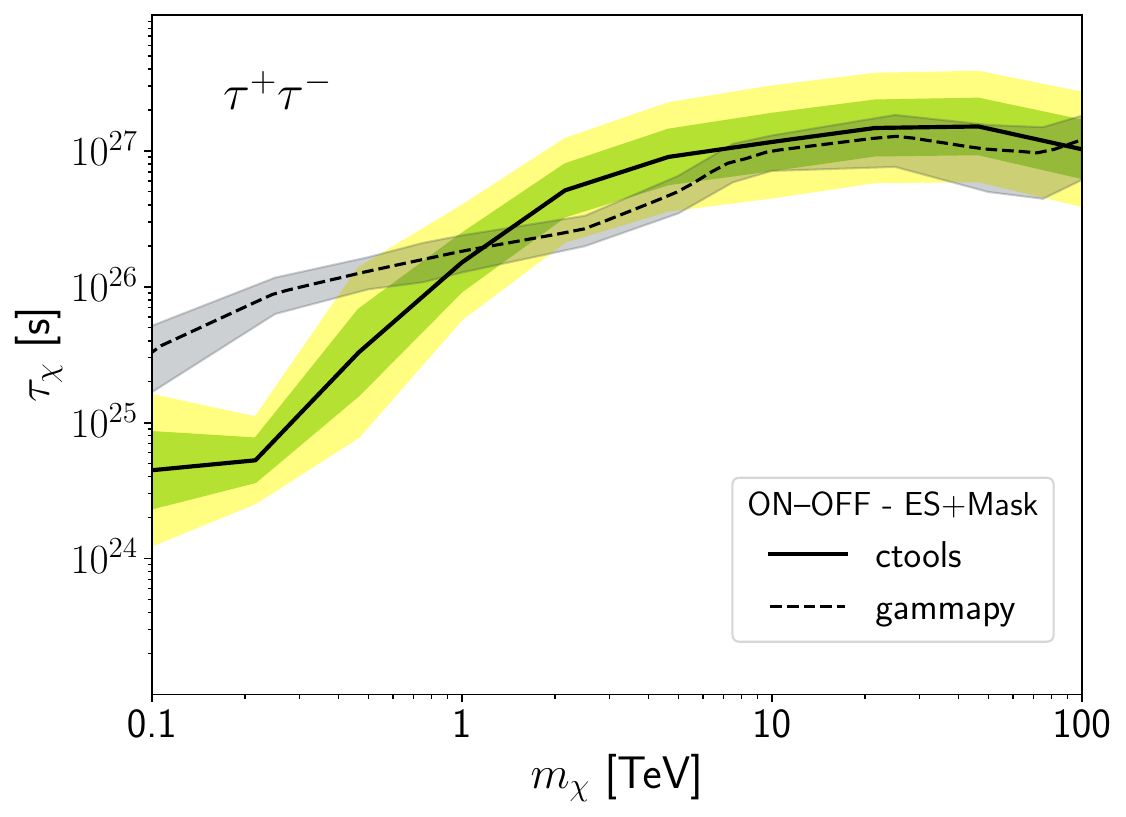}
	\includegraphics[width=0.49\textwidth]{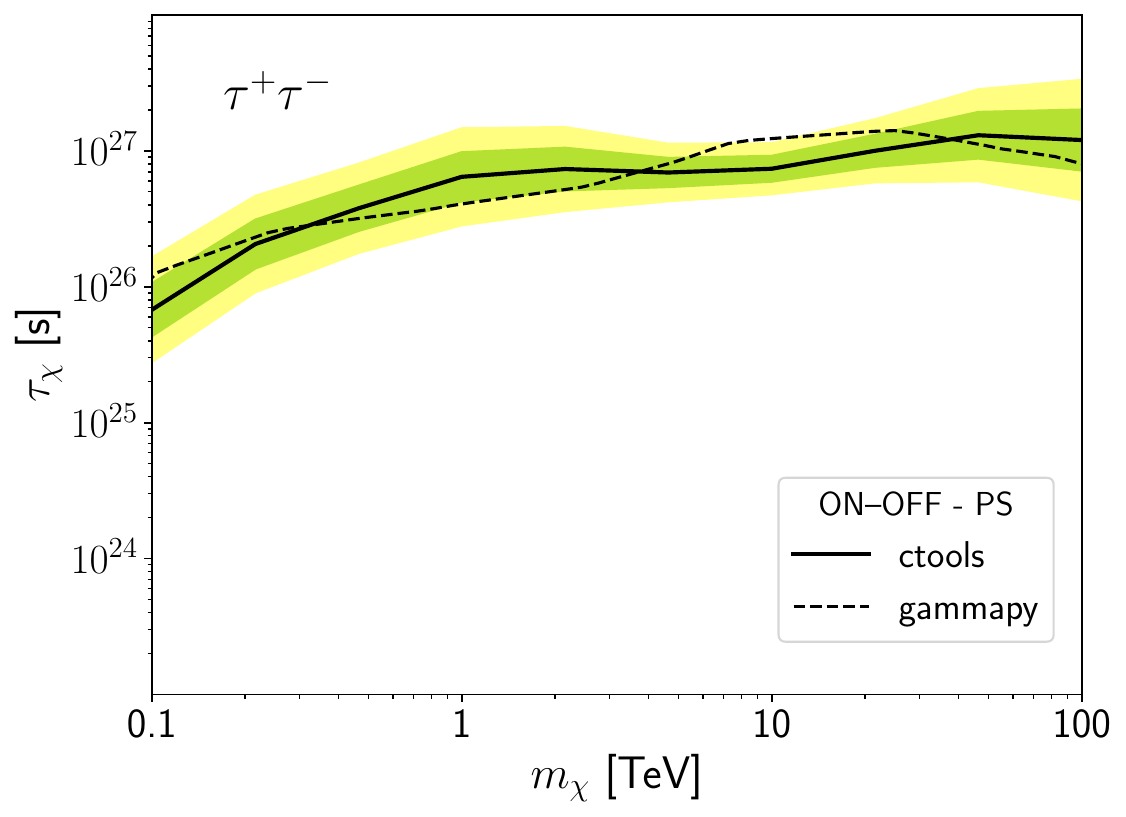}
	\caption{Sensitivity of CTA to a DM decay signal from the Perseus cluster, at 95\% C.L., in terms of the mean lower limits of the lifetime of the DM particle versus the DM mass, obtained via the two CTA analysis softwares used in this work, \texttt{ctools} (solid line) and \texttt{gammapy} (dotted line). The green (yellow) band shows the $1\sigma$ ($2\sigma$) scatter of the projected limits obtained with \texttt{ctools},and the gray bands (when visible) represent the $1\sigma$ and $2\sigma$ bands for \texttt{gammapy}.\textbf{Top left panel:} Mean lifetime lower limits for the $\tau^+\tau^-$ decay channel obtained with \texttt{ctools} (solid) and \texttt{gammapy} (dotted), in both cases adopting the template-fitting analysis approach (Sections~\ref{sec:DM_template_fitting} and \ref{app:ctools_DM_sens_tempfitting}). \textbf{Top right panel:} Mean lifetime lower limits for the $b\bar{b}$ channel for \texttt{ctools} (solid) and \texttt{gammapy} (dotted) for the template fitting (Sections~\ref{sec:DM_template_fitting} and \ref{app:ctools_DM_sens_tempfitting}). \textbf{Bottom panels:} Mean lifetime lower limits for the $\tau^{+}\tau^{-}$ channel in the DEC scenario, and ON-OFF 'ES + Mask' (\textbf{left panel}) and ON-OFF 'PS' analyses (\textbf{right panel}), as well as their comparison with the \texttt{gammapy} results (Section~\ref{app:DM_on_off}). 
	}
\label{fig:DM_decay_comparison}
\end{figure}

Finally, in this section we show the comparison between the results obtained via the two analysis pipelines used in this paper to search for a DM-induced $\gamma$-ray signal in the Perseus cluster. The results obtained with \texttt{ctools} are shown in Sections~\ref{app:ctools_DM_sens_onoff} and \ref{app:ctools_DM_sens_tempfitting}, while those with \texttt{gammapy} are presented in Sections \ref{sec:DM_result_annihil}, \ref{sec:DM_result_decay}, and \ref{app:DM_on_off}. Figures~\ref{fig:DM_anna_comparison} and \ref{fig:DM_decay_comparison} show the comparison for annihilating and decaying DM for the different observational setups considered in this work: template fitting, ON-OFF (ES+Mask), and ON-OFF (PS). In the following, we only focus the discussion in the comparison for the MED annihilation scenario. For the DEC scenario, as well as MIN and MAX annihilation models, conclusions are similar.

In the case of the template-fitting analysis, we observe that the results obtained with \texttt{ctools} are less restrictive in comparison to the limits obtained with \texttt{gammapy} for DM masses below 1 TeV (10 TeV) for $\tau^{+}\tau^{-}$ ($b\bar{b}$) annihilation channel. This difference is a consequence of the restrictive mask placed on NGC~1275 in the alternative setup used for \texttt{ctools}, which significantly decreases the CTA sensitivity at lower energies (i.e., DM masses). The larger difference for $b\bar{b}$ is due to the fact that in this case the annihilation spectrum has its maximum located at $\thicksim1/20$ of the considered DM mass, thus even still at high masses of around few TeV, the major contribution to the photon spectra comes from lower energies, where the CTA sensitivity is lower. Above 1 TeV or 10 TeV, depending on the annihilation channel, the results between \texttt{ctools} and \texttt{gammapy} are in good agreement, indeed being within the $2\sigma$ statistical fluctuations.

For the ON-OFF (ES+Mask) analysis, we observe that the projected limits obtained with \texttt{ctools} and \texttt{gammapy} are in good agreement for DM masses above 1 TeV, always within the $1\sigma$ bands. Similarly to the previous case, we also observe a decrease in the sensitivity with \texttt{ctools} for DM masses below 1 TeV, being the magnitude of this decrement comparable to the decrease of sensitivity observed in the template-fitting method. We find the same effect for $b\bar{b}$, in concordance with the results obtained for the template fitting as well. As discussed in previous subsections of this appendix, we believe that the size of the mask used for this analysis, of  $0.1~\text{deg}$ radius, is not sufficient to entirely cover the emission from NGC~1275, which impacts our sensitivity at the lowest energies, where the angular resolution is worse. We did not increase the the size of the mask placed on NGC~1275, as our goal is to compare the performance of the \texttt{ctools} and \texttt{gammapy} DM pipelines under the same observation setups.

Finally, for the ON-OFF (PS) analysis, we observe that the results from \texttt{ctools} and \texttt{gammapy} are in excellent agreement, within $1\sigma$ of statistical fluctuations, and independent of the assumed DM annihilation channel. 

As an overall conclusion from these exercises of comparison between \texttt{ctools} and \texttt{gammapy}, we find a good agreement between both analysis pipelines at high DM masses, either in basic or more complex analysis setups and scenarios. However, for lower DM mass, typically below 1 TeV, \texttt{ctools} is less sensitive to disentangling the emission from multiple components. For the Perseus cluster, the main reason is that NGC~1275 overshines or eclipses the contribution of the DM- and CR-induced $\gamma$-ray emissions, inducing a signal leaking in these templates that causes unstable, untrustable results. Indeed, when the emission of NGC~1275 is artificially decreased by placing a mask on it, conveniently chosen according to CTA's angular resolution properties, we are capable of recovering the sensitive obtained with \texttt{gammapy}.

\bibliography{biblio}

\end{document}